\documentclass[]{aastex631}

\usepackage{amsmath}
\usepackage{comment}

\usepackage{makecell} 

\definecolor{darkgreen}{rgb}{0.13, 0.55, 0.13}
\definecolor{brown}{rgb}{0.65, 0.16, 0.16}

\newcommand{\Msun}{\mbox{M$_{\sun}$}}
\newcommand{\Lsun}{\mbox{L$_{\sun}$}}
\newcommand{\Tbol}{\mbox{T$_{\rm{bol}}$}}
\newcommand{\Lbol}{\mbox{L$_{\rm{bol}}$}}
\newcommand{\Ltot}{\mbox{L$_{\rm{tot}}$}}

\newcommand{\dummyABCDE}{eHOPS-aql-18}
\newcommand{\dummyABCED}{eHOPS-aql-35}
\newcommand{\dummyABDCE}{eHOPS-aql-74}
\newcommand{\dummyABDEC}{eHOPS-aql-67}
\newcommand{\dummyABECD}{eHOPS-aql-48}
\newcommand{\dummyABEDC}{eHOPS-aql-2}
\newcommand{\dummyACBDE}{eHOPS-aql-44}
\newcommand{\dummyACBED}{eHOPS-aql-72}
\newcommand{\dummyACDBE}{eHOPS-aql-77}
\newcommand{\dummyACDEB}{eHOPS-aql-81}
\newcommand{\dummyACEBD}{eHOPS-aql-76}
\newcommand{\dummyACEDB}{eHOPS-aql-141}
\newcommand{\dummyADBCE}{eHOPS-aql-109}
\newcommand{\dummyADBEC}{eHOPS-aql-49}
\newcommand{\dummyADCBE}{eHOPS-aql-70}
\newcommand{\dummyADCEB}{eHOPS-aql-150}
\newcommand{\dummyADEBC}{eHOPS-aql-54}
\newcommand{\dummyADECB}{eHOPS-aql-46}
\newcommand{\dummyAEBCD}{eHOPS-aql-117}

\newcommand{\dummyAECBD}{eHOPS-aql-108}
\newcommand{\dummyAECDB}{eHOPS-aql-75}
\newcommand{\dummyAEDBC}{eHOPS-aql-9}
\newcommand{\dummyAEDCB}{eHOPS-aql-152}
\newcommand{\dummyBACDE}{eHOPS-aql-149}
\newcommand{\dummyBACED}{eHOPS-aql-110}
\newcommand{\dummyBADCE}{eHOPS-aql-38}
\newcommand{\dummyBADEC}{eHOPS-aql-56}
\newcommand{\dummyBAECD}{eHOPS-aql-166}
\newcommand{\dummyBAEDC}{eHOPS-aql-161}
\newcommand{\dummyBCADE}{eHOPS-aql-143}
\newcommand{\dummyBCAED}{eHOPS-aql-55}
\newcommand{\dummyBCDAE}{eHOPS-aql-164}
\newcommand{\dummyBCDEA}{eHOPS-aql-165}
\newcommand{\dummyBCEAD}{eHOPS-aql-171}
\newcommand{\dummyBCEDA}{eHOPS-aql-3}

\newcommand{\dummyBDCAE}{eHOPS-aql-36}
\newcommand{\dummyBDCEA}{eHOPS-aql-115}
\newcommand{\dummyBDEAC}{eHOPS-aql-29}
\newcommand{\dummyBDECA}{eHOPS-aql-93}
\newcommand{\dummyBEACD}{eHOPS-aql-78}
\newcommand{\dummyBEADC}{eHOPS-aql-154}

\newcommand{\dummyBECDA}{eHOPS-aql-53}
\newcommand{\dummyBEDAC}{eHOPS-aql-124}
\newcommand{\dummyBEDCA}{eHOPS-aql-27}

\defcitealias{Furlan16}{F16}

\shorttitle{eHOPS: Aquila}
\shortauthors{R. Pokhrel et al.}
\graphicspath{{./}{}}

\begin{document}

\title{Extension of HOPS\footnote{HOPS: Herschel Orion Protostar Survey} Out to 500 ParSecs (eHOPS). I. \\
Identification and Modeling of Protostars in the Aquila Molecular Clouds}

\correspondingauthor{Riwaj Pokhrel}
\email{Riwaj.Pokhrel@utoledo.edu, riwajpokhrel@gmail.com}

\author[0000-0002-0557-7349]{Riwaj Pokhrel}
\affiliation{Ritter Astrophysical Research Center, Department of Physics and Astronomy, University of Toledo, Toledo, OH 43606, USA}

\author[0000-0001-7629-3573]{S. Thomas Megeath}
\affiliation{Ritter Astrophysical Research Center, Department of Physics and Astronomy, University of Toledo, Toledo, OH 43606, USA}

\author[0000-0002-6447-899X]{Robert A. Gutermuth}
\affil{Department of Astronomy, The University of Massachusetts, Amherst, MA, 01003, USA}

\author[0000-0001-9800-6248]{Elise Furlan}
\affiliation{NASA Exoplanet Science Institute, Caltech/IPAC, Mail Code 100-22, 1200 E. California Blvd., Pasadena, CA 91125, USA}

\author[0000-0002-3747-2496]{William J. Fischer}
\affiliation{Space Telescope Science Institute, Baltimore, MD, US}

\author[0000-0002-6136-5578]{Samuel Federman}
\affiliation{Ritter Astrophysical Research Center, Department of Physics and Astronomy, University of Toledo, Toledo, OH 43606, USA}

\author[0000-0002-6195-0152]{John J. Tobin}
\affiliation{National Radio Astronomy Observatory, 520 Edgemont Rd., Charlottesville, VA, US}

\author[0000-0003-2300-8200]{Amelia M.\ Stutz}
\affiliation{Departamento de Astronom\'{i}a, Universidad de Concepci\'{o}n,Casilla 160-C, Concepci\'{o}n, Chile}

\author[0000-0003-1430-8519]{Lee Hartmann}
\affiliation{University of Michigan, Ann Arbor, MI, USA}

\author[0000-0002-6737-5267]{Mayra Osorio}
\affiliation{Instituto de Astrofísica de Andalucía, CSIC, Glorieta de la Astronomía s/n, E-18008 Granada, Spain}

\author[0000-0001-8302-0530]{Dan M. Watson}
\affiliation{Department of Physics and Astronomy, University of Rochester, Rochester, NY 14627-0171, USA}

\author[0000-0002-5812-9232]{Thomas Stanke}
\affiliation{Max-Planck-Institute f\"{u}r extraterrestrische PhysiK, Giessenbachstr. 1, 85748 Garching, Germany}

\author[0000-0002-3530-304X]{P. Manoj}
\affiliation{Department of Astronomy and Astrophysics, Tata Institute of Fundamental Research, Homi Bhabha Road, Colaba, Mumbai 400005, India}

\author[0000-0002-0554-1151]{Mayank Narang}
\affiliation{Department of Astronomy and Astrophysics, Tata Institute of Fundamental Research, Homi Bhabha Road, Colaba, Mumbai 400005, India}

\author[0000-0002-4026-126X]{Prabhani Atnagulov}
\affiliation{Ritter Astrophysical Research Center, Department of Physics and Astronomy, University of Toledo, Toledo, OH 43606, USA}

\author[0000-0002-2667-1676]{Nolan Habel}
\affiliation{Ritter Astrophysical Research Center, Department of Physics and Astronomy, University of Toledo, Toledo, OH 43606, USA}

\author[0000-0001-9030-1832]{Wafa Zakri}
\affiliation{Department of Physics, Jazan University Jazan, Saudi Arabia}








\begin{abstract}

We present a Spitzer/Herschel focused survey of the Aquila molecular clouds ($d \sim 436$~pc) as part of the eHOPS (extension of HOPS Out to 500 ParSecs) census of nearby protostars. For every source detected in the Herschel/PACS bands, the eHOPS-Aquila catalog contains 1-850~$\mu$m SEDs assembled from 2MASS, Spitzer, Herschel, WISE, and JCMT/SCUBA-2 data. Using a newly developed set of criteria, we classify objects by their SEDs as protostars, pre-ms sequence stars with disks, and galaxies.  A total of 172 protostars are found in Aquila, tightly concentrated in the molecular filaments that thread the clouds. Of these, 71 (42\%)  are Class 0 protostars, 54 (31\%) are Class I protostars,  43 (25\%) are flat-spectrum protostars, and 4 (2\%) are Class II sources. Ten of the Class 0 protostars are young PACS Bright Red Sources similar to those discovered in Orion. We compare the SEDs to a grid of radiative transfer models to constrain the luminosities, envelope densities, and envelope masses of the protostars. A comparison of the eHOPS-Aquila to the HOPS protostars in Orion finds that the protostellar luminosity functions in the two star-forming regions are statistically indistinguishable, the bolometric temperatures/envelope masses of eHOPS-Aquila protostars are shifted to cooler temperatures/higher masses, and the eHOPS-Aquila protostars do not show the decline in luminosity with evolution found in Orion. We briefly discuss whether these differences are due to biases between the samples, diverging star formation histories, or the influence of environment on protostellar evolution.

\end{abstract}



\section{Introduction} \label{sec:intro}

Protostars are young stellar objects (YSOs) surrounded by infalling envelopes of gas and dust \citep[e.g.,][]{Dunham14}. In a simple picture of a protostellar system, material from the envelope falls onto the circumstellar disk that surrounds the protostar.  Material from the disk then accretes onto the central protostar \citep[e.g.,][]{Hartmann2016}. The same disks that mediate accretion also launch jets and winds that drive outflows \citep[e.g.,][]{Frank14,Lee2020}. These outflows carve through the molecular gas and remove mass from the envelopes \citep[e.g.,][]{Bally16, Habel21}. The final masses of protostars - and the initial masses of the stars they form  - are a consequence of gas infall, accretion, and outflow over a $\sim$0.5 Myr timespan \citep{Dunham14,Dunham15}.

Studies of protostars have been hampered by the relatively small samples of these deeply embedded and rapidly evolving objects. The deployment of the Spitzer Space Telescope, with its high sensitivity, 2-6$\arcsec$ angular resolution, and capability for wide-field surveys in the mid-IR, enabled relatively complete surveys of protostars in nearby molecular clouds \citep{Megeath04, Allen04, Hartmann05, Allen07, Gutermuth09, Evans09, Kryukova12, Dunham15, Megeath16}. 
These surveys are most complete in clouds within 0.5~kpc where low luminosity and/or densely clustered protostars can be detected and distinguished. The large sample of protostars discovered by Spitzer poses a need for the robust characterization of their properties. The peak of the protostellar emission, however, is at far-IR wavelengths. The SEDs of young protostars show rising slopes in the near to mid-IR, peak in the far-IR region at $\sim$100 $\micron$, and fall in the sub-mm region with Rayleigh-Jeans type spectra modified by the dust properties (e.g., \citealt{Enoch09, Furlan16}). Furthermore, the most deeply embedded protostars can only be detected in the far-IR and longer wavelengths \citep{Stutz13}.

The combination of mid and far-IR observations is a powerful tool for studying protostars.
The luminosity emitted by a protostar is scattered and absorbed by the surrounding disk and envelope, which re-radiates the luminosity at mid-IR to far-IR wavelengths. This makes protostars faint or undetectable at visible wavelengths and motivates the need for mid-IR and far-IR observations with relatively high sensitivity and an angular resolution sufficient to separate protostars from neighboring sources and cloud structures. Even if the protostars are spatially unresolved, spectral energy distributions (SEDs) spanning the infrared regime can constrain the fundamental properties such as their luminosities and the densities of their infalling envelopes. Furthermore, absorption by silicate grains in the envelopes cause broad absorption features at 10 and 18 $\micron$ as well as additional ice features from their mantles, and spectra in the infrared can probe the composition of ices and dust being delivered to the central disks by infall (e.g., \citealt{Boogert08, Pontoppidan08, Poteet11, Poteet13}). Far-IR emission lines can be used to trace the effect of outflows \citep[e.g.,][]{Manoj13,Karska18}.

The mid to far-IR SEDs of protostars also provide a window into the evolution of protostars. 
The SEDs of protostars can be roughly divided into two regimes.  The 1-10~$\mu$m regime is dominated by light from the central protostars and inner disks, often scattered by dust grains in the envelope \citep[e.g.,][]{Fischer13,Habel21}. The $> 10~\micron$ regime, in contrast, is dominated by thermal dust emission from the envelopes \citep[e.g.,][]{Furlan16}. As protostars evolve, their envelopes are partially accreted onto the protostars and partially dispersed by winds and jets. The outflow cavities may grow and the envelopes thin \citep{Fischer17, Habel21}. This will result in an increase in the 1-10~$\mu$m regime of the SED, and a corresponding decrease in the $> 10~\micron$ regime as radiation from the central protostars escapes without being reprocessed by the envelopes. Furthermore, the peak of the thermal dust emission shifts to shorter wavelengths as the density of the envelopes decreases and the temperature increases \citep{Furlan16}. Thus, the SEDs evolve as envelopes are dispersed. The SEDs, however, are also dependent on the inclination of the protostellar envelopes, disks, and their outflow cavities. To characterize the SEDs of protostellar systems in different stages of evolution and inclinations, radiative transfer modeling from the near-IR to sub-mm is required for all evolutionary stages.

The far-IR and sub-mm detectors PACS and SPIRE on the Herschel Space Observatory provided unprecedented sensitivity and dynamic range at wavelengths around the peak of the SEDs of protostars \citep[e.g.][]{Bontemps10,Sewilo10}. Using PACS, the Herschel Orion Protostar Survey (HOPS) created a uniformly analyzed catalog of protostars using SEDs from near-IR ($\sim$1 $\micron$) to sub-mm ($\sim$1 mm) wavelengths \citep{Manoj13, Stutz13, Furlan16, Fischer17, Fischer20, Habel21}. HOPS targeted the rich population of protostars in the Orion molecular clouds, combining Herschel far-IR observations with Spitzer mid-IR photometry and spectra, 2MASS and HST near-IR imaging, and APEX sub-mm mapping. It performed the most detailed study to date of the 330 protostars inhabiting various regions in the Orion A and Orion B clouds \citep[][hereafter \citetalias{Furlan16}]{Furlan16}. 

The success of HOPS motivates similar studies in other nearby star-forming regions within 0.5 kpc, where the protostars can also be characterized in detail. Such surveys allow us to compare protostars in different star-forming environments and further constrain protostellar properties by constructing a larger sample of protostars. In this work, we extend the HOPS survey to the star-forming regions in the Aquila cloud complex as a part of the extension of HOPS Out to 500 ParSecs (eHOPS) program. Aquila is the second most protostar-rich star-forming region in the nearest 0.5 kpc (after Orion). Results for other molecular clouds (Ophiuchus, Perseus, Auriga, Cepheus, Chameleon, Corona-Australis, Lupus, Musca, and Pipe) will be released in future papers.

In this paper, we present the eHOPS protostar catalog for the Aquila molecular clouds. 
In section \ref{sec:obs}, we overview the sources of observational data in the $1 - 850 \,\micron$ wavelength region. We discuss the SEDs in section \ref{sec:sed} and present YSO identification and classification techniques in section \ref{sec:ps}. The spatial distribution of identified YSOs and galaxies is presented in Sec.~\ref{sec:dist}.
Sec.~\ref{sec:modeling} describes model fits to the SEDs using an existing grid of radiative transfer models. 
Sec.~\ref{sec:hopsvsehops} discusses a comparative study of the properties of protostars in the Aquila and Orion molecular clouds. Finally, we list the conclusions in section \ref{sec:conclusion}.

\subsection{Aquila molecular clouds}

\begin{figure}
\centering
\includegraphics[scale=0.95]{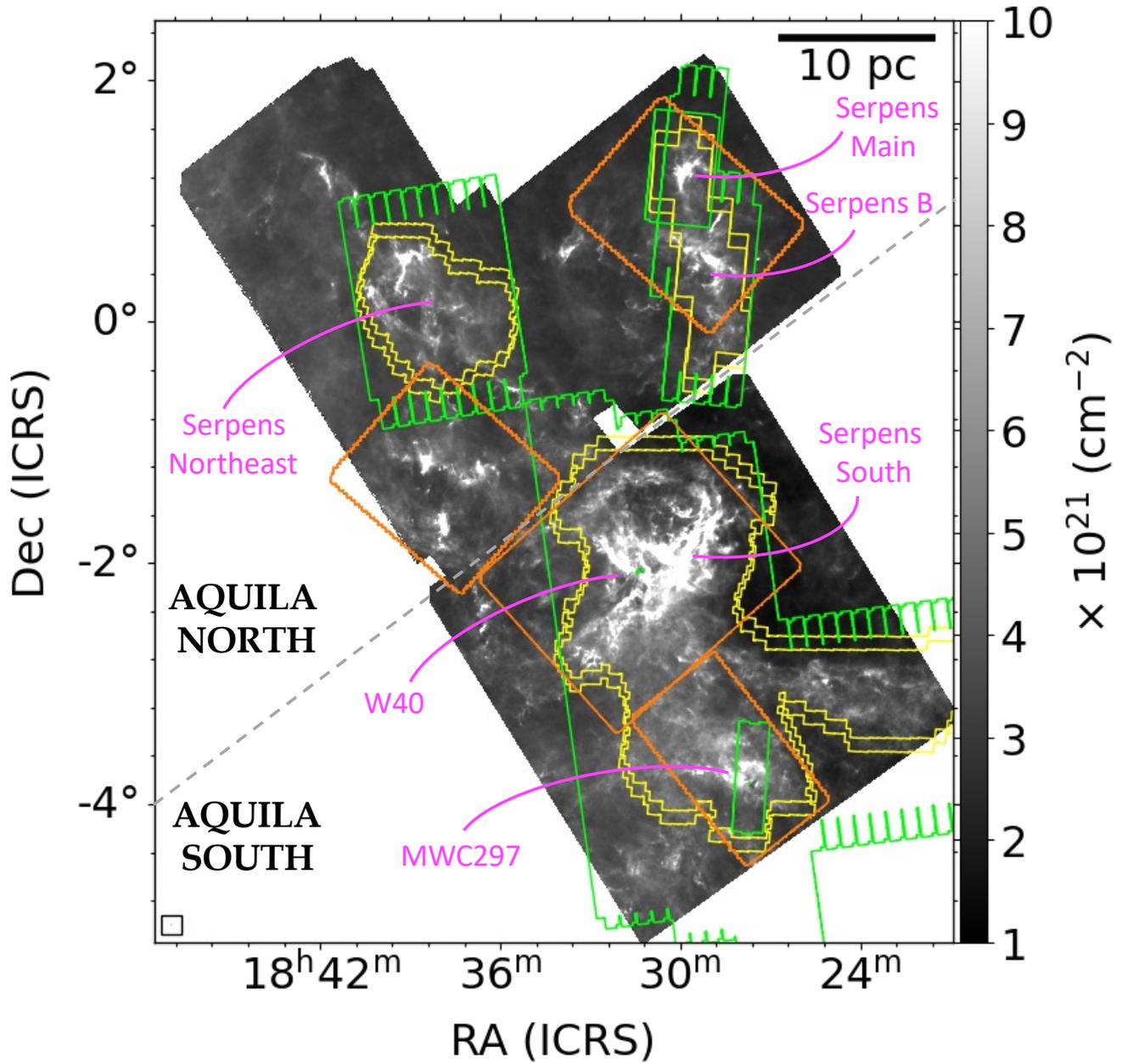}
\vspace{-.8in}
\caption{Herschel-derived column density map of the Aquila molecular clouds from the Herschel Gould Belt Survey \citep{Andre10}. A grey dashed line separates Aquila North and Aquila South. The spatial coverage of the IRAC, MIPS, and PACS 100 $\micron$ maps are shown by yellow, green, and orange contours, respectively. PACS 70 and 160 $\micron$, and SPIRE 250, 350, and 500 $\micron$ maps cover the entire displayed region. The major star-forming regions in Aquila North and Aquila South are labeled.
\label{fig:aquila_NH2}}
\end{figure}

The Aquila cloud complex (c.f. Figure \ref{fig:aquila_NH2}) contains several star-forming clumps concentrated in two distinct regions, Aquila-North and Aquila-South (e.g. \citealt{Reipurth08}). The Aquila-North region is located at $b=2^{\circ}-6^{\circ}$ and $l=29^{\circ}-34^{\circ}$ and harbors several star-forming regions including the Serpens Main and Serpens B cluster \citep{Winston07, Harvey07, Dunham15}. Aquila-North contains $\sim$3.5 $\times$ 10$^4$ M$_{\odot}$ gas in dense regions ($>$ 1 $A_{\rm{V}}$) and harbors $\sim$400 YSOs \citep{Pokhrel20, Pokhrel21}. Aquila-South is located in the southern part of the Aquila Rift at $b=2^{\circ}-5^{\circ}$ and $l=26^{\circ}-30^{\circ}$. There are three major star-forming regions in Aquila-South: Serpens South, W40, and MWC297 \citep{Dunham15, Winston18}. Discovered by the Spitzer Gould Belt survey \citep{Gutermuth08a},  Serpens-South is a young, protostar-rich cluster forming from a dense filamentary cloud  \citep{Nakamura11, Kirk13}. W40 is another young stellar cluster that is associated with an HII region that has cleared much of the natal molecular gas \citep{Kuhn10, Broos13, Pirogov13}. Herschel observations show $\sim5 \times 10^4$ M$_{\odot}$ gas at $> 1A_{\rm{V}}$ regions in Aquila-South \citep{Pokhrel20, Pokhrel21}. 

Past ambiguities in the distance to Aquila are now being resolved by parallax measurements (see discussions by \citealt{Bontemps10, Kirk13, Dunham15, Konyves15}). While stellar photometry previously suggested a distance of $\sim$255 pc \citep{Straizys03}, X-ray and HR diagram analysis suggested a distance $>$ 350 pc \citep{Winston10}. VLBI observations measured trigonometric parallaxes to stars in the Serpens Main cluster and obtained a distance of $\sim$415 pc \citep{Dzib10}. The method used by \cite{Straizys03} is sensitive toward the front wall of the extinction region with cluster-forming clumps at larger distances (e.g., \citealt{Bontemps10, Dzib10}). In more recent years, further observations with VLBI and Gaia converged on even larger distances. As a part of Gould's Belt Distances Survey (GOBELINS), \cite{Ortiz17} used VLBI observations that covered an 8-year observational span to find that the individual cluster forming clouds such as Serpens Main, W40, and Serpens South are physically associated and form a single cloud structure at an average distance of $436 \pm 9$~pc. The VLBI distance is also consistent with the more recent Gaia observations (see \citealt{Ortiz18} and \citealt{Anderson22} for the details of Gaia observations of Aquila). For the purpose of this work, we adopt a single distance of 436 pc for all the protostars in Aquila.

\begin{figure}[ht]
\centering
\includegraphics[scale=0.95]{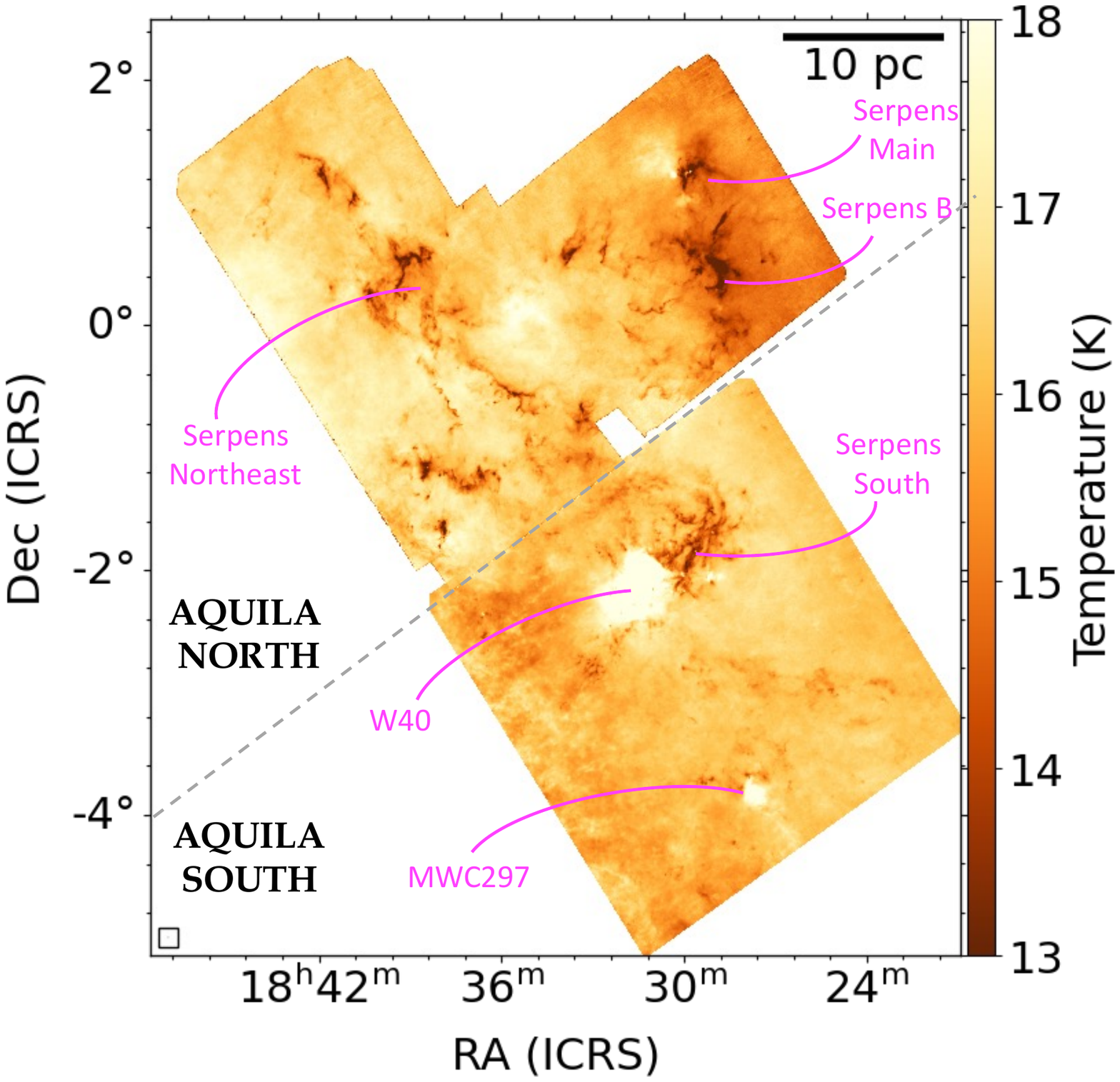}
\vspace{-.8in}
\caption{Herschel-derived dust temperature map of the Aquila molecular clouds from the Herschel Gould Belt Survey \citep{Andre10}. A grey dashed line separates Aquila North and Aquila South. The major star-forming region in Aquila North and Aquila South are labeled.
\label{fig:aquila_temp}}
\end{figure}

The Herschel Gould Belt Survey \citep{Andre10} modeled thermal dust emission from Herschel with a modified blackbody function to obtain the column density and temperature maps. Figure \ref{fig:aquila_NH2} shows a column density map of the Aquila molecular clouds from the Herschel Gould Belt Survey \citep{Konyves15, Fiorellino21}. Figure \ref{fig:aquila_temp} shows the dust temperature map for Aquila. The protostar-rich regions such as Serpens Main, Serpens B, and Serpens South have high column densities and low temperatures. Similarly, more evolved HII region W40 have higher dust temperatures than the surrounding regions.

\section{Archival Data} \label{sec:obs}

This paper presents SEDs assembled from photometry and spectra covering a wavelength range of 1-850~$\mu$m. These come from multiple catalogs.
The photometry from 1 - 24 $\micron$ is from the SESNA (Spitzer Extended Solar Neighborhood Archive) catalog compiled by R. Gutermuth et al. (2022, in preparation). SESNA is a uniform treatment of $>$90 deg$^2$ of archival Spitzer cryo-mission surveys of nearby star-forming regions. An additional $\sim$16 deg$^2$ of observations are used to remove the extragalactic contamination. SESNA combines near-IR (1.24, 1.67, and 2.16 $\micron$) photometry from the Two Micron All-Sky Survey point source catalog (2MASS; \citealt{Skrutskie06}) with Spitzer  photometry extracted from images made with  IRAC \citep{Fazio04} at 3.6, 4.5, 5.8, and 8.0 $\micron$ and with MIPS \citep{Rieke04} at 24 $\micron$. The SESNA catalogs are uniform in terms of their observing parameters, data treatment, source extraction algorithms, and photometric measurement techniques. SESNA mitigates the discrepancies seen in previous Spitzer YSO surveys that were analyzed by disparate groups emphasizing differing approaches and primary science goals.
Details will be described in the SESNA data release paper (R. Gutermuth et al. in prep.). The SESNA data products are publicly available through {\em http://bit.ly/sesna2021}.

We use observations from the Herschel Space Observatory \citep{Pilbratt10} to extract photometry in the 70-500 $\micron$ wavelength region. We obtain the observations from the Herschel Science Archive. We provide the observational details such as observation IDs, observed wavelengths, proposal names, and scan velocities for these observations in Appendix \ref{app:obsid} (Tables \ref{tab:pacs} and \ref{tab:spire}). The Photoconductor Array Camera and Spectrometer (PACS; \citealt{Poglitsch10}) detector on Herschel provides point source photometry at 70, 100, and 160 $\micron$. We use level-2.5 PACS data products with combined in-scan and cross-scan mode, i.e. in orthogonal scan directions to help with the mitigation of scanning artifacts. The data products labeled ``JSCANAM" are used for extracting PACS photometry since the JScanam map-maker method uses multiple sky passages with different scanning directions to remove noise that affects PACS maps while simultaneously preserving point source and real extended emission \citep{Gracia15}. We use PACS detection as a necessary criterion for selecting protostars. We then use these PACS selected sources to extract photometry at 250, 350, and 500 $\micron$ using data from the Spectral and Photometric Imaging REceiver (SPIRE; \citealt{Griffin10}) aboard Herschel. We use the Level 3 processed SPIRE maps to extract SPIRE photometry. In Section \ref{sec:sed}, we will describe in detail our source extraction method and photometry measurements.

The contours in Figure \ref{fig:aquila_NH2} show the coverage regions for different Spitzer and Herschel detectors: yellow, green, and orange contours showing the mapping area of IRAC 3.5 and 4.6 $\micron$, MIPS 24 $\micron$ and PACS 100 $\micron$ surveys, respectively. The PACS 70 and 160 $\micron$, and SPIRE 250, 350 and 500 $\micron$ coverage areas are not shown because they cover the entire region displayed in Figure \ref{fig:aquila_NH2}. 

The 850 $\micron$ photometry is from maps made with the Submillimeter Common-User Bolometer Array-2 (SCUBA-2; \citealt{Holland99}) instrument on the James Clerk Maxwell Telescope (JCMT). We use the 850 $\micron$ flux maps from the JCMT Gould Belt Survey data release repository\footnote{https://www.canfar.net/citation/landing?doi=18.0005} \citep{Kirk18}. Similar to SPIRE, we select sources for SCUBA-2 photometry based on the detection in at least one PACS wavelength, as described in section \ref{sec:sed}. 

When available, we complement the Spitzer photometry with the spectroscopic measurements from the Infrared Spectrograph (IRS; \citealt{Houck04}) on Spitzer using the Short-Low (SL; 5.2–14 $\micron$) and Long-Low (LL; 14–38 $\micron$) modules, both having a low spectral resolution of about 90 (see, e.g., \citealt{Kim16} for a description of IRS data reduction). We obtain the IRS observation from the Combined Atlas of Sources with Spitzer IRS Spectra (CASSIS\footnote{https://cassis.sirtf.com/}). We utilize the low-resolution IRS data for the purposes of this study \citep{Lebouteiller11}. 

We also utilize mid-IR photometry from the Wide-field Infrared Survey Explorer (WISE; \citealt{Wright10}). WISE surveyed the entire sky in four bands, W1 (3.4 $\micron$), W2 (4.6 $\micron$), W3 (12 $\micron$), and W4 (22 $\micron$), with angular resolutions of 6.1$''$, 6.4$''$, 6.5$''$, and 12$''$, respectively. The WISE photometry is retrieved from the AllWISE catalog in the NASA/IPAC Infrared Science Archive\footnote{https://irsa.ipac.caltech.edu}.
We primarily use Spitzer observations in the 3-25 $\micron$ wavelength range for the purpose of generating SEDs and estimating bolometric properties, and we use WISE observations only if Spitzer photometry is unavailable. Due to its better angular resolution and sensitivity, we use Spitzer photometry exclusively in the 3.5-24 $\micron$ wavelengths for model fitting, similar to \citetalias{Furlan16}.

\section{Spectral Energy Distributions} \label{sec:sed}

The identification, classification, and analysis of the Aquila protostars require SEDs spanning the infrared and sub-mm regimes.  These  SEDs are assembled from the archival data described in Sec.~\ref{sec:obs}.  In this section, we provide the details of this assembly. This includes an overview of the photometric extraction from the Spitzer, Herschel, and SCUBA-2 data. It also details the merging of the photometric data and  Spitzer/IRS spectra into SEDs for each of the sources detected in the Herschel/PACS bands.


\subsection{2MASS, Spitzer and IRS}

The SESNA catalog provides photometry for all point sources identified in the Spitzer images using the IDL-based interactive photometry visualization tool $PhotVis$ (see \citealt{Gutermuth08b, Gutermuth09} for the details of the automated search and photometry using {\em PhotVis}).  For each of these sources, the photometry for J (1.2 $\micron$), H (1.6 $\micron$), and Ks (2.2 $\micron$) are adopted from the 2MASS point source catalog using the NASA/IPAC Infrared Science Archive (IRSA\footnote{https://irsa.ipac.caltech.edu/}). Spitzer photometry is obtained by aperture photometry on point sources, using $PhotVis$. A positional tolerance of 1$\arcsec$ is used for matching sources in 2MASS and Spitzer/IRAC, and 1.3$\arcsec$ to match them with Spitzer/MIPS 24 $\micron$ sources. For the Spitzer/IRAC 3.6, 4.5, 5.8, and 8.0 $\micron$ images, the aperture size is 2.4$''$ with an annulus of 2.4-7.2$''$ for background subtraction. Similarly for the Spitzer/MIPS 24 $\micron$ data, the aperture size is 6.35$''$ with an annulus of 7.6-17.8$''$ for background subtraction. 

We adopt the same photometric calibration zero point magnitudes (Vega-standard magnitudes for 1 DN s$^{-1}$) as in \cite{Gutermuth08b}: 19.455, 18.699, 16.498, and 16.892 for 3.6, 4.5, 5.8, and 8.0 $\micron$ bands, respectively. These values are derived from the calibration effort presented in \cite{Reach05}, adjusted by standard corrections for the aperture sizes we adopt here. The 24 $\micron$ photometric calibration value that we adopt is that of \cite{Gutermuth08b} and adjusted in \cite{Gutermuth15} for the current aperture size selections, i.e. 14.525 mag for 1 DN s$^{-1}$ within the apertures we adopted for this work.

SESNA includes point source detection completeness mapping at 30$\arcsec$ resolution over much of its coverage and all five Spitzer bandpasses, and the SESNA source catalog provides the nearest 90\% differential completeness value to each source's position for all bands covered at that position.  We use these latter values as upper limits when a source is not detected in all the available bands (see Sec.~\ref{sec:criteria}).

For sources with Spitzer photometry and Herschel/PACS detections, 59 have an IRS low-res spectrum associated with them within the positional uncertainty of 2$\arcsec$, all retrieved from CASSIS (Sec.~2). We include these in the SEDs when available. If any flux mismatches were present in the short-low and long-low segments of the IRS spectrum, the short-low segment of the IRS data was usually scaled to match the flux level at the LL segment; the justification for which is provided in \citetalias{Furlan16}.

\subsection{Herschel/PACS} 


We adapt the procedure and software developed for the SESNA catalog to the PACS 70, 100, and 160~$\mu$m maps. The Herschel maps are scanned in both the parallel and Large Map modes. PACS 70 $\micron$ maps are scanned in the parallel mapping mode with Fast scanning speed (60$\arcsec$/s). PACS 100 $\micron$ maps are scanned in the Large Map mapping mode with Medium scanning speed (20$\arcsec$/s). PACS 160 $\micron$ maps are mapped in both mapping modes and are scanned at Fast as well as Medium speeds. We average the flux densities in both maps to obtain photometry at 160 $\micron$. Similar to the SESNA data, we performed automated point source detection and aperture photometry using {\em PhotVis} with the search parameters optimized for the PACS data.

To obtain calibrated photometry, we follow the recipe provided in the PACS Technical Report\footnote{http://archives.esac.esa.int/hsa/legacy/ADP/PSF/PACS/PACS-P/bolopsf\_22.pdf} and the associated Release Notes\footnote{http://archives.esac.esa.int/hsa/legacy/ADP/PSF/PACS/PACS-P/PACS-P\_PSF\_Release\_Notes.pdf} for the PACS maps. We adjust the aperture sizes according to our science needs and recalculate the aperture correction factor. 
Specifically, we use aperture photometry on the FITS images of the Vesta PACS observations\footnote{http://archives.esac.esa.int/hsa/legacy/ADP/PSF/PACS/PACS-P/}, for the scanning speeds and observing modes that match our data,  to compare the signal in our apertures to those with tabulated encircled energy fractions.  
The use of the FITS files with appropriate scanning speeds and observing modes is crucial, especially for the 70 and 100 $\micron$ images. This is particularly necessary for the blue filters in the parallel mode data, where the increase of the onboard frames averaging in the blue (70 $\micron$) and green (100 $\micron$) filters results in an elongation of the point spread functions (PSFs) in the scan direction compared to the prime mode (especially at the fast scanning velocity of 60$''$/s). In contrast, for the red camera (160 $\micron$) the sampling is identical to the Prime mode, and thus the PSFs are the same (see the PACS release notes for further details). The values for the aperture and annulus sizes and aperture correction factors are provided in Table \ref{tab:apsizes}.  We calculate the celestial coordinates of the sources using the astrometry in the PACS image headers; these provide adequate positional accuracy, as shown in Appendix~\ref{app:positions}. The uncertainties are calculated as the RMS flux on the background annulus using IDL's \emph{aper}.

\subsection{SPIRE and SCUBA}
\label{sec:mad}

We do not run a source identification algorithm on the SPIRE and SCUBA-2 maps due to their coarse angular resolution and their sensitivity to the extended,  structured dust emission in molecular clouds (Figure \ref{fig:aquila_NH2_T}). Instead, we search for sources at the known locations of PACS sources using the following process.

\begin{deluxetable*}{ccccccc} 
\tablecaption{Aperture sizes and Correction factors \label{tab:apsizes}} 
\tablewidth{0pt} 
\tablehead{ 
\colhead{Wavelength} & \colhead{Detector} & \colhead{Angular} & \colhead{Aperture} & 
\colhead{Inner} & \colhead{Outer} & \colhead{Aperture} \\ 
\nocolhead{} & \nocolhead{} & \colhead{Resolution} & \colhead{Radius} & 
\colhead{Annulus} & \colhead{Annulus} & \colhead{Correction} \\ 
\colhead{($\micron$)} & \colhead{} & \colhead{($''$)} & \colhead{($''$)} & 
\colhead{($''$)} & \colhead{($''$)} & \colhead{} 
} 
\decimalcolnumbers 
\startdata 
70 & PACS & 6 & 9.6 & 12.8 & 25.6 & 0.798 \\ 
100 & PACS & 8 & 8.0 & 9.6 & 16.0 & 0.729 \\ 
160 & PACS & 13 & 12.8 & 16.0 & 22.4 & 0.735 \\ 
250 & SPIRE & 18 & 18.0 & 24.0 & 36.0 & 0.658 \\ 
350 & SPIRE & 24 & 30.0 & 40.0 & 60.0 & 0.801 \\ 
500 & SPIRE & 36 & 42.0 & 56.0 & 84.0 & 0.771 \\ 
850 & SCUBA-2 & 15 & 15.0 & 18.0 & 30.0 & 0.85 \\ 
\enddata 
\end{deluxetable*}

At the location of each known source, we extract postage stamp images from the SPIRE and SCUBA-2 maps with diameters $\sim$3 times the outer annulus radius centered on the source, where the outer annulus radii are given in Table \ref{tab:apsizes}. The postage stamps contain the whole aperture and annulus regions with some extra space for visualizing the surrounding environment. 
For each postage stamp image, we create a radial profile plot that shows the variation of flux with distance for all pixels. We bin the radius axis of the radial profile plot by one pixel and compute the median flux in every bin. The median fluxes are then interpolated using a spline interpolation of third order (cubic interpolation) to show the smoothed behavior in the radial profile plot. We compute the background flux in the annulus region by first clipping fluxes above the three sigma RMS flux in the annulus regions defined in Table \ref{tab:apsizes}, followed by an MMM background\footnote{https://buildmedia.readthedocs.org/media/pdf/photutils/v0.3/photutils.pdf}($F_{\rm{bkg}}$) estimator of the form (3 $\times$ median) - (2 $\times$ mean). 

Our goal is to identify compact SPIRE and SCUBA-2 point sources that are at most only marginally more extended than point sources.   To do this, we compute the peak flux ($F_{\rm{peak}}$) in the interpolated data near the central source and define a compact source using the following criteria:

\[
\text{Source} = 
\begin{cases}
    \text{Compact source},& \text{if } F_{\rm{peak}} > (3 \times \rm{MAD} + F_{\rm{bkg}})\\
    \text{Extended source or non-detection},              & \text{otherwise.}
\end{cases}
\]

\begin{figure}
\centering
\includegraphics[scale=0.33]{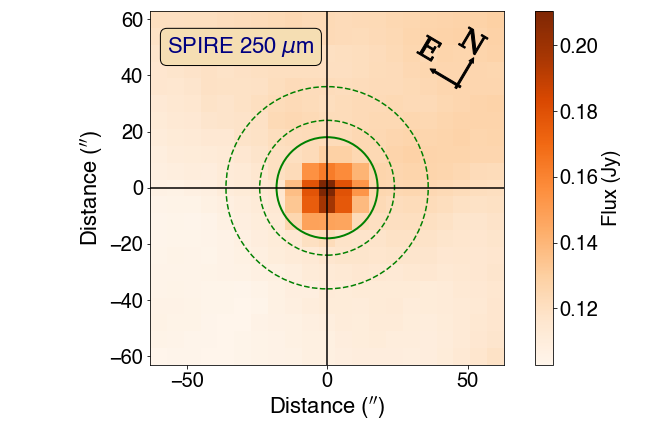}
\includegraphics[scale=0.33]{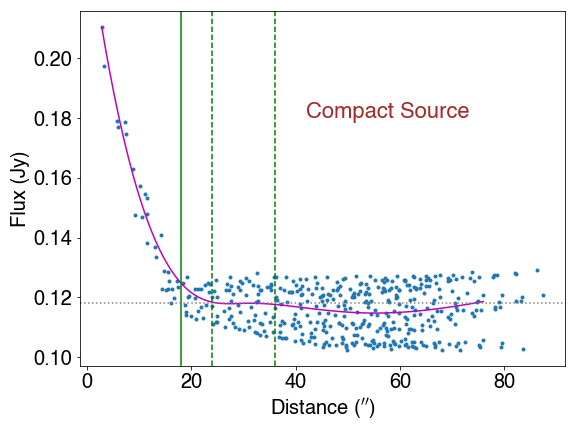}
\includegraphics[scale=0.33]{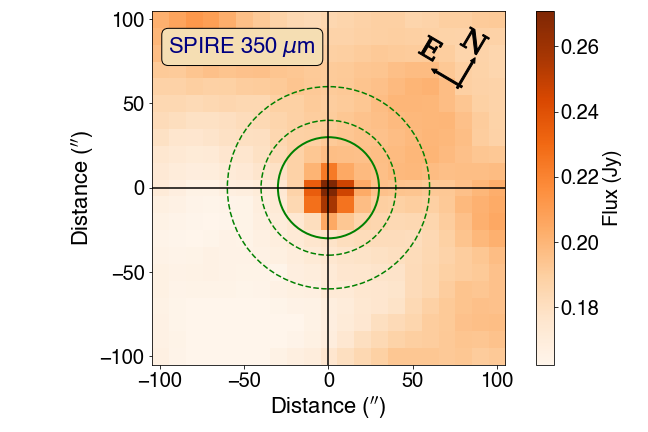}
\includegraphics[scale=0.33]{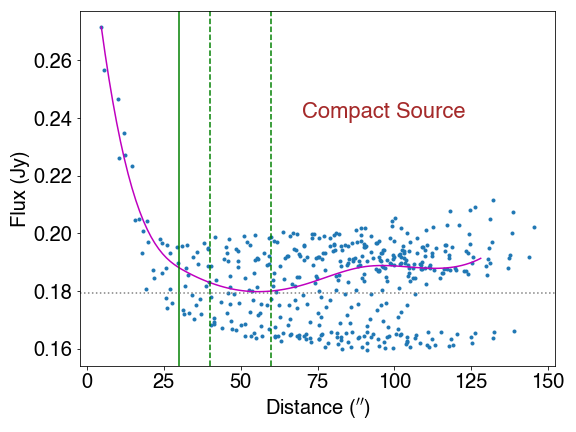}\\
\includegraphics[scale=0.33]{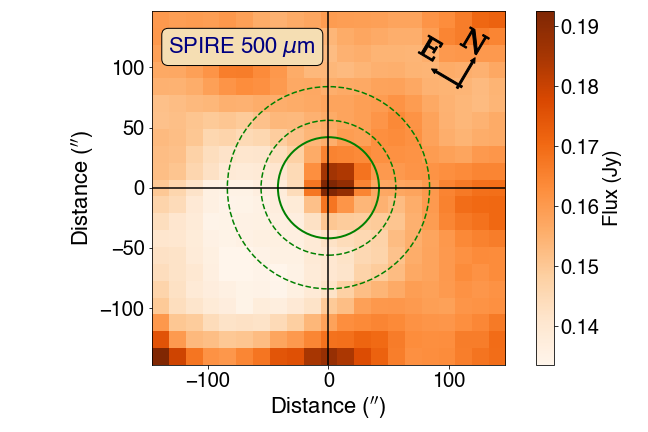}
\includegraphics[scale=0.33]{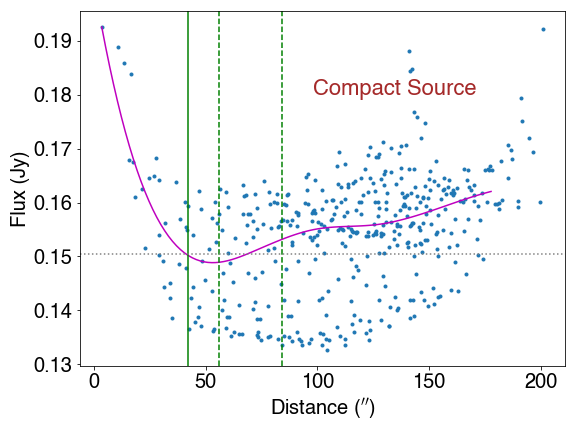}
\hspace*{.2cm}\includegraphics[scale=0.33]{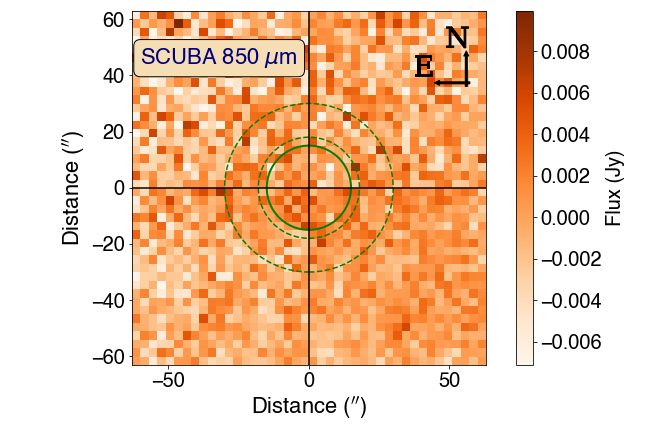}
\includegraphics[scale=0.33]{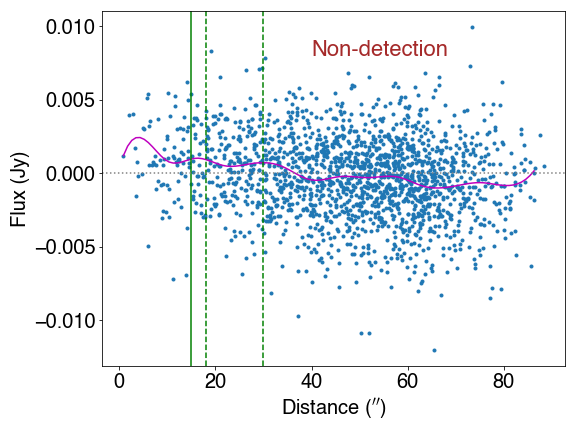}
\caption{An example of compact source identification in SPIRE and SCUBA-2 maps. In each panel, the left image shows the source, and green circles show aperture and annulus sizes for 250, 350, 500, and 850 $\micron$ maps, respectively. Dashed green circles represent the annulus over which we estimate the background flux. The right image shows the radial profile plot for the image on the left. Vertical lines show aperture and annulus sizes that correspond to the left image. The sky value is shown by a grey dashed horizontal line. The magenta lines shows the interpolated radial profile}. The source is classified as a compact source at 250, 350, and 500 $\micron$ but not at 850 $\micron$ according to the MAD analysis.
\label{fig:mad}
\end{figure}

\noindent
Here, MAD is the median absolute deviation of the fluxes in the central aperture. MAD is the median of the absolute deviation from the median value, providing a non-parametric measure of dispersion about the median and is more robust than other commonly used dispersion measures (see \citealt{Feigelson12} for a discussion on using MAD in astronomy). 
As an example illustration of the MAD technique,  we show in Figure \ref{fig:mad} one of the PACS-detected sources in the SPIRE maps in 250, 350, and 500 $\micron$ and in the SCUBA-2 850 $\micron$ map. This source is detected in all three SPIRE bands but is not detected by SCUBA-2.

We perform aperture photometry on the compact sources using IDL's $aper$ package from the IDL Astronomy Users' Library \citep{Landsman93}. We use the aperture and annulus sizes from Table \ref{tab:apsizes} for each respective wavelength band and divide by the corresponding aperture correction factors to account for the flux outside the aperture. For extended sources, we use an aperture size equal to the beam size and do not perform background subtraction. We treat such photometry as upper limits to the actual fluxes in that wavelength and do not include them in any calculations such as $\Lbol$ and $\Tbol$ (defined in section~\ref{sec:ps}). The upper limit photometry in the far-IR, however, plays an important role in selecting the best fit models (described in detail in Sec.~\ref{sec:modeling}).

For the SPIRE maps, the SPIRE Data Reduction Guide\footnote{http://herschel.esac.esa.int/hcss-doc-15.0/print/spire\_drg/spire\_drg.pdf} provides the details of aperture photometry. The guide also provides the FITS images for the SPIRE maps of the estimated encircled energy fraction for different aperture sizes\footnote{http://archives.esac.esa.int/hsa/legacy/ADP/PSF/SPIRE/SPIRE-P/}. The recommended values of aperture sizes for 250, 350, and 500 $\micron$ maps are 22$''$, 30$''$ and 42$''$ respectively, with a background annulus of 60-90$''$ for all three SPIRE maps. However, for the nearby ($<$500 pc) clouds such annulus sizes contain contributions from extended emission from nearby sources. We choose aperture sizes similar to the beam sizes and use a narrower annulus that is closer to the central aperture. The aperture correction factor for the adjusted sizes is recalculated using SPIRE maps of Neptune following the SPIRE Data Reduction Guide. The specific values are provided in Table \ref{tab:apsizes}.

We follow \cite{Dempsey13} for selecting aperture size and the aperture correction factor for the SCUBA-2 850 $\micron$ maps and refer the readers to the paper for a detailed discussion on aperture photometry of SCUBA-2 maps. In short, \cite{Dempsey13} uses a large sample of primary (Mars and Uranus) and secondary calibrator observations to investigate the instrument beam shape and photometry methods to deduce flux conversion factors for different aperture sizes. The aperture correction factors in Table \ref{tab:apsizes} are estimated following the routine in \cite{Dempsey13} using the curve of growth calculations (see Figure 6 and Table 3 in \citealt{Dempsey13}).

 The photometry uncertainties in the SPIRE and SCUBA-2 maps are obtained using a quadrature sum of the measured fluctuation (standard deviation) in the background annulus and flux uncertainty in the aperture calculated using flux uncertainty maps. This accounts for both formal uncertainties and uncertainties due to the background subtraction in fluctuating background. The signal-to-noise ratio in the SPIRE and SCUBA-2 photometry must be greater than 3. If not, the estimated flux is used as an upper limit.

We plot the fluxes of the SEDs in units of Jy. The flux units for Spitzer, Herschel, and JCMT are available in terms of Jy, however, 2MASS and WISE photometry are in units of magnitudes. To convert fluxes from magnitudes to Jy for 2MASS and WISE we use conversion factors based on Vega-based magnitudes with zero-point fluxes from \cite{Cohen03, Wright10} and  \cite{Jarrett11}.


\section{Source Classification} \label{sec:ps}

Once the SEDs have been constructed for the detected sources, each source must be classified and high-confidence protostars identified. 
Since the motivation of our work is to construct SEDs of protostars that span the mid and far-regimes, sources selected for further study must be detected in at least one of the PACS bands with a signal to noise ratio (SNR) of five or greater. 
The selected sources include objects other than protostars, such as external galaxies and pre-main sequence (pre-ms) stars with disks. In the following sections, we will explain the subsequent filtering techniques that we use to select the final sample of eHOPS protostars.

Previous attempts to identify and classify protostars often relied on mid-IR data in the $\le$ 24 $\micron$ range \citep{Megeath04, Allen07, Gutermuth09, Kryukova12}. Some studies also used coarser resolution and lower sensitivity Spitzer/MIPS 70 $\micron$ data or ground-based sub-mm data to identify protostars (e.g., \citealt{Evans09,Dunham13}).  
The HOPS program obtained deep 70~$\mu$m imaging targeting individual protostars identified with $3-24$~$\micron$ Spitzer photometry by \citet{Megeath12} and \citet{Megeath16}. As explained in \cite{Fischer20} and \citetalias{Furlan16}, the targeted protostars had MIPS 24 $\micron$ detections above a threshold flux and predicted PACS 70 $\micron$ fluxes $> 42$~mJy. Protostars that did not satisfy these criteria yet serendipitously fell into the HOPS fields were later added to the catalog. In particular,  deeply embedded protostars were discovered in the Herschel/PACS imaging that were faint or undetected in the  MIPS 24 $\micron$  band; these were added to the HOPS catalog as PACS Bright Red Sources \citep[PBRS,][]{Stutz13, Tobin15, Tobin16}. 

By relying on wide field, far-IR surveys covering entire clouds, eHOPS takes a different approach toward identifying protostars than HOPS and other programs. For the eHOPS sample,  we start by including all compact sources in the cloud surveys that have at least one detection in the Herschel/PACS wavelengths. We use a combination of mid and far-IR criteria, as well as the entire SEDs, to characterize and filter the sources. These criteria are applied over multiple steps to build a sample of protostars and reject contaminants. The end results are catalogs of protostars, galaxies, and pre-ms stars with disks, all with PACS detections.

\subsection{Source selection criteria} \label{sec:criteria}

We find 1102 sources that have detection in at least one of the Herschel/PACS wavebands with SNR $>$ 5. 
Out of the 1102 PACS detected sources, 272 sources are detected in only one wavelength in the $1-850~\micron$ wavelengths. Comparing the postage images of such sources at each wavelength, we find them to be mostly artifacts instead of a real source and we filter them from our catalog. For the remaining 830 sources with at least two detections in $1-850~\micron$, we devise a method to detect and classify protostars as well as identify pre-ms stars with disks and galaxies. 

Since our classification technique relies on mid-IR observations, Spitzer photometry and spectroscopy is an important prerequisites in identifying protostars. If a source is observed by Herschel but not mapped by Spitzer, the emission behavior in the mid-IR is unknown and hence the source cannot be classified. The most reliable classifications are for the sources that have both Spitzer and Herschel coverages. To account for different degrees of mapping coverage, we sort the 830 PACS detected sources into five different groups.

\begin{itemize}
    \item [A.] \underline{IRAC + MIPS + Herschel}: 492 sources are mapped both by Herschel and by Spitzer with IRAC and MIPS. Of these, 59 also have Spitzer IRS spectra. 

    \item [B.] \underline{WISE + MIPS + Herschel}: 51 additional sources are observed by Herschel and MIPS, are not observed by IRAC, yet are detected by WISE to replace the IRAC (3.6 and 4.5 $\micron$) bands.
    
    \item [C.] \underline{MIPS + Herschel}: 25 sources are only covered by MIPS and Herschel, but are not detected by WISE due to their faintness or due to confusion.
    
    \item [D.] \underline{WISE + Herschel}: 167 sources are not covered by the Spitzer maps. For these, we use WISE and Herschel detections. 
    
    \item [E.] \underline{Herschel-only}: 95 sources are detected in Herschel observations only ($\geq$70 $\micron$), are not covered by Spitzer, and are not detected by WISE.
    
\end{itemize}

In the remainder of the paper, we classify the sources that are in group A as they are the only ones that can be properly classified using mid-IR photometry and for which the far-IR properties can be constrained. Furthermore, in Sec.~\ref{sec:modeling} we only perform model fits for the protostars in group A since the model parameters cannot be constrained for sources that lack mid-IR Spitzer coverage. Table \ref{tab:aquila_phot} lists the 2MASS, Spitzer, Herschel, and JCMT photometry of all the group A sources classified as protostar. The column ``Object" in Table  \ref{tab:aquila_phot} denotes the eHOPS designation according to their increasing Right Ascension. The column ``Comments" provides the respective ``step" where the source is classified as a protostar along with brief reasoning behind its classification. In addition, we provide photometry of pre-ms stars with disks in Appendix \ref{app:disk} and galaxies in Appendix \ref{app:gal}. We provide photometry of sources in groups B, C, D, and E in Appendix \ref{app:out}.


\subsubsection{SED Based Diagnostics of IR Sources}

The characterization and classification of the protostars is based on the spectral indices, bolometric temperatures, and to a smaller extent, bolometric luminosities.  We define these quantities and devise source classification criteria similar to the HOPS survey papers (\citealt{Megeath12, Stutz13}; \citetalias{Furlan16}; \citealt{Fischer17}). The full SED spans $1-850 ~\micron$ photometry, including the IRS spectrum where available. We use the photometry with SNR $>$ 5 in all wavebands and exclude the upper limits for our calculations. If either of the Spitzer 3.6, 4.5 or 24 $\micron$ photometry is not available for any source, then we use the corresponding WISE 3.4, 4.6 or 22 $\micron$ photometry in our calculations.

We define the spectral index using the mid-IR SED slope as it probes the IR emission from the disk and inner envelope. Similar to HOPS, we use the spectral index between Spitzer/IRAC 4.5 $\micron$ and Spitzer/MIPS 24 $\micron$ to classify YSOs. Bands centered at wavelengths shorter than 4.5 $\micron$ are not used to estimate the spectral index to minimize the effects of extinction (also see \citealt{Anderson22} for a discussion on disfavoring the use of shorter wavelengths to estimate the spectral index). Since the SED between the two bands is not well approximated by a power law, we calculate the spectral index $\alpha$ from the two endpoints,

\begin{equation} \label{eq:alpha}
    \alpha_{4.5, 24} = \frac{d\,\rm{log}(\lambda F_{\lambda})}{d\, \rm{log}(\lambda)} = \frac{\rm log[24~F_{\lambda}(24~\mu m)]-log[4.5~F_{\lambda}(4.5~\mu m)]}{\rm log(24)-log(4.5)} .
\end{equation}

\noindent 
Similarly, for YSOs with IRS spectrum, we also estimate the spectral index ($\alpha_{\rm{IRS}}$) using the fluxes at 5 and 25 $\micron$. We use $\alpha_{\rm{IRS}}$ to find high-confidence protostars and pre-ms stars with disks in Section \ref{sec:step1}.

\cite{Myers93} defines $\Tbol$ as the effective temperature of a blackbody with the same flux-weighted mean frequency as in the observed SED. We follow \cite{Myers93} to calculate $\Tbol$, as is used by many previous literature works (e.g., \citealt{Dunham15}; \citetalias{Furlan16}):

\begin{equation} \label{eq:Tbol}
    \Tbol~\rm{[K]} = 1.25 \times 10^{-11} \frac{\int_0^\infty \nu F_{\nu} d\nu}{\int_0^\infty F_{\nu} d\nu}.
\end{equation}

\begin{longrotatetable} 
\begin{splitdeluxetable*}{c|cc|cc|cc|ccc|ccc|ccc|ccc|cccBcc|cc|cc|cc|ccc|ccc|ccc|cccBccc|ccc|ccc|c} 
\tablecaption{Observed fluxes (in mJy) for the eHOPS protostars in Aquila. \label{tab:aquila_phot}} 
\setlength{\tabcolsep}{4pt} 
\tablehead{ 
\nocolhead{} & \multicolumn{6}{|c|}{2MASS} & \multicolumn{15}{|c}{Spitzer} & \multicolumn{8}{c|}{WISE} & \multicolumn{18}{|c}{Herschel} & \multicolumn{3}{|c|}{JCMT} & \nocolhead{} \\ 
\hline 
\nocolhead{} & \multicolumn{6}{|c|}{} & \multicolumn{12}{|c|}{IRAC} & \multicolumn{3}{|c}{MIPS} & \multicolumn{8}{c|}{} & \multicolumn{9}{|c|}{PACS} & \multicolumn{9}{|c}{SPIRE} & \multicolumn{3}{|c|}{SCUBA2} & \nocolhead{} \\ 
\hline 
\multicolumn{1}{c|}{Object} & \colhead{$F_{\rm{J}}$} & \colhead{$\Delta F_{\rm{J}}$} & \multicolumn{1}{|c}{$F_{\rm{H}}$} & \colhead{$\Delta F_{\rm{H}}$} & \multicolumn{1}{|c}{$F_{\rm{K}}$} & \colhead{$\Delta F_{\rm{K}}$} & \multicolumn{1}{|c}{$F_{3.6}$} & \colhead{$\Delta F_{3.6}$} & \colhead{$F_{3.6}^C$} & \multicolumn{1}{|c}{$F_{4.5}$} & \colhead{$\Delta F_{4.5}$} & \colhead{$F_{4.5}^C$} & \multicolumn{1}{|c}{$F_{5.8}$} & \colhead{$\Delta F_{5.8}$} & \colhead{$F_{5.8}^C$} & \multicolumn{1}{|c}{$F_{8.0}$} & \colhead{$\Delta F_{8.0}$} & \colhead{$F_{8.0}^C$} & \multicolumn{1}{|c}{$F_{24}$} & \colhead{$\Delta F_{24}$} & \colhead{$F_{24}^C$} & \multicolumn{1}{c}{$F_{3.4}$} & \colhead{$\Delta F_{3.4}$} & \multicolumn{1}{|c}{$F_{4.6}$} & \colhead{$\Delta F_{4.6}$} & \multicolumn{1}{|c}{$F_{12}$} & \colhead{$\Delta F_{12}$} & \multicolumn{1}{|c}{$F_{22}$} & \colhead{$\Delta F_{22}$} & \multicolumn{1}{|c}{$F_{70}$} & \colhead{$\Delta F_{70}$} & \colhead{$F_{70}^S$} & \multicolumn{1}{|c}{$F_{100}$} & \colhead{$\Delta F_{100}$} & \colhead{$F_{100}^S$} & \multicolumn{1}{|c}{$F_{160}$} & \colhead{$\Delta F_{160}$} & \colhead{$F_{160}^S$} & \multicolumn{1}{|c}{$F_{250}$} & \colhead{$\Delta F_{250}$} & \colhead{$F_{250}^S$} & \multicolumn{1}{c}{$F_{350}$} & \colhead{$\Delta F_{350}$} & \colhead{$F_{350}^S$} & \multicolumn{1}{|c}{$F_{500}$} & \colhead{$\Delta F_{500}$} & \colhead{$F_{500}^S$} & \multicolumn{1}{|c}{$F_{850}$} & \colhead{$\Delta F_{850}$} & \colhead{$F_{850}^S$} & \multicolumn{1}{|c}{Comments} 
} 
\startdata 
eHOPS-aql-1 & --- & --- & --- & --- & --- & --- & 6.4 & 0.1 & 2.2 & 16.3 & 0.2 & 5.2 & 23.7 & 0.3 & 3.8 & 30.3 & 0.4 & 10.9 & 411.6 & 0.7 & 519.4 & --- & --- & --- & --- & --- & --- & --- & --- & 1725.9 & 96.7 & 44.1 & --- & --- & --- & 2793.3 & 142.3 & 30.9 & 3788.7 & 155.7 & U & 6622.5 & 324.8 & U & 5902.5 & 189.3 & U & --- & --- & --- & Step 3--- [3.6]-[4.5] $>$ 0.65, $\alpha_{4.5,24} > -0.3$ \\ 
eHOPS-aql-2 & --- & --- & 5.8 & 0.2 & 35.3 & 0.9 & 90.8 & 1.2 & 2.9 & 146.2 & 1.9 & 3.0 & 171.9 & 2.3 & 13.3 & 175.4 & 2.3 & 32.1 & 571.2 & 5.3 & 1023.4 & 53.6 & 1.2 & 135.2 & 2.5 & 193.0 & 2.8 & 707.3 & 13.7 & 1802.3 & 95.7 & 32.2 & --- & --- & --- & 7713.3 & 387.8 & 46.9 & 6401.0 & 195.3 & 208.0 & 6743.3 & 371.8 & U & 5720.9 & 149.6 & U & --- & --- & --- & Step 3--- [3.6]-[4.5] $>$ 0.65, $\alpha_{4.5,24} > -0.3$ \\ 
eHOPS-aql-3 & 46.3 & 1.0 & 62.7 & 1.3 & 181.9 & 3.9 & --- & --- & --- & --- & --- & 13.9 & --- & --- & --- & --- & --- & 2901.5 & --- & --- & --- & 691.1 & 44.6 & 1398.9 & 78.6 & 21219.3 & 469.1 & 48082.6 & 132.9 & 10816.2 & 582.8 & 217.2 & --- & --- & --- & 2314.8 & 120.1 & 35.1 & 557.5 & 35.4 & 78.7 & 159.9 & 61.6 & U & 208.2 & 57.8 & U & --- & --- & --- & Step 3--- [3.6]-[4.5] $>$ 0.65, $\alpha_{4.5,24} > -0.3$ \\ 
eHOPS-aql-4 & 58.7 & 1.2 & 126.6 & 2.7 & 161.4 & 3.1 & 179.1 & 2.4 & 10.9 & 210.6 & 2.8 & 9.0 & 240.1 & 3.1 & 6.0 & 350.1 & 4.6 & 71.4 & --- & --- & 248.6 & 275.4 & 7.1 & 345.3 & 7.0 & 508.1 & 6.6 & 1113.8 & 16.4 & 1571.9 & 87.0 & 36.8 & 1755.1 & 88.4 & 10.4 & 1578.0 & 81.1 & 15.0 & 1144.4 & 85.2 & 175.7 & 607.0 & 205.1 & U & 5208.0 & 8.8 & U & 110.5 & 7.9 & -0.2 & Step 3--- [3.6]-[4.5] $>$ 0.65, $\alpha_{4.5,24} > -0.3$ \\ 
eHOPS-aql-5 & 0.7 & --- & 6.5 & 0.2 & 18.5 & 0.4 & 44.0 & 0.6 & 2.3 & 65.1 & 0.9 & 3.0 & 82.0 & 1.1 & 2.2 & 100.6 & 1.5 & 9.4 & 413.9 & 5.1 & 745.1 & 37.8 & 0.9 & 69.3 & 1.4 & 91.4 & 4.7 & 261.6 & 12.0 & 1614.8 & 87.4 & 33.4 & 1984.1 & 99.5 & 7.5 & 7966.7 & 399.1 & U & 16716.3 & 36.4 & U & 13903.5 & 34.2 & U & 11414.2 & 21.9 & U & 202.9 & 11.2 & U & Step 3--- [3.6]-[4.5] $>$ 0.65, $\alpha_{4.5,24} > -0.3$ \\ 
\enddata 
\tablecomments{For each Spitzer wavebands, the third column labeled as ``F$^C$" denotes the completeness flux limit. For Herschel and JCMT wavebands ``F$^S$" denotes 1-$\sigma$ background flux. The letter ``U" implies that the flux is used as an upper limit when fitting with models. The column ``Comments" states the step in which the protostar is identified with a brief explanation. Only 5 sources are included in the table. Full list of sources is available online.}
\end{splitdeluxetable*} 
\end{longrotatetable}

\noindent
Similarly, the bolometric luminosity ($\Lbol$) of a source is calculated by integrating the observed flux, adopting a distance and assuming the source radiates equally over 4 $\pi$ sr:

\begin{equation} \label{eq:lbol}
    \Lbol~[\Lsun] = 4\pi D^2 \int_0^\infty F_{\nu} d\nu ,
\end{equation}

\noindent
where $D$ is the distance to the source. We use the trapezoidal summation rule to integrate over the finitely sampled SEDs in equation \ref{eq:Tbol} and \ref{eq:lbol}, similar to \cite{Dunham15} and \citetalias{Furlan16}.

\subsubsection{Step 1: Selection of high-confidence YSOs} \label{sec:step1}

High-confidence sources are those identified and classified as either a protostar, a pre-ms star with disk, or a galaxy with higher certainty using the full suite of MIPS, IRAC, IRS, and Herschel data. The goal of step 1 is to catalog only the high-confidence protostars and pre-ms stars with disks; we classify the remaining sources in subsequent steps. We use the high confidence sources to establish criteria in section \ref{sec:step2} for removing the extragalactic contaminants from our sample.

\paragraph{Protostars} The sample of protostars targeted in the HOPS survey (\citealt{Megeath12}; \citetalias{Furlan16}) were first identified using Spitzer photomery. HOPS protostars were characterized by a steeply rising SED between 3.6 and 4.5 $\micron$ such that the color [3.6] - [4.5] $>$ 0.65 (e.g., \citealt{Megeath04, Kryukova12}) and a flat or rising SED between 4.5 and 24 $\micron$ such that $\alpha_{4.5, 24}$ $\geq$ -0.3 \citepalias[also see][]{Furlan16}. For the high confidence protostars, we adopt these criteria and require a rising spectrum  ($\alpha_{4.5, 24}$ $\geq$ 0.3), and exlcude flat-spectrum sources.

In addition, the Spitzer IRS spectra provide a viable means of identifying deeply embedded protostars  \citepalias{Furlan16}. Many protostars are characterized by a silicate absorption feature at $\sim$10 $\micron$ along with ice features in the $5-8~\micron$ range. In protostars observed at a nearly edge-on inclination, there is instead a strong dip at about 10 $\micron$ which is between the scattered light at $< 10$~$\micron$  and the thermal emission at $> 10$~$\micron$ \citep{Whitney03a}. Note that the presence of a silicate absorption feature is not a unique signature of a protostar and is present in other highly reddened sources. Absorption features accompanied by ice lines are a stronger indication of a protostar as the ice features are found in only cold dense regions, but we do not require those in our current criteria.

We thus require that the high confidence protostars as those with a positive slope between 5 and 25 $\micron$ in the IRS spectra ($\alpha_{\rm{IRS}}$), indicative of a rising SED, and minima at $\sim 10$~$\micron$ due to silicate absorption or high inclinations.  By excluding  flat-spectrum sources, we minimize the chances of incorporating non-embedded sources in our list of high-confidence protostars. In summary, for a source with IRAC, MIPS, IRS and PACS data to be a high-confidence protostar, it should satisfy the following requirements:

\begin{equation} \label{eq:bon_ps}
\begin{split}
\alpha_{\rm{IRS}} \ge 0~\rm{and/or~absorption}&\rm{/minimum~at~} \lambda \approx 10 \micron\\
    \rm{and} \\
      [3.6] - [4.5] &> 0.65\\
    \rm{and} \\
      \alpha_{4.5, 24} &> 0.3
\end{split}
\end{equation}

\begin{figure}[b!]
\includegraphics[scale=0.3]{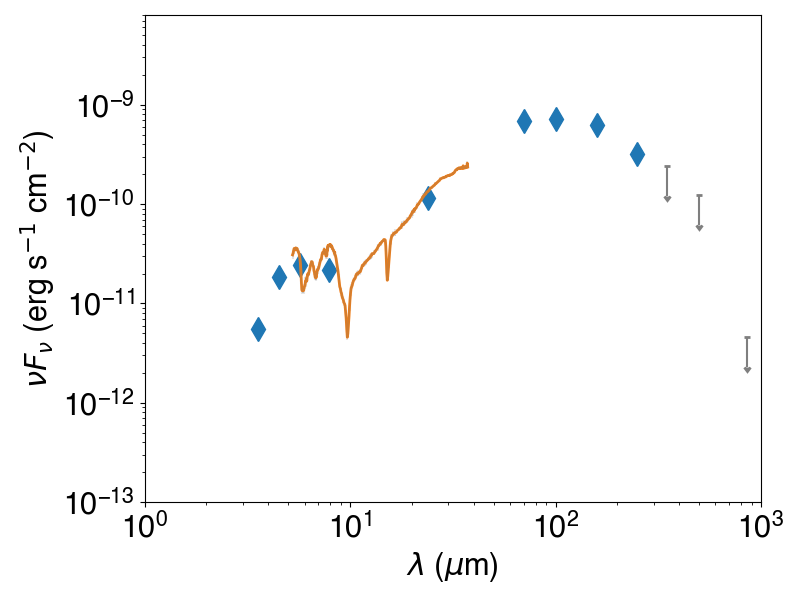}
\includegraphics[scale=0.3]{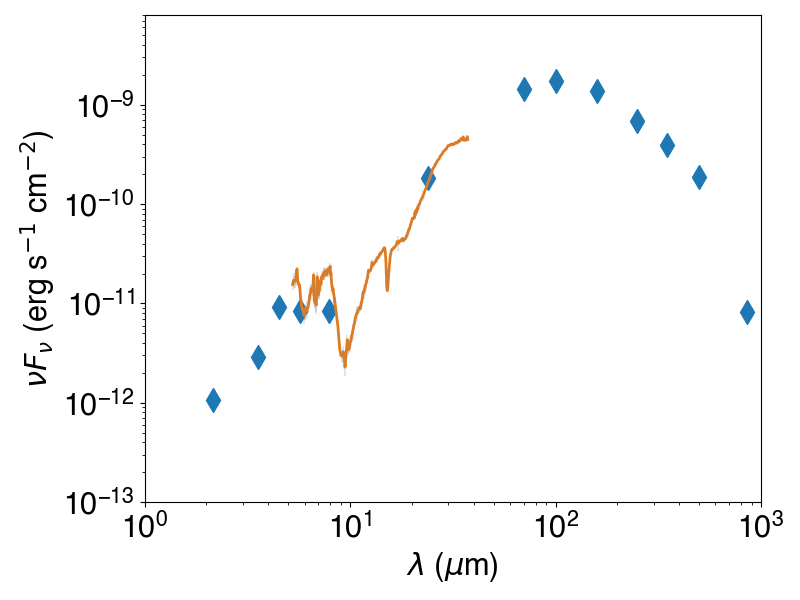}
\includegraphics[scale=0.3]{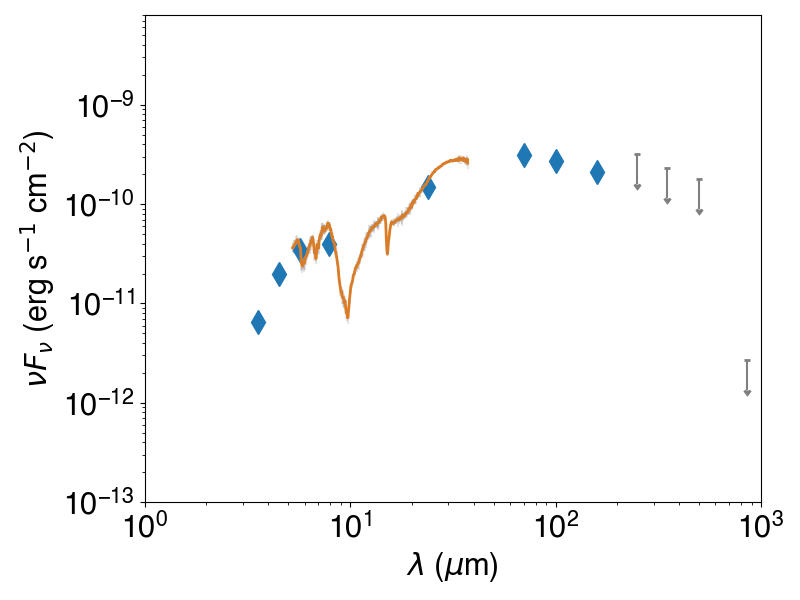}
\caption{Three examples of high-confidence protostars that are selected using the requirements in section \ref{sec:step1}. The protostars are \dummyABCDE, \dummyABCED ~and \dummyABDCE ~from left to right. All sources exhibit prominent silicate absorption features at 10 and 20 $\micron$, and rising SEDs in the $\sim~1-100~\micron$ wavelength range. 
\label{fig:ps_bf}}
\end{figure}

We classify 29 high confidence protostars using Equation \ref{eq:bon_ps}. In Figure \ref{fig:ps_bf}, we show 3 example SEDs of protostars out of 29: \dummyABCDE, \dummyABCED ~and \dummyABDCE. After we identify and filter the extragalactic contamination, we relax the above conditions to incorporate other protostars in Sec.~\ref{sec:step3}.

\paragraph{Pre-ms stars with disks} To distinguish the pre-ms stars with dusty disks from protostars, we require a declining SED slope between 3.6 and 4.5 $\micron$. This requirement may not be satisfied by highly reddened pre-ms stars which we will discuss in Sec.~\ref{sec:step3}. Additionally, we restrict ourselves to the sources that have $\alpha_{4.5, 24}$ $<$ -0.3, thereby excluding flat-spectrum sources. For pre-ms stars with a disk, the silicate feature at $\sim 10$~$\micron$ is often apparent in emission. The silicate emission is produced by hot dust grains in the upper layers of protostellar disks. Similar to protostars, we combine all three requirements for selecting high-confidence pre-ms stars with disks:

\begin{equation} \label{eq:bon_disk}
\begin{split}
\alpha_{\rm{IRS}} < 0~\rm{and/or}\rm{~silicate}&\rm{~emission~at~} \lambda \approx 10 \micron\\
    \rm{and} \\
      [3.6] - [4.5] &< 0.65\\
    \rm{and} \\
      \alpha_{4.5, 24} &< -0.3
\end{split}
\end{equation}

\begin{figure}[ht!]
\includegraphics[scale=0.3]{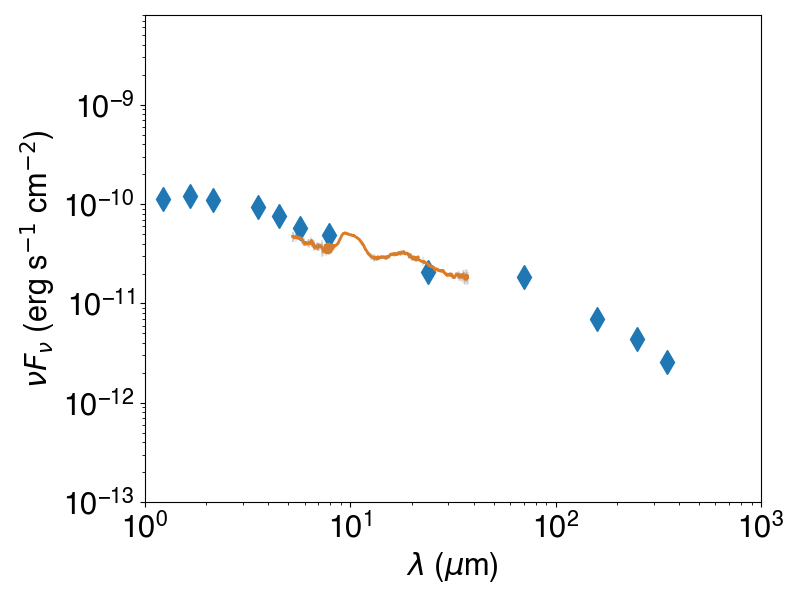}
\includegraphics[scale=0.3]{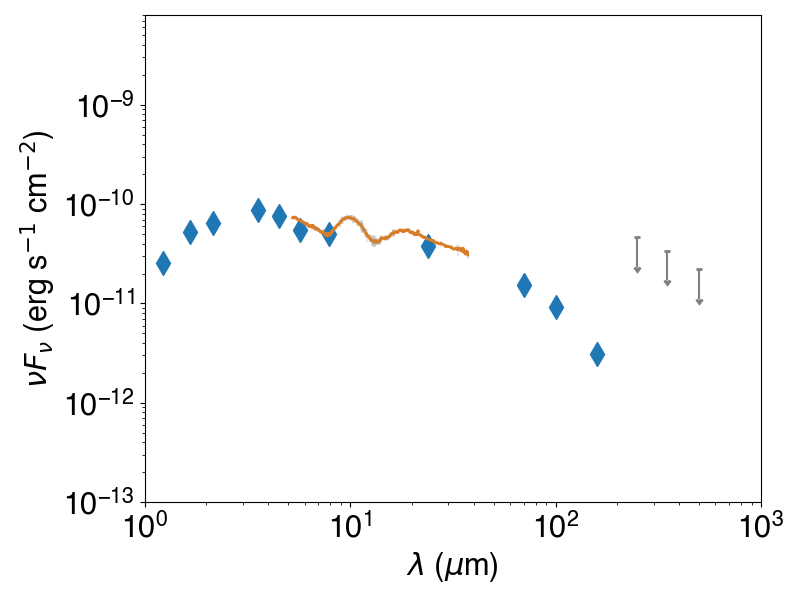}
\includegraphics[scale=0.3]{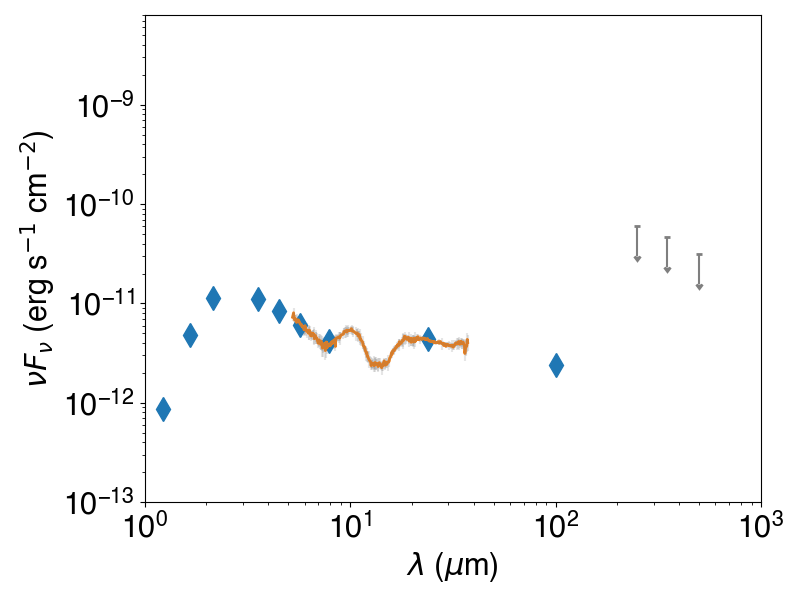}
\caption{Examples of high-confidence pre-ms stars with disks that are selected using the requirements in Sec.~\ref{sec:step1}. The sources are \#30016, \#50600, and \#88493 from left to right (c.f. Table \ref{tab:aquila_phot_disk}). 
\label{fig:disk_bf}}
\end{figure}

We classify 13 high-confidence pre-ms stars with disks using Equation \ref{eq:bon_disk}. Figure \ref{fig:disk_bf} shows 3 example SEDs of the high-confidence pre-ms stars with disks.

\subsubsection{Step 2: Identification of galaxies} \label{sec:step2}

In section \ref{sec:step1}, we identify 42 high-confidence YSOs that require IRS spectra for classification (29 protostars and 13 pre-ms stars with disks). 
In this section, we discuss different techniques that we use to classify galaxies in the remaining 450 sources to account for extragalactic contamination in our total sample. First, we search the literature for well-studied galaxies that may be present in our sample. Second, we search the high-resolution visible and IR databases to identify nearby galaxies that are spatially resolved such that the spiral arm or elliptical morphologies of galaxies are explicit. Third, we use IRAC colors to detect star-forming galaxies, a technique used by \cite{Stutz13}. Finally, we apply a method based on fitting the observed SEDs with extragalactic model templates. This method robustly identifies the galaxies that do not show strong PAH emission from star-forming regions and are not resolved in the high-resolution visible/IR images.

We search the Set of Identifications, Measurements, and Bibliography for Astronomical Data (SIMBAD) database\footnote{http://simbad.u-strasbg.fr/simbad/} to find galaxies that are studied previously in the literature. We find 3 galaxies that were studied previously in \cite{Oliveira10}. These galaxies are noted as ``Literature" in column ``Comments" in Table  \ref{tab:aquila_phot_gal}. Then we search the Panoramic Survey Telescope and Rapid Response System (Pan-STARRS) images \citep{Chambers16} to find galaxies that are resolved at optical wavelengths. We also search the UKIRT Infrared Deep Sky Survey (UKIDSS) database \citep{Lawrence07} in the near-IR. UKIDSS has higher angular resolution than 2MASS and thus is helpful to morphologically distinguish galaxies that are resolved at near-IR wavelengths. We use a tolerance of 2$\arcsec$ to match our catalog with SIMBAD, UKIDSS and Pan-STARRS. In appendix \ref{app:morph}, Figure \ref{fig:morph} shows a few examples of morphologically identified galaxies. We identify 34 such galaxies in our sample. Morphologically identified galaxies are noted as ``Morphology" in column ``Comments" in Table  \ref{tab:aquila_phot_gal}. Using these two techniques, we still cannot identify the galaxies that are farther away and not spatially resolved.

We follow \cite{Stutz13} to detect star-forming galaxies based on their Spitzer/IRAC color-color plots, specifically using the [3.6]-[4.5] and [5.8]-[8.0] colors, or equivalently, $\alpha_{3.6, 4.5}$ and $\alpha_{5.8, 8.0}$. Star-forming galaxies are often characterized by polycyclic aromatic hydrocarbon (PAH) emission around the 5-8 $\micron$ regime. \cite{Stutz13} found that the sources with $\alpha_{5.8, 8.0}$ $\geq$ 3 and $\alpha_{3.6, 4.5}$ $\leq$ 0.5 show characteristics of star-forming galaxies where bright PAH emission results in high values of $\alpha_{5.8, 8.0}$ but the $\alpha_{3.6, 4.5}$ range is dominated by starlight. This $\alpha_{5.8, 8.0}$ criterion is higher than the criterion used by \cite{Gutermuth08b}, but that only ensures that we have star-forming galaxies with high confidence when we apply the criterion of \cite{Stutz13}. In addition, we cross-check the SEDs of all the sources that satisfy the \cite{Stutz13} criteria to further confirm they show PAH emission at $\sim8~\micron$. In Table  \ref{tab:aquila_phot_gal}, the IRAC-color criterion is mentioned in the column ``Comments" for the galaxies identified using IRAC colors.

\begin{figure}[ht!]
\includegraphics[scale=0.3]{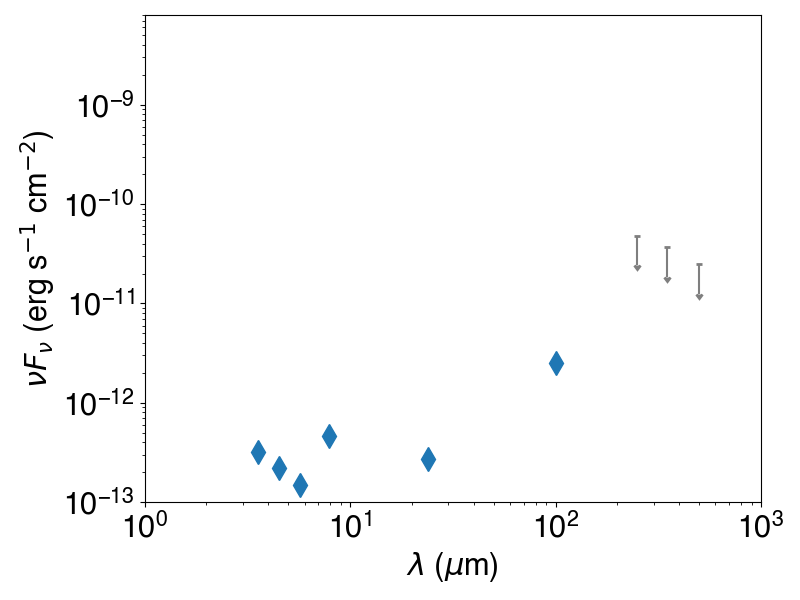}
\includegraphics[scale=0.3]{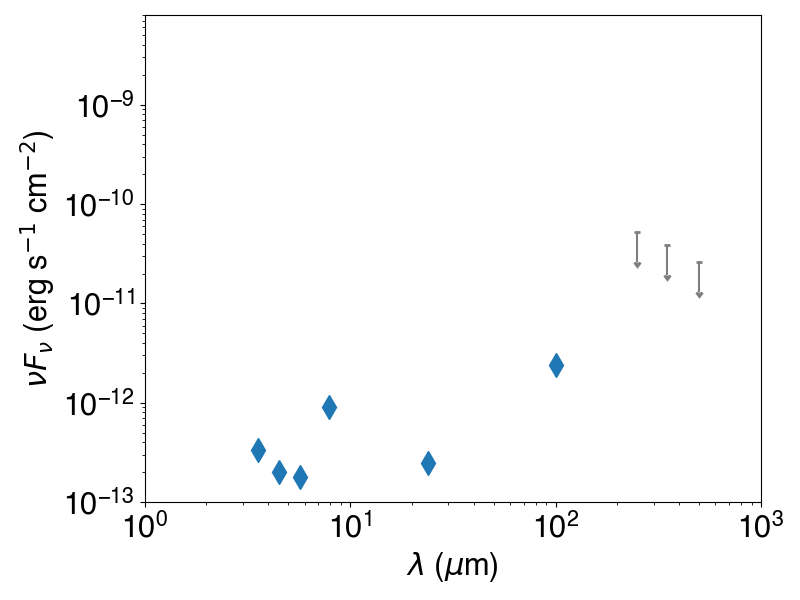}
\includegraphics[scale=0.3]{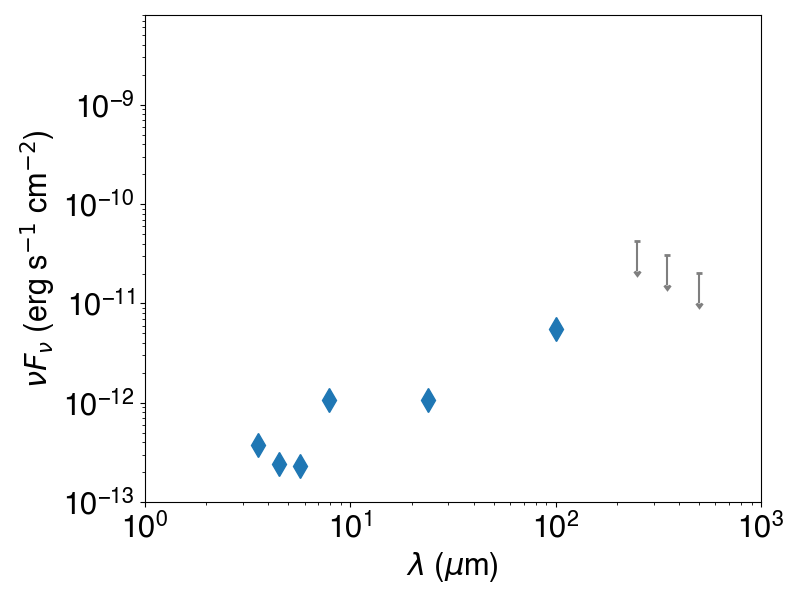}
\caption{SEDs for some star-forming galaxies with prominent PAH emission at $\sim8\micron$ that are found using the Spitzer/IRAC color criterion in \cite{Stutz13}. The sources are \#1288709, \#1186350, and \#1218688 from left to right (c.f. Table \ref{tab:aquila_phot_gal}). These are identified as a part of Step 2.
\label{fig:sed_sfg}}
\end{figure}

Figure \ref{fig:sed_sfg} shows SEDs for three of the star-forming galaxies with prominent PAH emission at $\sim$ 8 $\micron$ that are selected using the criterion from \cite{Stutz13}.  We refine our selection using the conditions in Sec.~\ref{sec:step4} after selecting other galaxies in the sample. We find 41 star-forming galaxies in our sample using the IRAC-color criterion from \cite{Stutz13}. Some galaxies are found using more than one criterion, for example from both the literature and IRAC colors. Altogether, we detect 63 high-confidence galaxies following the above-described processes.


\paragraph{Extragalactic SED template fitting}
The majority of the galaxies identified so far are either star-forming galaxies with PAH emission at 8~$\micron$ or the galaxies that are spatially resolved in the high-resolution near-IR images. However, other extragalactic contaminants such as AGN-dominated galaxies do not show PAH emission and can be missed in the previous steps. We devise a novel approach to reduce contamination by galaxies in the low luminosity sources using the model SED templates for external galaxies of \cite{Kirkpatrick15}. Templates encompass different star-formation properties, from actively star-forming to AGN-dominated to composite galaxies that are between actively star-forming and quenched galaxies. We use these extragalactic model templates to fit all the 492 sources of group A to identify other galaxies in our sample. In this way, we can identify actively star-forming, non-star-forming (AGN-dominated), and composite galaxies. We use the goodness-of-fit measure for SEDs called $R$ to measure how close the observed sources are to extragalactic templates.  Sources with low $R$ values have an SED similar that of a galaxy.  See Appendix \ref{app:extgalfit} for the details of best fit.

\begin{figure}
\includegraphics[scale=0.4]{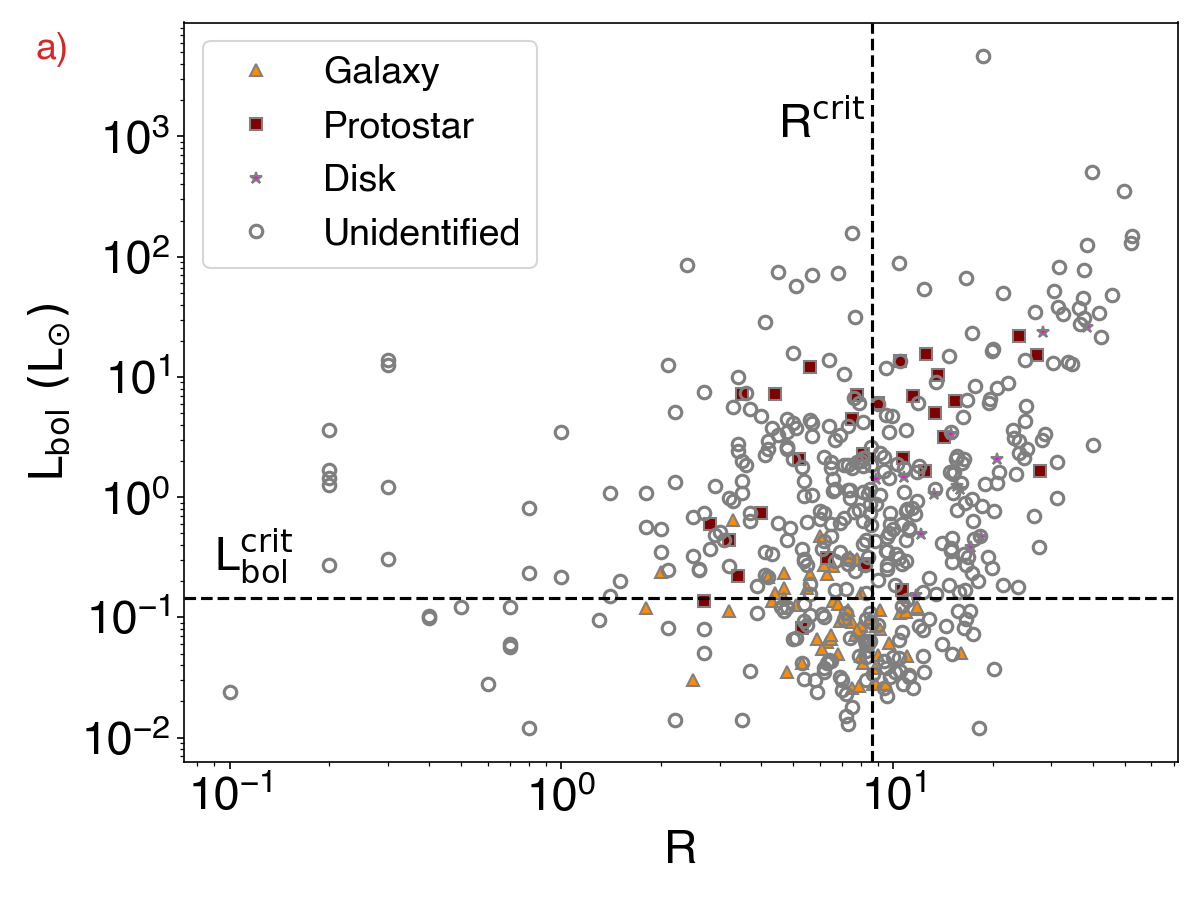}
\includegraphics[scale=0.4]{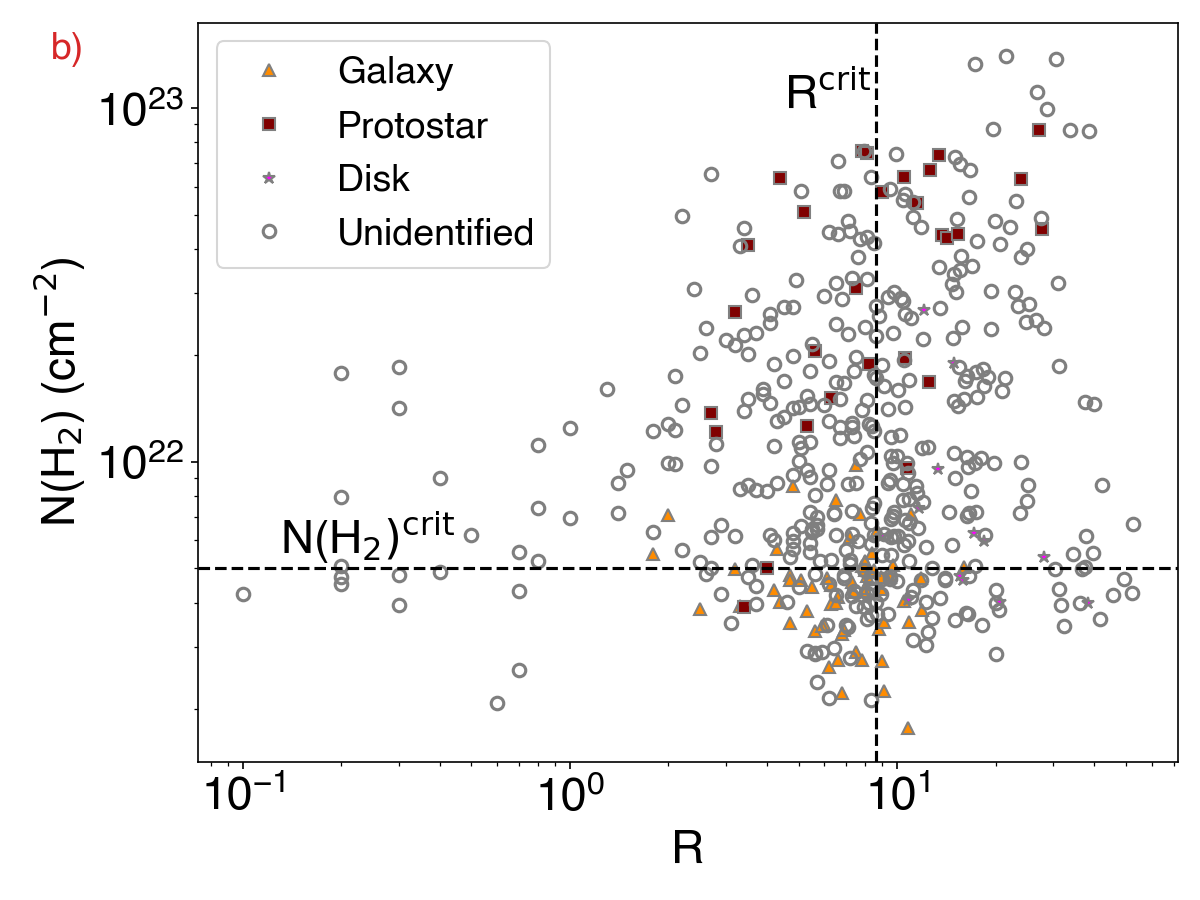}\\
\includegraphics[scale=0.4]{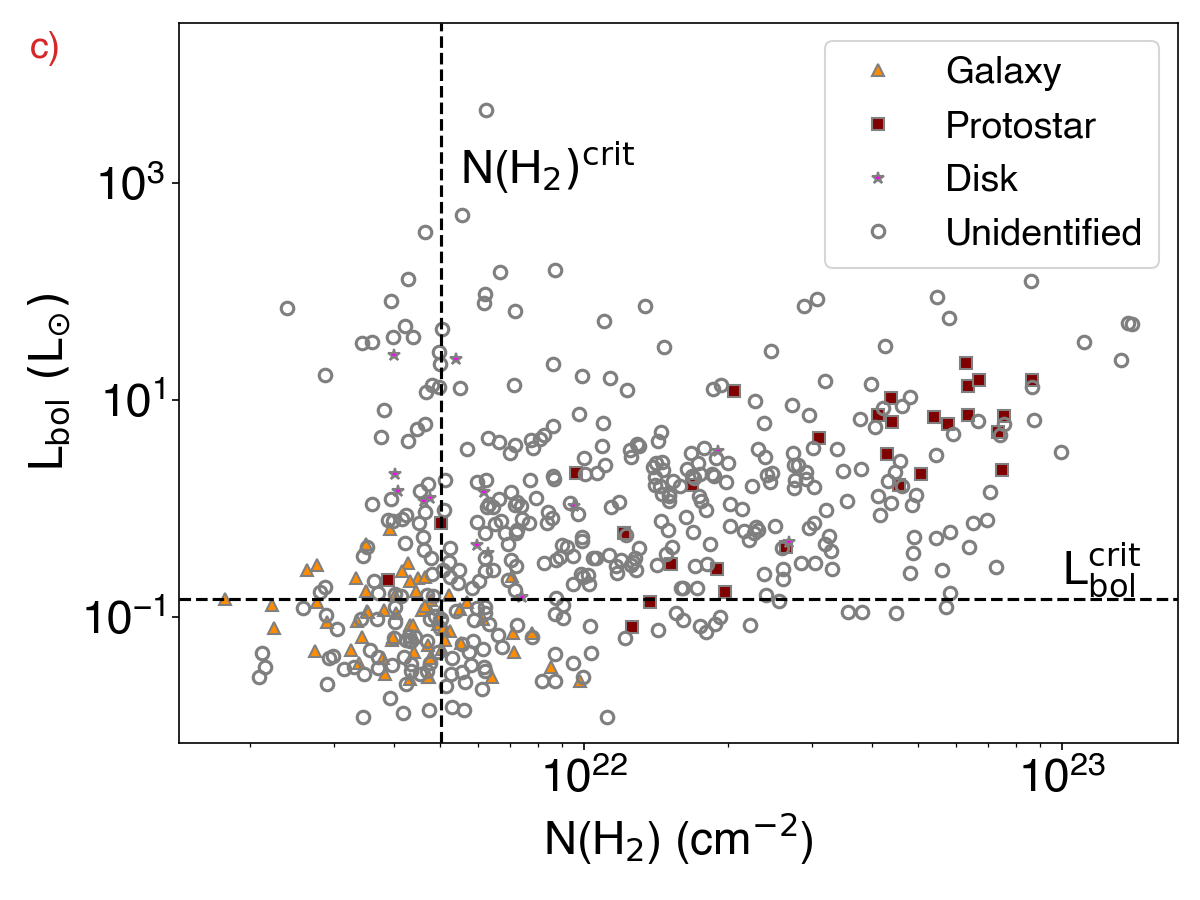}
\includegraphics[scale=0.4]{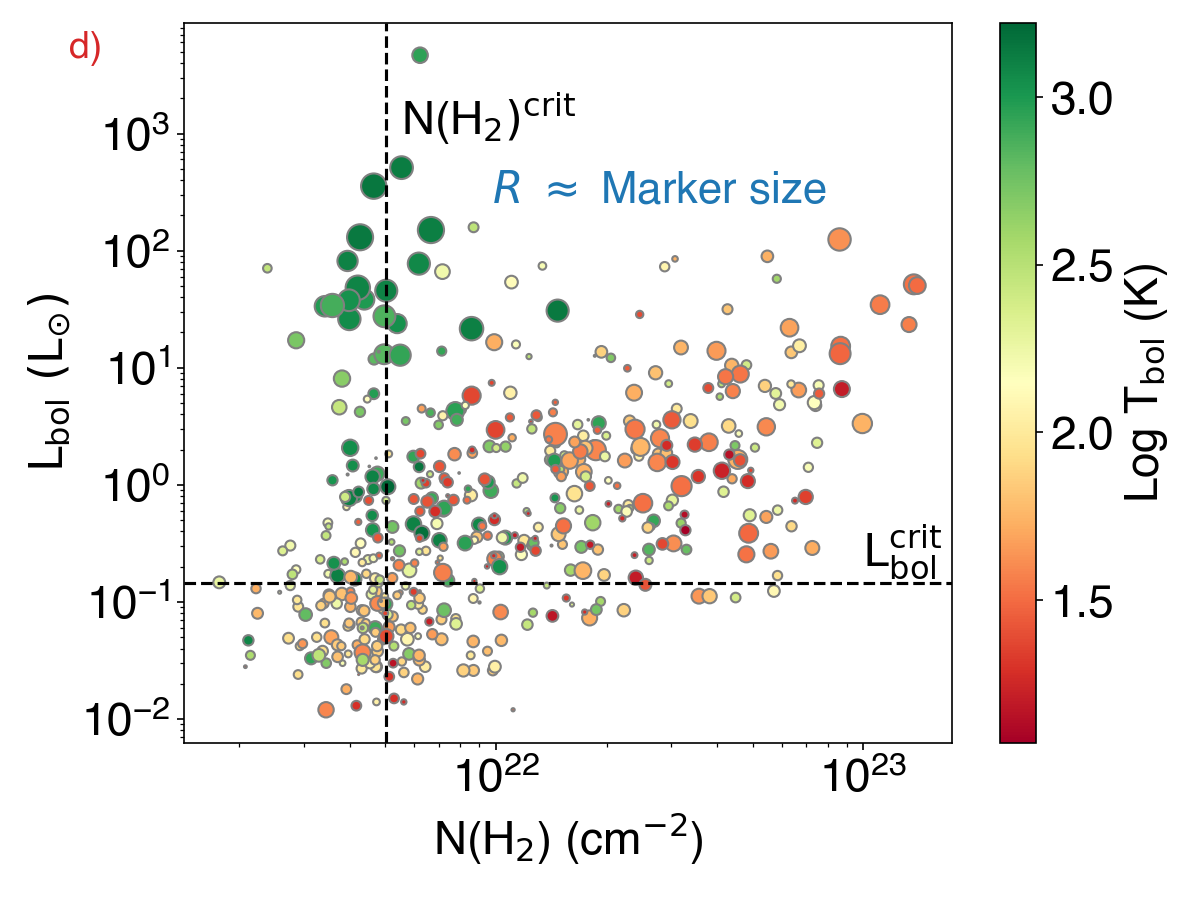}
\caption{Panels a), b) and c) show the scatter plots of \Lbol, N(H$_2$) and $R$, respectively for the high-confidence protostars, pre-ms stars with disks, and galaxies. Panel d) is a repetition of panel c but with a marker size proportional to $R$ and colors that depend on the \Tbol. In all panels, the horizontal and vertical lines correspond to the 75$^{\rm{th}}$ percentile of L$_{\rm{bol}}$, N(H$_2$), and $R$ for the high-confidence galaxies in our sample.
\label{fig:extGal}}
\end{figure}

Identification of galaxies based on $R$ alone is not effective. There may be protostars that appear similar to galaxies particularly when they are detected in only a few photometric bands. In these cases, sources can have lower $R$ values since they have fewer photometric bands in their SEDs.
Luminosity is another way to distinguish between galaxies and protostars; most galaxies are faint across all wavelengths and have a low luminosity for an assumed distance of 436~pc.  A low luminosity source with low $R$ may not necessarily be a galaxy, however, but a low-luminosity protostar such as a VeLLO \citep{Dunham15}. This motivates using the H$_2$ column density (N($\rm{H_2}$)) around the source as an additional criterion; low luminosity sources in regions of low gas column density are likely to be galaxies. Hence, we combine three conditions to identify galaxies in our catalog:

\begin{equation} \label{eq:gal}
\begin{split}
    L_{\rm{bol}}~&<~L_{\rm{bol}}^{\rm{crit}} \\
    \rm{} & \rm{and} \\
    N(\rm{H_2})~&<~ N(\rm{H_2})^{\rm{crit}} \\
    \rm{} & \rm{and} \\
    \rm{R}~&<~\rm{R}^{\rm{crit}}
\end{split}
\end{equation}

\noindent
The projected column density for each source is the Herschel derived column density, N(H$_2)$, measured at the position of the source and smoothed to the beam size of the 500 $\micron$ SPIRE map (Figure \ref{fig:aquila_NH2}), and L$_{\rm bol}$ assumes a distance of 436~pc. L$_{\rm{bol}}^{\rm{crit}}$, N(H$_2)^{\rm{crit}}$ and $R^{\rm{crit}}$ are estimated by studying the variation of L$_{\rm{bol}}$, N(H$_2$) and $R$ for all 492 sources. Figure \ref{fig:extGal} shows the variations of L$_{\rm{bol}}$, N(H$_2$) and $R$ for the sources in the Aquila star-forming clouds. The horizontal and vertical dashed lines mark the critical values below which we find probable galaxies, defined as the 75$^{\rm{th}}$ percentiles of L$_{\rm{bol}}$, N(H$_2$), and $R$ for the sample of previously identified galaxies. For Aquila, we find L$_{\rm{bol}}^{\rm{crit}}$, N(H$_2)^{\rm{crit}}$ and R$^{\rm{crit}}$ to be 0.15 L$_{\odot}$, 5 $\times$ 10$^{21}$ cm$^{-2}$ and 9, respectively. We choose conservative critical values for L$_{\rm{bol}}$, N(H$_2$), and $R$ to ensure that we do not falsely identify low-luminosity protostars as galaxies. 

Using these criteria in Equation \ref{eq:gal}, in the Aquila regions, we find 31 additional galaxies. The galaxies that are identified by fitting observed SEDs with extragalactic templates are designated ``Step 2" in the ``Comments" column in Table \ref{tab:aquila_phot_gal}. We inspect the SEDs of each of these galaxies to make sure there are no obvious YSOs.
Furthermore, we cross-match our galaxy catalogs with the Gaia Early Data Release 3 (EDR3) catalog\footnote{https://www.cosmos.esa.int/web/gaia/earlydr3} to ensure that there is no source in our galaxy catalog that has a reliable Gaia distance corresponding to the distance of Aquila.
We do not find any galaxies with a Gaia distance $<$ 1 kpc and distance/$\Delta$(distance) $>$ 3, confirming that no YSO with a known Gaia distance is misclassified as a galaxy.

\begin{figure}[ht!]
\includegraphics[scale=0.3]{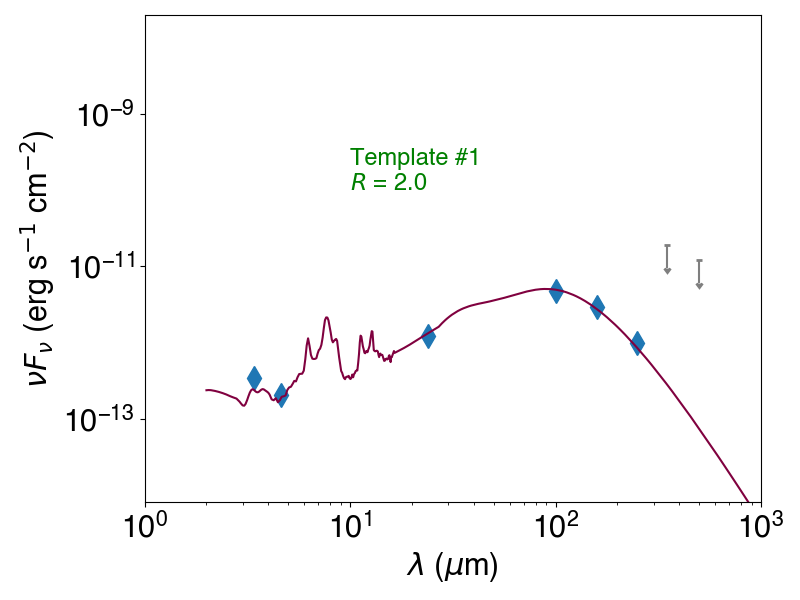}
\includegraphics[scale=0.3]{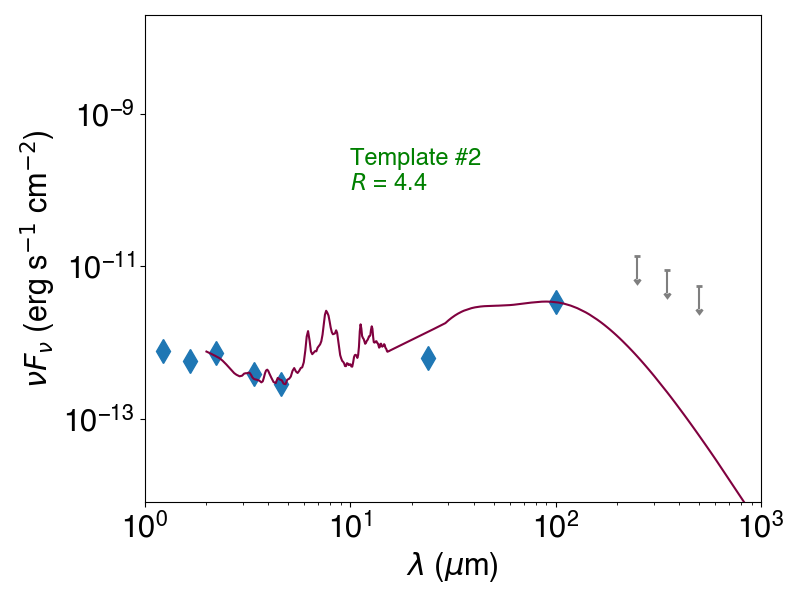}
\includegraphics[scale=0.3]{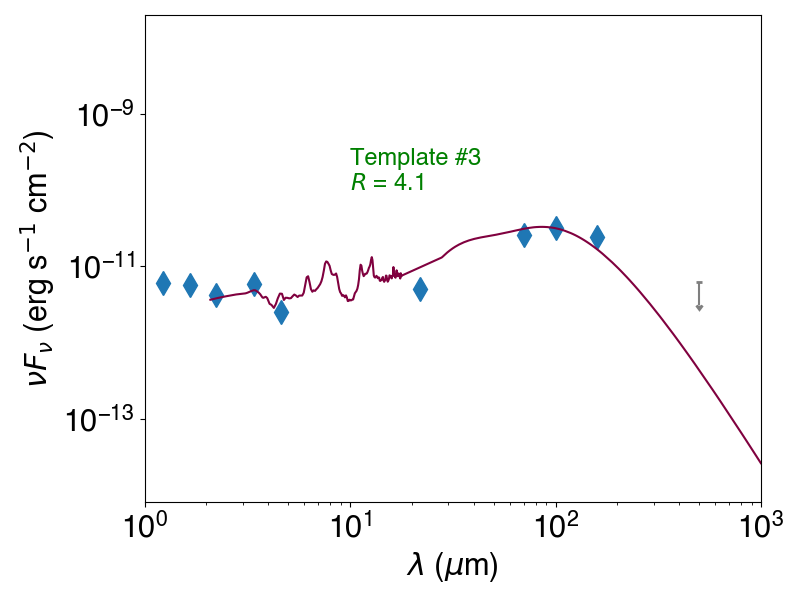}
\caption{Best-fit results with extragalactic templates overplotted to the SEDs of three galaxies in Aquila. The template number and $R$ value for the best fitting model are noted. Model template numbers from \#1 to \#8 refer to galaxies from actively star-forming to AGN dominated (see \citealt{Kirkpatrick15}). The sources are \#2067206, \#192641 and \#36616 from left to right (c.f. Table \ref{tab:aquila_phot_gal}). These galaxies are identified using the template fitting criteria in Step 2.
\label{fig:sed_sfg_fit}}
\end{figure}

Figure \ref{fig:sed_sfg_fit} shows three examples of SEDs of galaxies that are identified by the fitting process with the best fit of the extragalactic SED template. 
Our fitting technique successfully identifies external galaxies, whether they are star-forming, AGN-dominated, or composites. Furthermore, we find our criteria based on a combination of L$_{\rm{bol}}$, N(H$_2$), and $R$ are particularly helpful in identifying galaxies that otherwise would be missed if we use only Spitzer/IRAC colors. 
It should be noted that although our criteria successfully identify galaxies with apparently low L$_{\rm{bol}}$ located in less dense clouds, we may still miss galaxies that do not obey the criteria. For example, a galaxy with L$_{\rm{bol}}$ = 0.2 L$_{\odot}$ but N(H$_2$) $<$ N(H$_2)^{\rm{crit}}$ and R $<$ R$^{\rm{crit}}$ may escape our selection criteria. In subsequent steps, we deploy additional techniques to identify them.

\subsubsection{Step 3: Identification of remaining YSOs} \label{sec:step3}

After identifying the sources that are extragalactic in nature, we use Spitzer photometry to classify the remaining 356 sources of group A into four categories of YSOs: protostars, pre-ms stars with disks, reddened disks, and candidate protostars. To select protostars, we follow previous work by \cite{Megeath04, Megeath12, Gutermuth08b, Gutermuth09, Kryukova12, Dunham15} to set the necessary conditions based on mid-IR photometry. We identify protostars using the following criteria:

\begin{equation} \label{eq:ps1}
\begin{split}
    [3.6] - [4.5] &> 0.65 \\
    \rm{and}& \\
    \alpha_{4.5, 24} &> -0.3.
\end{split}
\end{equation}

\noindent
Similarly, we use the following criteria to identify pre-ms stars with disks:

\begin{equation} \label{eq:disk1}
\begin{split}
    [3.6] - [4.5] &< 0.65 \\
    \rm{and}& \\
    \alpha_{4.5, 24} &< 0.3.
\end{split}
\end{equation}

Note the overlapping region of -0.3 $<$ $\alpha_{4.5, 24}$ $<$ 0.3 between Equations \ref{eq:ps1} and \ref{eq:disk1}. This narrow range of the spectral index is indicative of flat spectrum sources that fall on the borderline between protostars and pre-ms stars with disks. Flat spectrum sources can be a mixture of protostars observed at more face-on inclination angles, protostars with thin envelopes, and a few highly reddened pre-ms stars with disks (\citealt{Calvet94}; \citetalias{Furlan16}; \citealt{Habel21}; \citealt{Federman22}. 
\cite{Heiderman15} report that only about half of the flat-spectrum sources are true protostars using envelope tracers in a large sample of Class 0/I and flat-spectrum sources. Other studies such as \cite{Groschedl19} report that a substantial fraction of the flat-spectrum sources is at a younger evolutionary phase compared to Class II YSOs. \citetalias{Furlan16} argued that most flat spectrum sources in Orion have far-IR emissions from thin, dusty envelopes. 
To minimize any ambiguity posed by the true stage of flat-spectrum sources, we additionally require [3.6] - [4.5] $>$ 0.65 for flat-spectrum sources in our protostellar catalog in addition to $\alpha_{4.5, 24} > -0.3$ (see also \citealt{Kryukova12, Megeath12}). Hence, using Equations \ref{eq:ps1} and \ref{eq:disk1}, a flat-spectrum source is either classified as a protostar or a pre-ms star with disk, but not both, by imposing the [3.6] - [4.5] color criteria.

We find 104 protostars and 56 pre-ms stars with disks, in addition to the high-confidence sources, using the criteria in Equations \ref{eq:ps1} and \ref{eq:disk1}, respectively. Out of 56 pre-ms stars with disks, five sources (Source\# 72009, 561496, 602852, 1525323, and 2062387 in Table \ref{tab:aquila_phot_gal} in Appendix \ref{app:gal}) have either no detection in one or more IRAC channels but detect elevated emission in the 8 $\micron$ channel based on the completeness limits in the  channels with non-detections, or show resolved emission in the IRAC channels. These five sources are reclassified as galaxies. The reason for their reclassification is mentioned in the column ``Comments" in Table \ref{tab:aquila_phot_gal}.

\begin{figure}[ht!]
\includegraphics[scale=0.3]{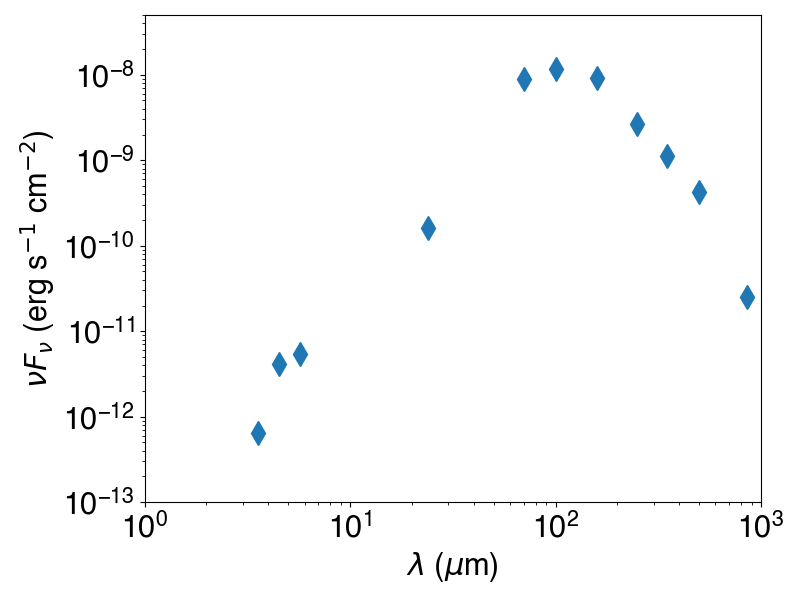}
\includegraphics[scale=0.3]{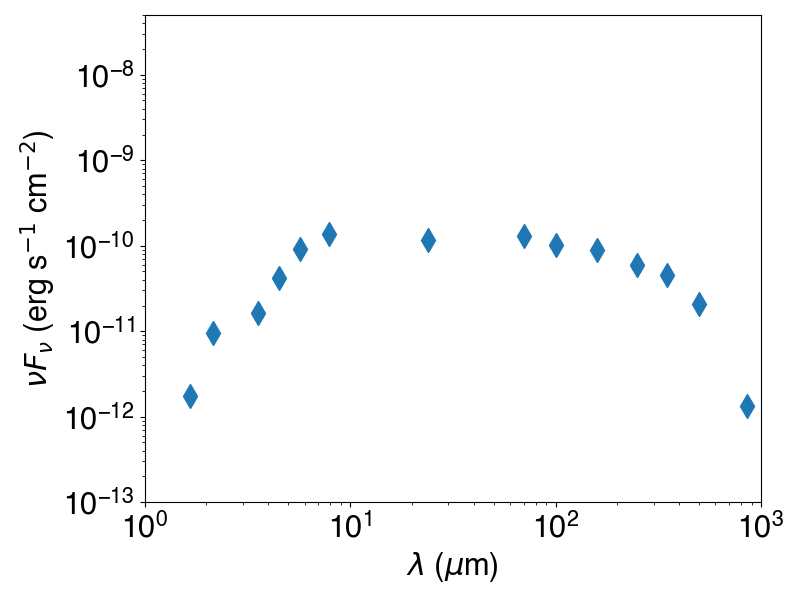}
\includegraphics[scale=0.3]{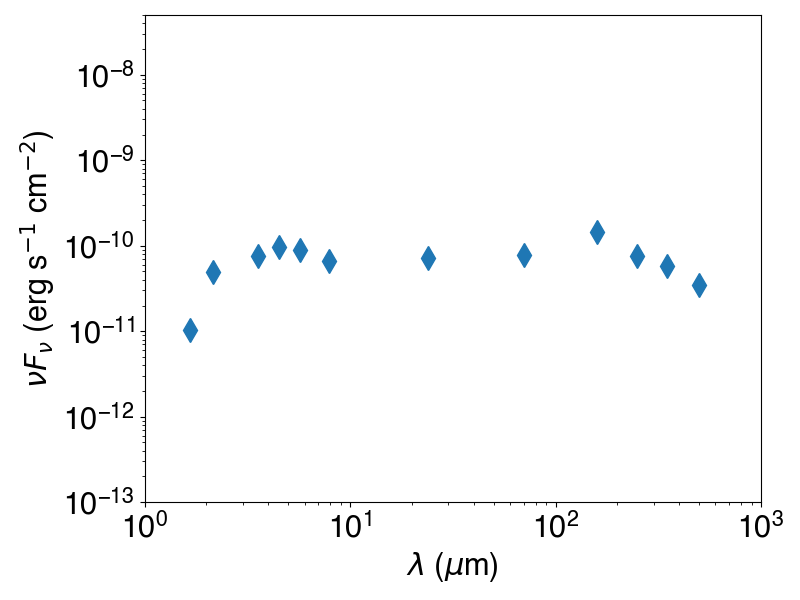}
\caption{A few examples of SEDs of protostars in Aquila regions that are identified using Equation \ref{eq:ps1} defined in section \ref{sec:step3}. The sources are \dummyABDEC ~(Class 0), \dummyABECD ~(Class I), and \dummyABEDC ~(flat-spectrum), respectively from left to right. These protostars are identified as part of Step 3.
\label{fig:sed_ps}}
\end{figure}

\begin{figure}[ht!]
\includegraphics[scale=0.3]{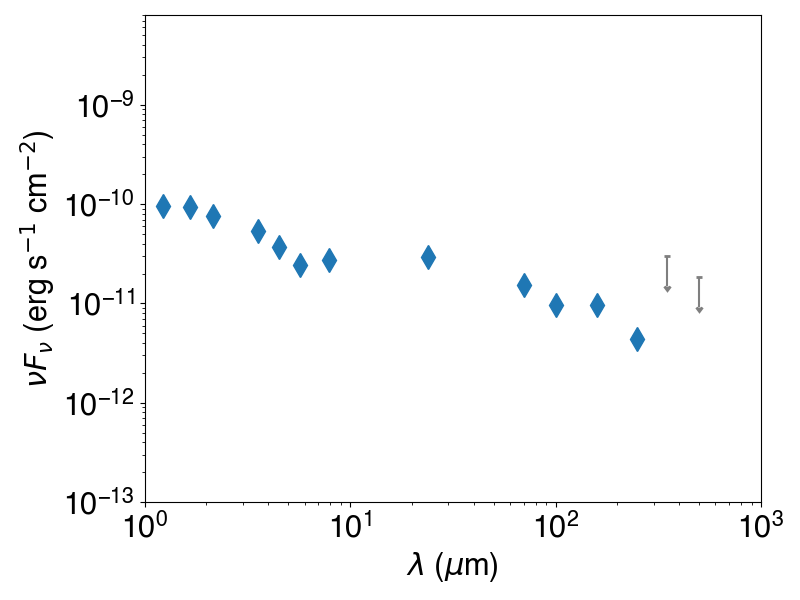}
\includegraphics[scale=0.3]{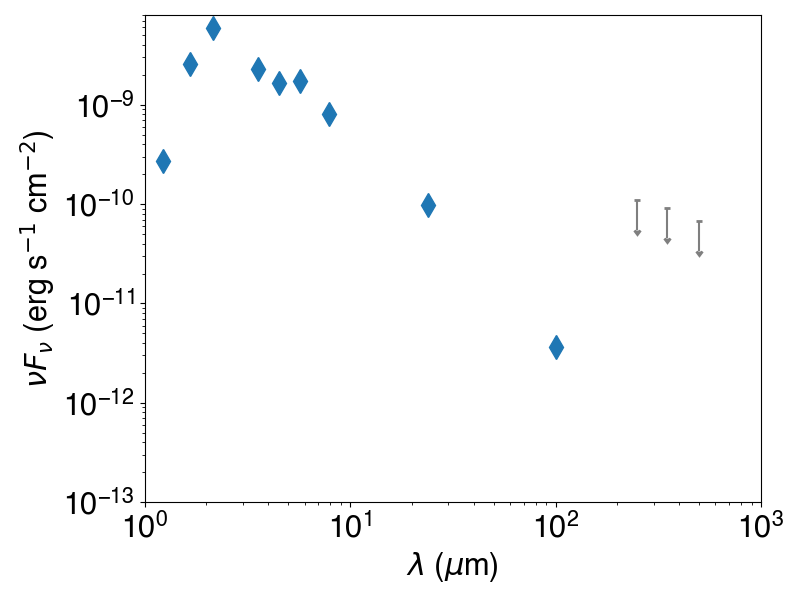}
\includegraphics[scale=0.3]{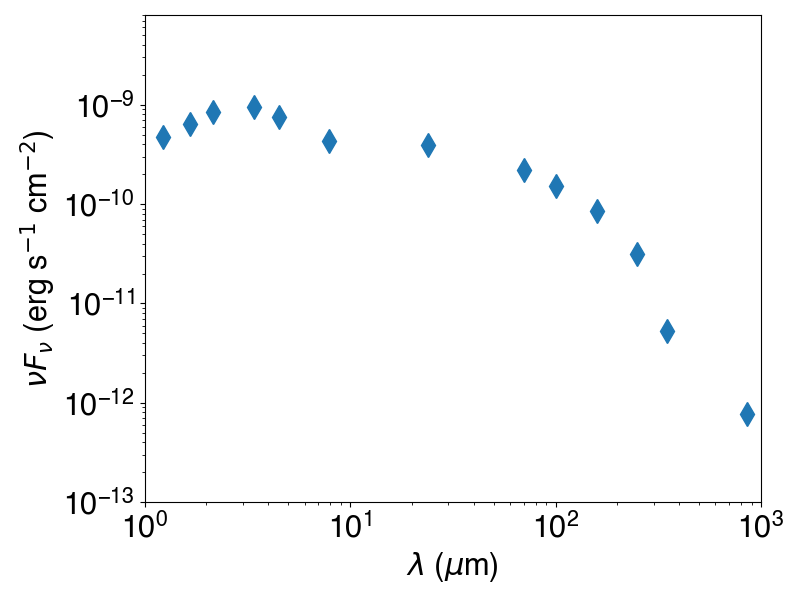}
\caption{A few examples of SEDs of pre-ms stars with disks in Aquila that are identified using Equation \ref{eq:disk1} defined in section \ref{sec:step3}. The sources are \#122731, \#1225768 and \#106653 from left to right (c.f. Table \ref{tab:aquila_phot_disk}). These pre-ms stars with disks are identified as part of Step 3.
\label{fig:sed_disk}}
\end{figure}

So far we detect 133 protostars, 64 pre-ms stars with disks, and 99 galaxies out of the total of 492 sources. A total of 196 sources are yet to be classified. 
Another category of sources is the pre-ms stars with disks that reside in high-extinction regions such that [3.6] - [4.5] $\nless$ 0.65. Our classification technique for pre-ms stars with disks in Equation \ref{eq:disk1} is biased towards the pre-ms stars that are located in the lower column density regions. In high column density regions, the near-IR emission from YSOs suffer from extinction and an increased [3.6] - [4.5] color. For the reddened pre-ms stars with disks, extinction may have a minimal effect on $\alpha_{4.5, 24}$ but [3.6] - [4.5] can be greater than our critical value of 0.65. We identify such reddened pre-ms stars with disks in the remaining 196 sources using the following criteria:

\begin{equation} \label{eq:redDisk}
\begin{split}
    [3.6] - [4.5] &> 0.65 \\
    \rm{and}& \\
    \alpha_{4.5, 24} &< -0.3.
\end{split}
\end{equation}

\noindent
We identify 32 reddened pre-ms stars with disks using Equation \ref{eq:redDisk}. Out of 32 reddened pre-ms stars with disks, one (Source\# 1371293 in Table \ref{tab:aquila_phot_gal} in Appendix \ref{app:gal}) has a peak at IRAC 8 $\micron$ but does not have photometry at 5.8 $\micron$ due to its extended morphology in the 5.8 $\micron$ map. Hence the source is not classified as a galaxy by our IRAC-color based criteria. The source is also missed by our extragalactic contamination  criteria (Equation \ref{eq:gal}) because it is located at N(H$_2$) $\sim$ 7.8 $\times$ 10$^{21}$ cm$^{-2}$, slightly greater than N(H$_2$)$^{\rm{crit}}$. We reclassify it as a galaxy.

\begin{figure}[ht!]
\includegraphics[scale=0.3]{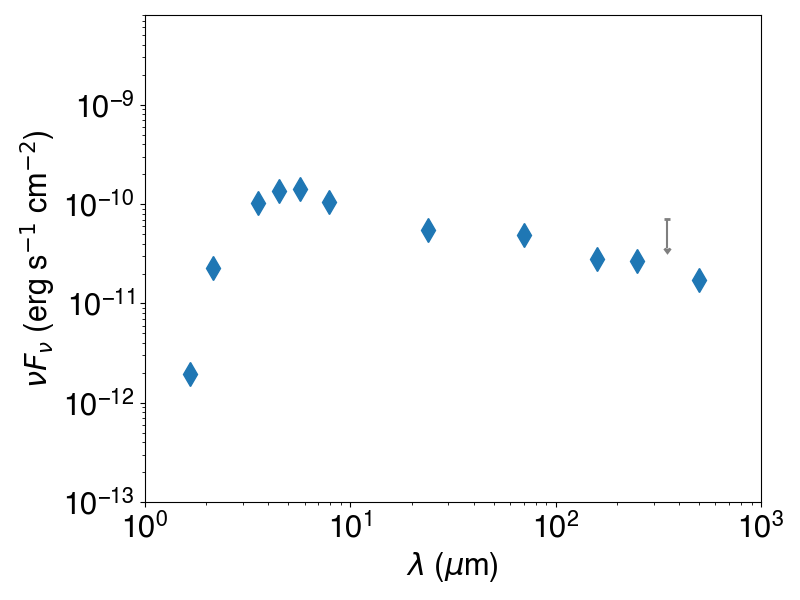}
\includegraphics[scale=0.3]{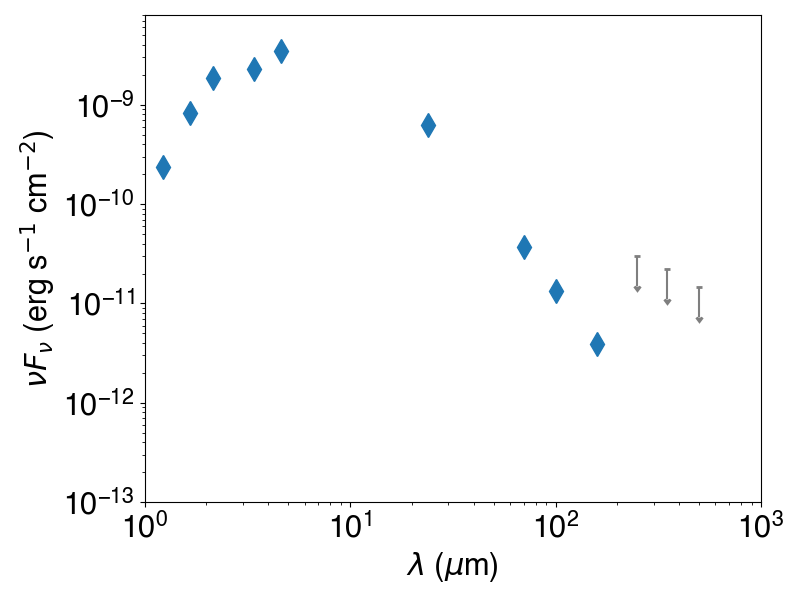}
\includegraphics[scale=0.3]{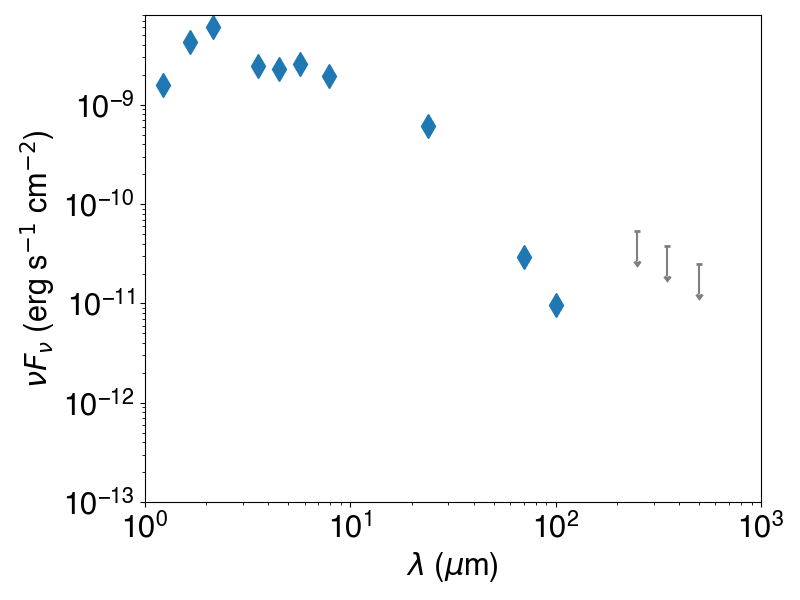}
\caption{Three examples of reddened pre-ms stars with disks in Aquila regions that are identified using criteria defined in section \ref{sec:step3}. The sources are \#280461, \#61202, and \#120110 from left and right (c.f. Table \ref{tab:aquila_phot_disk}). These are identified as part of Step 3.
\label{fig:sed_disk_red}}
\end{figure}

Out of the remaining 31 reddened pre-ms stars with disks, four sources (\dummyACBDE, \dummyACBED, \dummyACDBE ~and \dummyACDEB) are reclassified as protostars. These sources have $-0.5 < \alpha_{4.5, 24} < -0.3$ so were not previously classified as protostars as we require $\alpha_{4.5, 24} > -0.3$ for a protostar. The source \dummyACDEB ~has $\alpha_{4.5, 24} = -0.32$ but has an elevated far-IR emission. The sources \dummyACBDE ~and \dummyACBED ~have weak silicate absorption at $\sim$10 $\micron$. Another source, \dummyACDBE ~show scattered light emission in the near-to-mid IR maps. Although our strict set of classification criteria classifies these four sources as reddened pre-ms stars with disks, the above-mentioned features show their protostellar nature. Figure \ref{fig:sed_disk_red} shows three examples of SEDs of reddened pre-ms stars with disks that we identify using Equation \ref{eq:redDisk}.

A similar problem arises for classifying protostars. We require both [3.6] - [4.5] $>$ 0.65 and $\alpha_{4.5, 24} > -0.3$ to classify protostars. However, some protostars may have [3.6] - [4.5] slightly less than 0.65 but the SED rises beyond 24 $\micron$. The criteria based on Equation \ref{eq:ps1} will exclude them from being a protostar. Additionally, a declining [3.6] - [4.5] color and a rising $\alpha_{4.5, 24}$ value can also be the signature of a few of the outlier galaxies that are missed by our selection criteria. For now, we define all such sources as ``candidate protostars". We find candidate protostars (hereafter: CP) in the remaining 168 unidentified sources using the following criteria:

\begin{equation} \label{eq:cp}
\begin{split}
    [3.6] - [4.5] &< 0.65 \\
    \rm{and}& \\
    \alpha_{4.5, 24} &> 0.3
\end{split}
\end{equation}

We define CP as the sources that satisfy criteria for both the pre-ms stars with disks and for the protostars, thereby preventing a robust classification. We initially find 20 CP based on criteria in Equation \ref{eq:cp}. Further investigation of the mid-IR maps and SEDs show that four out of 20 are galaxies (Source\# 149381, 1154877, 1251530, and 1492533 in Table \ref{tab:aquila_phot_gal} in Appendix \ref{app:gal}). These four galaxies show a peak at $\sim$8 $\micron$ but are missed by our IRAC-color based criteria because of the incomplete IRAC detections. Another four sources are classified as protostars (\dummyACEBD, \dummyADBCE, \dummyACEDB, and \dummyADBEC) due to their rising SED in the far-IR wavelengths, scattered emission in the mid-IR maps, and/or the flux ratio plots described in Sec.~\ref{sec:step4}.

\begin{figure}[ht!]
\includegraphics[scale=0.3]{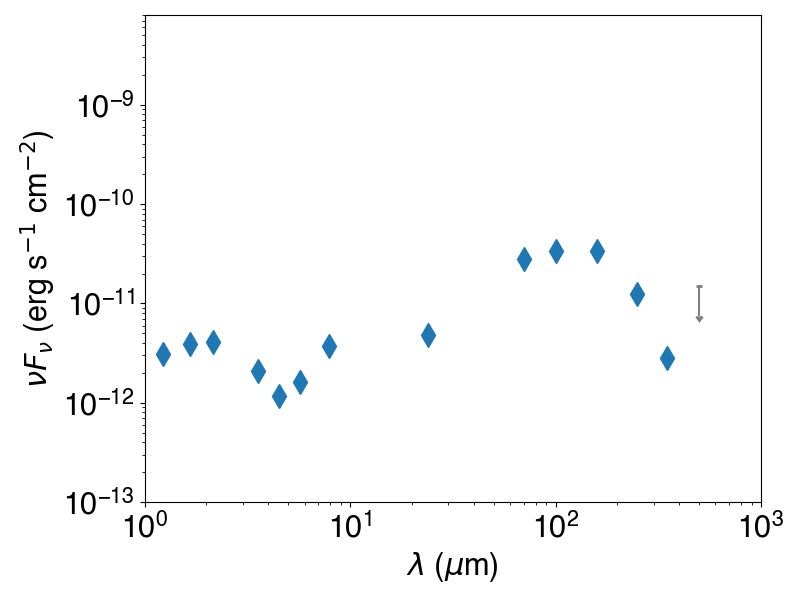}
\includegraphics[scale=0.3]{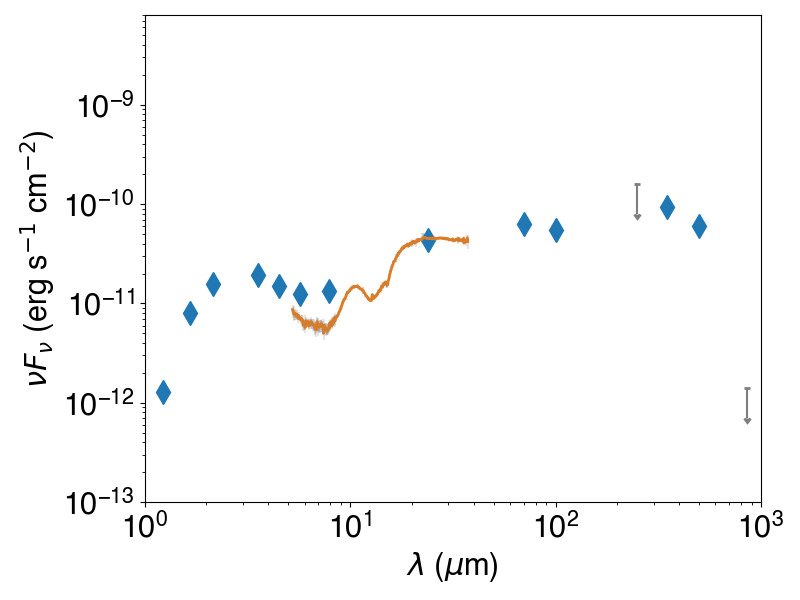}
\includegraphics[scale=0.3]{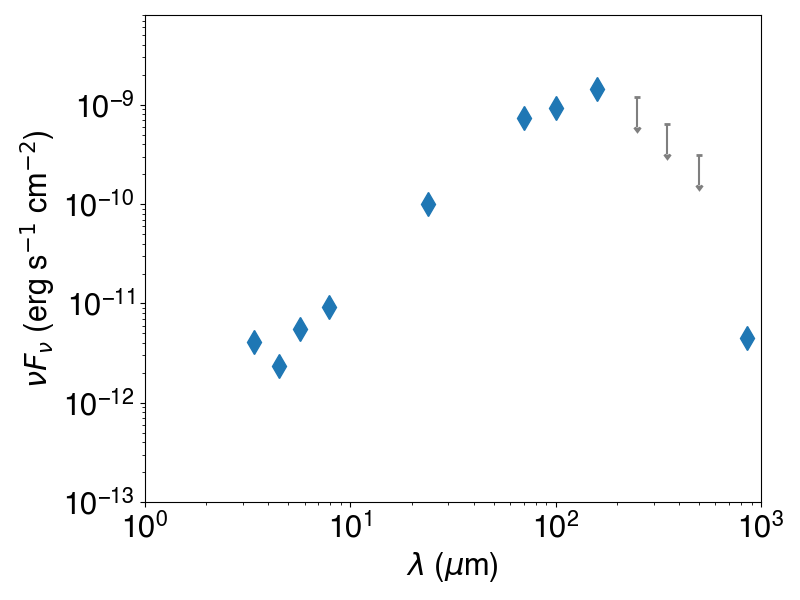}
\caption{Three examples of SEDs of candidate protostars (CP) in Aquila identified using the criteria defined in section \ref{sec:step3}. The SED in the left panel is a galaxy confirmed from near-IR morphology that escaped our selection criteria (\#78912 in Table \ref{tab:aquila_phot_gal}). The middle and right panels show SEDs of two protostars, \dummyACEBD, and \dummyACEDB, respectively. These are identified as part of Step 3.
\label{fig:sed_cp}}
\end{figure}

Figure \ref{fig:sed_cp} shows three example SEDs of CP. The SED in the first panel is of one of the four galaxies detected using scattered emissions in the near-IR maps. The remaining two are protostars, \dummyACEBD ~, and \dummyACEDB, respectively. At the end of this step, we have identified a total of 141 protostars, 104 galaxies, 64 pre-ms stars with disks, 27 reddened pre-ms stars with disks, and 12 CPs. There are still 144 unidentified sources which we will categorize in Sec.~\ref{sec:step4}.

\subsubsection{Step 4: Refinement based on color-color/color-magnitude plots} \label{sec:step4}

\begin{figure}[ht!]
\includegraphics[scale=0.44]{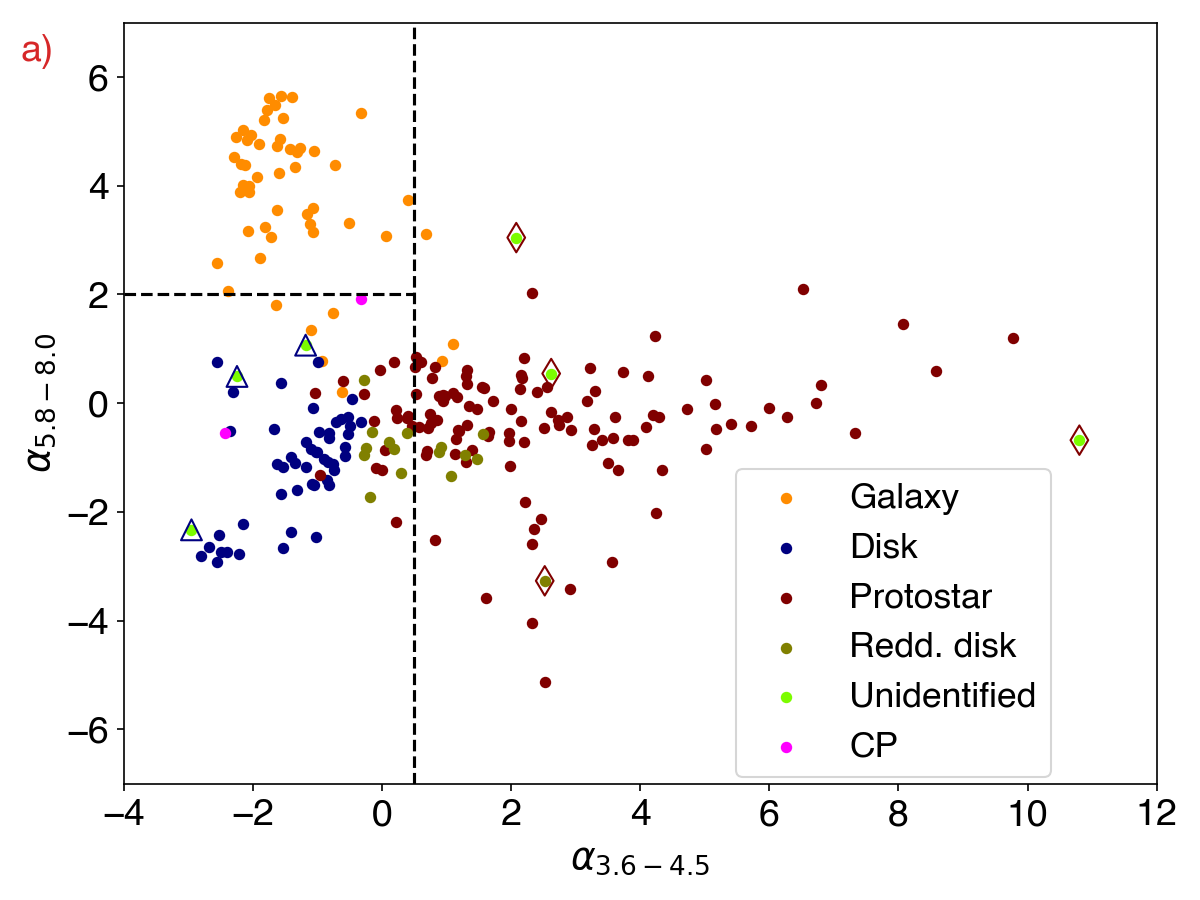}
\includegraphics[scale=0.44]{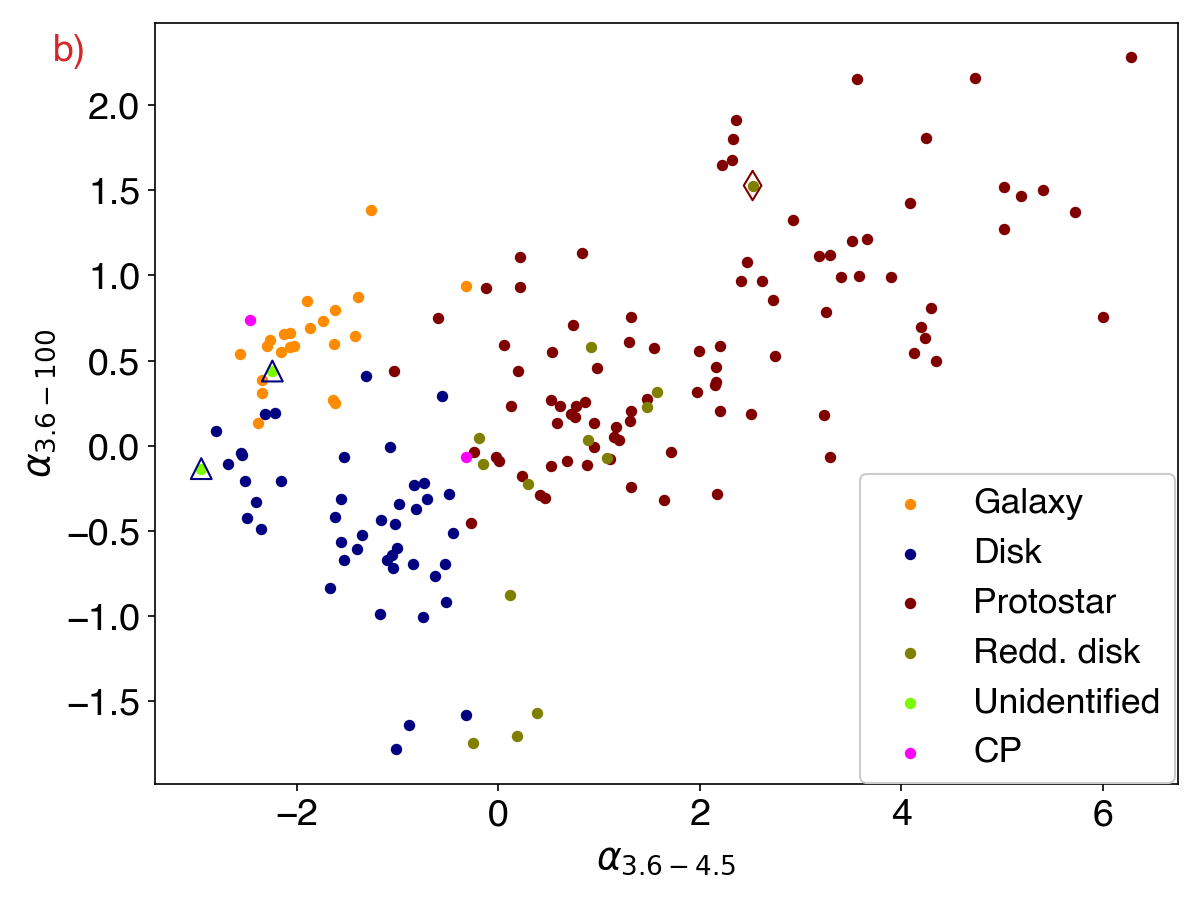}
\caption{a) Plot showing $\alpha_{5.8, 8.0}$ vs. $\alpha_{3.6, 4.5}$ and b) $\alpha_{3.6, 100}$ vs. $\alpha_{3.6, 4.5}$. Maroon diamonds and blue triangles denote reclassified protostars and pre-ms stars with disks, respectively. The horizontal and vertical dash lines in panel a denotes $\alpha_{5.8, 8.0} = 2$ and $\alpha_{3.6, 4.5} = 0.5$, respectively}. In the legend, CP stands for candidate protostars.
\label{fig:cc_cmd}
\end{figure}

In this step, we first assess and refine the classification of our sources using color-color and color-magnitude diagrams. In Figure \ref{fig:cc_cmd}a, we plot $\alpha_{5.8, 8.0}$ vs $\alpha_{3.6, 4.5}$ with the  classification from Steps 1-3. We see a cluster of star-forming galaxies in the upper left corner of the plot in the $\alpha_{5.8, 8.0} \geq 2$ and $\alpha_{3.6, 4.5} \leq 0.5$ region. Similarly, in their study of Orion protostars, \cite{Stutz13} previously reported the cluster of star-forming galaxies in the $\alpha_{5.8, 8.0} \geq 3$ and $\alpha_{3.6, 4.5} \leq 0.5$ region. The sources in the $\alpha_{5.8, 8.0} \leq 2$ and $\alpha_{3.6, 4.5} \leq 0.5$ region are mostly pre-ms stars with disks. We find most protostars in the $\alpha_{3.6, 4.5}$ $>$ 0.5 region. We scrutinize the interlopers in the vicinity of these clusters and investigate their SEDs to find if they have been misclassified. We find that three unidentified sources (\dummyADCBE, \dummyADCEB, and \dummyADEBC) have no MIPS photometry at 24 $\micron$, either due to saturation or confusion with a nearby brighter source. These sources, however, have an increasing mid-IR SED slope with the peaks of their emission in the far-IR wavelengths. Since our classification criteria require a detection at 24 $\micron$, these three sources are not yet classified as protostars. With the help of IRAC color criteria in Figure \ref{fig:cc_cmd}a, we classify them as protostars. Another source, \dummyADECB, has $\alpha_{4.5, 24} < -0.3$ so is not identified as a protostar but instead a reddened pre-ms star with disk. However, \dummyADECB ~has $\alpha_{3.6, 4.5} > 0.5$, shows scattered light nebulae in the mid-IR maps, and has an increasing SED slope in the far-IR wavelengths suggesting its protostellar nature. We reclassify \dummyADECB ~as a protostar. On the other hand, three unidentified sources at $\alpha_{3.6, 4.5} < 0.5$ region have declining mid-IR SED slopes, but due to missing 24 $\micron$ photometry, were not identified before as pre-ms stars with disks. We reclassify them as pre-ms stars with disks (Source\# 135046, 1230147, and 1230429 in Table \ref{tab:aquila_phot_disk} in Appendix \ref{app:disk}). 

Figure \ref{fig:cc_cmd}b shows the color-color diagram of $\alpha_{3.6, 100}$ versus $\alpha_{3.6, 4.5}$. The plot is similar to Figure 3 in \cite{Stutz13}, but instead of 160 $\micron$ we use 100 $\micron$ due to its higher angular resolution and lower contamination from extended emission. Three clusters of sources, corresponding to galaxies, protostars, and pre-ms stars with disks are again apparent. Again, we assess the individual SEDs and reclassify the sources that have been misclassified or unidentified. We identify two additional pre-ms stars with disks in this diagram (sources \#1230147 and 135046 in Table \ref{tab:aquila_phot_disk}).

In the next step, we search for the reddest protostars.
The most recently formed protostars are deeply buried in a dense envelope and can be undetected in some, or even all, of the Spitzer bands at $\le 24$~$\micron$. In many cases, they are detected only as extended emissions from scattered light nebulae or outflow jets instead of point sources in the 3.6 and 4.5~$\micron$ IRAC bands, and consequently, do not have measured magnitudes in the Spitzer source catalogs \citep{Stutz13}. In some cases, they are not detected in the IRAC bands or even at 24~$\micron$ \citep{Stutz13}.  Using Herschel data, \cite{Stutz13} identified extremely red protostars that were not previously identified in the Spitzer data.  Many of these are bright in the Herschel PACS wavebands and were classified as PACS Bright Red Sources (PBRS) by \citet{Stutz13}; the PBRS include the youngest protostars known in Orion \citep{Karnath20} (see Sec.~\ref{sec:pbr}). Since the selection criteria in steps 1-3 rely on the Spitzer mid-IR photometry, we need additional criteria for identifying these deeply embedded protostars, as well as any other protostars that have been missed by Spitzer. 

We adopt the following components of the technique described in \cite{Stutz13} to identify the protostars with the Herschel data. First, they included sources that were not detected at the shorter Spitzer wavelengths such as 3.6 or 4.5 $\micron$. We start by requiring a non-detection in at least one of these two bands; sources with detections in these bands would have already been identified in Steps 1 and 3. Second, they eliminated sources with Spitzer IRAC detections that had colors similar to galaxies. We do this in Step 2 and the first part of Step 4. Third, they eliminated the requirement of \citet{Megeath12} that the Spitzer 24 $\micron$ for protostars be brighter than 7 magnitudes; this requirement was adopted to reduce extragalactic contamination \citep{Kryukova12}. We never adopted a limiting magnitude at 24~$\mu$m and have identified galaxies through other means.  
Finally, they required that the SED slope should be increasing between Spitzer 24 $\micron$ and PACS 70 $\micron$. 
Since the eHOPS sample has more sensitive 100 $\micron$ photometry, we require a rising slope between Spitzer 24 $\micron$ and PACS 70 $\micron$ or Spitzer 24 $\micron$ and PACS 100 $\micron$.  We also use the PACS 100 $\micron$ flux limit below which we detect galaxies, F$_{100}^{\rm{gal}}$. We adopt F$_{100}^{\rm{gal}}$ as the 95$^{\rm{th}}$ percentile of the 100 $\micron$ flux for the 104 identified galaxies to assure that our estimate of F$_{100}^{\rm{gal}}$ is not biased by outliers.
Thus, for a Spitzer unidentified source to be identified as a protostar, the source should have

\begin{equation} \label{eq:redps}
\begin{split}
   \rm{No~detection~at~3.6}&{\rm~and/or~4.5~\micron} \\
    \rm{and}&\\
    F_{100} > &~F_{100}^{\rm{gal}}\\
     \rm{and}&\\
     \nu F_{\nu}(70)/\nu F_{\nu}(24) > 1~\rm{or}&~\nu F_{\nu}(100)/\nu F_{\nu}(24) > 1
\end{split}
\end{equation}

\noindent
For Aquila, we find F$_{100}^{\rm{gal}}$ = 0.6 Jy. We detect eight protostars using Equation \ref{eq:redps}-- \dummyAEDBC, \dummyBADCE, \dummyAECDB, \dummyAECBD, \dummyBACED, \dummyAEBCD, \dummyBACDE, and \dummyAEDCB. We defer the inclusion of sources without detections at 24  $\micron$ to Step 5.

\begin{figure}[ht!]
\includegraphics[scale=0.44]{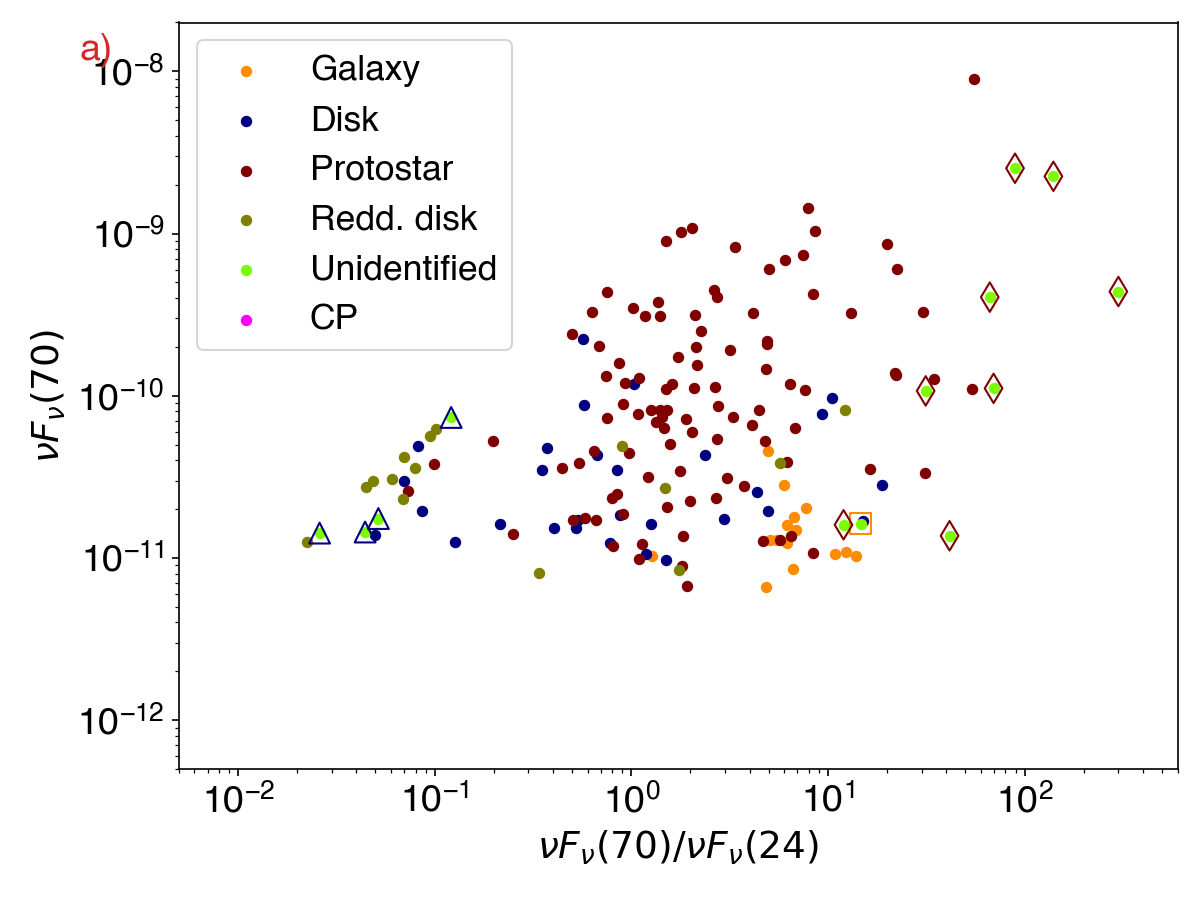}
\includegraphics[scale=0.44]{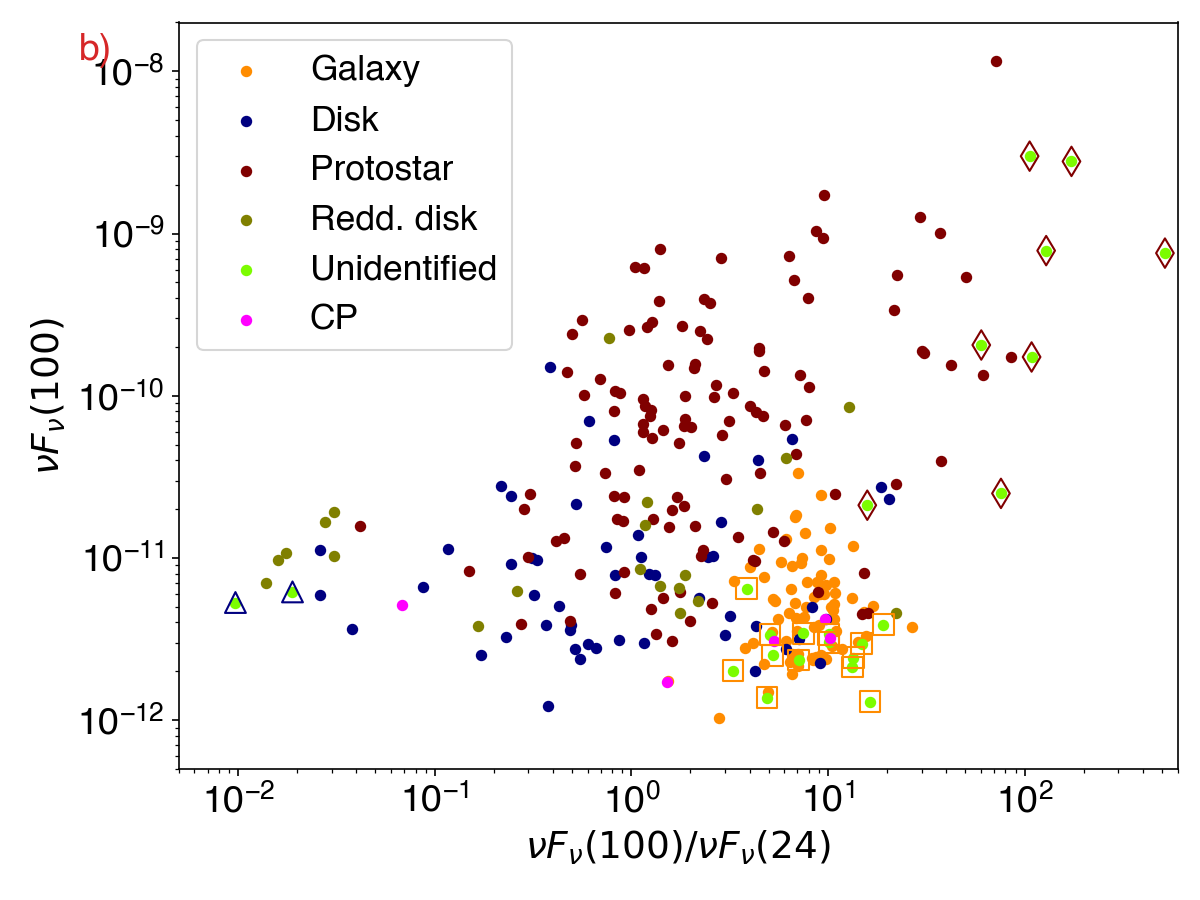}
\caption{Panel a) Variation of $\nu F_{\nu}(70)$ with $\nu F_{\nu}(70)/\nu F_{\nu}(24)$. Panel b) variation of $\nu F_{\nu}(100)$ with $\nu F_{\nu}(100)/\nu F_{\nu}(24)$. Orange squares denote newly identified galaxies, maroon diamonds denote newly identified protostars, and blue triangles are newly identified pre-ms stars with disks. 
\label{fig:cc_cmd1}}
\end{figure}

Figure \ref{fig:cc_cmd1}a shows the "color-magnitude" plot of $\nu F_{\nu}(70)$ with $\nu F_{\nu}(70)$/$\nu F_{\nu}(24)$. The different categories of sources are colored differently as shown in the legend. The eight newly detected protostars using Equation \ref{eq:redps} are shown by maroon diamonds. These eight sources have incomplete Spitzer/IRAC photometry due to their deeply embedded nature but the SEDs peak in the Herschel/PACS far-IR wavelengths. Additionally, we use the $\nu F_{\nu}(70)$ versus $\nu F_{\nu}(70)$/$\nu F_{\nu}(24)$ ratio plot to detect pre-ms stars with disks that have steeply declining mid-to-far IR slopes but were not previously detected due to sparse IRAC photometry. We find four pre-ms stars with disks in the $\nu F_{\nu}(70)$/$\nu F_{\nu}(24)$ $<$ 0.1 region (Source\# 267898, 290129, 409790, and 1303802 in Table \ref{tab:aquila_phot_disk} of Appendix \ref{app:disk}).

The eight new protostars using Equation \ref{eq:redps} are also shown in Figure~\ref{fig:cc_cmd1}b. 
An advantage of the $\nu F_{\nu}(100)$ versus $\nu F_{\nu}(100)$/$\nu F_{\nu}(24)$ plot is its ability to segregate faint galaxies due to the higher sensitivity at PACS 100 $\micron$ in our data. In Figure \ref{fig:cc_cmd1}b, a cluster of galaxies are apparent at a region enclosed by $\nu F_{\nu}(100)/\nu F_{\nu}(24) \gtrsim 3$ and $\nu F_{\nu}(100) \lesssim 2 \times 10^{-11}$ erg s$^{-1}$ cm$^{-2}$. We find 13 additional galaxies (Source\# 696942, 739146, 918673, 980308, 981917, 1116822, 1117119, 1155813, 1291988, 1419621, 1450123, 1524950, 2059229 and 2063039 in Table \ref{tab:aquila_phot_gal} in Appendix \ref{app:gal}). 

At the end of ``Step 4", we identify a total of 153 protostars, 118 galaxies, 71 pre-ms stars with disks, 26 reddened pre-ms stars with disks, and 12 CP. There are still 115 sources that need to be classified. 

\subsubsection{Step 5: Classification of Sources with Incomplete Photometry} \label{sec:step5}

The classification criteria defined so far are based on the mid-IR Spitzer and Herschel/PACS photometry with ${\rm SNR} > 5$. There are, however, sources that may have a lower SNR in one of the wavebands of Spitzer or Herschel/PACS or may be missing a detection in one of the 3.6, 4.5, or 24 $\micron$ bands, for example, due to either lower sensitivity (for a faint source) or saturation (for a bright source). Our criteria will fail to catch such sources and we may miss some protostars, except in the case of the most deeply embedded protostars with bright PACS emission (Sec.~\ref{sec:step4}).

If a source is not detected at a  wavelength required for classification, we use the corresponding completeness limit at that wavelength. As an example, if a source is not detected at 4.5 $\micron$ then $\alpha_{4.5, 24}$ cannot be calculated, missing out on a crucial component for the source classification. In that case, we use the completeness limit as upper limit of the 4.5 $\micron$ maps at the source position to compute $\alpha_{4.5, 24}$. Hence, in the case of pre-ms stars with disks,

\begin{equation} \label{eq:disklim0}
\begin{split}
    \{[3.6] - [4.5]\}^{\rm{lim}} &< 0.65 \\
    \rm{and}& \\
    \alpha_{4.5, 24}^{\rm{lim}} &< 0.3
\end{split}
\end{equation}

\begin{figure}[ht!]
\includegraphics[scale=0.31]{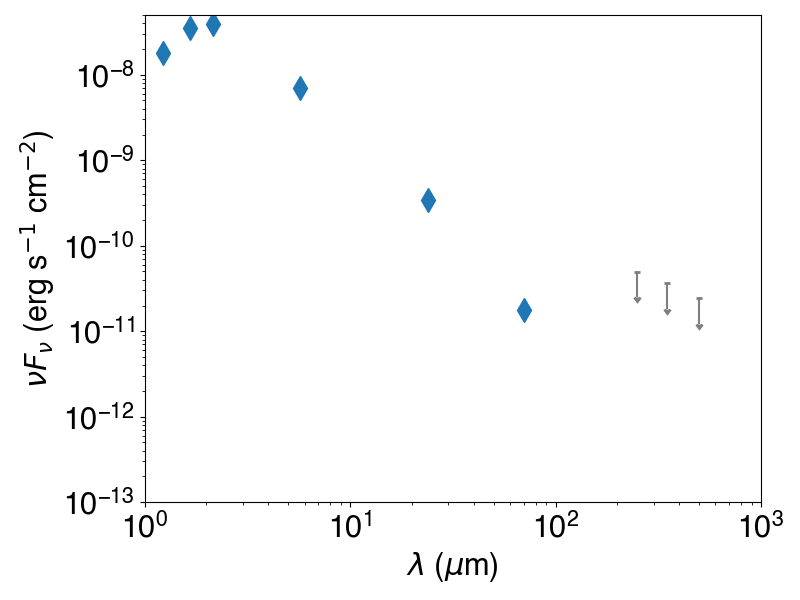}
\includegraphics[scale=0.31]{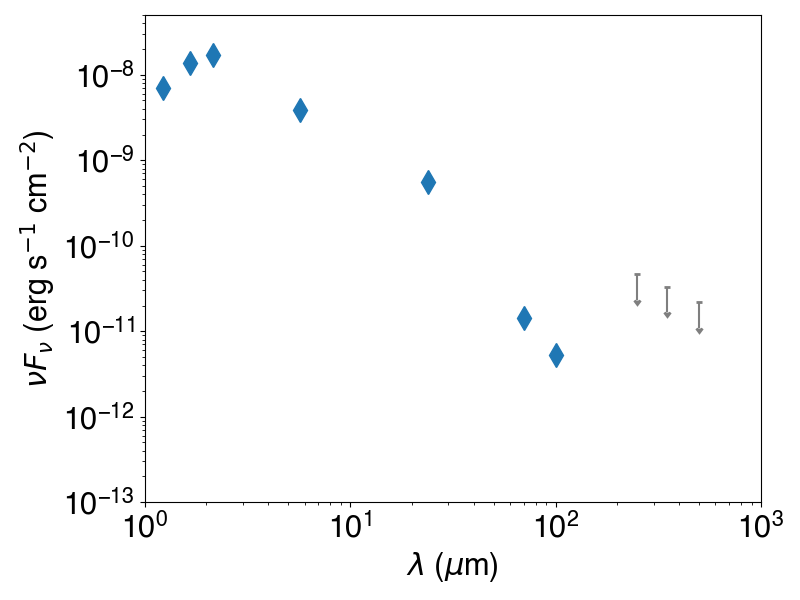}
\includegraphics[scale=0.31]{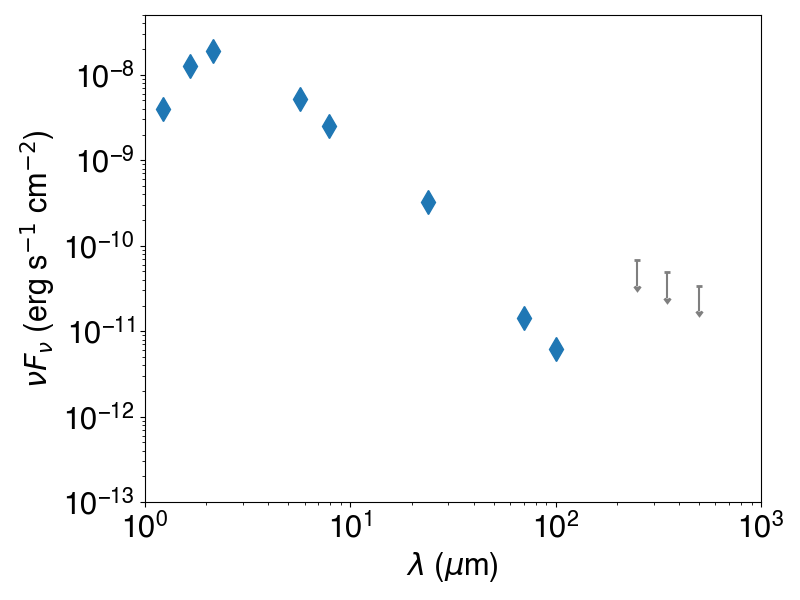}
\caption{Examples of SEDs of pre-ms stars with disks that show steeply falling mid-to-far IR emission but are not identified on the basis of the mid-IR photometry due to missing photometry. The sources are \#290129, \#409790, and \#1303802 from left to right (c.f. Table \ref{tab:aquila_phot_disk}). These are identified using the completeness limits in Step 5.
\label{fig:sed_disklim}}
\end{figure}

\noindent
Note that from Equations \ref{eq:disklim0} to \ref{eq:ps_lim}, the superscript ``lim" on middle brackets denotes that a completeness limit is used in any one of the two photometric points inside the middle brackets, but not on both. The calculation is meaningless if upper limits on both photometric points are used simultaneously. Furthermore, there are pre-ms stars with disks with steeply falling $\alpha_{4.5, 24}^{\rm{lim}}$ and also rapidly falling IR emissions from 24 to 160 $\micron$. Some of those sources are not detected in more than one wavelength among 3.6, 4.5, and 24 $\micron$, and hence they cannot be identified using the above conditions. We define additional criteria to identify them using the completeness limits for the far-IR wavelengths. For the bright pre-ms stars with disks that are saturated in the Spitzer/IRAC wavelengths, a steeply declining SED slope has ratios of PACS to MIPS fluxes less than one. For any two out of three PACS wavelengths, $\nu F_{\nu}(\rm{PACS})/\nu F_{\nu}(24)^{\rm{lim}} < 1$. Figure \ref{fig:sed_disklim} shows a few examples of pre-ms stars with disks that are identified using the following condition:

\begin{equation} \label{eq:disk_lim}
\begin{split}
\{\nu F_{\nu}(\rm{PACS})/\nu F_{\nu}(24)\}^{\rm{lim}} < 1 & \rm{~for~at~least~two~PACS~bands}.
\end{split}
\end{equation}

\noindent
Similar to the pre-ms stars with disks, we identify other candidate protostars using the completeness limits for missing photometry. For the sources where one of the required Spitzer photometry for classifying a CP is not available, we again use the rising far-IR flux ratio as a proxy for a candidate protostar. Figure \ref{fig:sed_cp} shows a few examples of such CPs that are identified using the following conditions invoking completeness limits:

\begin{equation} \label{eq:cp_lim}
\begin{split}
    \{[3.6] - [4.5]\}^{\rm{lim}} < 0.65 &\rm{~and~} \alpha_{4.5, 24}^{lim} > -0.3\\
    &\rm{~~or} \\
    \{[3.6] - [4.5]\}^{\rm{lim}} < 0.65 &\rm{~and~} \{\nu F_{\nu}(\rm{PACS})/\nu F_{\nu}(24)\}^{lim} > 1.
\end{split}
\end{equation}

In Equation \ref{eq:cp_lim}, the condition $\nu F_{\nu}(\rm{PACS})/\nu F_{\nu}(24)^{lim} > 1$ must hold true for at least one PACS wavelengths.
CPs may be protostars, but they may also be non-protostellar sources such as galaxies that were not removed by our filtering process. 

In Sec.~\ref{sec:step4}, we discuss the deeply embedded protostars with red SEDs in the far-IR wavelengths that do not show emission in shorter Spitzer wavelengths such as 3.6 and/or 4.5 $\micron$. Using Equation \ref{eq:redps}, we find eight such protostars but the criteria fail to identify the sources that are not detected at 24 $\micron$. To find such sources, we use the 24 $\micron$ flux completeness limit and require $\{\nu F_{\nu}(70)/\nu F_{\nu}(24)\}^{\rm{lim}} > 1$ or $\{\nu F_{\nu}(100)/\nu F_{\nu}(24)\}^{\rm{lim}} > 1$ in Equation \ref{eq:redps}. We identify six additional protostars using the completeness limit as the upper limit on the 24 $\micron$ photometry-- \dummyBEDCA, \dummyBDEAC, \dummyBDCAE, \dummyBECDA, \dummyBDCEA, and \dummyBEDAC.

For the remaining sources, we further update the selection criteria for protostars from section \ref{sec:step3} by including their completeness limits:

\begin{equation} \label{eq:ps_lim}
\begin{split}
    \{[3.6] - [4.5]\}^{\rm{lim}} > 0.65 &\rm{~and~} \alpha_{4.5, 24}^{lim} > -0.3 \\
    &\rm{~~or} \\
    \{\nu F_{\nu}(\rm{PACS})/\nu F_{\nu}(24)\}^{lim} > 1 & \rm{~for~at~least~two~PACS~bands}.
\end{split}
\end{equation}



\begin{figure}[ht!]
\centering
\includegraphics[scale=0.5]{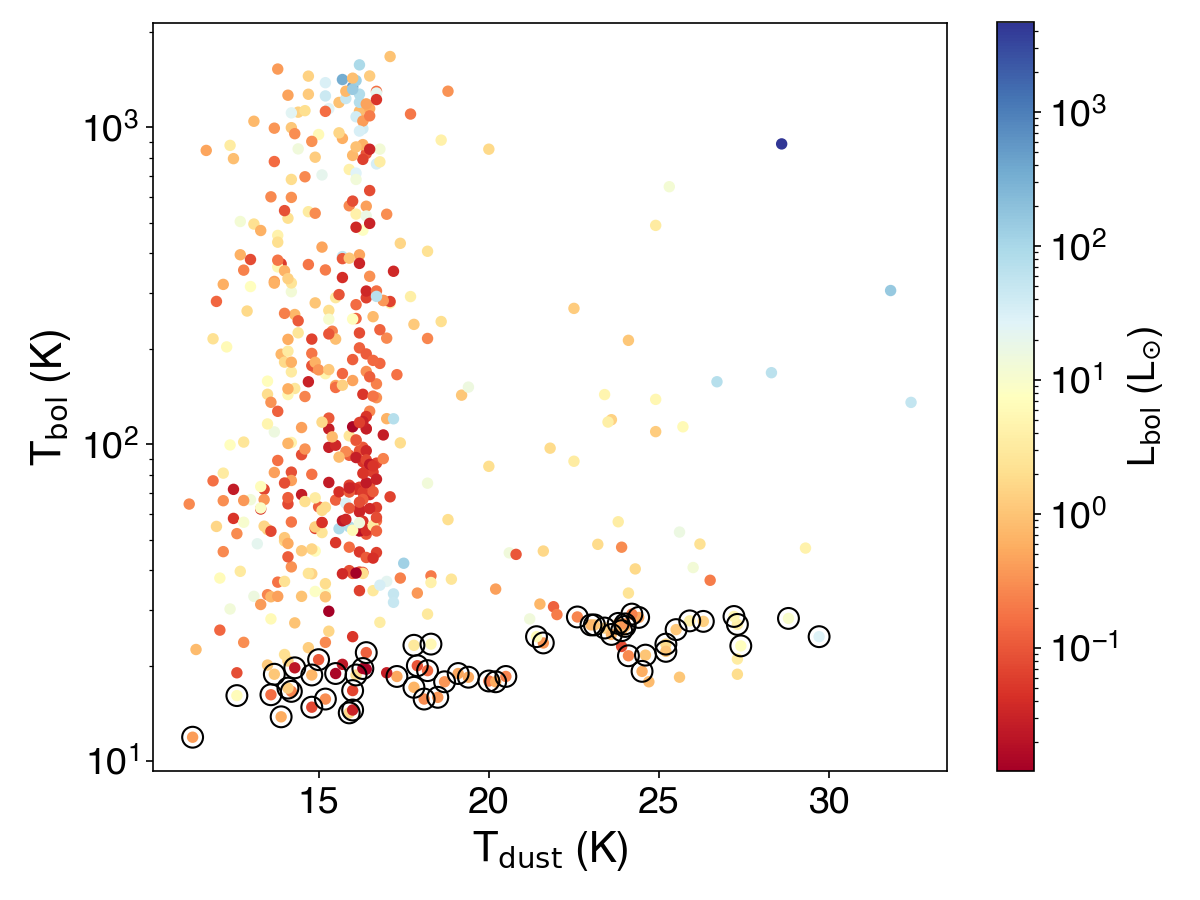}
\caption{Bolometric temperatures (\Tbol) for SEDs of sources vs. Herschel-derived dust temperature (T$_{\rm{dust}}$) at their position (c.f. Figure \ref{fig:aquila_NH2_T}). This diagram is used to identify starless cores in our sample. Sources in empty black circles are starless cores. This is analysis is part of Step 5.
\label{fig:stc}}
\end{figure}

The emission in the far-IR wavelength region can be used to detect and identify very young protostars with weak or no mid-IR emission such as the PBRS. But the selection criteria purely based on far-IR emission and completeness limits are also prone to select starless cores or other over-densities that do not harbor protostars. These are mostly the sources that show no emission at $\leq$100 $\micron$. Due to the absence of an internal heating source, the bolometric temperature of these sources is similar to the surrounding dust temperature. We identify such sources by calculating the difference between T$_{\rm{bol}}$ (calculated using the SED) and T$_{\rm{dust}}$ (obtained from Herschel observations, Figure \ref{fig:aquila_temp}). For any source, if the difference is less than 5 K (typical uncertainty in T$_{\rm{dust}}$) and there is no emission at $\leq$100 $\micron$, we classify them as starless cores.

We find a total of 19 protostars, one galaxy, and five pre-ms stars with disks using the completeness limits in Step 5. At the end of Step 5, there are still five sources that are not yet classified using our classification methods. These sources have sparsely sampled photometry and cannot be robustly classified. From the remaining unidentified sources, we remove the ones that do not have a detection in the 24-100 $\micron$ range as they are unlikely to be protostars. The sources that do have at least one detection in the 24-100 $\micron$ region are classified as candidate protostars and individually examined for signs of protostar such as scattered light nebulae in the mid-IR region; however, no additional protostars are identified.

After following the steps from Sec.~\ref{sec:step1} to \ref{sec:step5}, we find a total of 172 protostars (Table \ref{tab:aquila_phot}), 73 pre-ms stars with disks (Table \ref{tab:aquila_phot_disk}), 24 reddened pre-ms stars with disks (Table \ref{tab:aquila_phot_disk}), 118 galaxies (Table \ref{tab:aquila_phot_gal}), and 12 candidate protostars (Table \ref{tab:aquila_phot_cp}).

\subsection{AGB Contamination} \label{sec:agb}

Besides extragalactic contamination, another prominent source of contamination is background stars with infrared excesses, such as Asymptotic Giant Branch (AGB) stars. Previous studies (\citealt{Cieza10, Romero12, Dunham15}) show that AGB contamination is lower for the clouds that are farther away from the Galactic plane, such as Orion A ($b$=$\sim$-20$^{\circ}$), and higher for the clouds that are closer to the Galactic plane such as Aquila ($b$=2-6$^{\circ}$). For example, using optical spectroscopy \cite{Romero12} found an AGB contamination rate of $>40\%$ in Class III sources in Lupus V and VI ($b$=6$^{\circ}$). The rate decreases to $\sim20-25\%$ for Ophiuchus at $b$=$\sim$16$^{\circ}$. However, such studies also show that AGB contamination is a significant issue only for more evolved YSOs. \cite{Dunham15} found that the AGB contamination is almost negligible as the spectral index decreases until the Class II/III boundary, after which the AGB contamination rate steeply rises to $>50\%$ for Class IIIs in their sample. In the Aquila Main region, \cite{Oliveira09} reported $\sim 62\%$ of AGB contamination in Class III sources, with a sharp decrease to $\sim 5\%$ for Class II.
As the focus of this work is to produce a protostellar catalog and the envelope properties using protostellar models, AGB contamination is negligible in our protostellar catalog. However, AGB contamination can be significant for the pre-ms stars with disk sample that we present in Table \ref{tab:aquila_phot_disk}, especially for the sources that have a high L$_{\rm{bol}}$ and low spectral index \citep{Dunham15}.

\subsection{Comparison to previous catalogs} \label{sec:compare}

Previous studies identifying protostars in Aquila primarily used mid-IR photometry with Spitzer, often with a combination of near-IR or X-ray data \citep{Winston07, Gutermuth09, Enoch09, Kryukova12, Dunham15, Sun22}. Recent ALMA observations have also provided mm/sub-mm sources (for example \citealt{Plunkett18}) and their properties such as protostellar disk mass (for example \citealt{Anderson22}) that have enriched the protostar catalog in Aquila. In this section, we present a detailed comparison between the protostellar catalog that we identify in this study and five other widely used catalogs in the literature. The catalogs are from \cite{Winston07, Gutermuth09, Enoch09, Kryukova12} and \cite{Dunham15}. Table \ref{tab:allPrevious} lists whether a particular eHOPS source in Aquila is detected (``Y") or not (``N") in these catalogs. We use a matching tolerance of 2$\arcsec$ to compare the positions of eHOPS protostars to the catalogs in literature.

\cite{Winston07} use Spitzer and Chandra observations to find YSOs in the Serpens Main region (c.f. Figure~\ref{fig:aquila_NH2}). They use Spitzer/IRAC and Spitzer/MIPS to detect thermal emission from circumstellar disks and envelopes to classify YSOs using color-color diagrams and SEDs. They also use Chandra X-ray observations to study the effects of circumstellar disks on stellar X-ray properties and to identify young stars without disks or envelopes. Table 4 in \cite{Winston07} lists their identified 137 YSOs, out of which 37 sources are either Class 0/I or flat-spectrum sources. We detect 17 out of 37 sources in the eHOPS catalog. Out of the remaining 20, 19 do not have PACS detection and are excluded by our selection criteria. We inspected the SEDs and postages in the 1-850 $\micron$ images and found that most of these 19 sources are reddened pre-ms stars with disks with no significant emission (SNR $>$ 5) in the $>$24 $\micron$ wavelength bands. A source, S-ID 28 in Table 4 of \cite{Winston07} is confused with the neighboring protostar but appears to be a reddened pre-ms stars with disk. The final remaining source that does have PACS detection is classified as a pre-ms star with disk in the eHOPS catalog. 

Based on the mid-infrared imaging and photometric survey with 2MASS and Spitzer (1-24 $\micron$), \cite{Gutermuth09} surveyed two regions in Aquila. These regions include star-forming clusters in Serpens Main and MWC 297 (c.f., Figure \ref{fig:aquila_NH2}). In the two regions, \cite{Gutermuth09} report 27 protostars out of which 17 are identified in the eHOPS survey. The remaining 10 sources do not have a PACS detection and are not included by the eHOPs selection criteria. We inspected these 10 sources and found mostly reddened pre-ms stars with disks with no emission in the $>$ 24 $\micron$ wavelengths. 

\cite{Enoch09} present a catalog of protostars in Ophiuchus, Perseus, and Serpens region of Aquila-North by combining large-scale 1.1 mm Bolocam continuum and Spitzer Legacy surveys. They identify protostars based on their mid-IR properties at the positions predetermined by the 1.1 mm core positions. They determine the luminosities and \Tbol for these protostars using low angular resolution Spitzer data for the 70 and 160~$\mu$m fluxes. Thus there are differences in both the sample of identified protostars and the properties of those protostars.  In Serpens, they identify 34 protostars. 
Out of 34 protostars from \cite{Enoch09}, 29 are identified as protostars in our eHOPS survey. The median luminosity for the matched 29 protostars is 4.5 L$_{\odot}$ from both \cite{Enoch09} and from our eHOPS estimation. The median T$_{\rm{bol}}$ for the matched 29 sources is 110 K from \cite{Enoch09} and 81 K from our study. Out of the remaining five sources in \cite{Enoch09} that do not match with our catalog, four sources do not have detections in any of the PACS wavebands (Ser-emb 23, Ser-emb 24, Ser-emb 27, and Ser-emb 32 in \citealt{Enoch09}), and one source is reclassified as a reddened pre-ms stars with disk (Ser-emb 4). 

Another survey of protostars in the nearest kpc was by \cite{Kryukova12} using 2MASS and Spitzer in the 1-24 $\micron$ wavelengths. \cite{Kryukova12} reports 40 protostars in the two clusters in the Serpens Main region, out of which 29 match with the eHOPS catalog. The remaining 11 sources do not have PACS detection despite being included inside the PACS coverage therefore they get excluded from the eHOPS catalog.

\cite{Dunham15} provides a catalog of YSOs in 18 molecular clouds using the Spitzer ``Cores to Disks (C2D)" and ``Gould Belt" Legacy surveys. They compute the spectral index ($\alpha$) using a linear least-squares fit to all available 2MASS and Spitzer photometry between 2 and 24 $\micron$. In Aquila, the \cite{Dunham15} catalog contains 200 sources that have extinction corrected $\alpha$ $>$ -0.3, out of which 115 sources match our eHOPS survey catalog. \cite{Dunham15} used $d$ = 260 pc for Aquila and $d$ = 429 pc for the Serpens region. We correct the distance in calculating  L$_{\rm{bol}}$ for the 115 sources that are matched in both \cite{Dunham15} and our eHOPS catalog. We find the median luminosity for these 115 matched sources of 0.7  L$_{\odot}$ from Table 2 of \cite{Dunham15} and 1.8 L$_{\odot}$ from the eHOPS estimates. 
The median T$_{\rm{bol}}$ is 255 K and 119 K from  \cite{Dunham15} and our eHOPS estimates, respectively, for the matched 115 sources. Among the 85 unmatched sources between \cite{Dunham15} catalog and our eHOPS catalog, 75 sources do not have at least one detection in our PACS survey despite being included inside the PACS coverage regions. Three sources are outside the PACS coverage region. Among the remaining seven sources, five are reclassified as reddened pre-ms stars with disks, one as a pre-ms star with disk, and one is a candidate protostar.

\begin{deluxetable*}{ccccccc} 
\tablecaption{Source comparison between eHOPS catalog and other catalogs from the literature for Aquila clouds \label{tab:allPrevious}} 
\tablewidth{0pt} 
\tablehead{ 
\colhead{eHOPS source} & \colhead{Other names} & \colhead{W07} & \colhead{E09} & \colhead{G09} & \colhead{K12} & \colhead{D15} 
} 
\decimalcolnumbers 
\startdata 
eHOPS-aql-1 & 2MASS J18251133-0258532 & N & N & N & N & Y \\ 
eHOPS-aql-2 & 2MASS J18251332-0259549 & N & N & N & N & Y \\ 
eHOPS-aql-3 & --- & N & N & N & N & N \\ 
eHOPS-aql-4 & IRAS 18245-0342 & N & N & N & N & N \\ 
eHOPS-aql-5 & 2MASS J18275019-0349140 & N & N & Y & N & Y \\ 
\enddata 
\tablecomments{Full table is available in the online version.} 
\end{deluxetable*}

A detailed comparison between eHOPS catalog and each of the \cite{Winston07, Gutermuth09, Enoch09, Kryukova12} and \cite{Dunham15} are presented in Appendix \ref{app:prevCat} in Tables \ref{tab:winston07}, \ref{tab:gutermuth09}, \ref{tab:enoch09}, \ref{tab:kryukova12} and \ref{tab:dunham15}, respectively. If a certain source in one of the previous catalogs is missing in the eHOPS catalog, Tables \ref{tab:winston07} through \ref{tab:dunham15} also explains the reason behind its exclusion in column ``Comments". We find new protostars with Herschel that were not previously identified by Spitzer which shows that the inclusion of far-IR data is essential to making a robust protostar catalog.
We note that there are sources in the literature that are not identified by eHOPS because of their lacking $>$ 24 $\micron$ emission. Possible explanations that would cause the sources to be below the PACS detection limits include having a very low luminosity ($<$0.05 \Lsun), being a more evolved YSO such as a reddened pre-ms star with disk, or being a residual galaxy or AGN. Categorizing each such source that is not included in the eHOPS catalog is out of the scope of this paper.

\section{Results: The Spatial Distributions  of YSOs and Dense Gas} \label{sec:dist}

\begin{figure}[ht!]
\centering
\includegraphics[scale=0.8]{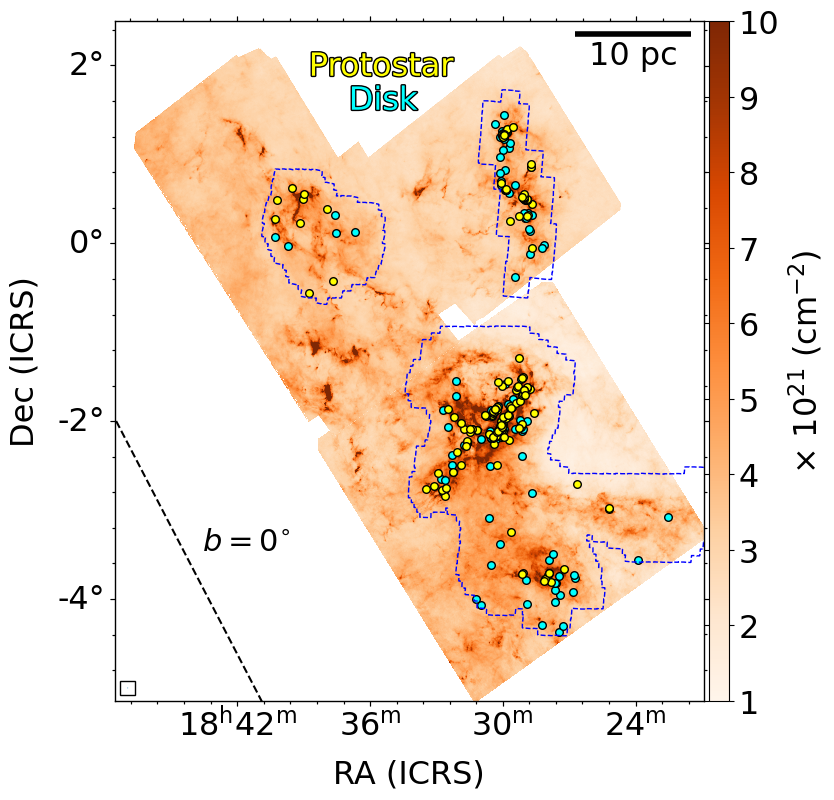}
\caption{Protostars (yellow circles) and pre-ms stars with disks (cyan circles) overplotted on the column density map of Aquila star-forming clouds. The central galactic latitude line ($b$=0$^{\circ}$) is shown for reference and blue dashed lines show the surveyed regions. The protostars are mostly concentrated towards the dense regions of the cloud. Blue dash lines denote the region mapped by both Spitzer and Herschel.}
\label{fig:aquila_map_ps}
\end{figure}

\begin{figure}[ht!]
\centering
\includegraphics[scale=0.8]{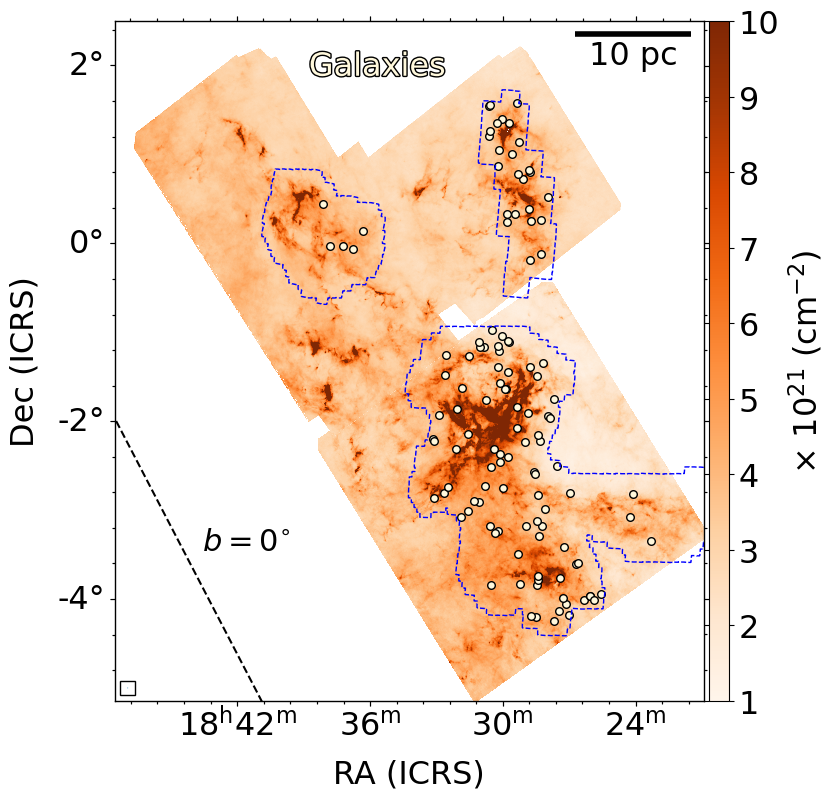}
\caption{Similar to Figure \ref{fig:aquila_map_ps} but galaxies are overplotted on the column density map. Unlike protostars, galaxies are mostly found in less dense regions.
\label{fig:aquila_map_gal}}
\end{figure}

In Figure \ref{fig:aquila_map_ps}, we display the spatial distribution of protostars and pre-ms stars with disks on the Herschel-derived column density map to show the spatial arrangement of the YSOs. Similarly, in Figure \ref{fig:aquila_map_gal}, we overplot the identified galaxies on the column density map. The regions with combined Spitzer and Herschel coverage are outlined to show the extent of our survey; in these regions we have the multi-band coverage needed to identify YSOs. The protostars and pre-ms stars with disks are both concentrated in the filamentary regions with high gas column density.  The fainter galaxies, in contrast, are preferentially found in regions of low gas column density where they are less obscured by dust extinction.   The density of galaxies increases in regions with 100~$\mu$m and Spitzer coverage due to the higher sensitivities of the 100~$\mu$m data compared to the PACS 70~$\mu$m maps (see Figure~\ref{fig:aquila_NH2} for the extent of 100~$\mu$m maps). Since the 100~$\mu$m fields target regions of high gas density, we also see a higher concentration of protostars in these fields. Yet, galaxies are spread throughout the observed  100~$\mu$m fields and not concentrated on the highest column density gas like the YSOs. 

\begin{figure}[t!]
\includegraphics[scale=0.6]{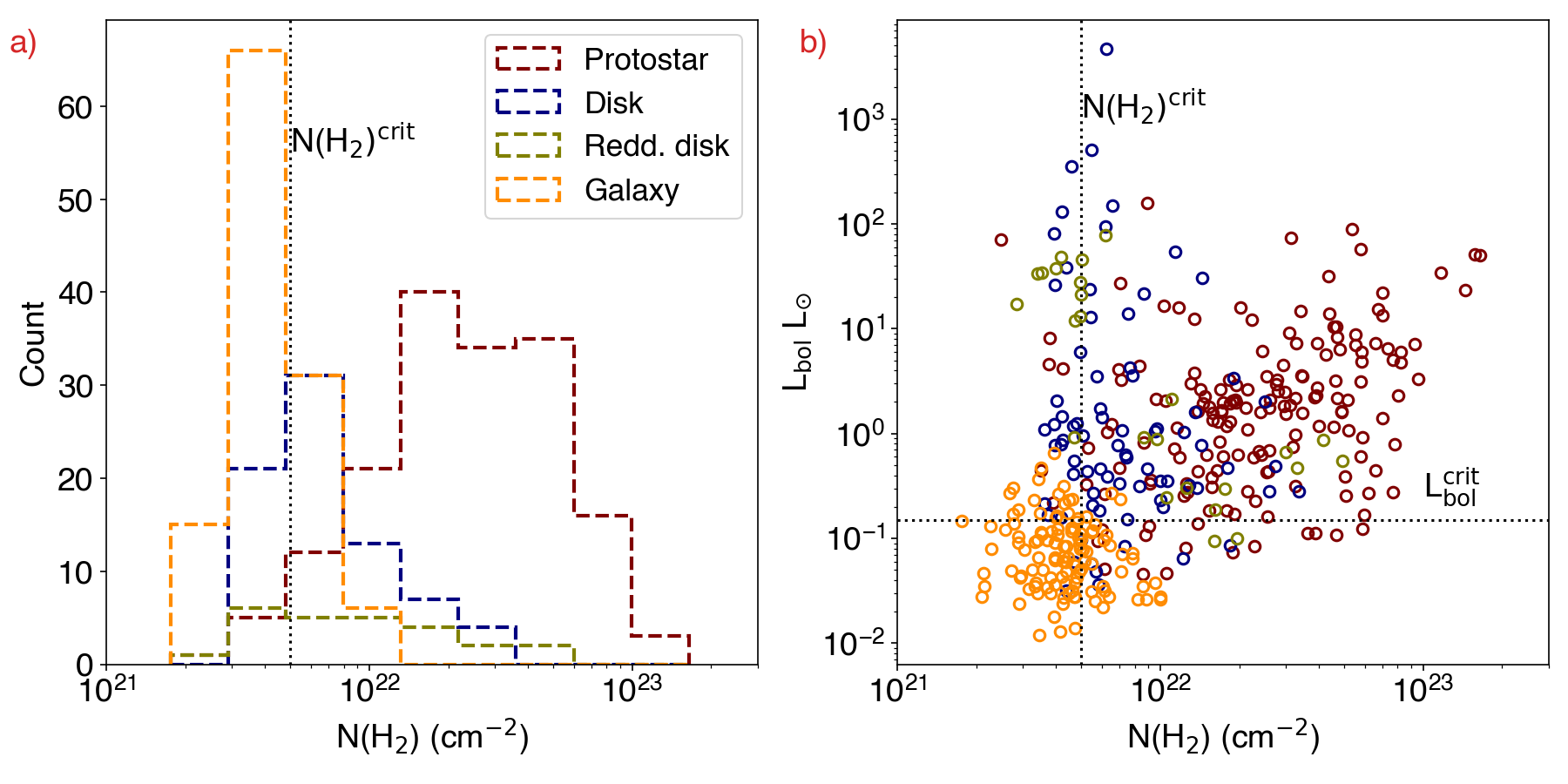}
\caption{Panel a): Distribution of projected column densities for protostars, pre-ms stars with disks, and galaxies in Aquila. Galaxies are mostly concentrated in lower N(H$_2$) and protostars in higher N(H$_2$) regions, respectively. Panel b): Variation of L$_{\rm{bol}}$ with N(H$_2$) for the same sources as in Panel a. Color labels in panel b are the same as in panel a. In addition to being found in lower density regions, galaxies also have lower \Lbol.
\label{fig:lbolNH2}}
\end{figure}

These spatial arrangements between sources and gas are illustrated in Figure \ref{fig:lbolNH2}. Figure \ref{fig:lbolNH2}a shows histograms of N(H$_2$)  around each type of source. The protostars are mostly in the high column density bins, consistent with Figure~\ref{fig:aquila_map_ps}. In contrast, as expected from Figure~\ref{fig:aquila_map_gal}, the galaxies are predominantly found in the low column density bins. The pre-ms stars with disks, which closely follow the distribution of protostars in Figure \ref{fig:aquila_map_ps}, are found at a broad range of column densities in Figure \ref{fig:lbolNH2}a. These pre-ms stars, as well as the more evolved protostars, have dispersed their natal gas and are mostly found in lower density regions. 
Figure \ref{fig:lbolNH2}b shows the projected N(H$_2$) vs L$_{\rm{bol}}$  for all of the identified sources. Galaxies are concentrated  in the L$_{\rm{bol}}$ $<$ L$_{\rm{bol}}^{\rm{crit}}$ and N(H$_2$) $<$ N(H$_2)^{\rm{crit}}$ region; this partly reflects the selection criteria for galaxies in Equation \ref{eq:gal}. Galaxies are randomly distributed in the low-density survey region. As our primary selection criterion is based on Herschel/PACS detection, extinction only plays a minor role in hiding galaxies behind dense gas as we impose the N(H$_2)^{\rm{crit}}$ based criterion only in the later steps. A few galaxies that are identified at higher column density regions (although still $<$ 10$^{22}$ cm$^{-2}$) may be the result of chance projection. Protostars dominate high N(H$_2)$ regions whereas pre-ms stars with disks cover a broad range of L$_{\rm{bol}}$ and N(H$_2)$. Some of the pre-ms stars with disks at high L$_{\rm{bol}}$ and low N(H$_2)$ may be other contaminants such as the AGB stars, as discussed in Sec.~\ref{sec:agb}.

We note a small number of galaxies concentrated in regions of high gas column density or protostars in regions of low gas column density that are suspicious and merits future investigation. We find 31 galaxies with projected N(H$_2$) $>$ N(H$_2)^{\rm{crit}}$. These galaxies are located in the 5 $\times$ 10$^{21}$ cm$^{-2}$ $<$ N(H$_2$) $<$ 1 $\times$ 10$^{22}$ cm$^{-2}$ region. Out of 31 galaxies, 19 are star-forming galaxies identified using bright PAH emission at Spitzer/IRAC 8 $\micron$, six are morphologically confirmed to be galaxies from near-to-mid IR high-resolution images, and the remaining 6 are low luminosity galaxies from extragalactic SED template fits. Thus, the eHOPS galaxies are robustly identified and there is a minimum chance, if any, of their misclassification. On the other hand, we find six protostars (\dummyBCEDA, \dummyBCAED, \dummyBCDAE, \dummyBCDEA, \dummyBAECD, and \dummyBCEAD) in the N(H$_2$) $<$ N(H$_2)^{\rm{crit}}$ region. These sources have increasing SED slope in the 1-24 $\micron$ wavelengths and thus are classified as protostars using Equation \ref{eq:ps1}. A peculiar feature of all these protostars is a declining SED at $>$ 24 $\micron$ wavelengths. This suggests the possibility that these sources are reddened pre-ms stars with disks, galaxies, or even planetary nebulae if there is a peak at $\sim$ 24 $\micron$ (see \citealt{Ueta06}). However, due to the lack of conclusive evidence to classify them otherwise, and an increasing mid-IR slope that satisfies Equation \ref{eq:ps1}, we classify them as protostars in our catalog. 

\section{Results: Protostellar SEDs}
\label{sec:modeling}

In this section, we focus on characterizing the protostars in the eHOPS-Aquila catalog by their SEDs.  We start by classifying protostars based on their spectral index and \Tbol. We follow the empirical SED classification system adopted by \citetalias{Furlan16} that is based on $\alpha_{4.5, 24}$ and \Tbol\ to classify protostars into different evolutionary classes as follows:

\begin{itemize}
    \setlength\itemsep{-0.1cm}
    \item Class 0 $\longrightarrow$  $\alpha_{4.5, 24}$ $>$ 0.3 \& $\Tbol$ $<$ 70 K 
    \item Class I $\longrightarrow$  $\alpha_{4.5, 24}$ $>$ 0.3 \& $\Tbol$ $>$ 70 K 
    \item Flat-spectrum source $\longrightarrow$  -0.3 $<$ $\alpha_{4.5, 24}$ $<$ 0.3 
    \item Class II $\longrightarrow$ -1.6 $<$ $\alpha_{4.5, 24}$ $<$ -0.3 
    \item Class III $\longrightarrow$ $\alpha_{4.5, 24}$ $<$ -1.6
\end{itemize}

The spectral indices, \Lbol, \Tbol, and evolutionary classes are found in Table \ref{tab:aquila}. 
After classifying protostars, we fit their SEDs to a grid of radiative transfer models.  The best fit values for the primary parameters of the models are tabulated in Table~\ref{tab:aquila}. Specifically, we give the total luminosity (\Ltot), the centrifugal radius of the disk ($R_{\rm disk}$), the envelope density at 1000~au ($\rho_{1000}$), the mass of the envelope within 2500~au ($M_{\rm env}$), the outflow cavity half-angle ($\theta$), the inclination ($i$), and the amount of foreground reddening ($A_v$).  We also show the scale factor applied to the model SEDs and the goodness of fit parameter ($R$).  These are described in more detail below.

\subsection{A Grid of Radiative Transfer Models}

A physical model that includes all the crucial components of a protostellar system is required to infer the fundamental physical properties of protostars from their SEDs. We use the model grid that was developed by \citetalias{Furlan16} using the \cite{Whitney03a, Whitney03b} radiative transfer code after incorporating the improvements by \cite{Ali10} and \cite{Stutz13}. The \cite{Whitney03a, Whitney03b} model of a protostellar system consists of a central luminosity source, flared disk, bipolar outflow cavities, and a rotating, collapsing envelope. In these models, the density distribution of the disk follows power laws in both the radial and vertical directions. The envelopes are described by TSC models (\citealt{Terebey84}) where the density distribution corresponds to a rotating collapsing cloud core with a constant infall rate (also see \citealt{Ulrich76}). The outflow cavity in the envelope follows a polynomial shape. \citetalias{Furlan16} used the dust opacities from \cite{Ormel11}, which include scattering cross-sections that are required by the modified Henyey-Greenstein function in the \cite{Whitney03a} models. This opacity law was chosen since it reproduced the mid-IR extinction law measured in molecular clouds by \cite{McClure09}.   We adopt a gas-to-dust ratio of 100 in the models. 
The grid sampling and approach is carefully detailed in \citetalias{Furlan16}. Our models differ from the \cite{Robitaille06, Robitaille07} and \cite{Robitaille17} grids in the choices of grid sampling and the opacity law, which were optimized in our models for low to intermediate luminosity protostars in nearby clouds.

The \cite{Whitney03a} Monte Carlo radiative transfer code has many input parameters. These parameters are related to stellar properties (stellar mass, stellar temperature, stellar radius), disk properties (for example disk mass, disk outer radius, disk-to-star accretion rate), and envelope properties (for example envelope density, cavity opening angle). Other parameters of interest are intrinsic stellar luminosity, total (stellar+accretion) luminosity, and inclination angle. In the models, an inclination of 0 degrees is a face-on view of the protostellar system. Because of the degeneracies involved and computational complexity, \citetalias{Furlan16} fixed some of the parameters at a constant value and vary others that have the most impact on protostellar SEDs. For example, \citetalias{Furlan16} fixed the stellar mass at 0.5 M$_{\odot}$, stellar effective temperature at 4000 K, and envelope outer radius at 10000 au. The parameters that \citetalias{Furlan16} vary are stellar radius (which determines the intrinsic stellar luminosity), disk outer radius (which sets the flattening of the envelope in TSC models), disk-to-star accretion rate for different $R_{\rm{star}}$ (which determines the accretion luminosity and together with the intrinsic luminosity sets the total luminosity), envelope density at 1000 au ($\rho_{1000}$), cavity opening angle, and inclination angle. The specific values of the model parameters are included in Table 3 of \citetalias{Furlan16}, and the justification for the adopted initial parameter values are provided in Sec.~4.1 of \citetalias{Furlan16}. 

The model grid contains 3040 main models that are parameterized for eight total (intrinsic+accretion) luminosities, nineteen envelope densities (which for a given stellar mass, determines the infall rate), four disk radii, and five cavity opening angles. Each model is further calculated for 10 different inclination angles. In total, 30,400 different model SEDs fit the observed SEDs. In this section, we provide a brief discussion of the major components of the modeling. Readers should refer to \citetalias{Furlan16} for a more detailed description of the modeling procedure, including the choice of parameter space, degeneracies and biases in the fits, and uncertainties in model parameters.

\subsection{Modifications from the HOPS Grid }\label{sec:modification}

The model grids used by the HOPS collaboration \citepalias{Furlan16} were designed specifically for the protostars in Orion. Since we are using the same model grid to fit the protostars in other molecular clouds, it is important to modify the models in places where the parameters are specific to Orion, such as distance to the cloud. In this section, we explain the general changes that we apply to the models so that they can be used to fit the protostellar SEDs in Aquila.

The distance affects the observed fluxes calculated from the models, which in turn affects the scaling factors of the model SEDs used to determine the luminosities of the protostars. It also affects the physical sizes of the apertures. The models were produced for 24 different apertures, from 1\arcsec\ to 24\arcsec. At the distance of the Orion clouds, 420 pc, the apertures correspond to 420-10,080 au in steps of 420 au. The use of discrete aperture sizes in models also means that there may not be an accurate model flux that corresponds exactly to the aperture of the observed flux. For example, at the distance of Aquila, IRAC detector has a FWHM beam of 1046 au, while the model fluxes close to 1046 au are available for apertures of 840 and 1260 au. In such cases, we do a linear interpolation of the fluxes at 840 and 1260 au to approximate the model flux at 1046 au. The interpolation ensures a more accurate comparison of the observed and modeled fluxes. For IRS fluxes we interpolate to a 5.3$''$ aperture. Since most of the sources appear as point sources in the Spitzer and PACS 70 and 100 $\micron$ data, the aperture size typically does not have a significant effect on the flux. 

We have SPIRE and SCUBA-2 data points in our observed SEDs that were not included in the model grids in \citetalias{Furlan16}. Similarly, HOPS SEDs contained APEX/SABOCA and APEX/LABOCA observations that are not included in our eHOPS SEDs.  We compute the model fluxes for SPIRE and SCUBA-2 wavebands using their respective spectral response functions. For the sources in Aquila, SPIRE 350 and 500 $\micron$ observations correspond to $\sim$10,500 and $\sim$15,700 au, respectively. Both values are greater than the maximum aperture size in the model. The models assume a 10,000 au outer envelope radius, so there is no emission from the protostar at larger radii. However, the observation will capture the actual emission at radii larger than 10,000 au, which can be from a larger envelope than the model assumes or extended emission from dense gas structures in the Aquila cloud. For consistency between model and observed SEDs, we use the observed photometry at SPIRE 350 and 500 $\micron$ wavelengths as upper limits and treat them as such when fitting the observed SEDs with models. Furthermore, for sources with \Lbol~$<$ 0.1 \Lsun, the emission at the far-IR wavelengths is susceptible to contamination from outer envelopes. For such cases, we use the observed photometry for $\ge$160 $\micron$ as upper limits.

\subsection{Fitting technique} \label{sec:fittingTechnique}

We use a customized fitting routine initially developed by \cite{Ali10} and \citetalias{Furlan16} to fit the observed SEDs of protostars in Orion. Here we summarize the important aspects of the fitting technique and ask the readers to consult Sec.~5 of \citetalias{Furlan16} for more details. We use photometry for the entire SED: 2MASS, IRAC, MIPS, PACS, SPIRE, SCUBA-2, and the spectrum from Spitzer/IRS (where available) to fit with the model SEDs. For many sources, photometry covering the entire near-IR to sub-mm SED is available, but for some sources, there are upper limits in the SPIRE and SCUBA-2 fluxes.
Of our 172 modeled protostars in Aquila, 35 have IRS observations. We do not include any additional data from the literature to maintain uniformity in the data used in the fits.

For the sources with IRS data, the spectrum tends to dominate the fit because of the many data points in the spectrum. This causes the fit to be highly biased towards the mid-IR region when IRS data are present. Furthermore, there are ice absorption features in the IRS spectrum in the 5-8 $\micron$ region and at $\sim$15.2 $\micron$ which are not included in the model opacities. To mitigate these issues, we follow the approach of \citetalias{Furlan16} and rebin the IRS spectrum in 16 wavelengths that avoid the ice absorption region but trace silicate absorption features at $\sim$10 and 20 $\micron$. A source with a finely sampled SED can have up to 15 photometric data points from 1-850 $\micron$ and 16 spectroscopic data points from the IRS. In such cases, the mid-IR remains the most finely sampled region of the SEDs, and the IRS spectra strongly influence the resulting fit. This is particularly valuable in determining the extinction toward each protostar.

We use spectral response functions for each detector and wavelength to calculate model fluxes for the same aperture sizes as in the observations, ensuring a direct comparison between observed and model fluxes. The use of models to predict Herschel fluxes is described by \cite{Ali10}. For the model data corresponding to the IRS spectrum, the model fluxes are linearly interpolated to the same 16 wavelengths as in the observed IRS data. 

The fitting is done in two steps. First, we determine a scale factor and extinction. Our model grid contains only eight values for model luminosities and does not initially account for extinction. To account for variations in luminosity and extinction,  we scale the models by a scale factor, $s$, that usually ranges between 0.5 and 2, and apply our adopted extinction law. The modeled and observed SEDs are then related as

\begin{equation} \label{eq:fit}
    F_{\rm{obs}, \lambda} = sF_{\rm{mod}, \lambda}10^{-0.4A_{\lambda}},
\end{equation}

\noindent
where $A_{\lambda}$ is foreground extinction to the protostar at wavelength $\lambda$.  
We use the Herschel-derived column density maps, smoothed to the beam size of the SPIRE 500 $\micron$ maps,  to calculate the projected column density at the position of a protostar. We convert the column density to a total extinction through the cloud using the conversion factor of 1 $A_V$ = 10$^{21}$ cm$^{-2}$ \citep{Winston10, Pillitteri13}. Setting this Herschel-derived extinction as an upper limit, a least-squares fit  simultaneously determines $s$ and $A_{\lambda}$ using Equation \ref{eq:fit}. See \citetalias{Furlan16} for the details.

Second, after a best fit scale factor and the extinction are applied to each model, we assess which model best reproduces the observed SED.  The proximity of the model SED to the observed SED is measured using a goodness-of-fit parameter, $R$, similar to the method used for fitting the galaxy SEDs calculated using Equation \ref{eq:Rfactor} \citep[\citetalias{Furlan16}, ][]{Fischer14}.  The justification for using $R$ as the goodness-of-fit and the inverse of the approximate fractional uncertainty as weights for the photometric data points are given in Appendix \ref{app:extgalfit}. In short, $R$ is approximately equal to the distance of the model from the data in units of the fractional uncertainty. As a general rule of thumb, the smaller the value of $R$, the better the model fit, and we identify a best-fit model for a single source as that with the minimum value of $R$. 

In this approach, a discrepant data point does not affect the fit as strongly as it would using a least-squares fit; this is important given the potential for contamination of the photometry beyond 100~$\mu$m.   Determining the values of $R$ that are consistent with good fits and the comparison of $R_{\rm{min}}$ values for different sources are complex tasks, since $R$ depends on the number of data points in the observed SED. As with other goodness of fit statistics, an SED with fewer data points often gives a lower value of $R$, whereas an SED with more data points can give a larger value of $R$, but the fit may still be good. We further discuss these issues in the following sections, and we further address the robustness of the fits in Appendix~\ref{app:unc}.

\subsection{Model Fits  Results} \label{sec:best fit}

\begin{figure}[ht]
\centering
\includegraphics[scale=0.45]{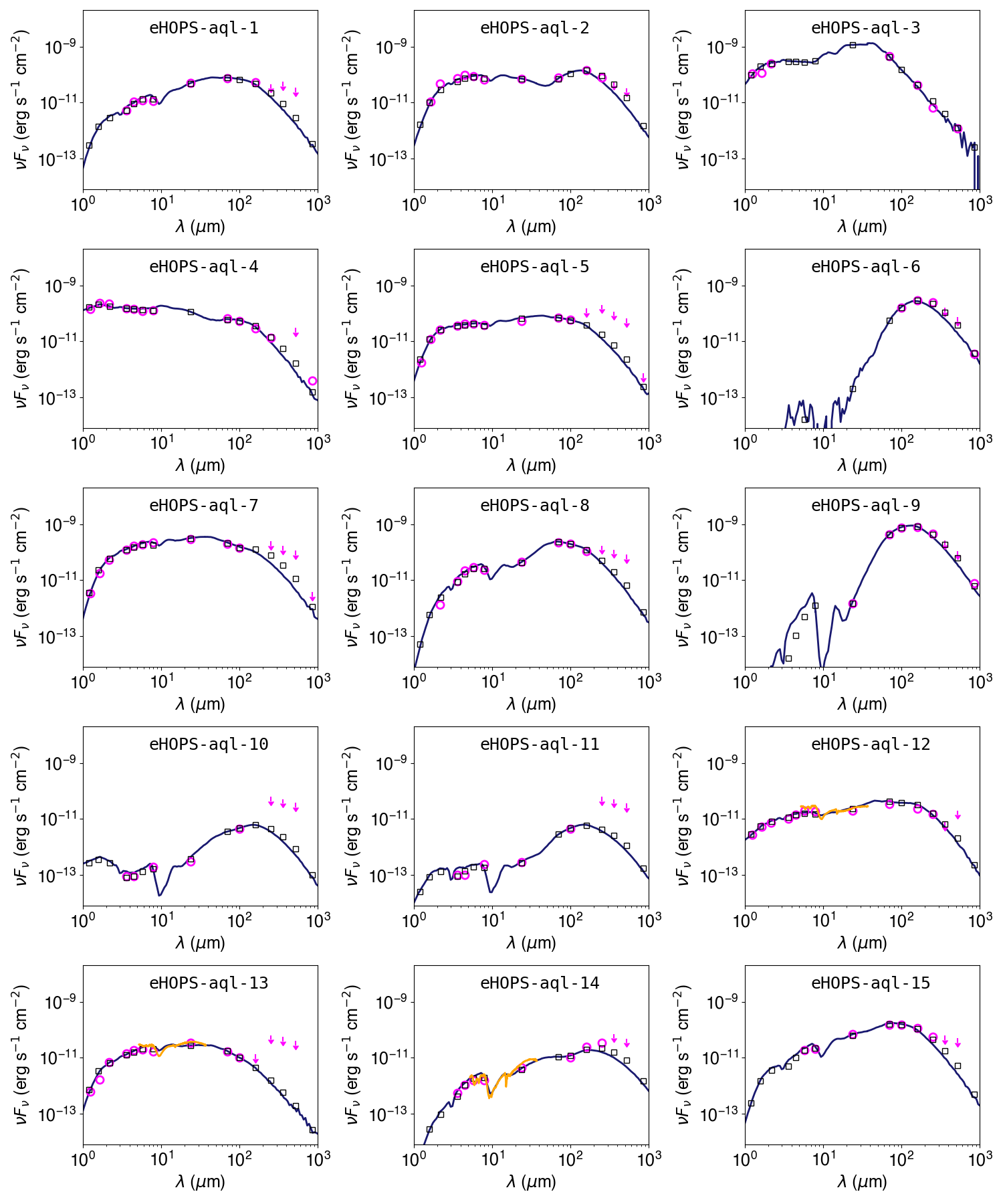}
\vspace{-0.1in}
\caption{SEDs of 15 protostars in Aquila clouds with best fit models. For each SED, magenta circles represent observed data, arrows are upper limits, and the IRS spectrum is included where available in orange. Open squares are best-fit model photometry measured in the same apertures and bandpass as in the data, except for SPIRE 350 and 500 $\micron$ for which we use the maximum model aperture of 24$\arcsec$ and treat the observed fluxes as upper limits.} The best fit model for each protostar is shown by a dark-blue line with fluxes taken from a 4$''$ aperture for $\lambda$ $<$ 8 $\micron$, a 5 $''$ aperture for $\lambda$=8-37 $\micron$ and a 10$''$ aperture for $\lambda$ $>$ 37 $\micron$. The complete figure set for the remaining sources is available online.
\label{fig:bestfit}
\end{figure}

Figure \ref{fig:bestfit} shows the best fits for protostars from eHOPS-aql-1 to eHOPS-aql-15. 
The best-fit parameters for individual SEDs are included in Table \ref{tab:aquila}. Due to the large number of protostars and possible degeneracies in the model parameters, we discuss the statistical distribution of the model parameters in this section rather than individual protostellar properties. \citetalias{Furlan16} presented the statistical study of the best fit properties in a similar manner and found that certain properties such as envelope density and total luminosity are better constrained than others. We also find degeneracies between parameters such as disk radius, cavity opening angles, and inclination (see the model uncertainties and degeneracies section in Appendix \ref{app:unc}) in our study. Thus, in this section, we only discuss the distribution of the best-constrained model parameters, but we include all the fit parameters in Table \ref{tab:aquila} for completeness.

\begin{figure}[b!]
\centering
\includegraphics[scale=0.6]{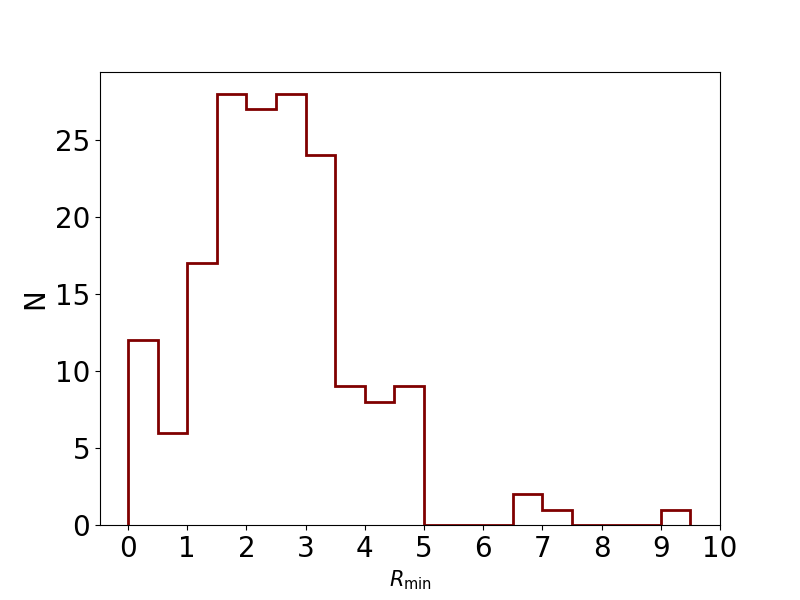}
\caption{Histogram of $R_{\rm{min}}$ values for the best fits of protostars in the eHOPS sample for Aquila. About 97 \% of the SEDs are well-constrained with $\geq$3 data points in their SEDs and $R_{\rm{min}} < 5$.}
\label{fig:R_hist}
\end{figure}

Figure \ref{fig:R_hist} shows the distribution of $R_{\rm{min}}$ values for the Herschel detected protostars in the Aquila clouds. Both the median and mean of the distribution are 2.4. This implies that, on average, the natural logarithm of the ratio between observed and modeled flux deviates by about 2.4 times the average fractional uncertainty. 
In general, a low $R_{\rm{min}}$ value implies that the modeled photometry is similar to the observed photometry. The models with fewer photometry points, however, are less restrictively constrained.  Thus, as discussed in Section \ref{sec:fittingTechnique}, a lower number of fitted data points can also yield a lower $R_{\rm{min}}$. We find 18 sources that have $R_{\rm{min}} < 1$, out of which 14 sources are Class 0, two sources are Class I and two sources are flat-spectrum sources. A total of 11 sources out of 18 with $R_{\rm{min}} < 1$ have only two to five data points (excluding the 350 and 500 $\micron$ data that we use as upper limits for fitting with models) and therefore yield systematically lower $R_{\rm{min}}$ values compared to other sources.

Similar to \citepalias{Furlan16}, the discrepancy between the observed SEDs and models increases substantially when $R_{\rm{min}} > 5$. We find four protostars with $R_{\rm{min}} > 5$. Among them, \dummyACDBE ~has a mismatch between the photometry obtained from the IRS and MIPS 24 $\micron$, suggesting that the source may be variable at $\sim$24 $\micron$. Also, \dummyACDBE ~has a lower observed PACS 70 $\micron$ photometry compared to the modeled 70 $\micron$ data because of the bright, structured emission in the background annulus, thereby raising the estimated $R_{\rm{min}}$. Another source, \dummyADECB~ has an $R_{\rm{min}}$ value of $\sim$7 because of the lower 24 $\micron$ photometry due to the presence of a bright source in the background annulus. The source \dummyADECB~ has scattered dust emission in the mid-IR wavebands that affect the Spitzer/IRAC photometry leading to a mismatch with the model photometry and an increased $R_{\rm{min}}$ of $\sim$ 7. Another protostar, \dummyBADEC~ has a similar discrepancy at 100 $\micron$ because of the presence of a bright source in the background, resulting in $R_{\rm{min}} \sim 7$. Finally, \dummyBAECD~ has $R_{\rm{min}} \sim 9$ due to the discrepancy between observed and modeled photometry in the Spitzer wavelengths. The slope of the SED for \dummyBAECD~ rises in the Spitzer/IRAC wavelengths, peaks at $\sim$ 24 $\micron$, and falls at $>24 \micron$. The source is also located in a lower density region, N(H$_2$) $\sim$ 3 $\times$ 10$^{21}$ cm$^{-2}$. These features suggest that the source may not be a protostar but a galaxy or planetary nebula that shows a peak around 24 $\micron$ due to the [O IV] line arising from highly ionized regions \citep[e.g., ][]{Ueta06}.  The SED classification criteria based on Spitzer photometry (Equation \ref{eq:ps1}) classifies \dummyBAECD~ as a protostar, but the observed SED deviates substantially from the best fit model SED leading to the largest $R_{\rm{min}}$ value in our sample of protostars. We currently leave this in our protostar sample; future observations are needed to determine the nature of this object.

\begin{longrotatetable} 
\begin{deluxetable*}{lllccccccccccccc} 
\tablecaption{Classification and Best-fit Model Parameters for the eHOPS sample in Aquila \label{tab:aquila}} 
\tablewidth{0pt} 
\tablehead{ 
\colhead{Object} & \colhead{R.A.} & \colhead{Decl.} & \colhead{Class} & \colhead{$L_{\rm{bol}}$} & \colhead{$T_{\rm{bol}}$} & \colhead{$\alpha_{4.5,24}$} & 
\colhead{$L_{\rm{tot}}$} & \colhead{$R_{\rm{disk}}$} & \colhead{$\rho_{1000}$} & \colhead{$M_{\rm{env}}$} & \colhead{$\theta$} & \colhead{$i$} & \colhead{$A_{\rm{V}}$} & \colhead{Scaling} & \colhead{$R$} \\ 
\nocolhead{} & \colhead{($^{\circ}$)} & \colhead{($^{\circ}$)} & \nocolhead{} & \colhead{($L_{\odot}$)} & \colhead{(K)} & \nocolhead{} & \colhead{($L_{\odot}$)} & \colhead{(au)} & 
\colhead{(g cm$^{-3}$)} & \colhead{($M_{\odot}$)} & \colhead{($^{\circ}$)} & \colhead{($^{\circ}$)} & \colhead{(mag)} & \colhead{Factor} & \nocolhead{} 
} 
\decimalcolnumbers 
\startdata 
eHOPS-aql-1 & 276.2975 & -2.9815 & Class 1 & 1.563 & 95.7 & 0.93 & 3.7 & 50 & 5.9 $\times$ 10$^{-19}$ & 0.0239 & 35 & 63.3 & 12.9 & 1.21 & 1.755 \\ 
eHOPS-aql-2 & 276.3056 & -2.9986 & Flat & 2.992 & 282.4 & -0.19 & 5.8 & 5 & 5.9 $\times$ 10$^{-18}$ & 0.1828 & 45 & 56.7 & 9.0 & 1.89 & 2.069 \\ 
eHOPS-aql-3 & 276.6675 & -2.7158 & Class 1 & 1.003 & 46.4 & 1.26 & 157.4 & 5 & 2.4 $\times$ 10$^{-20}$ & 0.0012 & 15 & 87.2 & 2.3 & 0.52 & 1.426 \\ 
eHOPS-aql-4 & 276.8053 & -3.6707 & Flat & 4.130 & 735.0 & 0.05 & 4.2 & 5 & 2.4 $\times$ 10$^{-19}$ & 0.0094 & 35 & 31.8 & 1.7 & 1.38 & 1.371 \\ 
eHOPS-aql-5 & 276.9592 & -3.8206 & Flat & 2.062 & 243.6 & 0.10 & 2.8 & 50 & 5.9 $\times$ 10$^{-19}$ & 0.0239 & 35 & 56.7 & 10.0 & 0.92 & 1.362 \\ 
eHOPS-aql-6 & 276.9762 & -3.7116 & Class 0 & 2.173 & 18.7 & --- & 34.9 & 100 & 2.4 $\times$ 10$^{-17}$ & 0.7619 & 45 & 81.4 & 16.2 & 1.16 & 0.134 \\ 
eHOPS-aql-7 & 276.9781 & -3.7107 & Class 1 & 7.313 & 248.0 & 0.35 & 9.6 & 50 & 5.9 $\times$ 10$^{-18}$ & 0.1863 & 45 & 18.2 & 14.5 & 0.95 & 1.278 \\ 
eHOPS-aql-8 & 277.0227 & -3.7833 & Class 1 & 2.619 & 104.4 & 0.42 & 2.6 & 5 & 2.4 $\times$ 10$^{-18}$ & 0.1249 & 15 & 31.8 & 15.3 & 0.85 & 2.313 \\ 
eHOPS-aql-9 & 277.0382 & -3.8032 & Class 0 & 8.829 & 33.8 & 1.06$^a$ & 10.6 & 5 & 2.4 $\times$ 10$^{-17}$ & 1.1197 & 25 & 49.5 & 30.7 & 1.05 & 0.355 \\ 
eHOPS-aql-10 & 277.1542 & -1.9074 & Class 0 & 0.046 & 56.6 & 0.71 & 0.2 & 500 & 1.2 $\times$ 10$^{-18}$ & 0.0535 & 35 & 69.6 & 0.0 & 0.53 & 2.526 \\ 
eHOPS-aql-11 & 277.1700 & +0.4428 & Class 0 & 0.047 & 58.3 & 0.63 & 0.1 & 5 & 2.4 $\times$ 10$^{-18}$ & 0.1120 & 25 & 69.6 & 3.4 & 0.89 & 3.343 \\ 
eHOPS-aql-12 & 277.1745 & -0.0560 & Flat & 0.736 & 256.3 & 0.19 & 1.5 & 500 & 5.9 $\times$ 10$^{-19}$ & 0.0219 & 45 & 63.3 & 1.7 & 1.51 & 2.507 \\ 
eHOPS-aql-13 & 277.1835 & +0.8939 & Class 1 & 0.598 & 319.0 & 0.35 & 1.3 & 5 & 1.2 $\times$ 10$^{-20}$ & 0.0006 & 15 & 18.2 & 12.1 & 1.26 & 2.409 \\ 
eHOPS-aql-14 & 277.1867 & +0.8571 & Class 0 & 0.280 & 64.7 & 0.84 & 0.6 & 50 & 2.4 $\times$ 10$^{-17}$ & 0.7453 & 45 & 56.7 & 15.7 & 2.00 & 2.802 \\ 
eHOPS-aql-15 & 277.1877 & +0.8674 & Class 1 & 2.560 & 81.7 & 0.66 & 5.3 & 50 & 2.4 $\times$ 10$^{-18}$ & 0.0957 & 35 & 63.3 & 9.0 & 0.53 & 0.781 \\ 
eHOPS-aql-16 & 277.1991 & -1.6356 & Class 0 & 2.062 & 65.9 & 1.73 & 19.0 & 50 & 2.4 $\times$ 10$^{-18}$ & 0.0745 & 45 & 75.6 & 0.0 & 1.88 & 3.244 \\ 
eHOPS-aql-17 & 277.2135 & +0.3242 & Class 1 & 0.136 & 253.1 & 0.51 & 0.3 & 5 & 1.8 $\times$ 10$^{-20}$ & 0.0010 & 5 & 56.7 & 13.8 & 1.06 & 2.153 \\ 
eHOPS-aql-18 & 277.2253 & +0.4914 & Class 0 & 10.469 & 56.7 & 1.1 & 6.0 & 500 & 5.9 $\times$ 10$^{-18}$ & 0.3374 & 5 & 18.2 & 21.0 & 1.98 & 2.515 \\ 
eHOPS-aql-19 & 277.2287 & +0.4979 & Class 1 & 3.183 & 81.0 & 1.52 & 6.1 & 500 & 2.4 $\times$ 10$^{-18}$ & 0.1070 & 35 & 63.3 & 5.4 & 2.00 & 3.745 \\ 
eHOPS-aql-20 & 277.2288 & -1.6241 & Class 0 & 0.185 & 56.8 & 1.21 & 2.0 & 100 & 5.9 $\times$ 10$^{-19}$ & 0.0190 & 45 & 81.4 & 0.0 & 2.00 & 4.604 \\ 
\enddata 
\tablecomments{R.A. and Decl. represent mean of the coordinates from 2MASS, Spitzer and Herschel/PACS positions. $(a)$ denotes $\alpha_{4.5,24}$ values using 4.5 $\micron$ completeness limit.\        $(b)$ denotes $\alpha_{4.5,24}$ values with 24 $\micron$ completeness limit.\        Full Table is available online.} 
\end{deluxetable*} 
\end{longrotatetable}

\begin{figure}[t!]
\centering
\includegraphics[scale=0.48]{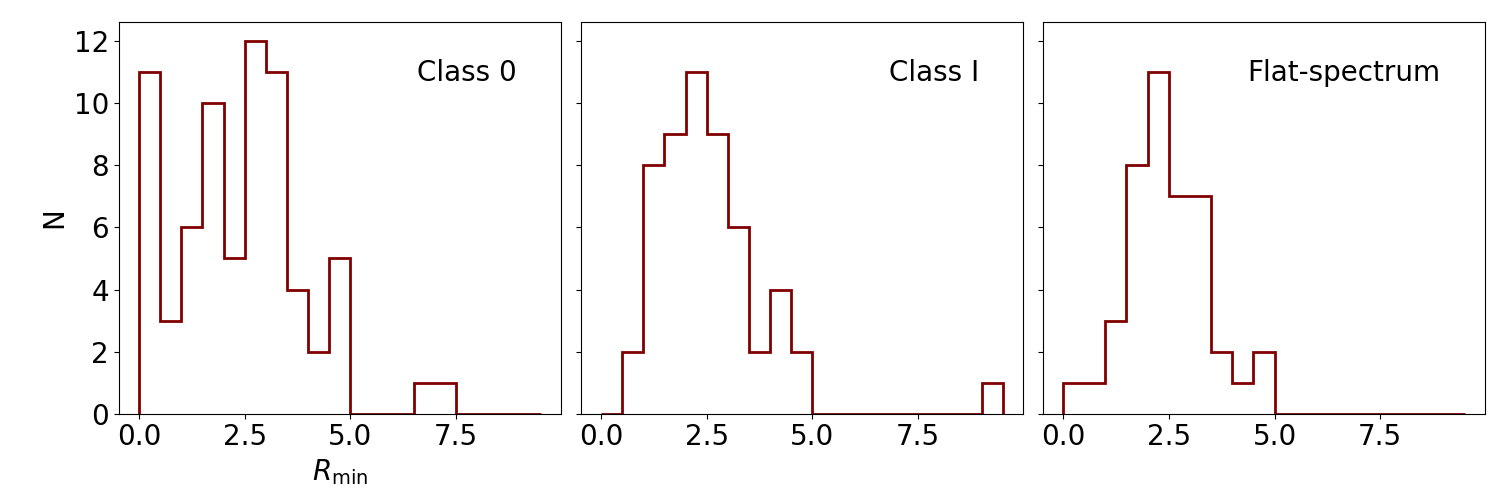}
\caption{Histogram of $R_{\rm{min}}$ values for the best fits of protostars in the eHOPS sample for Aquila,  shown separately for the different evolutionary SED classes. 95\% of Class 0, 98\% of Class I, and all flat-spectrum sources have reasonably good fits with $R_{\rm{min}} < 5$ and $\geq3$ data points in their SEDs.}
\label{fig:R_hist_cl}
\end{figure}

In Figure \ref{fig:R_hist_cl}, we show the distribution of $R_{\rm{min}}$ separately for different classes of protostars. The median $R_{\rm{min}}$ for the 3 classes are similar: 2.5, 2.4, and 2.5 for Class 0, Class I, and flat-spectrum sources, respectively.
About 95\% of Class 0, 98\% of Class I, and all flat-spectrum sources have reasonably good fits with $R_{\rm{min}} < 5$ with a minimum of three data points in their SEDs. The slightly lower fraction of Class 0 sources with reliable $R_{\rm{min}}$ values could be because of more uncertain SEDs, especially in the near-IR region, noisy IRS spectra, their location in extended structures such as filaments, and variability. The higher envelope density of Class 0 protostars also places them closer to the limit of parameter space probed by the model grid. 

The fits with a smaller number of photometry points may be reliable but are not as well constrained. Examining the SEDs of the protostars and their fits, particularly for the reddest protostars without mid-IR detections, we conclude that a minimum of three data points are required in the SED for a reliable fit. Out of 172 protostars, only one source, \dummyBECDA, contains less than three (two) data points in its SED so we caution that the model parameters for \dummyBECDA may not be constrained. All other 171 protostars have a minimum of three data points in their SEDs. Furthermore, there are four sources with $R_{\rm{min}} > 5$ for which the  models in the grid do not provide an adequate fit.
If we consider only the best fits that have a minimum three data points and $R_{\rm{min}} < 5$, we ascertain that 167 (97\%) of the protostars have reliable and well constrained best fit models. We only include these 167 protostars when discussing the parameter distribution in the rest of the section.

\subsection{Best Fit Parameters}
\label{sec:best_fit}

In this section, we discuss the best fit values for the most fundamental parameters of the models. Because of degeneracies between parameters and model pitfalls (see \citetalias{Furlan16}), we focus on the bulk properties of the best-fit parameters and refer the readers to Table \ref{tab:aquila} for the best fit values for individual protostars. We concentrate on the properties that primarily determine the shape and amplitude of the SEDs: the envelope density/mass and the total protostellar luminosity. We also discuss the inclination of the envelope and foreground reddening, which have a large impact on the fits. Since previous work measuring the sizes of outflow cavities using HST imaging shows that these are not well constrained by SED fitting \citep{Habel21}, we omit a discussion of the cavities. 

\subsubsection{Inclination Angle} \label{sec:incl}

\begin{figure}[t!]
\centering
\includegraphics[scale=0.6]{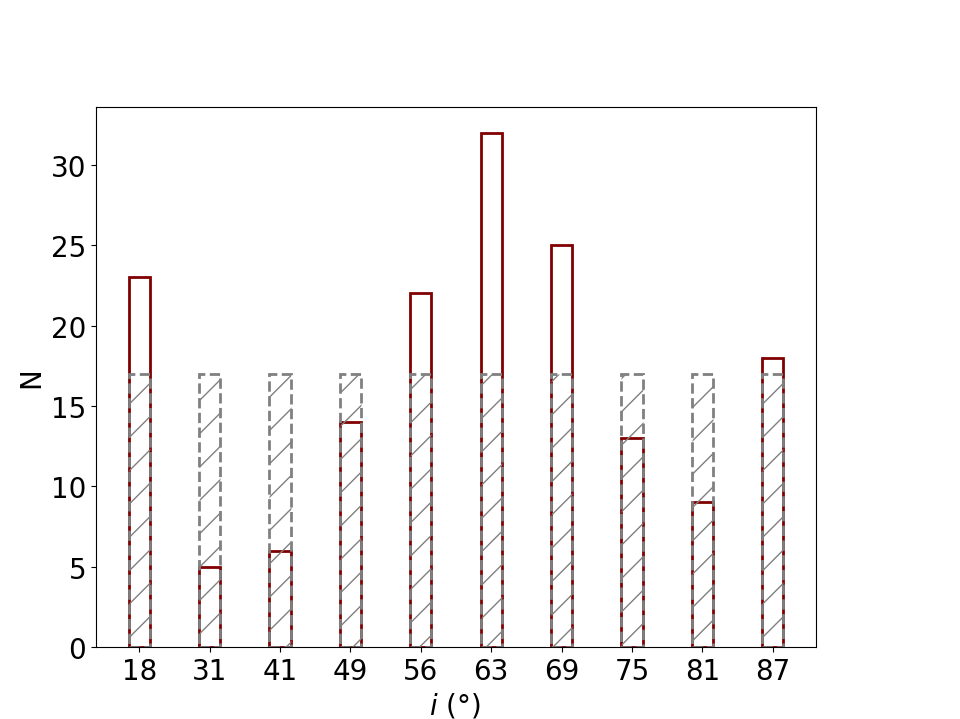}
\caption{Histogram of the best-fit inclination angles of the Aquila protostars. The grey dashed histogram shows the distribution of uniformly (randomly) distributed inclination angles. The inclination angles for the protostars in Aquila are not distributed randomly but peaked at $\sim$63$^{\circ}$.
\label{fig:incl}}
\end{figure}

We show the histogram of the inclination angles from the best fits in Figure \ref{fig:incl}; the bins correspond to the ten different inclinations in the model grid.  The inclinations of the grid are distributed uniformly in cos($i$) so that a sample with randomly distributed inclinations would show a flat distribution, as illustrated by the grey dashed histogram. We expect protostars to be randomly inclined in the sky with an equal number of protostars in each bin. Instead, we find the distribution peaking around $55-70^{\circ}$.  The median $i$ in Aquila is 63$^{\circ}$, which is close to the median value of 60$^{\circ}$ for isotropically distributed protostars. The histogram of inclination angle in Aquila is similar to the histogram for the HOPS protostars in Orion \citepalias{Furlan16}. Thus, in both clouds, the models do not recover the expected distribution of inclination. This likely indicates a bias in the models since degeneracies between inclination and other parameters can affect the determination of the fundamental parameters such as envelope density/mass and luminosity.  In addition, our identification of the protostars may be biased toward those at intermediate inclinations.  This may occur since sources at high inclinations may be too faint -- particularly in the near to mid-IR -- while sources at low inclinations may be misclassified as pre-ms stars or galaxies.

\begin{figure}[t!]
\includegraphics[scale=0.48]{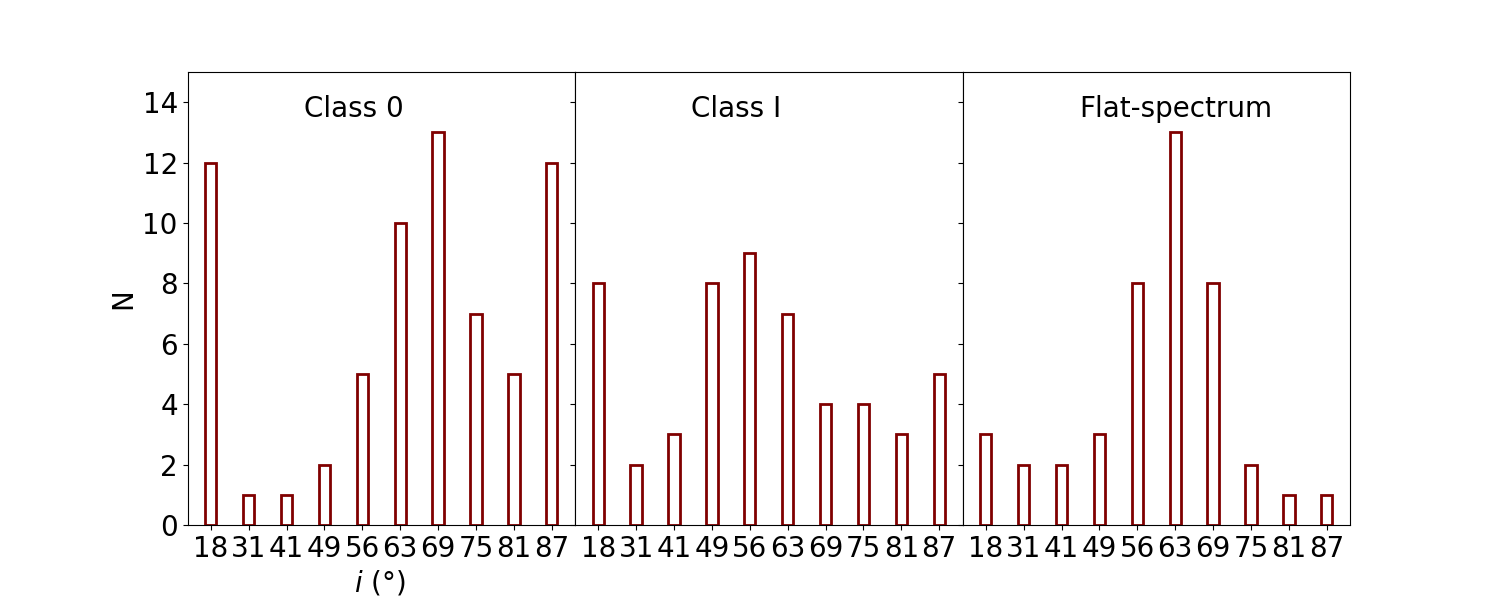}
\caption{Histogram of the inclination angles ($i$) for the best fits of protostars in the eHOPS sample for Aquila, shown for different SED classes.
\label{fig:incl_cl}}
\end{figure}

The different SED classes also show non-uniform distributions, in this case, due to degeneracies between evolutionary class and inclination.  
Figure \ref{fig:incl_cl} shows the distribution of inclination angles for different SED classes. We use the Anderson-Darling (AD) test to determine whether the inclination distribution for different SED classes is similar. Other non-parametric statistics commonly used in astronomy are the Kolmogorov-Smirnov (KS) test and the Cramer - von Mises (CvM) test. \cite{Feigelson12} mentions the AD test as the most sensitive of the three empirical distribution functions (also suggested independently by \citealt{Hou09}). 
The AD test shows that the distribution of inclination angles is significantly different between Class 0 and Class I protostars, and also between Class 0 protostars and flat-spectrum sources, but not between Class I protostars and flat-spectrum sources. 

There seems to be a deficit of lower inclination angles for Class 0 sources, except for $i$=18\degr; this is expected since protostars with face-on inclinations are likely to have higher values of \Tbol. Conversely, there are relatively fewer high-inclination angle sources classified as flat-spectrum protostars; such high inclination protostars are likely to be classified as a Class I or 0 protostar due to the concentration of mass around the midplane (e.g. \citealt{Fischer14, Habel21}). The Class 0 and Class I protostars both show a peak in the $i$=18$^{\circ}$ bin. Most of these sources are deeply embedded protostars that have fewer than average data points to fit. Likely, the inclination angle is not constrained for such sources.

\subsubsection{Foreground extinction}
\label{sec:av}

\begin{figure}[b!]
\centering
\includegraphics[scale=0.44]{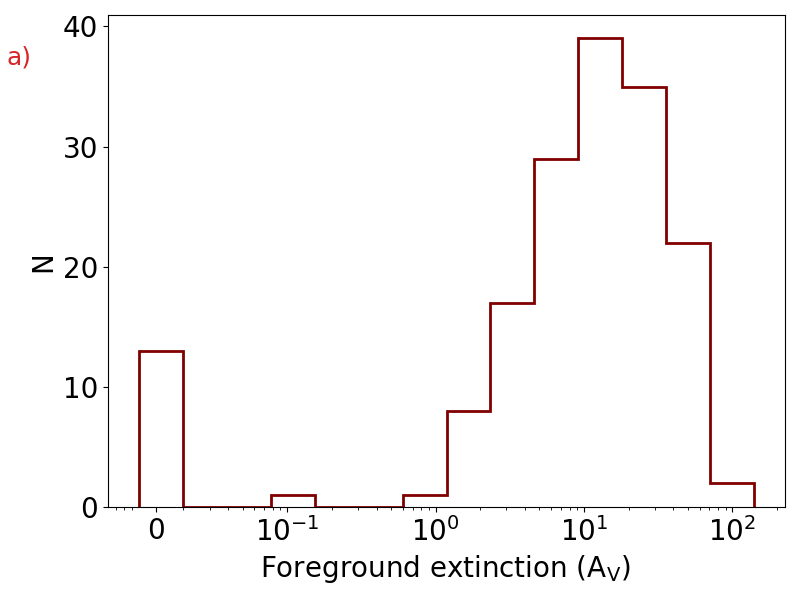}
\includegraphics[scale=0.44]{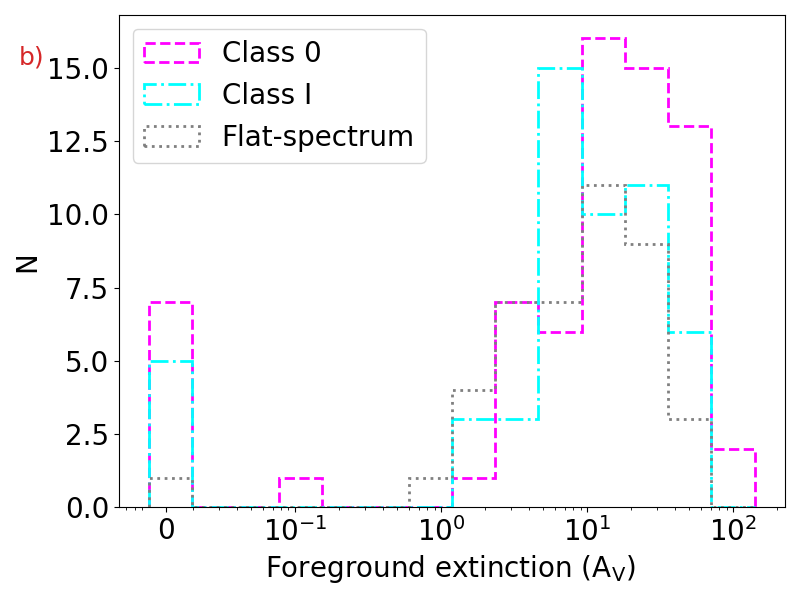}
\caption{Panel a) Histogram of the model-derived foreground extinction (A$_{\rm{V}}$) for the best fits of protostars in the eHOPS sample for Aquila. Panel b) Histogram of A$_{\rm{V}}$ separated by different SED classes. AD test shows a significant difference in the distribution of Class 0 and flat-spectrum sources.
\label{fig:av}}
\end{figure}

Figure \ref{fig:av}a shows the distribution of the best fit foreground extinction $A_{\rm{V}}$ for all the protostars in Aquila. There are 13 sources that are best fit by models that do not require foreground extinction. These sources have sparse mid-IR photometry so $A_{\rm{V}}$ is likely not constrained. A similar result is seen in \citetalias{Furlan16} for the sources with sparse or no mid-IR data. For the remaining protostars, the distribution peaks around $10-20$ mag, with a median at $\sim$ 13 mag. The distribution is slightly skewed towards the higher $A_{\rm{V}}$ region, which is expected due to the high column densities of the Aquila molecular cloud and since Aquila is close to the galactic plane with high foreground extinction. In Figure \ref{fig:av}b, we show $A_{\rm{V}}$ as a function of the SED classes. Out of 13 sources with $A_{\rm{V}}$ = 0, seven are Class 0 sources, a majority of which have a [3.6] - [4.5] $>$ 0.65 and declining $\alpha_{4.6,24}$ but the emission increases sharply beyond 24$\micron$. This is likely due to the large contribution from scattered light at $\lambda < 8$ $\micron$ in the Class 0 protostars. We find a decreasing median $A_{\rm{V}}$ as a function of SED Class, with the median $A_{\rm{V}}$ of 15.9, 11.3, and 9.9 mag for Class 0, Class I and flat-spectrum sources, respectively. 
The AD test finds a significant difference in the distribution of $A_{\rm{V}}$ for Class 0 and flat-spectrum sources ($p<0.01$) and no significant difference in either Class I and flat, and Class 0 and Class I sources. In Sec.~\ref{sec:density} we discuss whether foreground reddening is biasing protostars to earlier classes.

\begin{figure}[t!]
\centering
\includegraphics[scale=0.52]{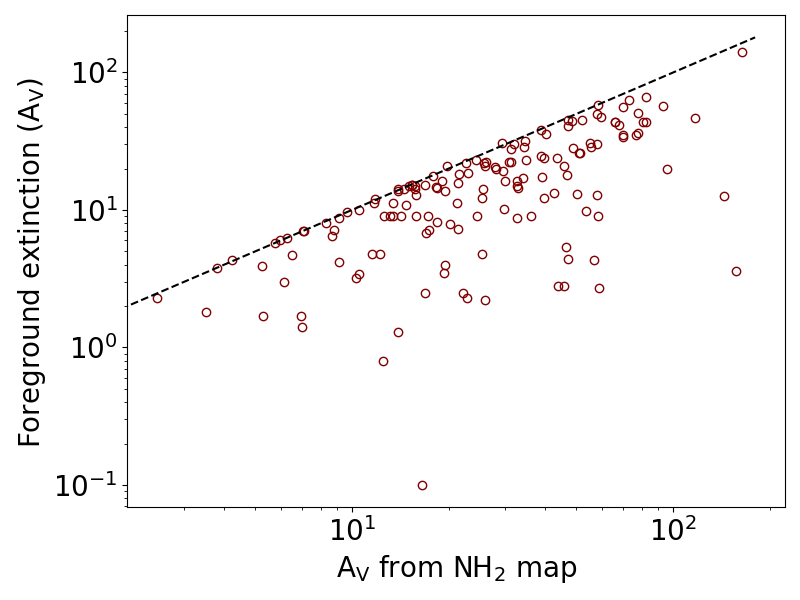}
\caption{Foreground extinction values (A$_{\rm{V}}$) from the best fit models vs.~the maximum A$_{\rm{V}}$ value determined from column density maps of Aquila. The dashed line indicates where the two A$_{\rm{V}}$ values are equal. The model derived A$_{\rm{V}}$ is typically less than the projected column density at the position of protostars, confirming that the protostars are embedded in the clouds rather than in the background.
\label{fig:avVsav}}
\end{figure}

Figure \ref{fig:avVsav} shows the variation of the foreground extinction A$_{\rm{V}}$ obtained from the best fits and the projected A$_{\rm{V}}$ values obtained from the Herschel maps. Similar to the HOPS results in \citetalias{Furlan16}, the majority of sources are fit with A$_{\rm{V}}$ lower than the Herschel-derived A$_{\rm{V}}$ values used as the maximum A$_{\rm{V}}$ allowed by the fitter. The ratio of the best fit A$_{\rm{V}}$ to the Herschel-derived A$_{\rm{V}}$ is $<$0.5 for $\sim$40\% of the sources. This is consistent with the protostars being inside the cloud, as opposed to background objects behind the cloud. There are sources for which the best fit foreground extinction is similar to the A$_{\rm{V}}$ from the Herschel maps, and also some sources for which the foreground extinction is close to zero.  These sources again point to a potential degeneracy between foreground extinction and the density of the envelope. We discuss this in the following section (Sec.~\ref{sec:density}).

\subsubsection{Envelope reference density} \label{sec:density}

The gas and dust density distributions for the collapsing protostellar envelopes models used in our model grid are described in the original TSC model paper (\citealt{Terebey84}; also see \citealt{Ulrich76} and \citealt{Cassen81}). In the TSC model, the initial cloud (envelope) has a solid body rotation with a radial density distribution corresponding to a singular isothermal sphere, $\rho$ $\propto$ $r^{-2}$. The collapse commences near the center of the protostellar system and then propagates outward at the sound speed but the material outside the collapsing region maintains hydrostatic equilibrium. In the collapsing region, the infall velocity approaches free-fall and the density distribution takes the form $\rho$ $\propto$ $r^{-3/2}$, which is the solution for a steady infall rate. In our grid of models, we assume that the entire envelope has this collapse profile, except when rotation and outflows cause deviations from spherical symmetry. In the TSC model, the material falling from the envelope accumulates in a disk that extends to the centrifugal radius ($R_{\rm{C}}$). In our grid, $R_{\rm{C}}$ is a measure of the angular momentum of the envelope, assuming an initial solid body rotation. It influences the SED by controlling the flattening of the inner envelope due to rotation. Although the disk can extend to larger radii, the model SEDs are relatively insensitive to the actual outer radius of the disk.   

To characterize the density at radii that are relatively unaffected by rotation, \citetalias{Furlan16} parametrize the envelope structure by the density at 1000 au.  The SED is strongly dependent on the envelope density, which is the fundamental physical parameter describing the envelope. For a given collapse model and central protostar mass, the envelope density gives the rate of mass infall. For a 0.5 M$_{\odot}$ protostar with steady, spherically symmetric infall, the infall rate is given by

\begin{equation} \label{eq:rho}
    \dot{M}_{\rm env}  =  \left(\frac{\rho_{1000}}{2.4 \times 10^{-18}~{\rm g\,cm^{-3}}}\right) \times 10^{-5}~\rm{M}_{\odot}~\rm{yr}^{-1}. 
\end{equation}

The values for $\rho_{1000}$ that are listed in Table 3 in \citetalias{Furlan16} correspond to $\dot{M}_{\rm{env}}$ from 5.0 $\times$ 10$^{-8}$ to 7.5 $\times$ 10$^{-5}$ M$_{\odot}$~yr$^{-1}$ for a protostar with 0.5 M$_{\odot}$. The reduction of the infalling mass due to removal by outflow cavities is not incorporated in these estimates of the infall rate. We also include one model with $\rho_{1000}=0$~g~cm$^{-3}$ (i.e., no envelope), which describes pre-ms stars that have dispersed their envelopes but have protostellar SEDs due to extinction.

\begin{figure}[t!]
\centering
\includegraphics[scale=0.44]{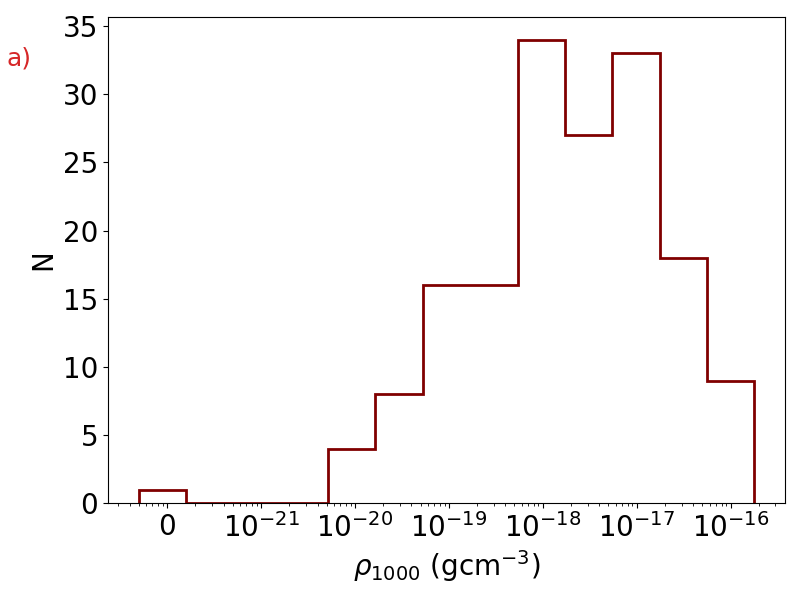}
\includegraphics[scale=0.44]{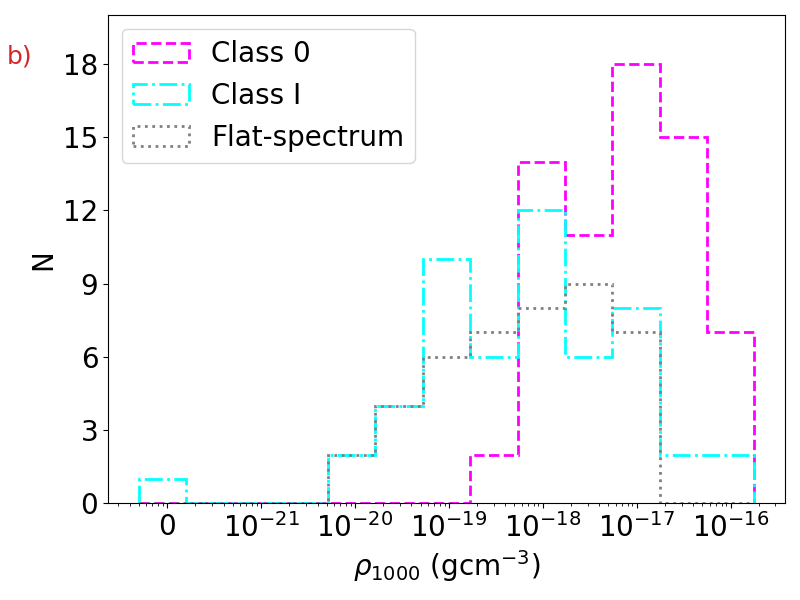}
\caption{Panel a) Histogram of the model-derived envelope reference density ($\rho_{1000}$) from the best fits of protostars in the eHOPS sample for Aquila. Panel b) Histogram of $\rho_{1000}$ separated by different SED classes. Class 0 protostars have the highest $\rho_{1000}$. AD tests show significant differences in the distribution of $\rho_{1000}$ between Class 0 and the remaining classes, but no significant difference between Class I and flat-spectrum sources.
\label{fig:rho}}
\end{figure}

The distribution of the best fit $\rho_{1000}$ values for all 172 protostars in Aquila are shown in Figure \ref{fig:rho}a. One Class I protostar, \dummyBAEDC, has the best fit with a model with no envelope ($\rho_{1000}=0$). For the remaining protostars, the median of the distribution is 1.8 $\times$ 10$^{-18}$ g cm$^{-3}$ with first and third quartiles at 2.4 $\times$ 10$^{-19}$ and 5.9 $\times$ 10$^{-18}$ g cm$^{-3}$. 
Between the first and the third quartiles, the sources are fairly uniformly distributed.

Figure \ref{fig:rho}b shows the distribution of the reference envelope density $\rho_{1000}$ for different SED classes. The median $\rho_{1000}$ for Class 0 sources is 5.9 $\times$ 10$^{-18}$ g cm$^{-3}$, which then decreases to 5.9 $\times$ 10$^{-19}$ g cm$^{-3}$ for both the Class I sources and the flat-spectrum sources. The mean $\rho_{1000}$ for Class 0, Class I and flat-spectrum protostars are  2 $\times$ 10$^{-17}$, 5.7 $\times$ 10$^{-18}$ and 2 $\times$ 10$^{-18}$, respectively. The mean  $\rho_{1000}$ thus decreases as a function of SED Class in the eHOPS Aquila protostars. The median $\rho_{1000}$ decrease from Class 0 to Class I by an order of magnitude, however, no such decrease in median $\rho_{1000}$ is noticed between Class I and flat-spectrum sources. Using the AD test, we find that the Class 0 sample is significantly different from the Class I and flat-spectrum protostars (p-value $<$ 0.01). The AD test, however, shows that we cannot reject the null hypothesis that the Class I and flat spectrum protostars are drawn from the same population. 

\begin{figure}[t!]
\includegraphics[scale=0.52]{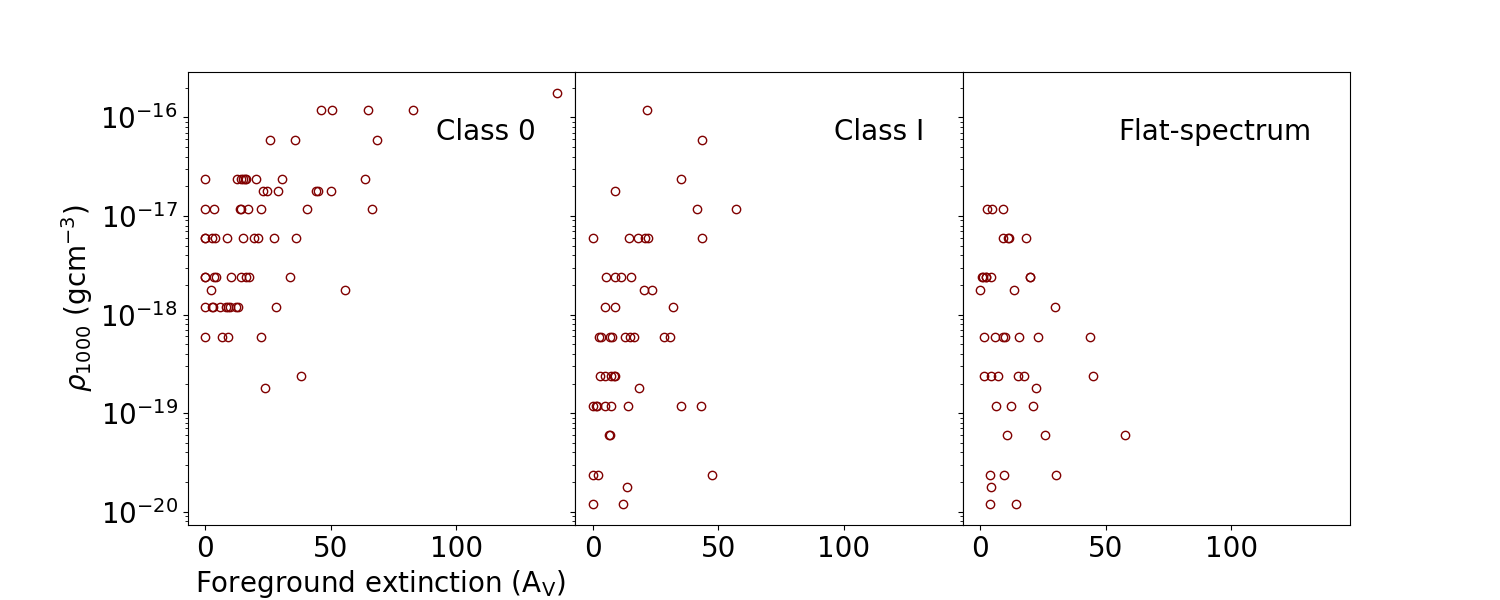}
\caption{Variation of the best fit envelope reference density $\rho_{1000}$ with foreground extinction $A_{\rm{V}}$ for different SED classes. 
\label{fig:rhovsav}}
\end{figure}

Since foreground extinction and envelope density can affect an SED in a similar way, we search for degeneracies between the  best fit envelope reference density $\rho_{1000}$ and foreground extinction $A_{\rm{V}}$. This will resolve the question posed in Sec.~\ref{sec:av}, does foreground extinction bias the classification of protostars and the determination of their envelope densities? If a degeneracy is present, we would expect a trade-off with foreground extinction increasing as envelope density decreases.   Instead, in Figure \ref{fig:rhovsav} we find that sources with higher $A_{\rm{V}}$ often have higher $\rho_{1000}$, particularly for the Class 0 and Class I protostars.  There are four Class 0 sources that have $\rho_{1000} > 10^{-16}$ g~cm$^{-3}$ and $A_{\rm{V}} > 50$, and no sources in either the Class I or flat-spectrum group in this region. Out of four Class 0 sources with high $\rho_{1000}$ and $A_{\rm{V}}$, three sources (\dummyBEACD, \dummyBDECA, and \dummyBACED) do not have measured emission shortward of 24 $\micron$ and thus are likely PBRS. For Class I protostars, the sources are mostly confined at $A_{\rm{V}} < 50$, with fewer sources in the high $\rho_{1000}$ regime compared to Class 0 protostars. Finally, the flat-spectrum sources are more confined in the lower $\rho_{1000}$ and lower $A_{\rm{V}}$ region compared to the other two classes.  This is consistent with the decreasing median values for both $\rho_{1000}$ and $A_{\rm{V}}$ with evolutionary class in Figures \ref{fig:rho}b and \ref{fig:av}b, respectively. In summary, we find no evidence for degeneracies between extinction and envelope density. On the contrary, the Class 0 protostars are not only embedded in higher density envelopes, but also appear to be concentrated in regions of higher cloud column density, and thus higher extinction. A similar trend is seen in parts of Orion \citep[e.g. \citepalias{Furlan16},][]{stutz2015,Megeath22}.

\begin{figure}[ht!]
\includegraphics[scale=0.52]{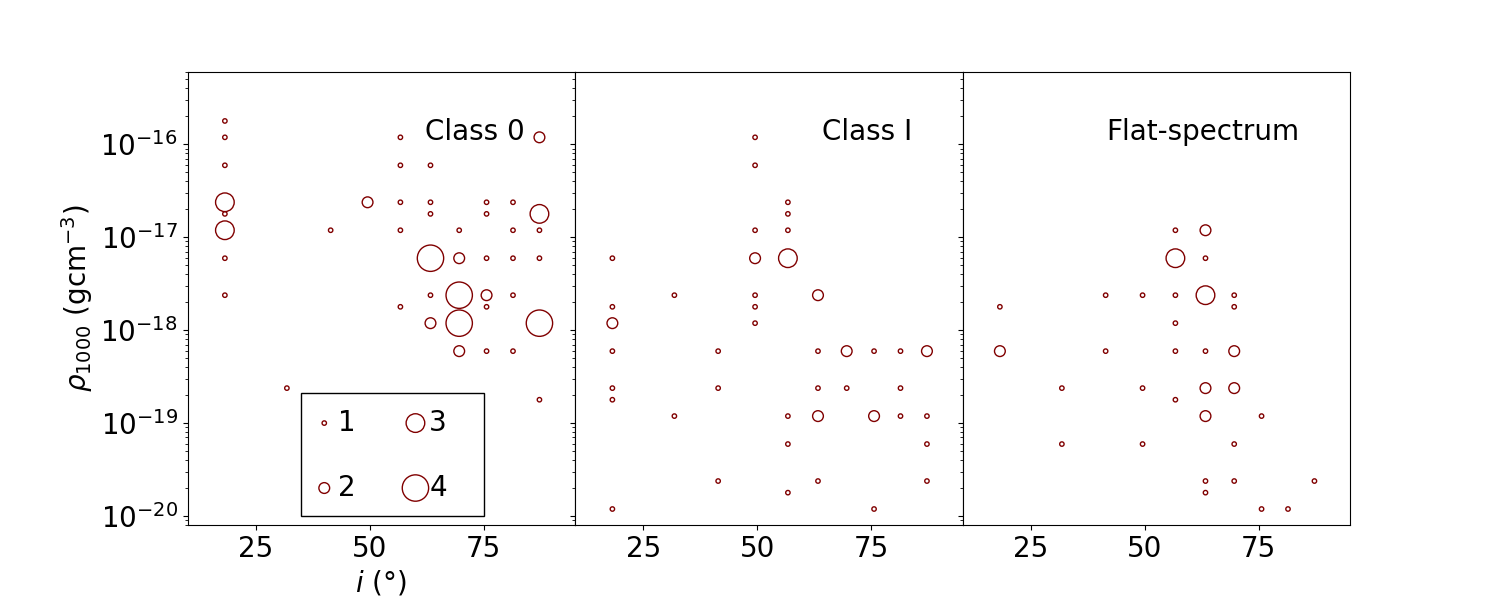}
\caption{Variation of the best fit envelope reference density ($\rho_{1000}$) with inclination angle ($i$) for different SED classes. The size of markers corresponds to the number of protostars with the same pair of ($\rho_{1000}$, i). The sizes that correspond to different numbers of protostars are shown in the lower region of the first panel. We find a majority of Class 0 protostars to be concentrated at intermediate to high inclination angles, whereas Class I protostars and flat spectrum protostars are roughly uniformly distributed.
\label{fig:rhovsincl}}
\end{figure}

A potential degeneracy, however, is found between inclination and envelope density.
In Figure \ref{fig:rhovsincl}, we plot one parameter as function of the other, to study their correlation. 
The Class 0 protostars often show high inclination angles when compared to later evolutionary classes (Figure \ref{fig:incl_cl}).  Within the sample of Class 0 protostars, the 13 Class 0 protostars that have the lowest possible inclination angle have high values of $\rho_{1000}$. The rest of the the Class 0 sample tend to have lower envelope density at higher inclination angles. A similar trend is found for Class I protostars, with those at high inclinations having lower envelope densities. Interestingly, a trend is not clearly apparent for flat spectrum protostars. This is in tension with HST 1.6~$\mu$m imaging of Orion protostars that shows that a large fraction of flat spectrum protostars has point source morphologies.  \citet{Habel21} explained this morphology by the flat spectrum protostars being populated by either protostar with low envelope densities seen at higher inclinations or protostars with high envelope densities seen at low inclinations through their outflow cavities. These correlations indicate that SED class is dependent on inclination, and other inclination independent methods are needed to more reliably sort sources by their evolutionary stage (e.g. \citealt{Federman22}).  Furthermore, these degeneracies and the lack of a uniform distribution of inclinations point to likely biases in our fits and sample. 

\subsubsection{Envelope Mass}
\label{sec:env_mass}

In Section \ref{sec:density}, we found that the envelope reference density decreases with evolutionary class in Aquila. Similar to the envelope density, early and deeply embedded protostar is expected to have a high envelope mass whereas an evolved protostar has dispersed most of its envelope with very little envelope left. The mass of the envelope inside some radius $r$ is a useful diagnostic that traces protostellar evolution. Furthermore, the envelope mass is an intrinsic property of a protostar that is independent of the inclination angle. 

The protostellar outflow carves out a section of the envelope with wider cavity angles carving out more envelope mass (see Figure 7 in \citealt{Fischer17}). As a result, the envelope mass calculated by merely converting $\rho_{1000}$ (Equation \ref{eq:rho}) to mass assuming spherical symmetry and steady infall rate is an overestimation of the actual envelope mass. For this reason, we adopt a more accurate modeling with axisymmetric envelope models where the envelopes have a rotational flattening and outflow cavities are included. The flattening component is characterized by the centrifugal radius ($R_C$), whereas outflow cavities are characterized by cavity opening angle ($\theta_C$). We assume a polynomial-shaped cavity with an exponent of 1.5. Note that in \cite{Whitney03a, Whitney03b} the cavity opening angles are measured from the axis of rotation, so they represent the half-angle instead of the full opening angle. The cavity opening angle is the angle from the pole to the cavity edge at a height above the disk plane equal to the envelope radius.

Since envelope mass is independent of inclination angle, the model estimates the envelope mass for a particular envelope density, centrifugal radius, and cavity opening angle for 3040 independent models. We then integrated the envelope mass out to a radius of 2500 au for two reasons. First, 2500 au is close to the FWHM of the PACS 160 $\micron$ beam at the distance of Aquila. In the observed SEDs, PACS 160 $\micron$ has the largest spatial extent near the expected SED peak and photometry at $>$200 $\micron$ data is often contaminated by cooler, extended dust emission, requiring us to use upper limits at these wavelengths. Second, to be consistent with the HOPS survey in Orion that uses 2500 au for estimating envelope mass in Orion (\citetalias{Furlan16}, \citealt{Fischer17}). In Sec.~\ref{sec:hopsvsehops}, we compare envelope masses for both HOPS sources in Orion and eHOPS sources in Aquila; using envelope masses within the same outer radius makes them easier to compare. 

\begin{figure}[ht!]
\centering
\includegraphics[scale=0.44]{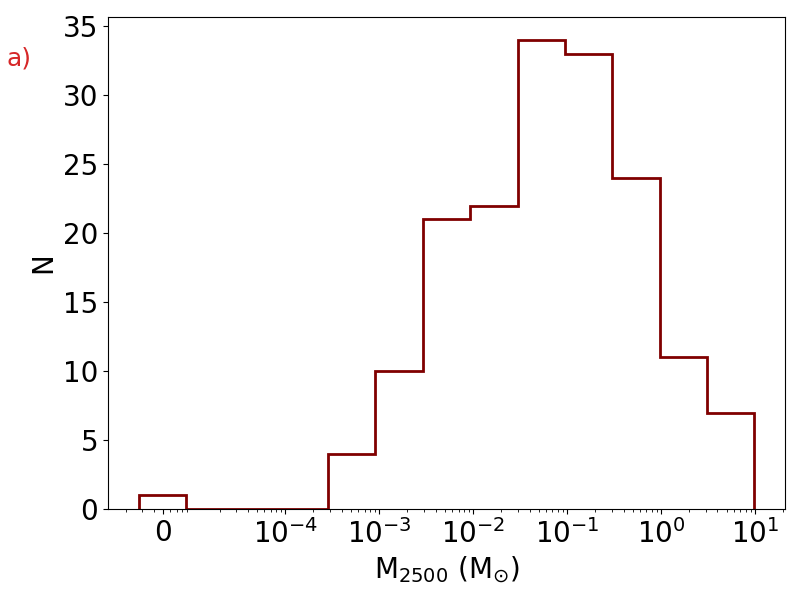}
\includegraphics[scale=0.44]{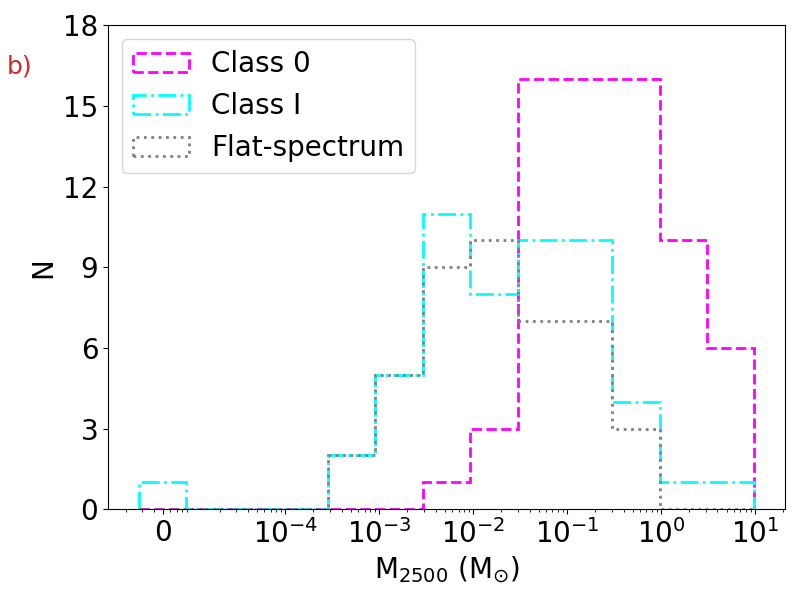}
\caption{Panel a) Histogram of the model-derived envelope mass within 2500 au (M$_{2500}$) from the best fits of protostars in the eHOPS sample for Aquila. Panel b) Histogram of M$_{2500}$ separated by different SED classes. Class 0 protostars have the highest M$_{2500}$. AD tests show significant differences in the distribution of M$_{2500}$ between Class 0 and the remaining classes, but no significant difference between Class I and flat-spectrum sources.
\label{fig:mass}}
\end{figure}

Figure \ref{fig:mass} shows the distribution of envelope masses of the best fit models. The median of the distribution is 0.07 M$_{\odot}$, with lower and upper quartiles at 0.01 and 0.4 M$_{\odot}$, respectively. Similar to $\rho_{1000}$, we find one protostar (\dummyBAEDC~) that has zero envelope mass. There are eight protostars that have extremely low envelope mass ($M_{2500} < 10^{-3}$), four of which are Class I and four are flat-spectrum sources. These eight protostars are likely in the later evolutionary stages where only a low-density envelope providing residual mass infall is present.  

Figure \ref{fig:mass}b shows the distribution of M$_{2500}$ for different SED classes. The median M$_{2500}$ for Class 0, I and flat-spectrum sources are 0.27, 0.02 and 0.01 M$_{\odot}$, respectively. Similarly, the mean M$_{2500}$ for Class 0, I and flat-spectrum sources are 0.87, 0.19 and 0.06 M$_{\odot}$, respectively. As expected, the envelopes around Class 0 protostars are significantly more massive than the other two classes. The AD test between the Class 0 and the other two classes shows that their distribution differs significantly ($p$-value $<< $ 1\%). The mean M$_{2500}$ for Class I protostars is a bit higher than those for flat-spectrum sources, however, the AD test between the Class I and flat-spectrum protostars show that there is no significant difference between the two distributions ($p$-value $\sim$ 70\%). This is consistent with an analysis by \cite{Federman22} of 870 $\micron$ continuum ALMA data which showed that that all SED classes contain a range of envelope masses, but that Class 0 protostars were typically the sources with the most massive envelopes.

\subsubsection{Total luminosity}
\label{sec:ltot}

\begin{figure}[ht!]
\centering
\includegraphics[scale=0.37]{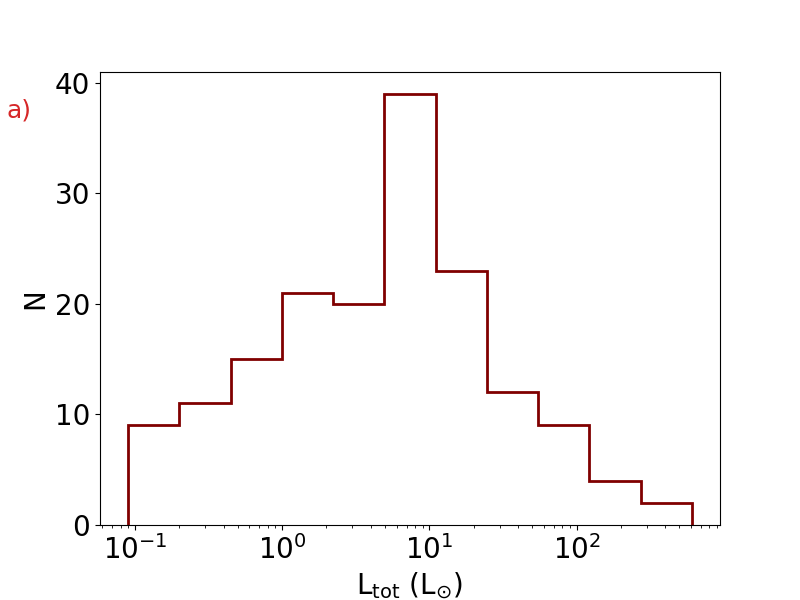}
\includegraphics[scale=0.37]{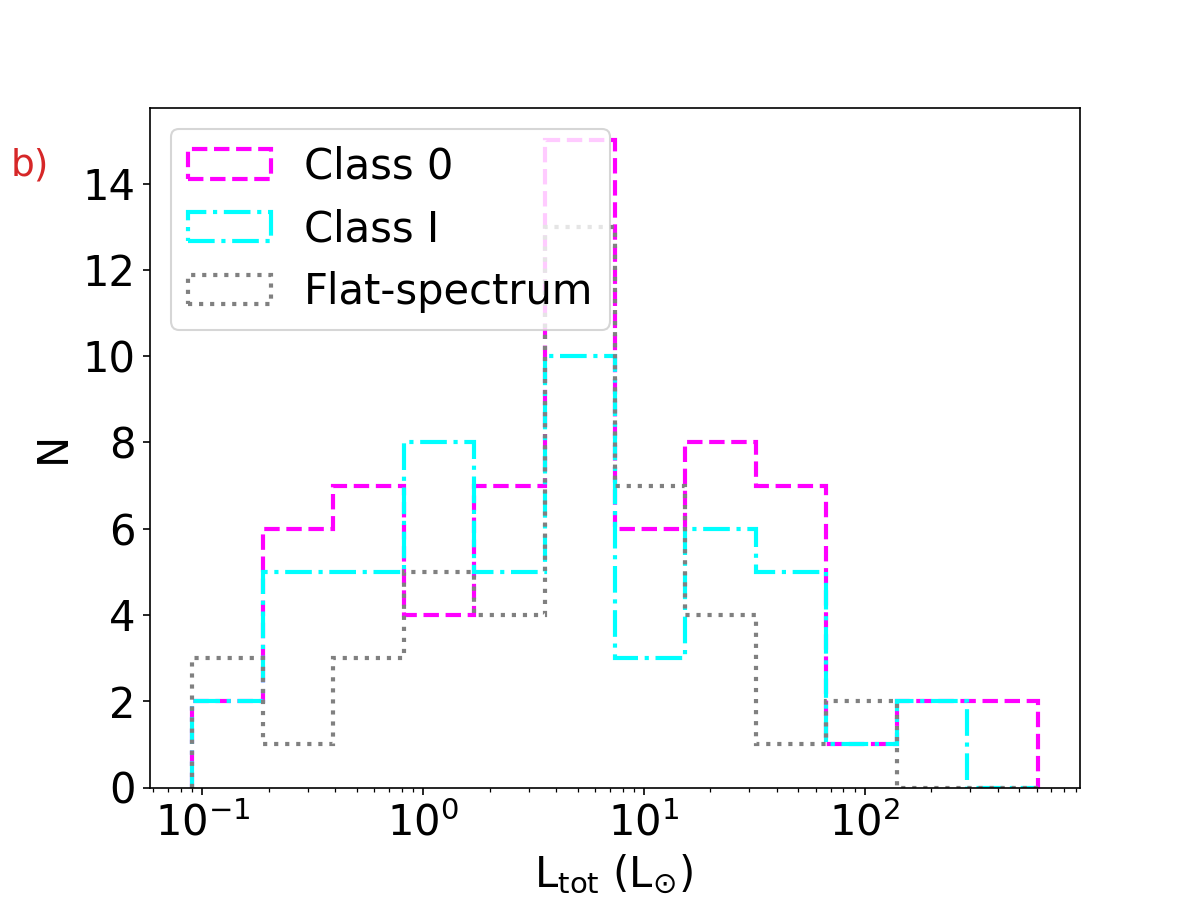}
\caption{Panel a) Histogram of the model-derived total luminosities (\Ltot) from the best fits of protostars in the eHOPS sample for Aquila. We find that \Ltot~peaks at $\sim$5.4 \Lsun. Panel b) Same as Panel a but for different evolutionary classes. AD test shows no significant difference in the distribution of \Ltot~for different SED classes.
\label{fig:ltot}}
\end{figure}

The total luminosity, or \Ltot, is the sum of the intrinsic photospheric luminosity from the central protostar and the accretion luminosity from accreting gas landing on the protostar.  \citep[There is also a contribution to the accretion luminosity by the disk itself, but this is typically negligible except during episodes of rapid accretion, e.g.][]{Hartmann2016}.  Figure \ref{fig:ltot}a shows the histogram of the total luminosity, L$_{\rm{tot}}$ obtained from the best fits.  The values for L$_{\rm{tot}}$ vary from 0.09 to 606.9 L$_{\odot}$, which is expected for maximum \Ltot\ as the model grid spans luminosities from 0.1 to 303.5 L$_{\odot}$ and the scaling factor varies from 0.5 to 2.
One protostar, \dummyBCADE\, hits the allowed luminosity ranges at the higher end of 606.9 L$_{\odot}$. The protostar \dummyBCADE\ is an extremely luminous Class I source with $L_{\rm{bol}} \sim 160$ L$_{\odot}$. It might be possible to obtain a better fit for \dummyBCADE\ by expanding the luminosity range in the model grid.
Overall, the histogram is strongly peaked with a median at $\sim$5.4 L$_{\odot}$, and the lower and upper quartiles at 1.5 and 15.7 L$_{\odot}$. We will compare L$_{\rm{tot}}$ for the eHOPS Aquila protostars with HOPS Orion protostars in Sec.~\ref{sec:hopsvsehops}. A comparison with all protostars in the nearest 500 pc region will be included in the next paper.

Upon inspecting L$_{\rm{tot}}$ for different classes of protostars in Aquila (Figure \ref{fig:ltot}b), we find median L$_{\rm{tot}}$ of 5.7, 3.9 and 5.8 L$_{\odot}$ for Class 0, I and flat-spectrum protostars, respectively. Due to the approximate symmetry of the distribution in log(L$_{\rm tot})$, the mean values of L$_{\rm{tot}}$ are much higher: 39.1, 30.9, and 11.9 L$_{\odot}$ for Class 0, I and flat-spectrum protostars, respectively.  The range of L$_{\rm{tot}}$ is similar for Class 0 and Class I protostars. For flat-spectrum protostars, the minimum L$_{\rm{tot}}$ is same as for other classes but the maximum L$_{\rm{tot}}$ is $\sim 115$ L$_{\odot}$ (instead of $\sim 607$ L$_{\odot}$ such as for Class 0 and Class I sources). Despite these differences, the $p$-value from the AD test indicates no significant difference in the distribution of L$_{\rm{tot}}$ for different SED classes.  We discuss the evolution of luminosity in Sec.~\ref{sec:ev_lum}.

\Ltot~for the best fit models typically differ from the observed bolometric luminosity (L$_{\rm bol}$) due to foreground extinction and the beaming of radiation out the outflow cavities \citep[\citetalias{Furlan16}, ][]{Whitney03b}.
L$_{\rm{bol}}$ is calculated by integrating the observed SEDs and assuming the protostar radiates isotropically in all directions. At high inclination angles, the protostellar luminosity is obscured by the intervening disk and envelope material, so less flux is received and therefore the total luminosity is higher than the bolometric luminosity. In contrast, at low inclination angle, the total luminosity is lower than the bolometric luminosity; this is due to the beaming of radiation through the outflow cavities. Since low inclination angles are unlikely for randomly oriented protostars, cases where $\Ltot < \Lbol$ are less common. Furthermore, while inclination can increase or decrease \Lbol, the foreground extinction always reduces the \Lbol.  This makes ratio L$_{\rm{tot}}$/L$_{\rm{bol}}$ less than one even rarer.  

\begin{figure}[t!]
\centering
\includegraphics[scale=0.42]{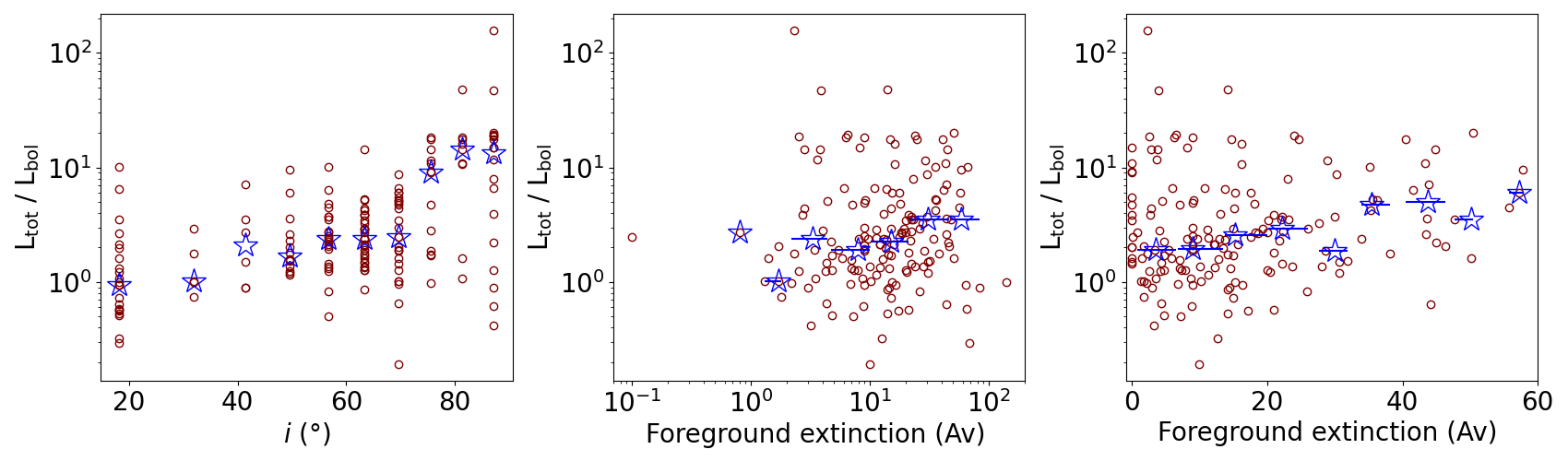}
\caption{Variation in the ratio of total luminosity to bolometric luminosity with inclination angle (left panel) and foreground extinction (middle panel) for the eHOPS protostars in Aquila. The right panel the the same as middle panel but in linear units of the foreground extinction. Blue stars mark the median L$_{\rm{tot}}$/L$_{\rm{bol}}$ ratio for specific bins of $i$ and $A_{\rm{V}}$ and horizontal line on stars show the range of $A_{\rm{V}}$ bin. As the inclination angle of protostars increases (towards edge-on configuration), the ratio \Ltot/\Lbol also increases. A similar trend is seen as an effect of foreground extinction.
\label{fig:lvsincl}}
\end{figure}

For the above reasons, the ratio of L$_{\rm{tot}}$ to L$_{\rm{bol}}$ is expected to increase with both inclination angle and foreground extinction, although with a large scatter. This is demonstrated in Figure \ref{fig:lvsincl}. The left panel of the figure shows the variation of the ratio L$_{\rm{tot}}$/L$_{\rm{bol}}$ with inclination angles obtained from the best fits. The median L$_{\rm{tot}}$/L$_{\rm{bol}}$ for each bin of inclination angle is shown by blue star. The median L$_{\rm{tot}}$/L$_{\rm{bol}}$ has an increasing trend with inclination angle. 
There are protostars with lower inclination angles that also have smaller cavity opening angles and therefore have higher Ltot/Lbol ratios. 
The middle panel shows the variation of L$_{\rm{tot}}$/L$_{\rm{bol}}$ with foreground extinction $A_{\rm{V}}$. The blue stars show the median L$_{\rm{tot}}$/L$_{\rm{bol}}$ in logarithmic bins of $A_{\rm{V}}$ with the bin ranges shown by horizontal blue lines on the top of the blue stars (median values). 
For the sources with best fit $A_{\rm{V}} > 1$ mag, we see mostly an increasing trend for L$_{\rm{tot}}$/L$_{\rm{bol}}$ with $A_{\rm{V}}$. 
The third panel is similar to the second panel, but it focuses on foreground extinction values up to 60 and shows them on a linear scale with linear bins. The trend of increasing L$_{\rm{tot}}$/L$_{\rm{bol}}$ with $A_{\rm{V}}$ is apparent in this panel too, where the median L$_{\rm{tot}}$/L$_{\rm{bol}}$ increases from 2.1 to 6.1 from the lowest to the highest $A_{\rm{V}}$ bins.

\begin{deluxetable*}{ccccccc} 
\tablecaption{Protostar classification in Aquila and Orion \label{tab:aql_orn}} 
\tablewidth{0pt} 
\tablehead{ 
\colhead{Cloud} & \colhead{PBRS} & \colhead{Class 0} & \colhead{Class I} & 
\colhead{Flat-spectrum} & \colhead{Class II} & \colhead{Total} 
} 
\decimalcolnumbers 
\startdata 
Aquila & 10 (6\%) & 61 (35\%) & 54 (31\%) & 43 (25\%) & 4 (2\%) & 172 \\ 
Orion & 19 (6\%) & 73 (22\%) & 125 (38\%) & 102 (31\%) & 11 (3\%) & 330 \\ 
\hline 
Orion A & 5 (2\%) & 55 (22\%) & 93 (37\%) & 88 (35\%) & 11 (4\%) & 252 \\ 
Orion B & 14 (18\%) & 18 (23\%) & 32 (41\%) & 14 (18\%) & 0 (0\%) & 78 \\ 
\enddata 
\tablecomments{Although PBRS and Class 0 are tabulated separately, we combine them together in the analyses in this paper, unless otherwise stated.} 
\end{deluxetable*}

\section{Discussion: Protostellar Evolution in Aquila and Orion}
\label{sec:hopsvsehops}

In this section, we conduct an initial comparison of the HOPS and eHOPS protostars. The HOPS sample is the protostars with Herschel/PACS 70~$\mu$m detections that inhabit the Orion~A and Orion~B clouds but excludes the Orion Nebula region which is too bright in the far-IR to extract reliable 1-870~$\mu$m SEDs. The eHOPS sample consists of the Herschel PACS detected protostars in Aquila North and Aquila South. Both the Orion and Aquila clouds are at similar distances: $\sim$420 pc for Orion \citepalias{Furlan16} and $\sim$436 pc for the Aquila clouds. Aquila has a similar gas mass to Orion; the Aquila clouds have a gas mass of 8.5 $\times$ 10$^{4}$ \Msun~ compared to 7.2 $\times$ 10$^{4}$ \Msun~ gas mass in the Orion clouds as measured inside the 1 A$_{\rm{V}}$ equivalent contour \citep{Pokhrel20}. The Orion clouds are about twice the size of Aquila clouds when measured in sky projection. The spatial size of the Aquila clouds as mapped by Herschel is 3 $\times$ 10$^{3}$ pc$^2$, whereas Orion clouds cover 6 $\times$ 10$^{3}$ pc$^2$ \citep{Pokhrel20, Pokhrel21}. Orion harbors rich star clusters with high-mass stars in the Orion Nebula Cluster (ONC) and the smaller NGC 2024 HII region. The Aquila clouds contain the W40 HII region which is similar to NGC~2024. 

Table \ref{tab:aql_orn} shows the number of protostars identified in the eHOPS-Aquila and HOPS samples, dividing the protostars into the SED-based evolutionary classes of protostars. We sub-divide Class 0 protostars into PBRS and more evolved Class~0 protostars; this will be discussed in Sec.~\ref{sec:pbr}. Several Class II objects are also included for both clouds.  
We categorize these as protostars and include them in the modeling exercise, the reasoning for which is provided in Sec.~\ref{sec:step3}. (Also see \citetalias{Furlan16} for a discussion of the HOPS Class II sources.) The HOPS protostars are further subdivided into those found in the Orion~A or Orion~B clouds. Despite having a similar gas mass concentrated in a region only half the projected size of Orion, Aquila harbors half as many protostars as Orion.  There are other differences.  For eHOPS-Aquila, 41\% of the total number of protostars are Class 0 protostars or PBRS. The remaining protostars are roughly equally split between Class I and flat-spectrum sources. In contrast, the HOPS-Orion protostars are roughly split equally between the three evolutionary classes \citepalias{Furlan16}.  

Despite our intention to analyze the HOPS and eHOPS data in a uniform manner to minimize biases, there are significant differences in the way each program identifies protostars and measures photometry in the wavelength range between 200 and 850 $\micron$. First, the HOPS observations used Herschel/PACS to observe protostars identified in the  Spitzer data by \citet{Megeath12}, \citet{Kryukova12}, and \citet{Megeath16} (See Sec.~\ref{sec:ps}).  These were supplemented by the Herschel identified PBRS and other protostars that serendipitously fell in the HOPS fields \citep[(Sec.~\ref{sec:pbr},]{Fischer20}. The HOPS program obtained medium-scan rate 70~$\mu$m data, and the protostar sample was required to have 70~$\mu$m detections.  In contrast,  eHOPS relies on 100~$\mu$m detections since the Aquila 70~$\mu$m data, which was obtained in fast-scan parallel observations,  has a lower sensitivity and angular resolution. The 100~$\mu$m data, in turn, has a lower angular resolution than the HOPS 70~$\mu$m data. Furthermore, in the long wavelengths range ($>$ 200 $\micron$), eHOPS-Aquila uses the low angular resolution ($>18''$) SPIRE data for data at 250, 350, and 500~$\mu$m, while HOPS used the lower sensitivity, higher angular resolution (7'') 350~$\mu$m APEX data. In the analysis of the eHOPS-Aquila data, the SPIRE photometry is used to calculate \Tbol\ and \Lbol\, but the two points at 350 and 500 $\micron$ are used as upper limits in the model fits. This may lead to biases in \Tbol\ and \Lbol\ due to extended dust emission in the SPIRE beams. Finally, almost all HOPS protostars have IRS spectra, while only a fraction of the eHOPS-Aquila protostars have spectra. This is particularly important for both classifying sources as protostars and constraining the fits. We will investigate potential biases in the evolutionary diagnostics and model fits in this discussion.

\subsection{Properties of PBRS in Aquila and Orion} \label{sec:pbr}

Within the HOPS sample, \citet{Stutz13} and \citet{Tobin15} identified 19 protostars in the Orion molecular clouds distinguished by their high 70~$\mu$m to 24~$\mu$m flux ratios.  Twelve of these protostars were not identified as protostars by the Spitzer data alone due to the faint or undetectable 24~$\mu$m emission. Most of these protostars appear to be young Class~0 protostars \citep{Stutz13,Tobin15,Tobin16}. Based on high angular resolution millimeter VLA and ALMA data, \citet{Karnath20} argued that four of these protostars were less than 10000 years old.  Of these four objects, two may be binary systems that include Class 0 protostars, one may contain a first-stage hydrostatic core, and the fourth may be in the process of forming a hydrostatic core. These are among the youngest protostars known. 

\begin{figure}[h]
\centering
\includegraphics[scale=0.43]{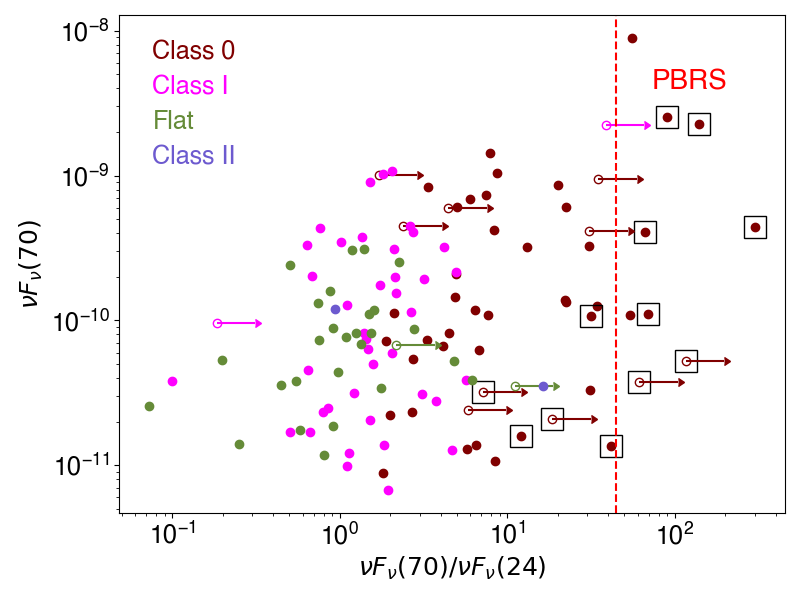}
\includegraphics[scale=0.43]{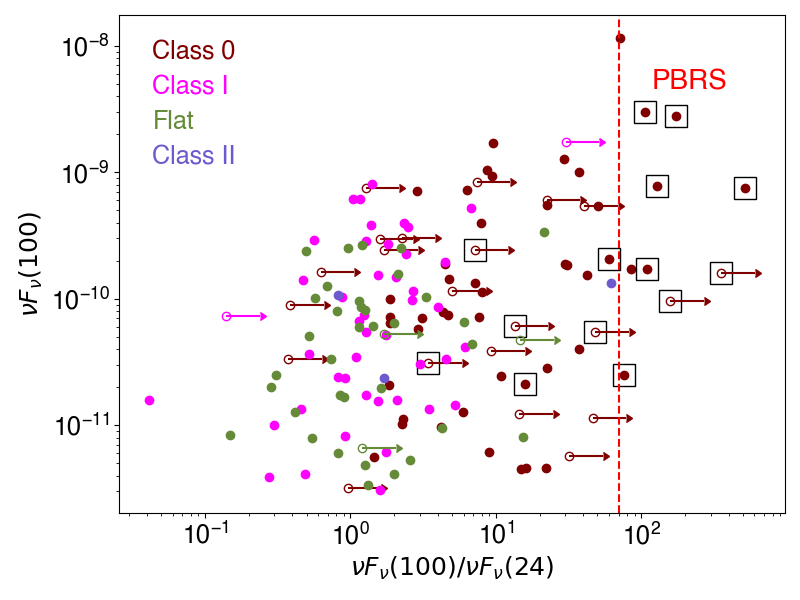}
\caption{Panel a) Variation of $\nu F_{\nu}(70)$ with $\nu F_{\nu}(70)/\nu F_{\nu}(24)$. Panel b) Variation of $\nu F_{\nu}(100)$ with $\nu F_{\nu}(100)/\nu F_{\nu}(24)$. In both panels, different colored symbols show different evolutionary classes as noted in the legend. Black open boxes show the extremely red protostars that are detected in Sec~\ref{sec:step4} and \ref{sec:step5}. Open circles denote that the protostars are not detected at 24 $\micron$ so the corresponding completeness limit is used as an upper limit flux. Filled circles denote the protostars where 24 $\micron$ photometry is available. The red vertical dash line in both panels mark the criteria that we use to find PBRS; $\nu F_{\nu}(70)$/$\nu F_{\nu}(24)$ = 10$^{1.65}$ for panel a (from \citealt{Stutz13}) and $\nu F_{\nu}(100)$/$\nu F_{\nu}(24)$ = 10$^{1.85}$ for panel b (see Sec.~\ref{sec:step4}).
\label{fig:pbr_70_100}}
\end{figure}

After Orion, Aquila is the 2nd molecular cloud complex that has been systematically searched for PBRS using the criterion established by \citet{Stutz13}. In Sec.~\ref{sec:step4}, we find eight extremely red protostars using Equation \ref{eq:redps} and six protostars of a similar nature using the completeness limit for the MIPS 24 $\micron$ flux in Sec.~\ref{sec:step5}.  \cite{Stutz13} defined the PBRS using the $\nu F_{\nu}(70)$/$\nu F_{\nu}(24)$ ratio; this was established using a grid of protostar models to determine the ratios that implied high envelope densities. We follow \cite{Stutz13} and define the PBRS are the protostars for which $\nu F_{\nu}(70)$/$\nu F_{\nu}(24) > 10^{1.65}$. Figure \ref{fig:pbr_70_100}a shows the  $\nu F_{\nu}(70)$ vs.~$\nu F_{\nu}(70)/\nu F_{\nu}(24)$  diagram.  The black open boxes mark the extremely red protostars that are detected using Equation \ref{eq:redps} in Sec.~\ref{sec:step4} and using completeness limits in Sec.~\ref{sec:step5}. In this figure, there are nine sources that satisfy the \cite{Stutz13} criterion and that are designated as PBRS.

We use the  PACS 100~$\micron$ data to find additional PBRS that were missed in the lower sensitivity 70 $\micron$ data. Figure \ref{fig:cc_cmd1}b shows the  $\nu F_{\nu}(100)$ vs. $\nu F_{\nu}(100)$/$\nu F_{\nu}(24)$ diagram for the eHOPS-Aquila protostars.  We determine the minimum $\nu F_{\nu}(100)$/$\nu F_{\nu}(24)$ ratio using the  PBRS identified in the 70 $\micron$ data. These PBRS satisfy the criteria $\nu F_{\nu}(100)/\nu F_{\nu}(24) > 10^{1.85}$.  Using this criteria, we identify one more PBRS in Figure \ref{fig:cc_cmd1}b.

In summary, using the $\nu F_{\nu}(70)$/$\nu F_{\nu}(24)$ and $\nu F_{\nu}(100)$/$\nu F_{\nu}(24)$ criteria in Figure \ref{fig:cc_cmd1}, we identify a total 10 PBRS in the eHOPS sample (\dummyAEDBC, \dummyBDEAC, \dummyBDCAE, \dummyBADCE, \dummyABDEC, \dummyAECDB, \dummyAECBD, \dummyBACED, \dummyAEDCB, and \dummyBEADC). Out of the 10 PBRS, two can be identified using Spitzer photometry alone (Equation \ref{eq:ps1}), six are newly identified extremely red protostars (Equation \ref{eq:redps}), and two are identified using the completeness limit at 24~$\micron$ in \ref{sec:step5}. 

\begin{figure}[t!]
\centering
\includegraphics[scale=0.3]{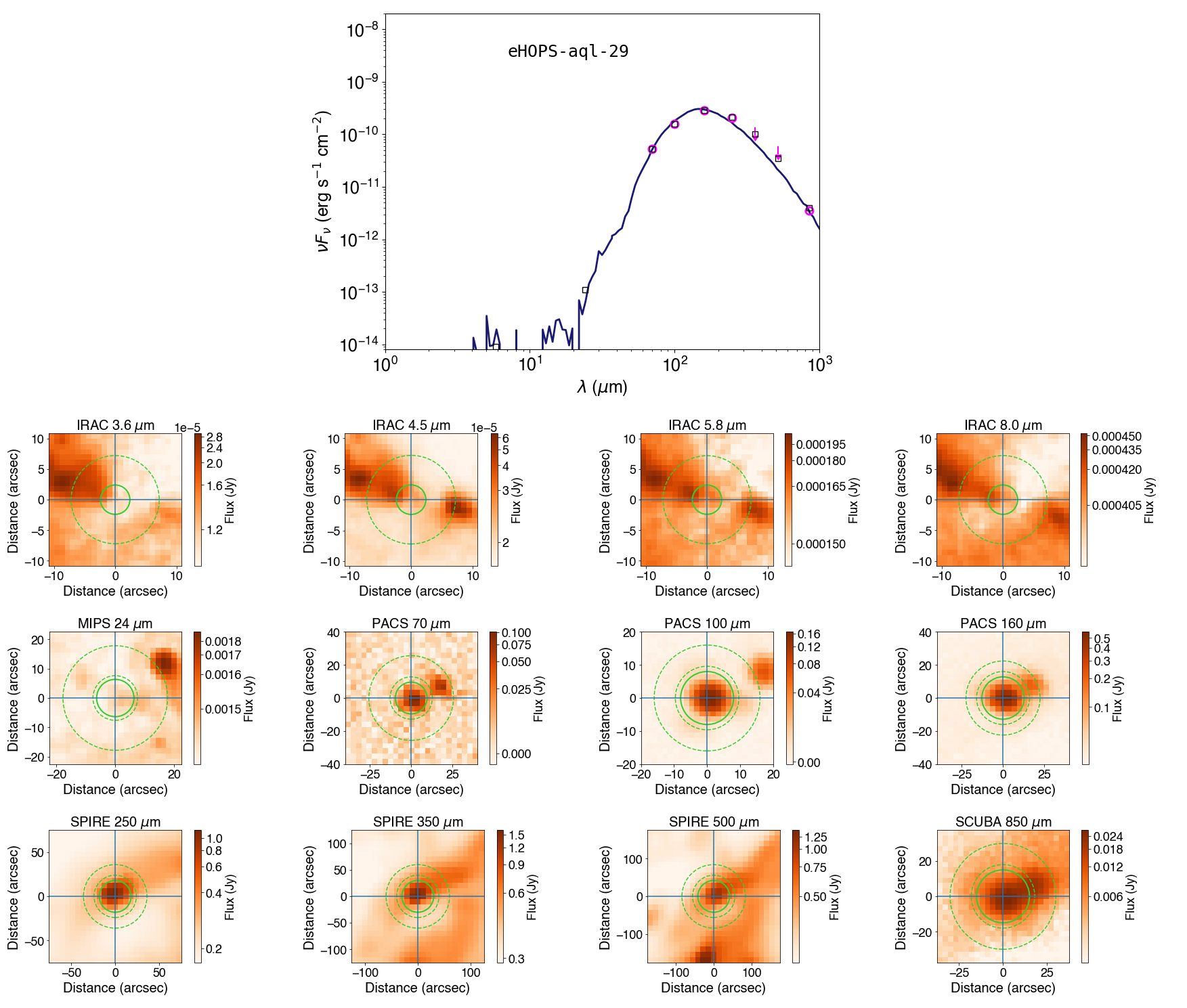}
\caption{The SED of a PBRS, \dummyBDEAC, along with the best fit model and image postages at different wavelengths from Spitzer/IRAC 3.6 $\micron$ to JCMT/SCUBA2 850 $\micron$. The color labels in the SED plot are the same as in Figure \ref{fig:bestfit}; magenta circles are detections and the best fit model photometry is overplotted with blue squares. The source \dummyBDEAC is detected only at $\geq$70 $\micron$. In the postage images, red and green circles show aperture and background annulus sizes used for aperture photometry, respectively. Similar figures for all the other PBRS are provided in the online version of the paper.
\label{fig:pbr}}
\end{figure}

As an example of a PBRS in the eHOPS Aquila catalog, Figure \ref{fig:pbr} shows the SED and postage stamp images of one of the PBRS in our sample, \dummyBDEAC. The source is not detected in any of the Spitzer wavebands in the $3.4-24 \micron$ wavelengths and is detected only by Herschel at $\geq70$~$\micron$.  This particular PBRS would not have been detected with just the Spitzer IRAC and MIPS detectors. The peak of the SED is between the PACS 100-160 $\micron$ bands. 

\begin{deluxetable*}{ccccc} 
\tablecaption{Properties of PBRS compared with Class 0 protostars in Aquila and Orion \label{tab:pbr}} 
\tablehead{ 
\colhead{Cloud} & \colhead{Aquila} & \colhead{Aquila} & \colhead{Orion} & \colhead{Orion} \\ 
\nocolhead{} & \colhead{Average} & \colhead{Median} & \colhead{Average} & \colhead{Median} \\ 
\colhead{Parameter} & \colhead{PBRS : Class 0} & \colhead{PBRS : Class 0} & \colhead{PBRS : Class 0} & \colhead{PBRS : Class 0} 
} 
\startdata 
T$_{\rm{bol}}$ (T$_{\odot}$) & 32.5 : 42.8 & 33.8 : 41.0&34.2 : 45.8&35.0 : 45.2 \\ 
L$_{\rm{bol}}$ (L$_{\odot}$) & 25.2 : 9.1 & 5.3 : 1.9&4.9 : 29.3&3.9 : 2.2 \\ 
L$_{\rm{tot}}$ (L$_{\odot}$) & 130.6 : 39.0 & 25.8 : 5.7&15.3 : 44.5&5.6 : 5.4 \\ 
$\rho_{1000}$ ($\times$ 10$^{-17}$ g~cm$^{-3}$) & 3.0 : 2.0 & 1.5 : 0.6&1.5 : 1.6&1.2 : 0.6 \\ 
M$_{2500}$ (M$_{\odot}$) & 1.4 : 0.9 & 0.6 : 0.3&0.7 : 0.6&0.7 : 0.2 \\ 
\enddata 
\tablecomments{Class 0 protostar sample in this table includes PBRS too.} 
\end{deluxetable*}

As shown in Table~\ref{tab:aql_orn},  Aquila has a rich population of PBRS.  The eHOPS-Aquila protostars have a similar fraction of PBRS as the HOPS protostars in Orion. In Orion, most of the PBRS are found in the Orion~B cloud \citep{Stutz13,Karnath20}.  The fraction of PBRS in Aquila is in between the low fractions of PBRS in Orion~A and the high fraction found for Orion~B.  To compare the properties of the HOPS and eHOPS PBRS, Table \ref{tab:pbr} lists the average and median values of the SED-based diagnostics and best-fit parameters for the PBRS and Class 0 protostars in Aquila and Orion. In Table \ref{tab:pbr}, the PBRS are included with the Class 0 protostars when calculating the properties of the Class 0 sample. 

In both clouds, the PBRS have a lower average and median \Tbol~compared to the whole sample of Class 0 protostars, as expected. The median values of \Tbol, \Lbol\, $\rho_{1000}$ and M$_{2500}$ are very similar between the PBRS in the Aquila and Orion clouds, and between the Class 0 protostars in those two clouds. Less agreement is seen for the average values; these may be affected by a few sources with very high luminosities or envelope masses.  Interestingly, the median values of \Ltot\ are higher in Aquila.
The higher \Ltot~values for Aquila PBRS may result from the different angular resolutions of the sub-mm data, which may favor fits with edge-on inclinations and higher \Ltot values for the eHOPS-Aquila sample. We will examine these potential biases in a future paper encompassing all of the molecular clouds in the eHOPS catalog. 

In general, we find very similar populations of PBRS in both clouds. The PBRS are distinguished from the typical Class~0 protostars by their lower \Tbol~values and higher envelope masses. These are consistent with the PBRS being young Class 0 protostars with dense envelopes and high infall/high accretion rates. Future ALMA and VLA observations are needed to search for very young ($< 10000$~yr) protostars similar to the four found in Orion.

\subsection{Total Luminosity vs. Bolometric Luminosity}

The total luminosity (\Ltot) of a protostar is the intrinsic luminosity of its photosphere summed with the luminosity generated from accretion onto the protostar. This can differ from the bolometric luminosity (\Lbol), which is calculated assuming that the protostar radiates isotropically with no correction for extinction.  As described in Sec.~\ref{sec:ltot}, the radiation of protostars is concentrated along the axis of the outflow cavities; this results in  $L_{\rm tot} > L_{\rm bol}$  for protostars observed a higher inclinations, and   $L_{\rm tot} < L_{\rm bol}$ for protostars observed near a face-on inclination.  In addition, foreground extinction lowers the value of \Lbol. In general, since face-on inclinations are less likely, and since foreground extinction always reduce \Lbol, we expect to see $L_{\rm tot} > L_{\rm bol}$ for most protostars (see Sec.~\ref{sec:ltot}).

\begin{figure}[t!]
\centering
\includegraphics[scale=0.44]{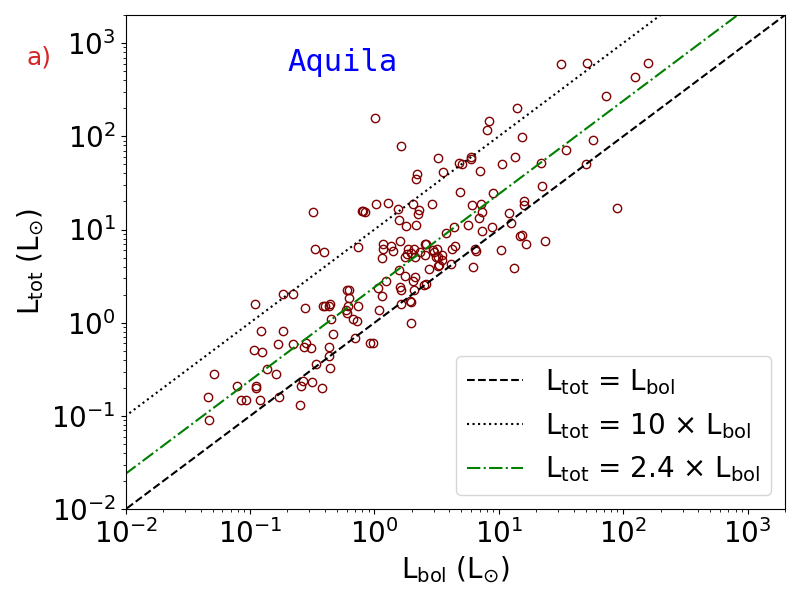}
\includegraphics[scale=0.44]{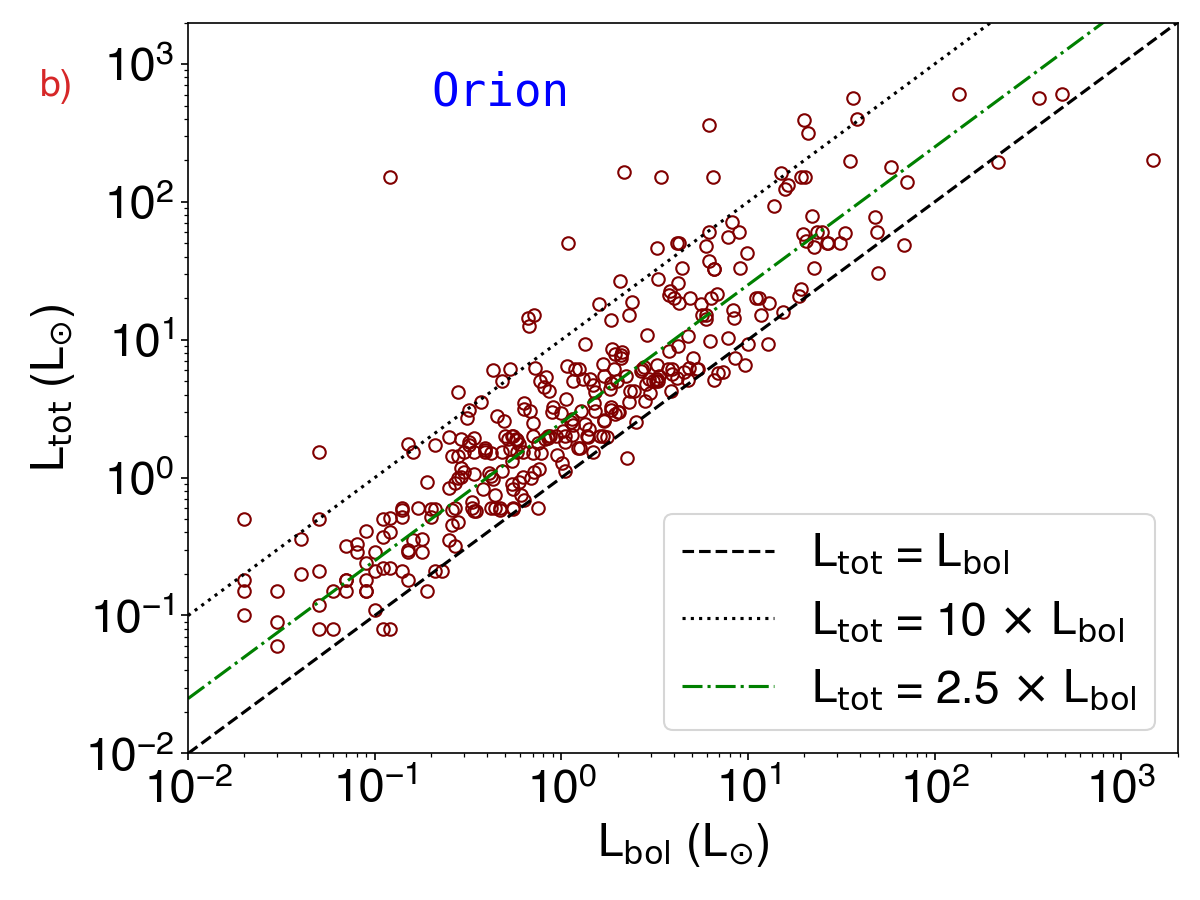}
\caption{Variation of bolometric luminosity (\Lbol) with total luminosity (\Ltot) for the eHOPS protostars in Aquila (Left panel) and HOPS protostars in Orion (Right panel). In both panels, green dash-dot line shows the median value of L$_{\rm{tot}}$/L$_{\rm{bol}}$, black dash line is the region where \Lbol\ and \Ltot\ are the same, and black dotted line represents \Ltot = 10 $\times$ \Lbol. The median L$_{\rm{tot}}$/L$_{\rm{bol}}$ is similar in Aquila (2.4) and Orion (2.5).
\label{fig:ltotvslbol}}
\end{figure}

Figure \ref{fig:ltotvslbol} compares the best fit  L$_{\rm{tot}}$ with the observed  L$_{\rm{bol}}$ for the protostars in the eHOPS-Aquila (left panel) and HOPS samples  \citep[right panel, from][]{Fischer17}. In both plots, the dashed line is the region where both L$_{\rm{tot}}$ and L$_{\rm{bol}}$ are equal. The median of L$_{\rm{tot}}$/L$_{\rm{bol}}$ are 2.4 and 2.5 for Aquila and Orion, respectively. The distribution of the sources is remarkably similar in both plots. In Aquila, the majority of the sources ($\sim$70\%) are between L$_{\rm{tot}}$/L$_{\rm{bol}}$ = 1 and  L$_{\rm{tot}}$/L$_{\rm{bol}}$ = 10. 
There are $\sim$15\% protostars for which L$_{\rm{tot}}$/$L_{\rm{bol}} < 1$; the median, best fit inclination angle for these protostars is $\sim$18$^{\circ}$ (more face-on). On the other end, there are another $\sim$15\% protostars for which L$_{\rm{tot}}$/L$_{\rm{bol}} > 10$; and their median,  best fit inclination angle is $\sim$82$^{\circ}$. These protostars are fit by models where the protostar is viewed nearly edge-on. The similarity of these L$_{\rm{tot}}$ vs L$_{\rm{bol}}$ diagrams suggest that there are no significant biases in the determination of luminosities due to the different bands used in the eHOPS and HOPS samples.

\begin{figure}[b!]
\includegraphics[scale=0.32]{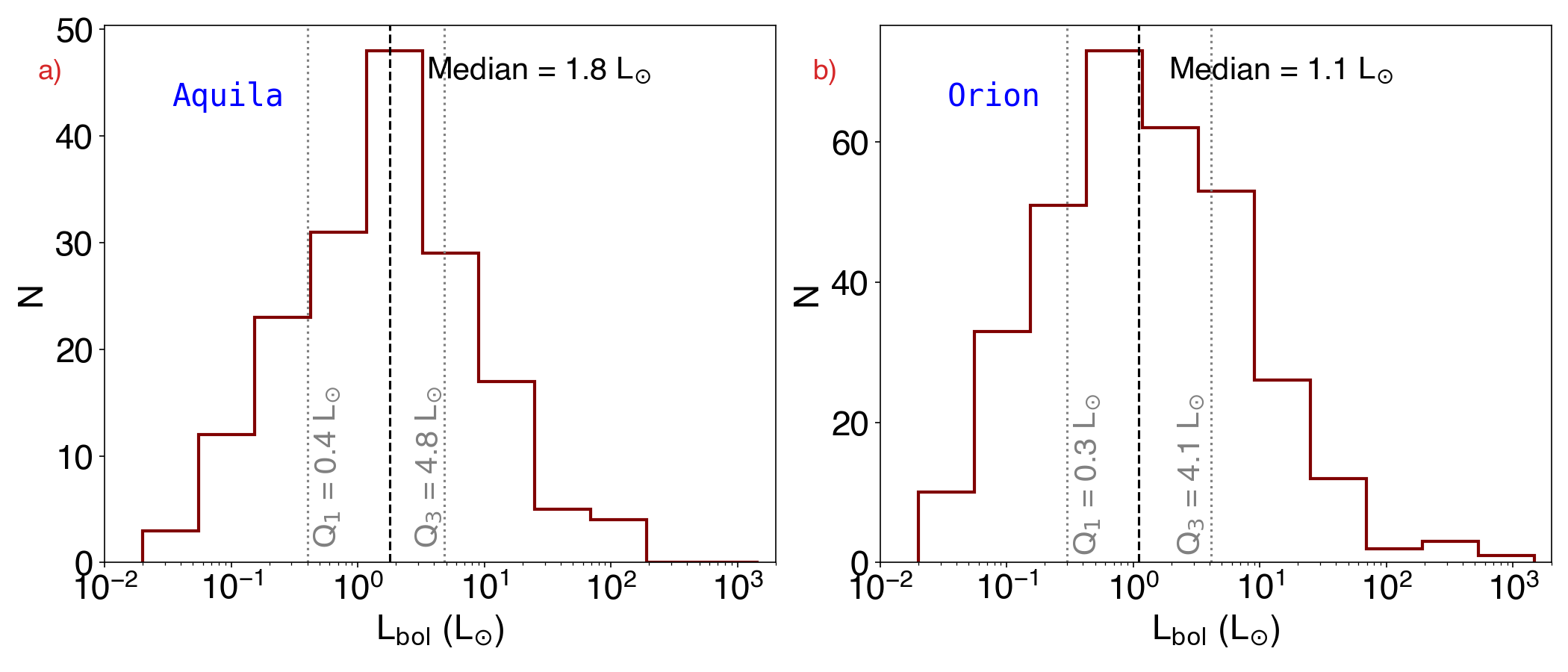}
\includegraphics[scale=0.32]{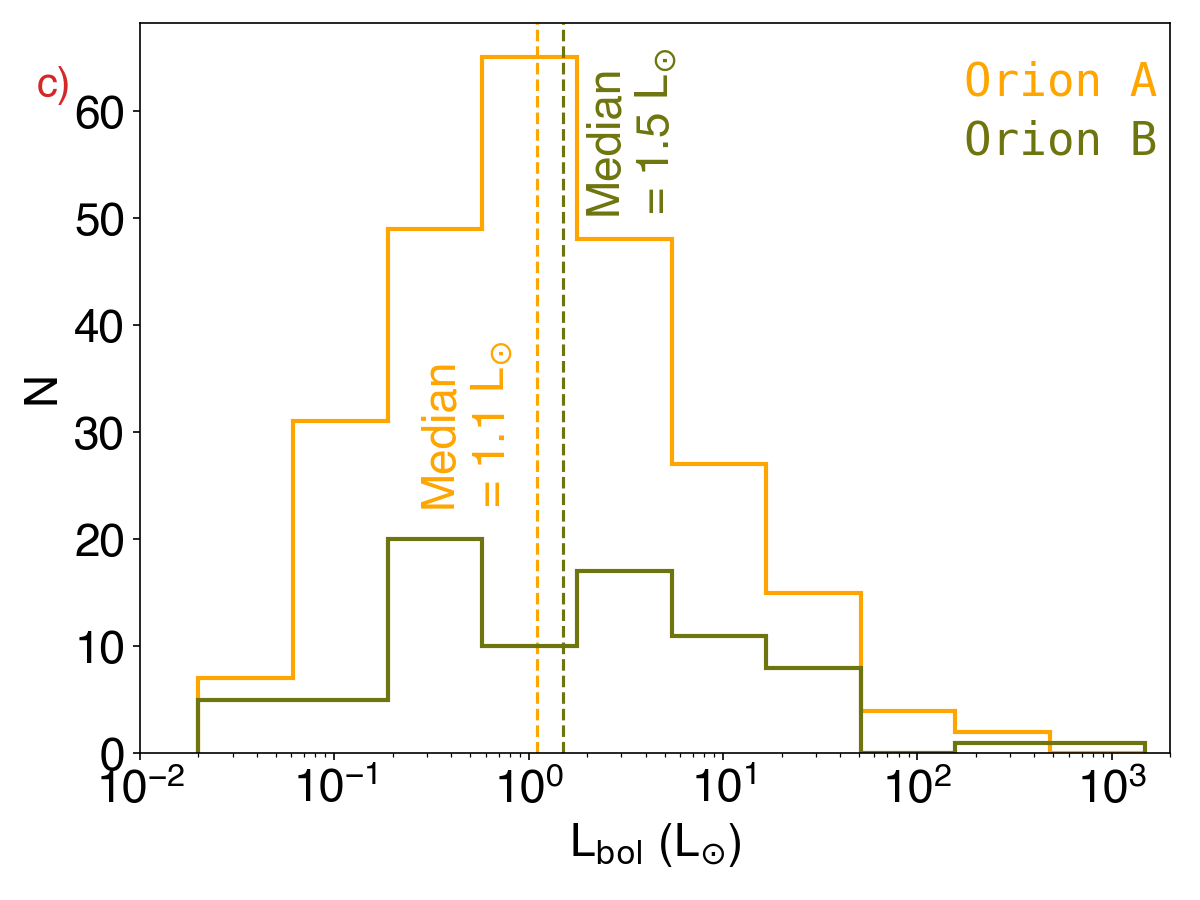}
\caption{The PLFs using L$_{\rm{bol}}$ for the eHOPS-Aquila protostars (panel a), HOPS protostars (panel b), and HOPS protostars separately for Orion A and Orion B (panel c). The median, the first quartile (Q1) and the third quartile (Q3) values of the distribution are given and indicated by vertical lines.
The AD test shows that the PLFs for Aquila and Orion B are likely drawn from the same parent distribution, but that this is unlikely for the Aquila and Orion A PLFs.
\label{fig:plf_1}}
\end{figure}

\subsection{The Protostellar Luminosity Function (PLF)} \label{sec:plf}

The protostellar luminosity function (hereafter:~PLF) places important constraints on the accretion history of protostars \citep{Dunham2010,Offner2011}. This function can be represented as the distribution of \Lbol\ or \Ltot\ for a population of protostars. Figures \ref{fig:plf_1}a, \ref{fig:plf_1}b, and \ref{fig:plf_1}c  show the distribution of \Lbol~for the eHOPS protostars in Aquila, the HOPS protostars in the entire Orion clouds (from \citetalias{Furlan16}), and separately for Orion A and Orion B, respectively. The PLFs in  Aquila and Orion have a similar form but with peak at different L$_{\rm{bol}}$. The PLF for Aquila peaks at $\sim 2$~L$_{\odot}$ but for Orion the PLF peaks at $\sim 1$~L$_{\odot}$. Similarly, the median L$_{\rm{bol}}$ for the eHOPS protostars in Aquila is 1.8 L$_{\odot}$, which is higher than the median L$_{\rm{bol}}$ of 1.1 L$_{\odot}$ for the HOPS sources in Orion. From Figure \ref{fig:plf_1}c, the median of the PLF for Orion A is the same as the median for the whole Orion clouds, 1.1 \Lsun\ while for Orion B L$_{\rm{bol}}$ is higher (1.5 \Lsun). However, Orion B is asymmetric and peaks at a luminosity $< 1$~\Lsun.

\begin{figure}[t!]
\includegraphics[scale=0.32]{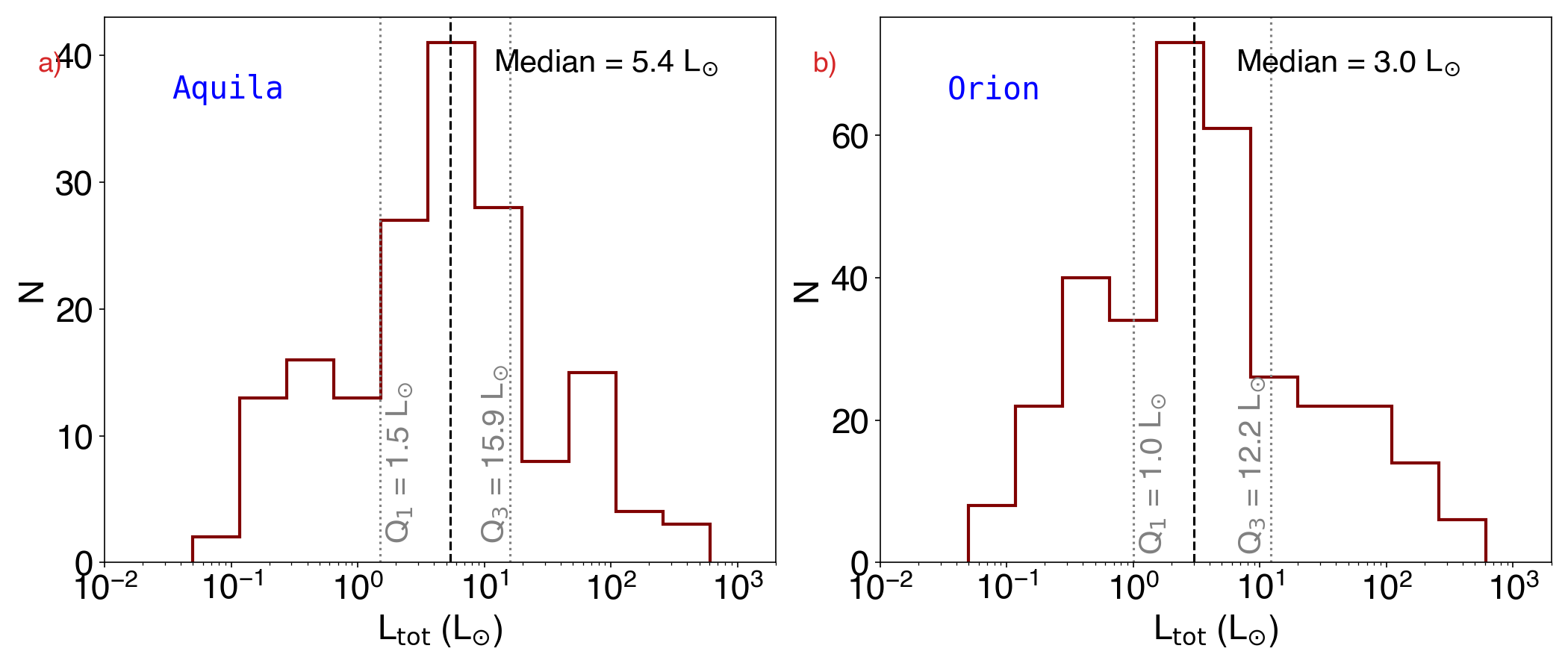}
\includegraphics[scale=0.32]{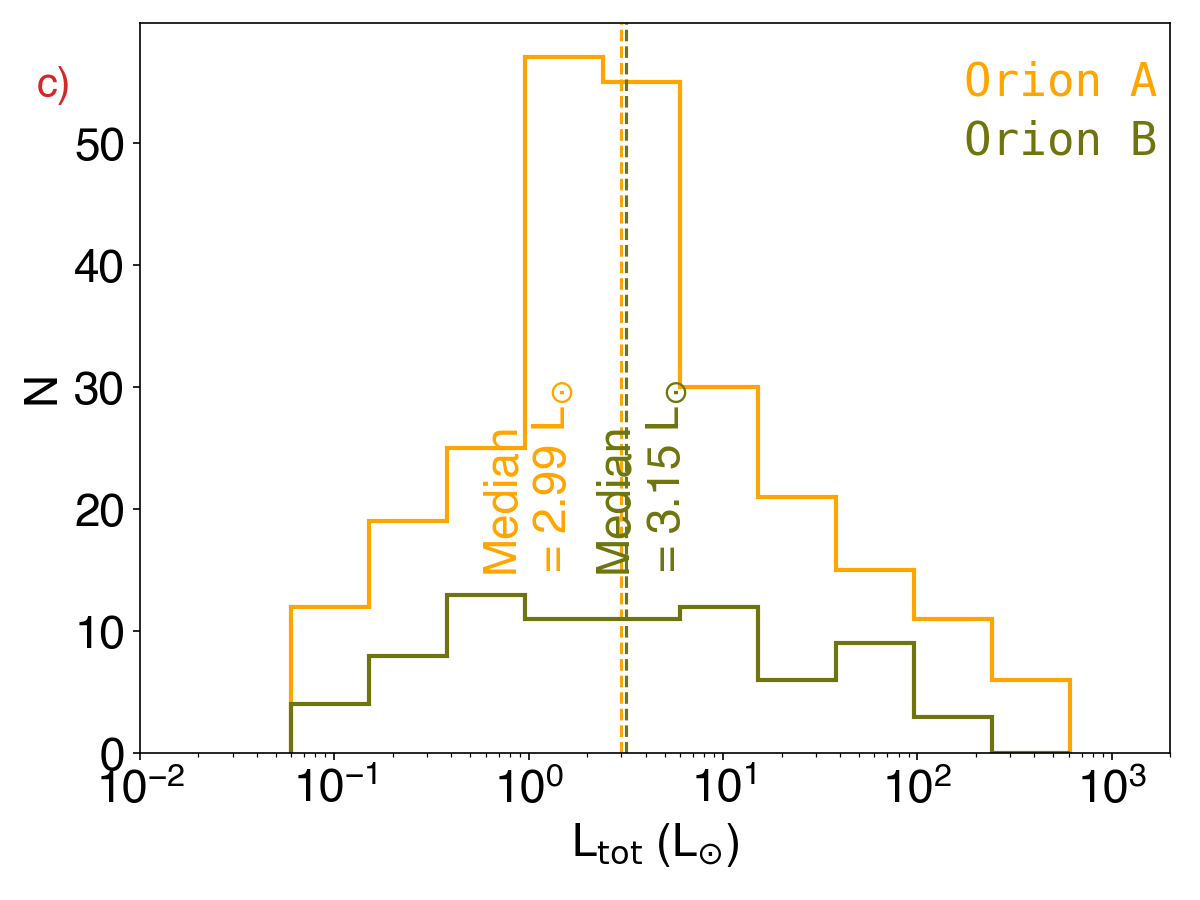}
\caption{The PLFs using L$_{\rm{tot}}$ for the eHOPS-Aquila protostars (panel a), HOPS protostars (panel b), and HOPS protostars separately for Orion A and Orion B (panel c). The median, the first quartile (Q1) and the third quartile (Q3) values of the distribution are given and indicated by vertical lines.
The AD test indicates that we cannot rule out they are drawn from the same parent population.
\label{fig:plf_2}}
\end{figure}

The differences in the PLFs are also evident when plotting in terms of the total luminosity, shown in Figure \ref{fig:plf_2}. \Ltot~takes into account the effects of inclination and extinction; however, it is also model dependent. Moreover, biases and degeneracies in the models can affect the best fit values. 
The median \Ltot\ for Aquila and Orion are 5.4~L$_{\odot}$ and 3.0~L$_{\odot}$, respectively. The ratio of the median \Ltot\ between Aquila and Orion is the same as the ratio of the median \Lbol\ between those clouds (1.8 and 1.9, respectively). The median \Ltot~for both Orion A and Orion B are 3~\Lsun.

\begin{figure}[b!]
\includegraphics[scale=0.45]{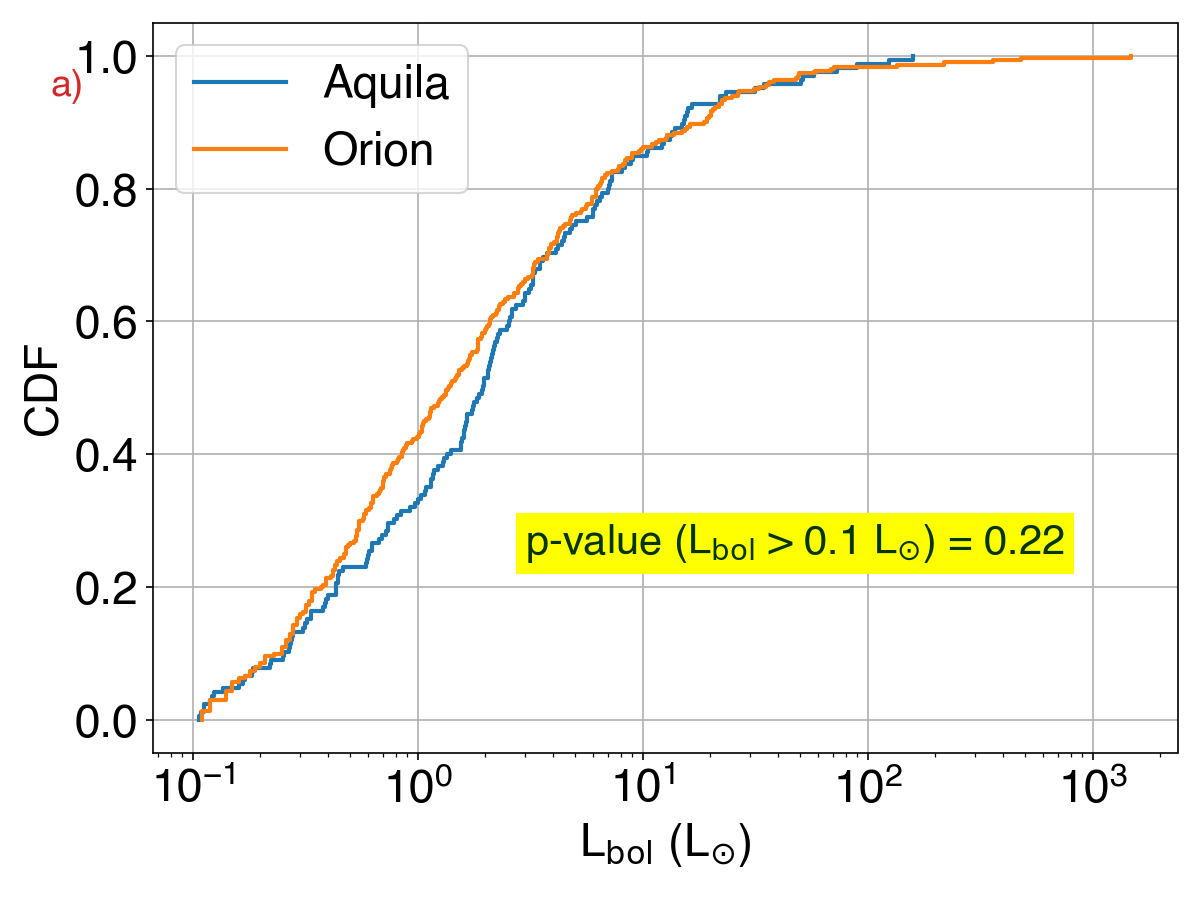}
\includegraphics[scale=0.45]{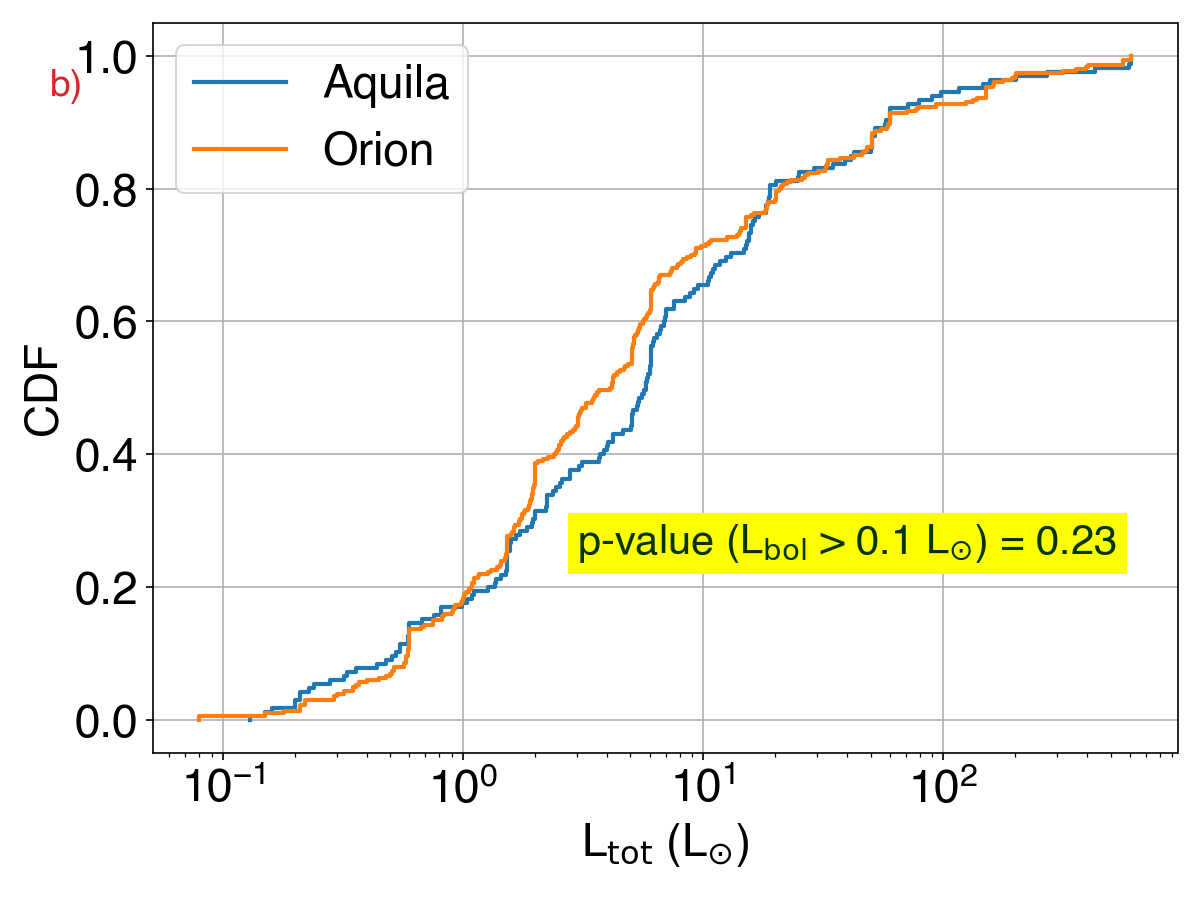}
\caption{Cumulative distribution function (CDF) of L$_{\rm{bol}}$ for protostars in Aquila and Orion molecular clouds in panel a. CDF for \Ltot~for the same clouds in panel b. AD test shows no significant difference in the distributions of \Lbol~and \Ltot~in Aquila and Orion for protostars with \Lbol~$>$ 0.1 \Lsun.
\label{fig:plf_ks}}
\end{figure}

Comparing the PLFS of the eHOPS-Aquila and HOPS samples, the AD test gives a p-value of 0.03 implying we can reject the null hypothesis that L$_{\rm{bol}}$ in Aquila and Orion are drawn from the same distribution at a 3\% significance level. However, there are potentially different completeness levels in the Herschel data used in HOPS and in the eHOPS-Aquila survey which must be considered. To mitigate differences in the completeness and to minimize any residual extragalactic contamination,  we perform the tests in protostars on Aquila and Orion for which L$_{\rm{bol}} > 0.1$ L$_{\odot}$. Figure \ref{fig:plf_ks}a shows the cumulative distribution function of L$_{\rm{bol}}$ for the protostars in Aquila and in Orion.  For the PLFs with L$_{\rm{bol}}$ for  $> 0.1$ L$_{\odot}$, Aquila and Orion are likely drawn from the same distribution with a $p$-value of 0.22 using the AD test. Thus, for the $> 0.1$ L$_{\odot}$ protostars, we cannot rule out that the distributions of L$_{\rm{bol}}$ for Aquila and Orion were drawn from the same parent sample. The AD test also shows that the distributions of \Lbol\  in Aquila and Orion B are likely drawn from the same distribution ($p$-value $\geq$ 0.25) but that Aquila and Orion A are not ($p$-value = 0.04). Also shown in Figure \ref{fig:plf_ks}b is the cumulative distribution function of \Ltot. Similar to \Lbol, we perform the AD test on protostars with \Ltot\ $> 0.1$~L$_{\odot}$ and find $p$-value = 0.23.  Furthermore, contrary to \Lbol, the distributions of \Ltot\ for Aquila, Orion A, and Orion B are likely drawn from the same distribution. 

In summary, we see higher medium and peak luminosities for both \Lbol~and \Ltot\ in Aquila compared to Orion. Statistical tests imply, however, that if we consider protostars above 0.1~\Lsun -- at which we have higher confidence that both data sets are complete -- we cannot reject the null hypothesis that the PLFs in Aquila and Orion are drawn from the same parent distribution. There is a possible exception in that the \Lbol~distributions of Aquila and Orion~A have a relatively small probability (4\%) of being drawn from the same population.

\begin{figure}[t!]
\includegraphics[scale=0.44]{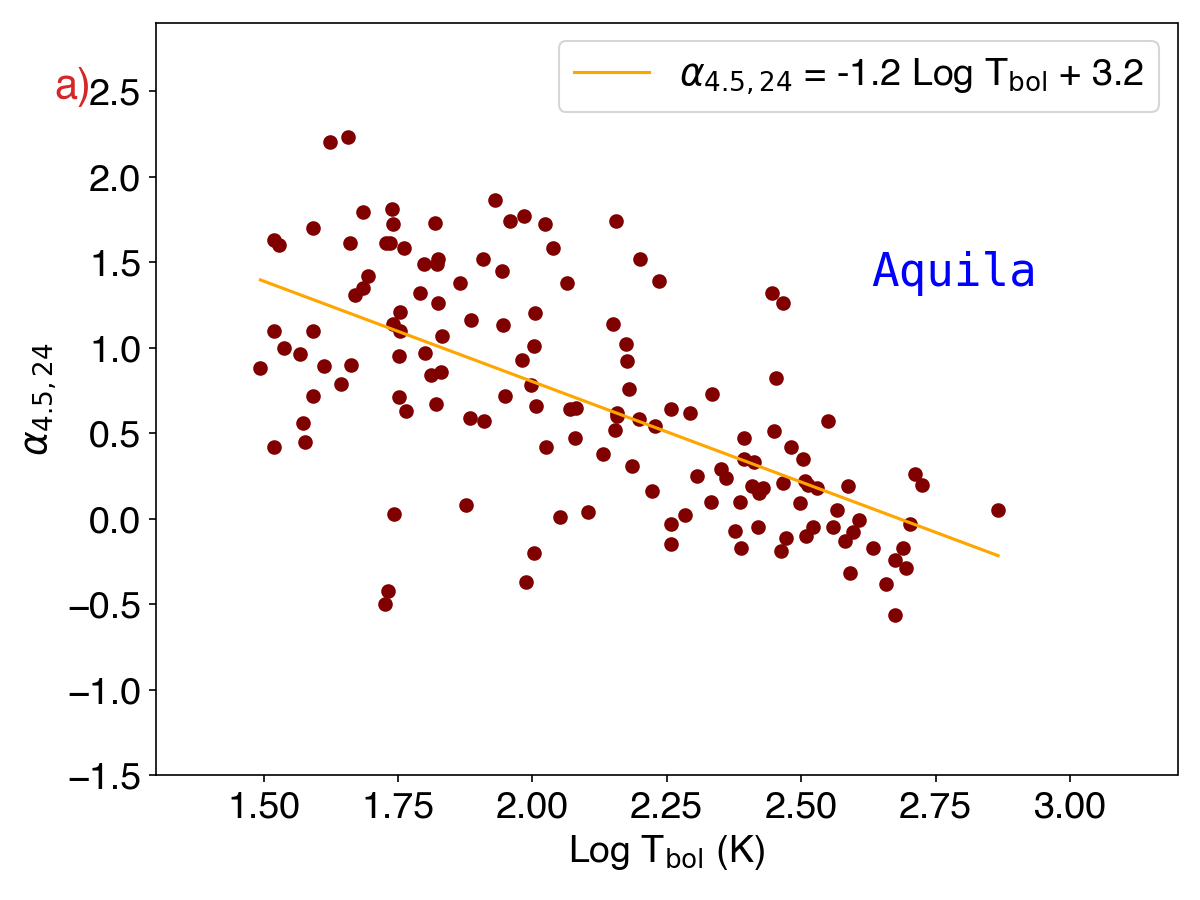}
\includegraphics[scale=0.44]{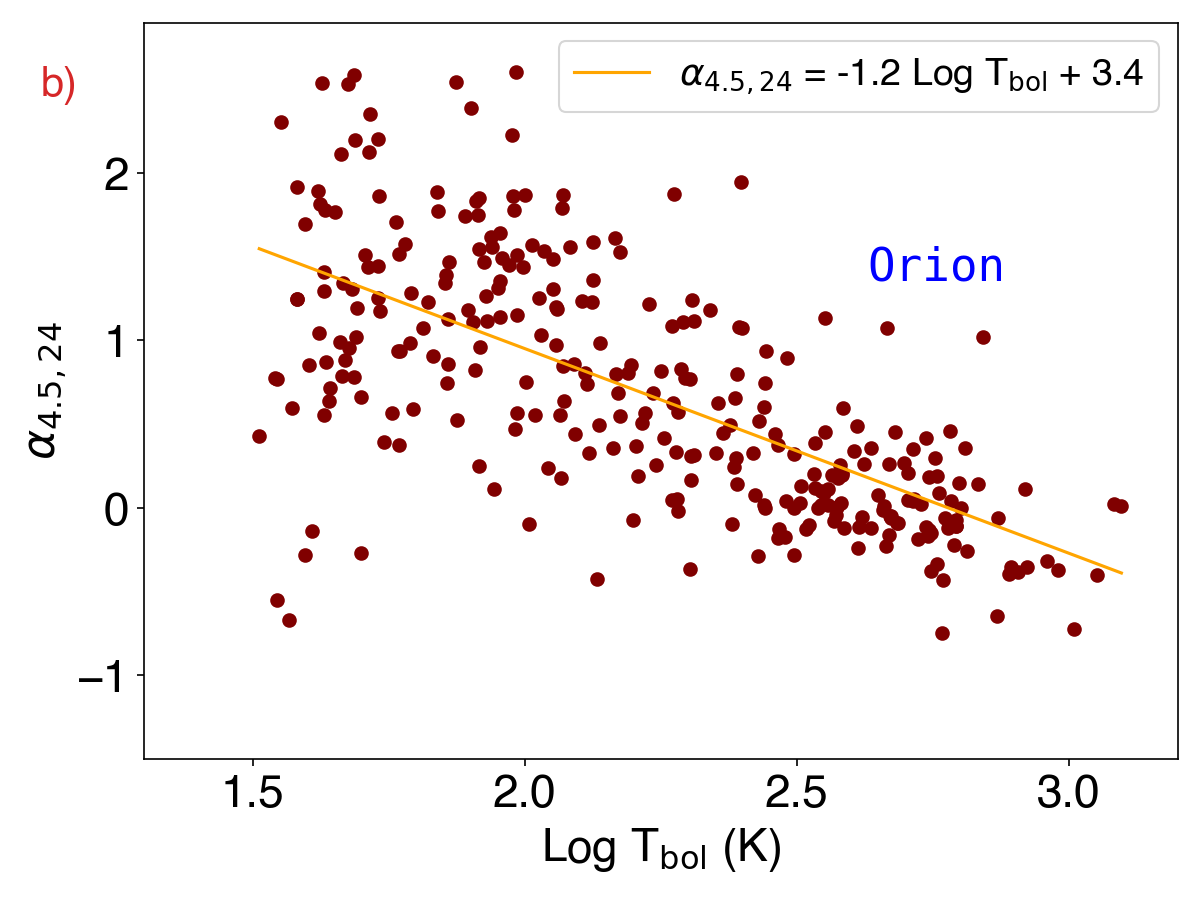}
\caption{The variation of the mid-IR $\alpha_{4.5, 24}$ with Log T$_{\rm{bol}}$ for Aquila (Panel a) and Orion (Panel b). Orange line represents the linear best fit line with the best fit relation shown in the legend.} We obtain a similar relation in both clouds.
\label{fig:plf_c}
\end{figure}

\subsection{Comparing Evolutionary Diagnostics}
\label{sec:ev_diag}

Several SED-based diagnostics are used to track the evolution of protostars as their envelopes are accreted/dispersed. These include the slope of the mid-IR SED \citep{Lada84,Greene94}, the bolometric temperature of the SED \citep{Myers93,Chen95}, and the envelope mass derived by model fits to the SED \citep{Fischer17}. Analyses based on Spitzer and Herschel data often use a combination of the slope and \Tbol~to place YSOs into an evolutionary context \citep[e.g.,][\citetalias{Furlan16}]{Evans09}. Each of these diagnostics, however, has limitations. The slope and \Tbol~are both strongly affected by inclination and foreground extinction \citep[e.g.,][]{Calvet94,Whitney03b,Fischer17}. Model fits to the SEDs can account for inclination and extinction; however, degeneracies between parameters can result in large uncertainties in the best fit values \citep[\citetalias{Furlan16},][]{Fischer17}. Recent work shows a strong, linear correlation between the best fit $M_{2500}$ and the envelope mass estimated from the 870~$\mu$m flux measured by ALMA \citep{Federman22}, although with significant scatter. 

One way to assess potential biases between the HOPS and eHOPS-Aquila data is to compare evolutionary diagnostics.  In Figure \ref{fig:plf_c} we compare the $\alpha_{4.5, 24}$ and \Tbol~for the two samples. To minimize the effect of foreground extinction, we measure the SED slope using only the mid-IR photometry between the 4.5 and 24 $\micron$ bands, similar to \citetalias{Furlan16}. Since the Spitzer data are similar between both data sets, the $\alpha_{4.5, 24}$ values for Aquila and Orion are calculated in a consistent manner. 

In both the HOPS and eHOPS-Aquila data, we find a  similar linear dependence between $\alpha_{4.5, 24}$ and Log T$_{\rm{bol}}$. (Figure \ref{fig:plf_c}).  A linear best fit plot in Figure \ref{fig:plf_c} reveals the relation $\alpha_{4.5, 24}$ = -1.2 Log T$_{\rm{bol}}$ + 3.2 for the protostars in Aquila, and an almost identical relation for the protostars in Orion, $\alpha_{4.5, 24}$ = -1.2 Log T$_{\rm{bol}}$ + 3.4. The dispersion from the linear relation is larger for the younger sources in both clouds. Since the $\alpha_{4.5, 24}$ data are dependent on Spitzer data alone while \Tbol~requires the entire SED, this similarity of these fits and diagrams indicates no substantial biases in \Tbol~between HOPS and eHOPS. 

\begin{figure}[t!]
\centering
\includegraphics[scale=0.51]{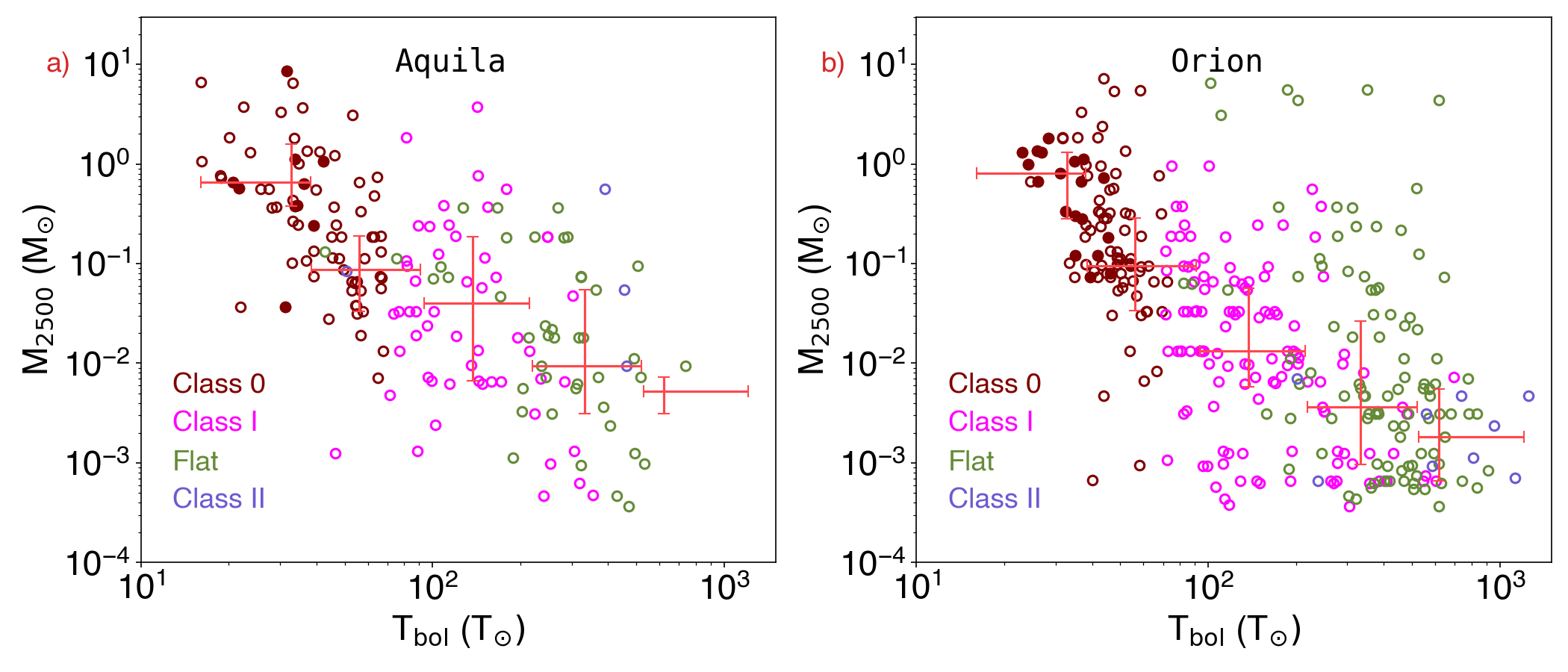}
\caption{Variation of the envelope mass within 2500 au (M$_{2500}$) with bolometric temperature (\Tbol) for Aquila (panel a) and Orion (panel b). Different SED classes are denoted by different colored open circles as labeled. Filled circles are the PBRS. The horizontal red bars show \Tbol~bins that are chosen to be equally spaced in log \Tbol. We use the same \Tbol~bins in both panels-- 16, 38, 92, 220, 523, and 1250 K. The vertical red bars denote the interquartile range of M$_{2500}$ for the protostars in a particular \Tbol~bin. 
\label{fig:MenvVsTbol}}
\end{figure}

In contrast, some differences are apparent when comparing envelope mass to \Tbol. Figure \ref{fig:MenvVsTbol} shows the variation of the envelope mass within 2500 au (M$_{2500}$) with T$_{\rm{bol}}$ for Aquila (panel a) and Orion (panel b). Note that the estimates for M$_{2500}$ are from model fits whereas we calculate T$_{\rm{bol}}$ directly from observations. 
We find the expected anti-correlation between T$_{\rm{bol}}$ and M$_{2500}$.  M$_{2500}$ is lower for the more evolved protostars that have higher T$_{\rm{bol}}$.  
The correspondence between the two clouds is strongest for the Class 0 protostars that populate the low \Tbol\ ($\lesssim$70 K) part of the plot.  On the other hand, for the sources with \Tbol\ $> 70$ K that are Class I, flats spectrum and Class II, we see that the interquartile ranges and median values are shifted to lower envelope masses in Orion.  This largely reflects the significant number of protostars with very low envelope masses ($\sim 10^{-3}$~M$_{\odot}$) in the HOPS sample. We also find differences at the extremes of the plot, with the Aquila data extending to lower \Tbol\ and a concentration of protostars in the highest \Tbol\ bin for HOPS which is not apparent for eHOPS-Aquila. We examine these differences in the distributions of \Tbol\ and envelope mass in the next sub-section.

\subsection{The Bolometric Temperature and Envelope Mass Distributions}

As protostars evolve, their bolometric temperatures increase and their envelopes masses decrease as the envelope is accreted and - simultaneously - dispersed by outflows. Thus, differences in the statistical distributions of \Tbol\ and M$_{2500}$ may indicate differences in star formation histories of these regions \citep{Fischer17}. This motivates a closer look at these distributions.

\begin{figure}[t!]
\includegraphics[scale=0.32]{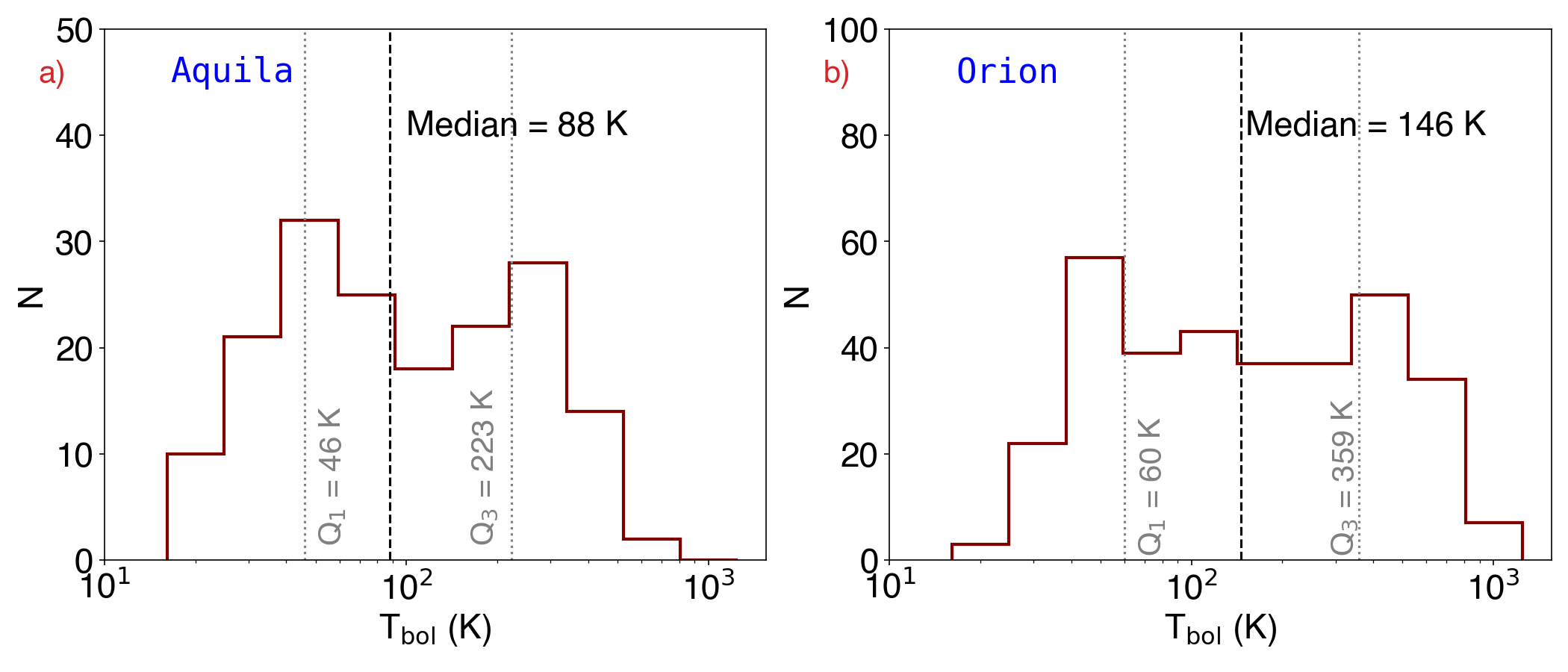}
\includegraphics[scale=0.32]{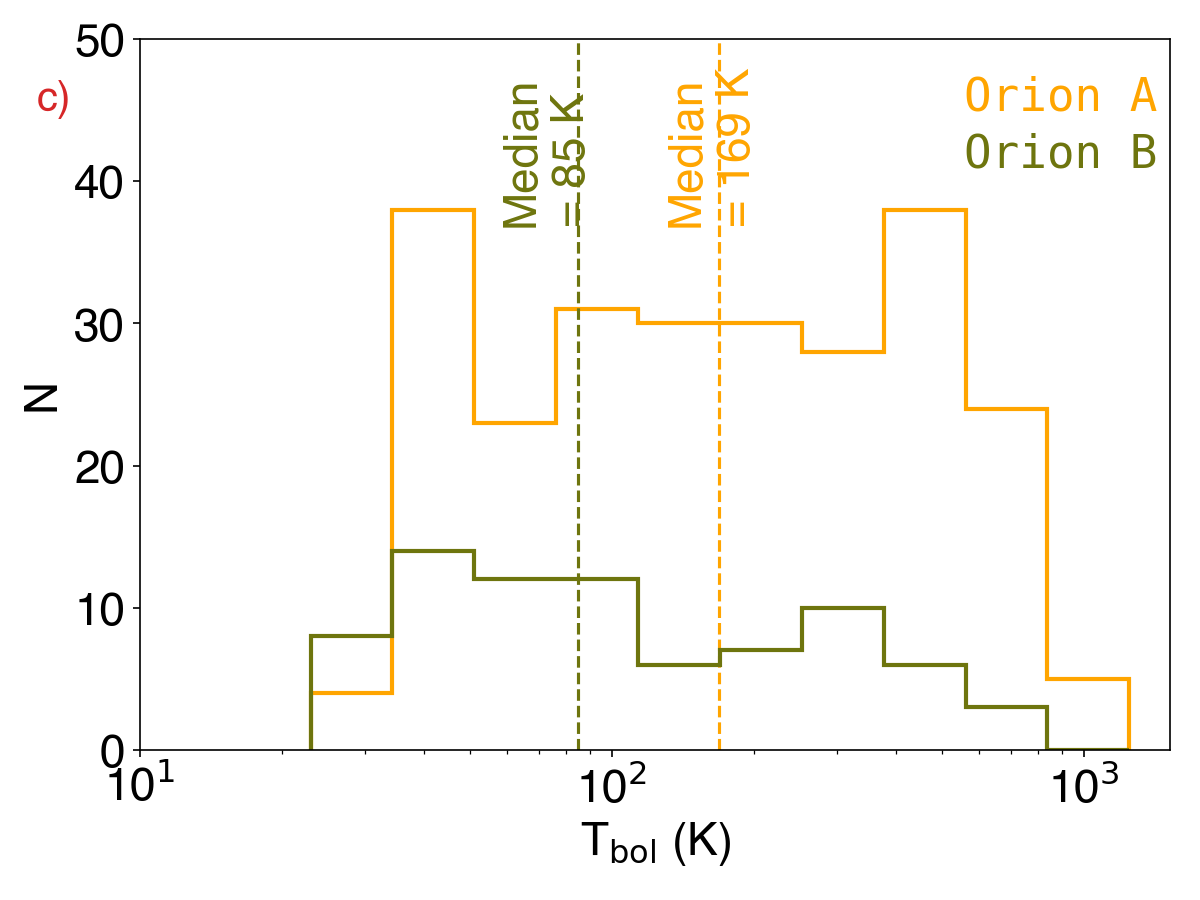}
\caption{Histogram of T$_{\rm{bol}}$ for eHOPS-Aquila (Panel a), HOPS (Panel b), and HOPS separately for Orion A and Orion B (Panel c). Median \Tbol~is shown by vertical dash lines and the first and third quartiles are shown by vertical dotted lines. AD test shows significant differences in the distribution of T$_{\rm{bol}}$ in Aquila and Orion when protostars in whole clouds are considered, but no significant difference in \Tbol~distribution between Aquila and Orion B.
\label{fig:plf_b}}
\end{figure}

Figure \ref{fig:plf_b} shows the histogram of T$_{\rm{bol}}$ for the protostars in the eHOPS-Aquila sample (Panel a) and the HOPS sample (Panel b). The histograms are broadly distributed for both clouds. The distribution for Aquila is shifted toward lower temperatures; the median T$_{\rm{bol}}$ of the eHOPS protostars in Aquila is 88 K compared to 146 K for the HOPS protostars in Orion. This is consistent with the higher fraction of PBRS and Class 0 protostars and a lower fraction of flat spectrum protostars in Aquila. The  AD test comparing the \Tbol~distribution of protostars in Aquila and Orion gives a $p$-value $<$ 0.01, thereby rejecting the null hypothesis that T$_{\rm{bol}}$ in Aquila and Orion are drawn from the same distribution. We also plot a histogram of \Tbol\ individually for Orion A and Orion B. For Orion~A, the AD test still shows a significant difference in \Tbol~distribution compared with Aquila with $p$-value $<<$ 0.01. However, for Orion~B, the AD test finds no significant difference in the distribution of \Tbol~when compared with Aquila with a $p$-value $>$ 0.9. In Orion B, the median \Tbol~is 85 K, again similar to that of Aquila. 

\begin{figure}[b!]
\includegraphics[scale=0.32]{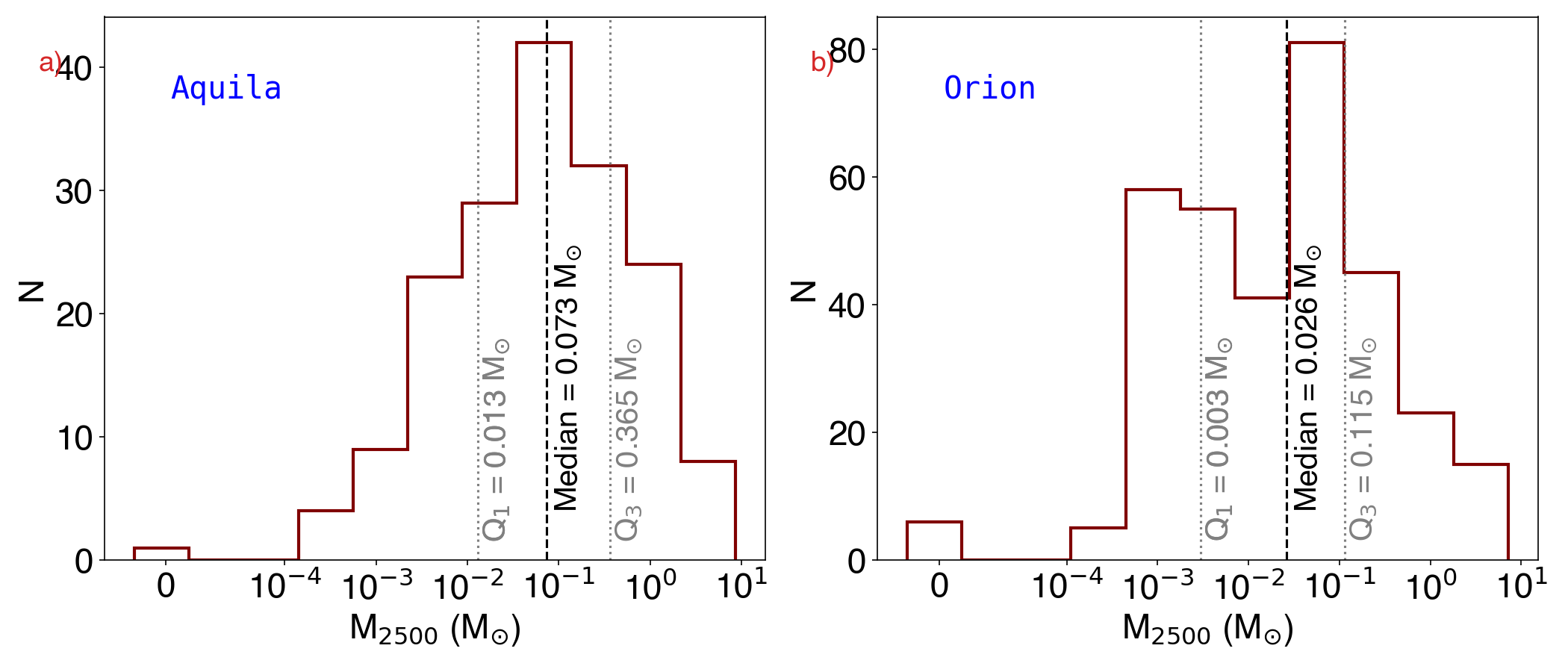}
\includegraphics[scale=0.32]{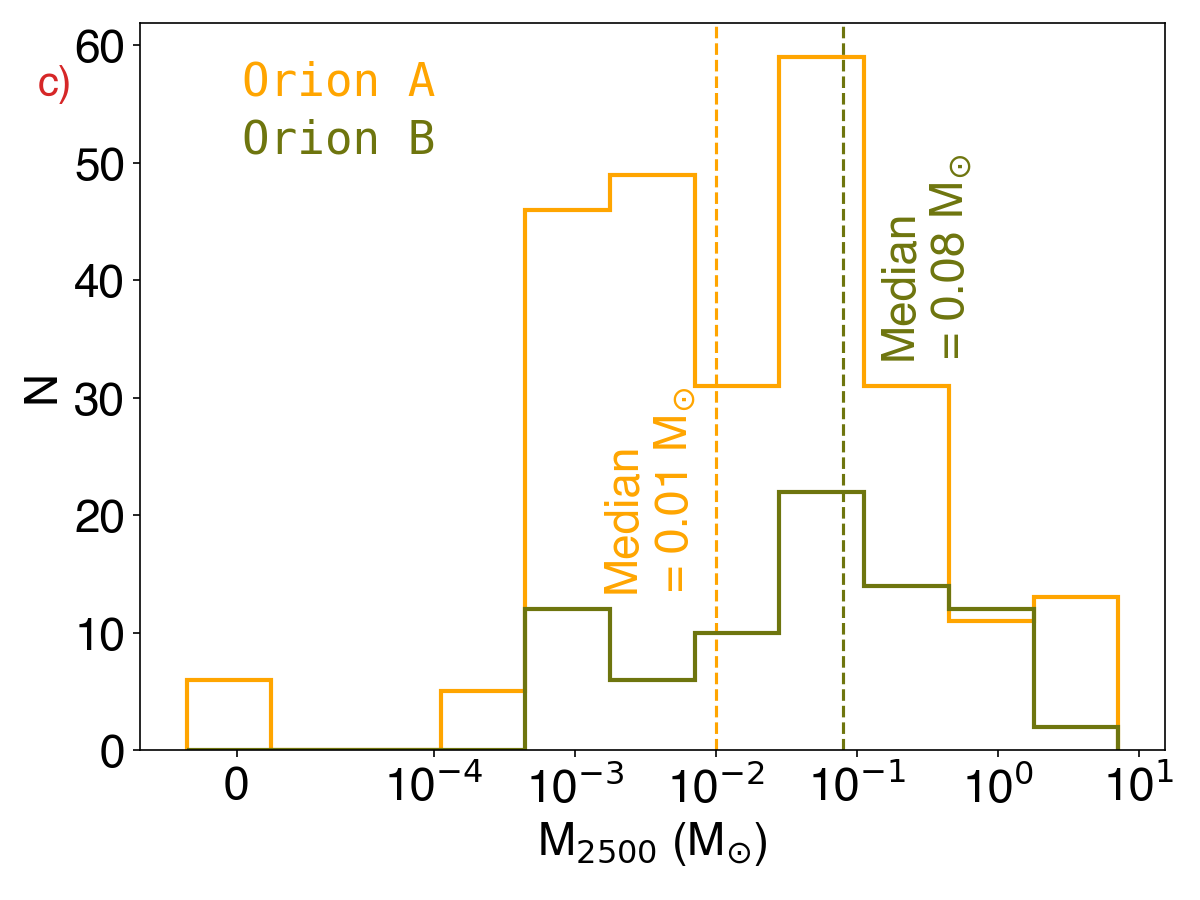}
\caption{Comparisons of the distributions of  M$_{2500}$ for protostars in eHOPS-Aquila (panel a), HOPS (panel b), and HOPS separately for Orion A and Orion B (panel c). The first and third quartiles are denoted as Q$_1$ and Q$_3$, respectively. The second quartile (median) is noted in all panels. AD test shows a significantly different distribution of M$_{2500}$ for protostars in Aquila and Orion. However, the protostars in Orion B have an M$_{2500}$ distribution similar to that in Aquila.
\label{fig:menv}}
\end{figure}
Similar results are found for the best-fit envelope masses to the eHOPS-Aquila and HOPS protostars.  Figure \ref{fig:menv} shows the distributions of M$_{2500}$ for the protostars in eHOPS-Aquila (panel a), HOPS, (panel b), and HOPS separately for Orion A and Orion B (panel c), respectively. As in the case of \Tbol, the distribution of M$_{2500}$ is shifted to higher masses for Aquila.   The median M$_{2500}$ in Aquila and Orion is 0.073 and 0.026 \Msun, respectively. The AD test shows a significant difference in the distribution of M$_{2500}$ in Aquila and in Orion with a $p$-value $<$ 0.01. Differences in the distribution of M$_{2500}$ are also found between the Orion A and Orion B clouds. The median M$_{2500}$ in Orion~B is 0.08~\Msun, similar to the median M$_{2500}$ in Aquila, while the median M$_{2500}$ in Orion~A is 0.01~\Msun, lower than the median M$_{2500}$ in the whole Orion cloud. The AD test shows that the distribution of M$_{2500}$~in Aquila is significantly different from Orion~A ($p$-value $<$ 0.01), but is similar to the M$_{2500}$ distribution in Orion~B ($p$-value $\geq$ 0.25). 

As we saw in Sec.~\ref{sec:ev_diag}, the HOPS sample contains a significant number of protostars with very low mass envelopes ($\sim 10^{-3}$~M$_{\odot}$) that are not present in eHOPS-Aquila. This is particularly true for Orion~A.  This difference may be due to differences in either the identification of protostars or their fits. The HOPS protostar sample may include more reddened pre-ms stars with disks or sources transitioning from protostars to pre-ms stars.  Alternatively, low mass protostars may have been removed by the galaxy template fits in Aquila, where the lack of IRS spectra would preclude the detection of silicate features indicative of protostars. Finally, the fits to the Orion SEDs may have resulted in more protostars with low envelope masses due to the different wavebands used in the fits. Establishing whether the low envelope mass sources are the result of biases, or whether they reflect real differences in the populations of protostars between these regions, requires a re-analysis of the HOPS data and is beyond the scope of this paper. 

Both eHOPS-Aquila and the HOPS protostars in Orion~B show a higher fraction of low \Tbol/high M$_{2500}$ protostars than Orion~A. The similarity between Orion~B and Aquila suggests these differences are not due to biases in the determination in \Tbol\ or M$_{2500}$. Furthermore, while \Tbol\ uses the Herschel/SPIRE band and may be affected by external dust emission from the surrounding cloud, the M$_{2500}$ values result from fits where the SPIRE data provide only upper limits. Thus, the difference in the M$_{2500}$ distributions cannot be explained by contamination of the SPIRE data. In total, these results suggest the differences between Orion~B/Aquila and Orion~A are real.

\subsection{The Evolution of Luminosity}
\label{sec:ev_lum}

\begin{figure}[t!]
\centering
\includegraphics[scale=0.5]{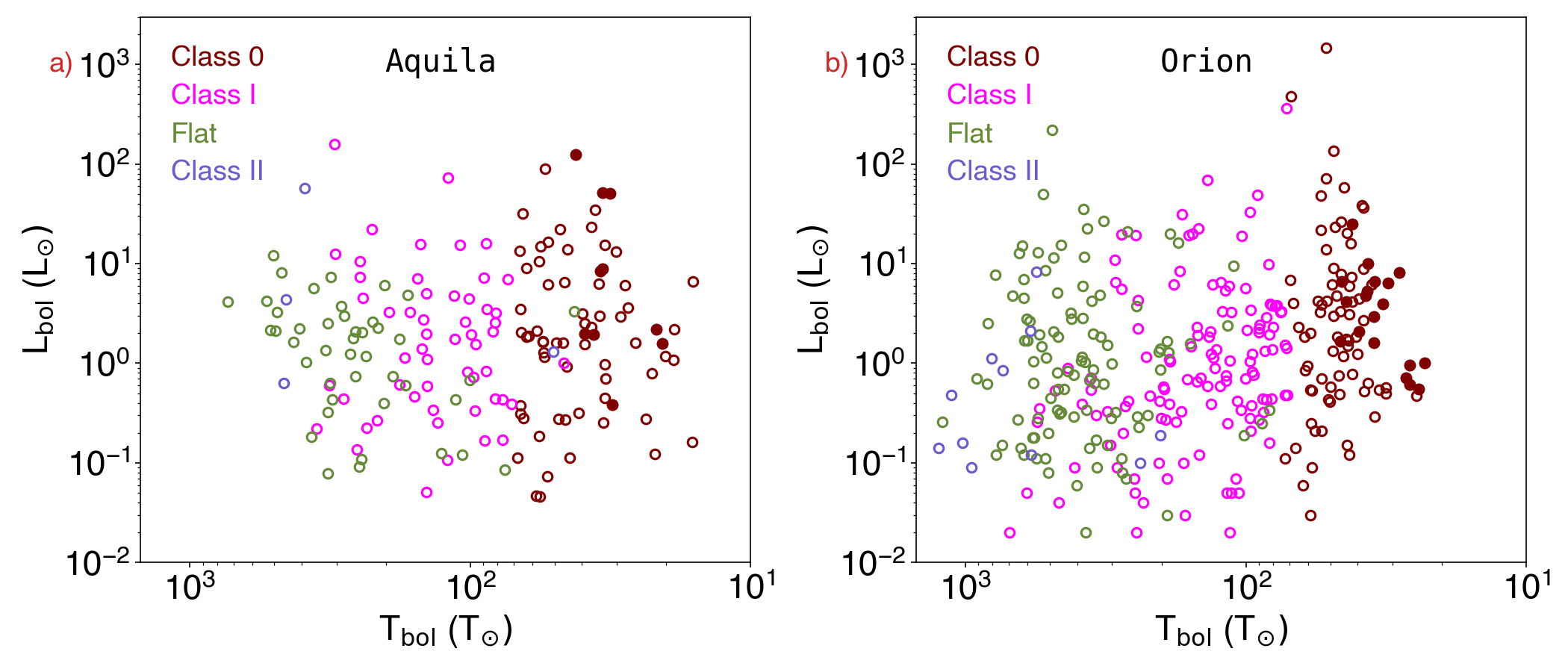}
\caption{The bolometric luminosity-temperature (BLT) diagram showing the variation of the bolometric luminosity with the bolometric temperature for the protostars in Aquila (panel a) and Orion (panel b). Colored open circles denote different SED classes as labeled. Filled circles denote the PBRS.
\label{fig:blt}}
\end{figure}

The Bolometric Luminosity versus Temperature (BLT) diagram is used to study the evolution of protostars \citep{Myers93,Evans09,Dunham15}. 
Figure \ref{fig:blt} shows the BLT diagrams for the eHOPS protostars in Aquila and HOPS protostars in Orion (c.f. \citealt{Fischer17}). The protostars in both Aquila and Orion span similar ranges in L$_{\rm{bol}}$. The maximum L$_{\rm{bol}}$ in the eHOPS Aquila catalog is 160 L$_{\odot}$. There are four protostars in the HOPS Orion catalog with L$_{\rm{bol}}$ $>$ 160 L$_{\odot}$: HOPS 361, HOPS 370, HOPS 376, and HOPS 384 (L$_{\rm{bol}}$ = 479, 361, 218 and 1478 L$_{\odot}$, respectively). 

\begin{figure}[b!]
\includegraphics[scale=0.5]{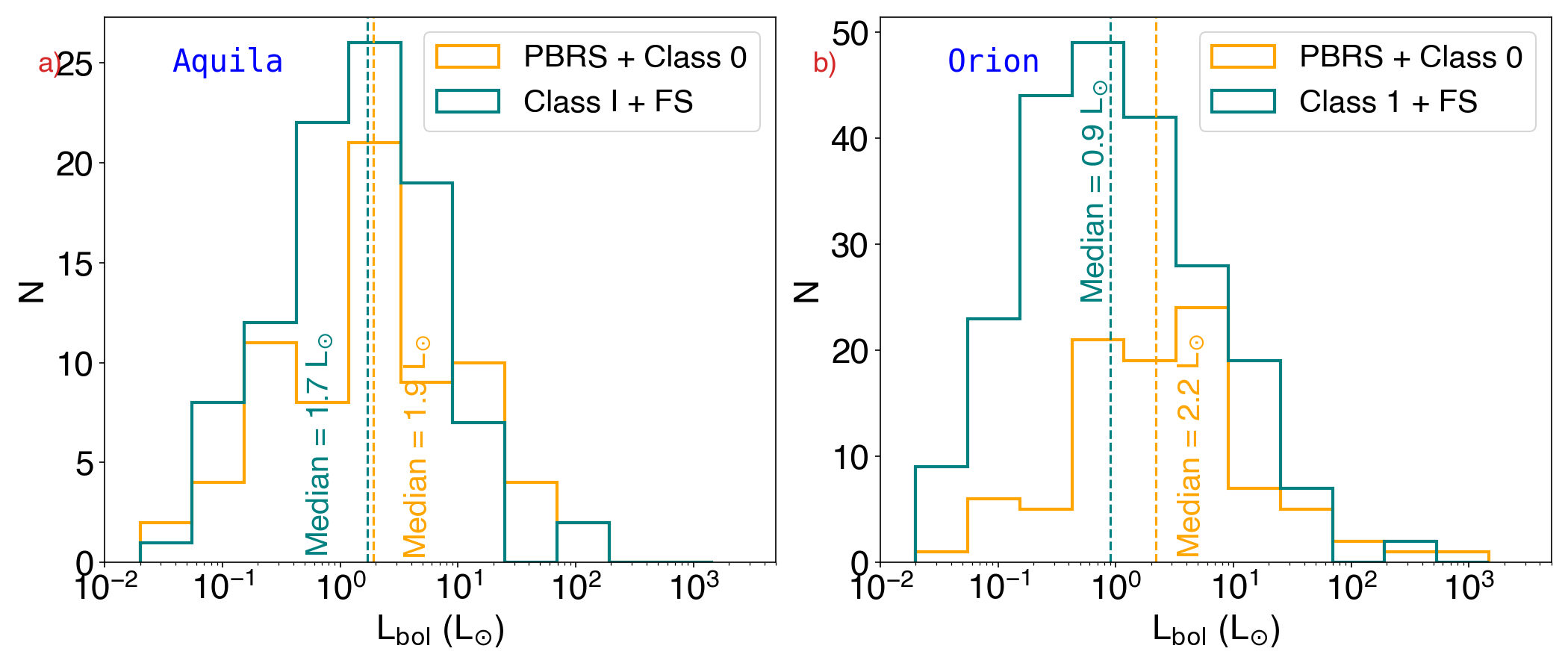}
\caption{Histogram of \Lbol~for Class 0 protostars (inclusive of PBRS) and a combined Class I and flat-spectrum sources for Aquila (panel a) and Orion (panel b). The median \Lbol~for Class 0 protostars in Orion is higher than for the combined Class I and flat-spectrum sources with statistically different distributions. In Aquila, the median \Lbol~for Class 0 is similar to the combined Class I and flat-spectrum sources, with statistically similar distributions.
\label{fig:plf_class}}
\end{figure}

The PLFs of the HOPS Class 0 protostars in Orion are shifted to higher \Lbol\ relative to more evolved protostars (see \citealt{Fischer17}).  This is true for both the Orion~A and B clouds. \citet{Fischer17} explain this shift as an evolution toward lower luminosities as the envelopes are depleted and infall/accretion decreases. In Figure \ref{fig:plf_class}, we compare the distribution of \Lbol~in eHOPS-Aquila and HOPS for deeply embedded protostars (PBRS + Class 0) and more evolved protostars (Class I + flat-spectrum sources). In Orion, the median \Lbol~for the  PBRS and Class 0  protostars are 2.2~\Lsun\ and for the  Class I and flat-spectrum protostars is 0.9 ~\Lsun. The AD test shows the distributions of \Lbol~for the combined PBRS and Class 0 protostars and the combined Class~1 and flat-spectrum protostars are unlikely to be drawn from the same parent distribution ($p$-value $<$ 0.01). However, we see no such behavior of \Lbol~in Aquila. The median \Lbol~of PBRS+Class 0 protostars in Aquila is 1.9 \Lsun, and for the later evolutionary phases (Class I + flat-spectrum sources) the median is 1.7~\Lsun. The AD test shows that \Lbol~distributions for the PBRS+Class~0 protostars and the  Class~1+flat-spectrum protostars are statistically indistinguishable ($p$-value $>$ 0.25). Thus, the eHOPS-Aquila protostars do not show the evolution toward lower luminosities observed in the HOPS protostars.

\begin{figure}[b!]
\centering
\includegraphics[scale=0.51]{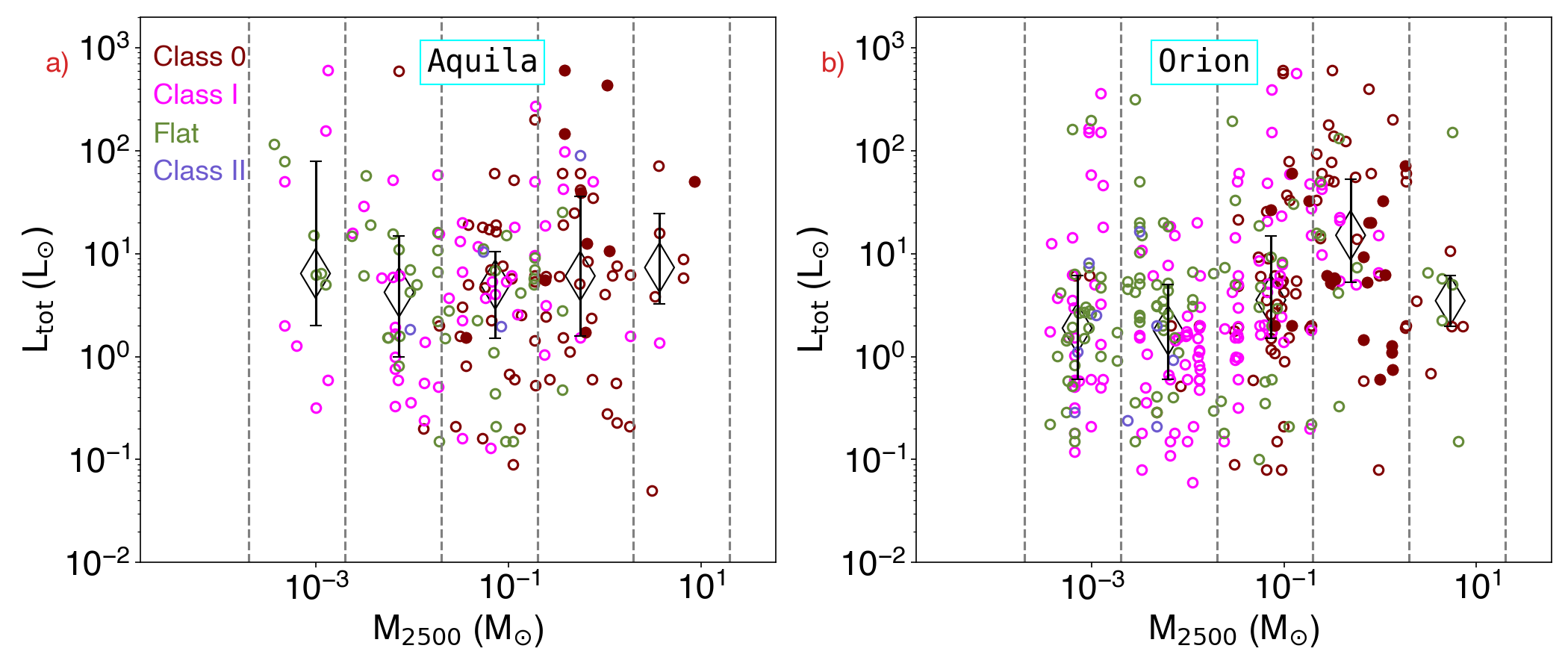}
\caption{Variation of total luminosity (\Ltot) with envelope mass within 2500 au  (M$_{2500}$) for the eHOPS protostars in Aquila (panel a) and HOPS protostars in Orion (panel b). Color labels are same as Figure \ref{fig:MenvVsTbol}. M$_{2500}$ are chosen to be equally spaced in logarithmic bins. We use the same M$_{2500}$ bins in both panels–  0.0002, 0.002, 0.02, 0.2, 2 and 20 M$_{\odot}$.
The vertical solid bars in each M$_{2500}$ bin denote the interquartile range of \Ltot with the median denoted by open diamonds. 
\label{fig:TLM}}
\end{figure}

\cite{Fischer17} also compared the total luminosity (\Ltot) with the envelope mass (M$_{2500}$) for the HOPS protostars in Orion as a means to study the evolution of luminosity of protostars. The plot is known as the Total Luminosity and Mass (TLM) diagram. The TLM diagram provides an alternative to the BLT diagram by directly comparing the physical protostellar properties from model fits. In Figure \ref{fig:TLM}, we provide a TLM diagram for the eHOPS protostars in Aquila and also recreate the TLM diagram from \cite{Fischer17} for the HOPS protostars in Orion. We divide the sample in equally spaced bins of Log M$_{2500}$ and compute the median \Lbol~(black diamonds) and upper and lower quartiles (bars on the diamonds) in each bin. 

For the HOPS protostars, \Ltot\ peaks around 0.3~M$_{\odot}$, after which \Ltot~ subsequently decreases as M$_{2500}$ decreases. This behavior was modeled by \citet{Fischer17} by assuming an exponentially decreasing envelope mass with time. 
AD test shows that the $p$-values for \Ltot\ in the consecutive M$_{2500}$ bins of 0.002, 0.02, 0.2, 2 and 20 M$_{\odot}$ are $<$0.01, implying that \Ltot\ is drawn from different distribution. Only for the lowest M$_{2500}$ bins of 0.0002 and 0.002 M$_{\odot}$, the $p$-value is $>$0.25, implying a statistically similar distribution. This behavior is apparent for both the Orion~A and Orion~B clouds \citep{Fischer17}. 
In the Aquila clouds, however, the AD $p$-values for \Ltot\ for each M$_{2500}$ bin are consistent with being drawn from the same distribution. To minimize differences due to incompleteness, we also study the variation of \Ltot~ with M$_{2500}$ for protostars with \Lbol~$>$0.1 L$_{\odot}$. We find that the AD-test results do not change by excluding \Lbol~$<$0.1 L$_{\odot}$ protostars.

In summary, in the eHOPS-Aquila sample, we do not find the evolution of luminosity present in the HOPS sample. This is due to lower average luminosities for the Class 0 protostars and higher average luminosities for the Class I/flat spectrum protostars in Aquila relative to Orion. In the next section, we briefly discuss the implications of this result.

\subsection{Differences between Aquila and Orion}

A primary goal in comparative studies of protostars in nearby clouds is to establish differences in the star-forming populations. These differences may arise due to different star formation histories, or due to the influence of the environment on the evolution of protostars.  Both can have significant implications. Variations in the star formation rate over the relatively short lifetimes of protostars (0.5~Myr) would provide new information on how star formation is induced and regulated in clouds. Variations with the environment would provide constraints on how environmental factors mediate protostellar evolution, which may prove useful in extrapolating our understanding of low-mass star formation in the nearest 0.5 kpc to the more active (and more representative) star-forming regions such as those found in the inner region of our galaxy  \citep{Megeath22}.   

In our cursory comparisons, we have established two differences. The first is the higher fraction of low \Tbol/high M$_{2500}$ protostars in Orion~B and eHOPS-Aquila. Orion~B has been known to have an unusual concentration of Class 0 protostars and PBRS \citep{Stutz13,Karnath20}. If these differences are real, the shift to lower \Tbol~and higher M$_{2500}$ in Aquila (and Orion B) may suggest that the star formation rate is increasing with time \citep{Fischer17}.  Consistent with this picture, the two clusters that dominate star formation in Aquila, the Serpens~Main and Serpens~South clusters, have high fractions of protostars compared to all YSOs \citep[57-71\%,][]{Gutermuth08a}.

The second difference is the lack of luminosity evolution in the eHOPS-Aquila. Such evolution in luminosity is apparent in both the Orion~A and B clouds.  Previous work has suggested that the luminosities of protostars depend on the environment \citep{Kryukova12, Kryukova14,Dunham14,Elmegreen14}.  These previous works, however, did not examine luminosity as a function of evolution. If the difference in luminosity evolution is real, there are a few possible explanations.  First, if the star formation rate is rapidly increasing, many of the Class I and flat spectrum sources may be younger protostars seen at favorable inclinations \citep{Federman22}.  Alternatively, the infall and accretion rate in Aquila may be systematically lower than those in Orion, lengthening the duration of the Class~0 phase. This, however, is in tension with the higher fraction of high M$_{2500}$ values in eHOPS-Aquila; protostars with higher  M$_{2500}$ values have higher envelope densities, shorter free-fall times, and higher infall/accretion rates. Alternatively, there may be differences in how much of the energy from accretion is radiated into space and the radii of the resulting protostars \citep[e.g.][]{Hosokawa11}.

Future work is needed to rigorously establish that such differences are not due to biases in the data. We will re-examine these differences in a future publication, in which we expand the eHOPS sample to include all the clouds within 500~pc except Taurus.

\section{Summary} \label{sec:conclusion}

We present a survey for protostars in the Aquila clouds (distance = 436~pc), the richest cloud complex within 500 pc after Orion. The goal of eHOPS-Aquila (eHOPS:extension of HOPS Out to 500 ParSecs) is to combine 2MASS, Spitzer, Herschel, and JCMT/SCUBA-2 data to create SEDs for all the protostars detected in the Herschel/PACS bands, and to model the SEDs to derive the general physical properties of the protostars in Aquila. This approach is motivated by the success of the HOPS (Herschel Orion Protostar Survey) program, which characterized 330 protostars in the Orion clouds with 2MASS, Spitzer, Herschel, and APEX data.  Due to differences in the available data - as well as due to lessons learned from HOPS - we have developed a new methodology for generating the SEDs and identifying protostars.  The primary achievements detailed in this paper are the following: 

\begin{enumerate}

\item{} For all the sources detected in at least one of the three Herschel/PACS bands (70, 100 and 160 $\micron$) in the Aquila complex, we have assembled SEDs spanning $1-850~\micron$. These combine near-IR photometry from 2MASS, mid-IR photometry and spectroscopy from Spitzer, far-IR photometry from Herschel/PACS, and sub-mm photometry from Herschel/SPIRE and JCMT/SCUBA-2.  WISE data is used when Spitzer data isn't available.  Tables of the photometry for all sources are presented. 

\item{} Using the SEDs, sources are separated into protostars, pre-ms stars with disks, or galaxies. An additional category of candidate protostars is established for sources that share some of the properties of pre-ms stars with disks. We accomplish this categorization with a five-step scheme that uses both the Spitzer and Herschel photometry; this includes finding galaxies using a novel template fitter.  We identify 172 protostars, 73 pre-ms stars with disks, 24 reddened pre-ms stars with disks, 118 galaxies, and 12 candidate protostars. The sample of pre-ms stars with disks is highly incomplete since most of those found by Spitzer do not have PACS detections.  

\item{} We compare the distribution of protostars, pre-ms stars with disks, and galaxies to Herschel maps of the molecular clouds. The protostars and pre-ms stars are concentrated in the filamentary clouds.  In contrast, galaxies are spread throughout the observed fields, as expected.  

\item{} The observed properties of the protostars -- evolutionary class, SED slope ($\alpha_{4.5,24}$), \Tbol, and \Lbol -- are determined from the SEDs. For most of the sources, the peak of the SED is in the Herschel/PACS bandpasses; thus the PACS data is required to reliably determine \Lbol\ and \Tbol.  Out of the 172 protostars in Aquila,  71 (42\%) are Class 0, 54 (31\%) are Class I, 43 (25\%) are flat-spectrum protostars. Another 4 (2\%) are Class II sources which likely have residual envelopes.

\item{}We fit the SEDs of the 172 protostars to a grid of 30,400 radiative transfer models originally generated to model the HOPS protostars.  The model grid is adapted to the distance of Aquila and the different wavelength bands used by eHOPS. We define the best fit models by using the $R_{\rm{min}}$ parameter (defined in Appendix \ref{app:extgalfit}) and we find that 93\% of the protostars have good fits with $R_{\rm{min}} < 5$ and more than three data points in their SEDs.

\item{We discuss degeneracies in determining the fundamental model parameters. We find that the envelope density ($\rho_{1000}$), envelope mass (M$_{2500}$), and total luminosity (\Ltot) are constrained by our models, whereas the disk outer radius (R$_{\rm{disk}}$) and cavity opening angle ($\theta_C$) are not constrained. The foreground extinction (A$_{\rm{V}}$) is not constrained for the sources that lack mid-IR photometry and/or spectrum in their SEDs; however, the lack of an extinction values does not strongly affect the  model fits to these very red SEDs.}

\end{enumerate}

With this new catalog of protostars, we perform an initial comparison between the eHOPS protostars in Aquila and HOPS protostars in Orion. The goal is to identify similarities and differences between the two samples.  Such differences may arise from variations in protostellar evolution or star formation histories between the two clouds, but they may also arise from distinctions both in the methods used for the identification of protostars and in the Spitzer/IRS or sub-mm data used for the two samples. The results from this comparison can be summarized as follows:

\begin{enumerate}

    \item{} In Aquila, we identify 10 PACS Bright Red Sources (PBRS); these very red sources comprise the same percentage of all protostars in both the eHOPS-Aquila and HOPS (6\%) samples.  The median values of the observed properties (\Lbol, \Tbol) and model-derived properties (\Ltot, $\rho_{1000}$, and M$_{2500}$) are similar in the two PBRS samples. These show that in both clouds, the PBRS are distinguished by lower \Tbol, higher envelope densities, and higher envelope masses compared with the rest of the protostars.
    
    \item The protostellar  luminosity function using $\Lbol$ for the eHOPS-Aquila protostars peaks at 2 L$_{\odot}$ (median \Lbol: 1.8~L$_{\odot}$) with  \Lbol\ ranging from 0.05 to 160 \Lbol. Although this value is higher than the 1~L$_{\odot}$  peak in Orion (median: 1.1~L$_{\odot}$), the Anderson-Darling (AD) test shows that the distribution of \Lbol~in Aquila is not statistically different than in Orion. The \Lbol\ distribution is similar in Aquila and Orion B, but statistically different between Aquila and Orion A. Looking at the distribution of the total luminosities, \Ltot, calculated from the best model fits in Aquila, the median value is also higher, 5.4 L$_{\odot}$, yet the AD test shows no significant difference in the luminosity function between Aquila, Orion A and Orion B.
    
    \item{} The eHOPS-Aquila sample shows a lack of protostars with low envelope masses ($< 10^{-3}$~M$_{\odot}$) compared to the HOPS-Orion sample. It is not clear if this is due to differences in the identification of protostars or in the wavelength bands used for the model fits to the SEDs, or whether this reflects real differences between Aquila and Orion.
    
    \item{} The statistical distribution of \Tbol(M$_{2500}$) is shifted to lower(higher) values for the eHOPS-Aquila protostars as compared to the HOPS protostars. The fraction of low \Tbol/high M$_{2500}$ protostars in Aquila is similar to that in Orion~B. This can result from a rising star formation rate in Aquila and Orion~B, as compared to Orion~A. 
    
    \item{} The HOPS protostars show a statistically significant decrease in luminosity between the deeply embedded PBRS + Class 0 protostars and the more evolved Class I and Flat Spectrum protostars. In contrast, there is no significant decline in the eHOPS-Aquila protostars. This may be due to differences in the star formation history or the effect of different environments on protostellar evolution, but we cannot currently rule out biases between the two protostar samples.

\end{enumerate}

This is the first publication of the eHOPS program, designed to present our methods and the sample for the most active cloud in the sample. Future work will present the eHOPS survey for all the clouds within 500~pc outside of Taurus. This work will extend our comparison to a larger sample of clouds and environments.  From this work, we will provide a more rigorous exploration of potential differences in the clouds and the physical implications.  The eHOPS catalogs will also be essential for planning future observations where individual properties such as protostellar accretion can be studied in much more detail, for example with ALMA and the JWST.

\begin{acknowledgments}

RP, STM, and SAF gratefully acknowledge the funding support for this work from the NASA/ADAP grant 80NSSC18K1564. 
STM and RP also acknowledge funding support from the NSF AST grant 2107827.
RAG acknowledges funding from the NASA/ADAP grant NNX17AF24G and the NSF AST grant 2107705. AS gratefully acknowledges support by the Fondecyt Regular (project
code 1220610), and ANID BASAL projects ACE210002 and FB210003. This work is based on observations made
with the Spitzer Space Telescope, which is operated by the Jet Propulsion Laboratory (JPL), California Institute of Technology (Caltech), under a contract with NASA; it is also based on observations made with the Herschel Space Observatory, a European Space Agency Cornerstone Mission with significant participation by NASA. The Herschel spacecraft was designed, built, tested, and launched under a contract to ESA managed by the Herschel/Planck Project team by an industrial consortium under the overall responsibility of the prime contractor Thales Alenia Space (Cannes), and including Astrium (Friedrichshafen) responsible for the payload module and for system testing at the spacecraft level, Thales Alenia Space (Turin) responsible for the service module and Astrium (Toulouse) is responsible for the telescope, with more than 100 subcontractors. The James Clerk Maxwell Telescope is operated by the East Asian Observatory on behalf of The National Astronomical Observatory of Japan; Academia Sinica Institute of Astronomy and Astrophysics; the Korea Astronomy and Space Science Institute; the National Astronomical Research Institute of Thailand; Center for Astronomical Mega-Science (as well as the National Key R\&D Program of China with No. 2017YFA0402700). Additional funding support is provided by the Science and Technology Facilities Council of the United Kingdom and participating universities and organizations in the United Kingdom and Canada. Additional funds for the construction of SCUBA-2 were provided by the Canada Foundation for Innovation. This publication also makes use of data products from the Wide-field Infrared Survey Explorer, which is a joint project of the University of California, Los Angeles, and the Jet Propulsion Laboratory/California Institute of Technology, funded by the National Aeronautics and Space Administration

\end{acknowledgments}


\software
~Plotly \citep{plotly}, DASH \citep{dash}, Astropy \citep{astropy:2013, astropy:2018, astropy:2022}, Aplpy \citep{aplpy2012,aplpy2019}, Matplotlib \citep{Hunter:2007}, Scipy \citep{Scipy01,SciPy20}, numpy \citep{vanderwalt, numpy}, TOPCAT \citep{Taylor17}

\appendix

\section{Herschel Observations} \label{app:obsid}

In this work, the primary selection criteria for identifying protostars is based on robust detection in the Herschel wavebands. We describe the observations in section \ref{sec:sed}. The Tables \ref{tab:pacs} and \ref{tab:spire} in this appendix contain more details of the Herschel/PACS and Herschel/SPIRE observations, respectively.

\begin{deluxetable*}{cccccc}
\tablecaption{Observation details for Herschel/PACS \label{tab:pacs}} 
\centering 
\tablehead{ 
\colhead{Wavelength} & \colhead{Observation ID} & \colhead{R.A.} & \colhead{Decl.} & \colhead{Proposal} & 
\colhead{Scan Velocity} \\ 
\colhead{($\micron$)} & \nocolhead{} & \colhead{($\deg$)} & \colhead{($\deg$)} & \nocolhead{} & 
\colhead{($\arcsec$/sec)} 
} 
\decimalcolnumbers 
\startdata 
70 & 1342186277,1342186278 & 277.6 & -2.73 & SDP\_pandre\_3 & 60 \\ 
70 & 1342206676,1342206695 & 277.77 & 0.72 & KPGT\_pandre\_1 & 60 \\ 
70 & 1342206694,1342206696 & 280.01 & 0.05 & KPGT\_pandre\_1 & 60 \\ 
100 & 1342228960,1342228961 & 279.46 & -1.3 & KPGT\_pandre\_1 & 20 \\ 
100 & 1342229079,1342229080 & 277.45 & 0.89 & KPGT\_pandre\_1 & 20 \\ 
100 & 1342193534,1342193535 & 277.83 & -2.09 & KPGT\_pandre\_1 & 20 \\ 
100 & 1342206702,1342206703 & 277.08 & -3.63 & KPGT\_pandre\_1 & 20 \\ 
160 & 1342228960,1342228961 & 279.46 & -1.3 & KPGT\_pandre\_1 & 20 \\ 
160 & 1342193534,1342193535 & 277.83 & -2.09 & KPGT\_pandre\_1 & 20 \\ 
160 & 1342206676,1342206695 & 277.77 & 0.69 & KPGT\_pandre\_1 & 60 \\ 
160 & 1342206702,1342206703 & 277.08 & -3.63 & KPGT\_pandre\_1 & 20 \\ 
160 & 1342186277,1342186278 & 277.59 & -2.73 & SDP\_pandre\_3 & 60 \\ 
160 & 1342206694,1342206696 & 280.01 & 0.05 & KPGT\_pandre\_1 & 60 \\ 
160 & 1342229079,1342229080 & 277.45 & 0.89 & KPGT\_pandre\_1 & 20 \\ 
\enddata 
\tablecomments{We use Level 2.5 processed Herschel/PACS observations that are selected from JSCANAM data products.} 
\end{deluxetable*}

\begin{deluxetable*}{cccccc}[ht] 
\tablecaption{Observation details for Herschel/SPIRE \label{tab:spire}} 
\centering 
\tablehead{ 
\colhead{Wavelength} & \colhead{Observation ID} & \colhead{R.A.} & \colhead{Decl.} & \colhead{Proposal} & 
\colhead{Scan Velocity} \\ 
\colhead{($\micron$)} & \nocolhead{} & \colhead{($\deg$)} & \colhead{($\deg$)} & \nocolhead{} & 
\colhead{($\arcsec$/sec)} 
} 
\decimalcolnumbers 
\startdata 
250 & \makecell{1342186111,1342186277,1342186278,\\ 1342204950,1342206676,1342206694,\\ 1342206695,1342206696} & 277.78 & -0.88 & \makecell{SDP\_okrause\_3,\\ SDP\_pandre\_3,\\ KPGT\_pandre\_1} & 30 \\ 
350 & \makecell{1342186111,1342186277,1342186278,\\ 1342204950,1342206676,1342206694,\\ 1342206695,1342206696} & 277.78 & -0.87 & \makecell{SDP\_okrause\_3,\\ SDP\_pandre\_3,\\ KPGT\_pandre\_1} & 30 \\ 
500 & \makecell{1342186111,1342186277,1342186278,\\ 1342204950,1342206676,1342206694,\\ 1342206695,1342206696} & 277.78 & -0.87 & \makecell{SDP\_okrause\_3,\\ SDP\_pandre\_3,\\ KPGT\_pandre\_1} & 30 \\ 
\enddata 
\tablecomments{We use Level 3 processed Herschel/SPIRE data products for this work.} 
\end{deluxetable*}

\section{A combined column density -- temperature map} \label{app:tdust}

\begin{figure}
\centering
\includegraphics[scale=.98]{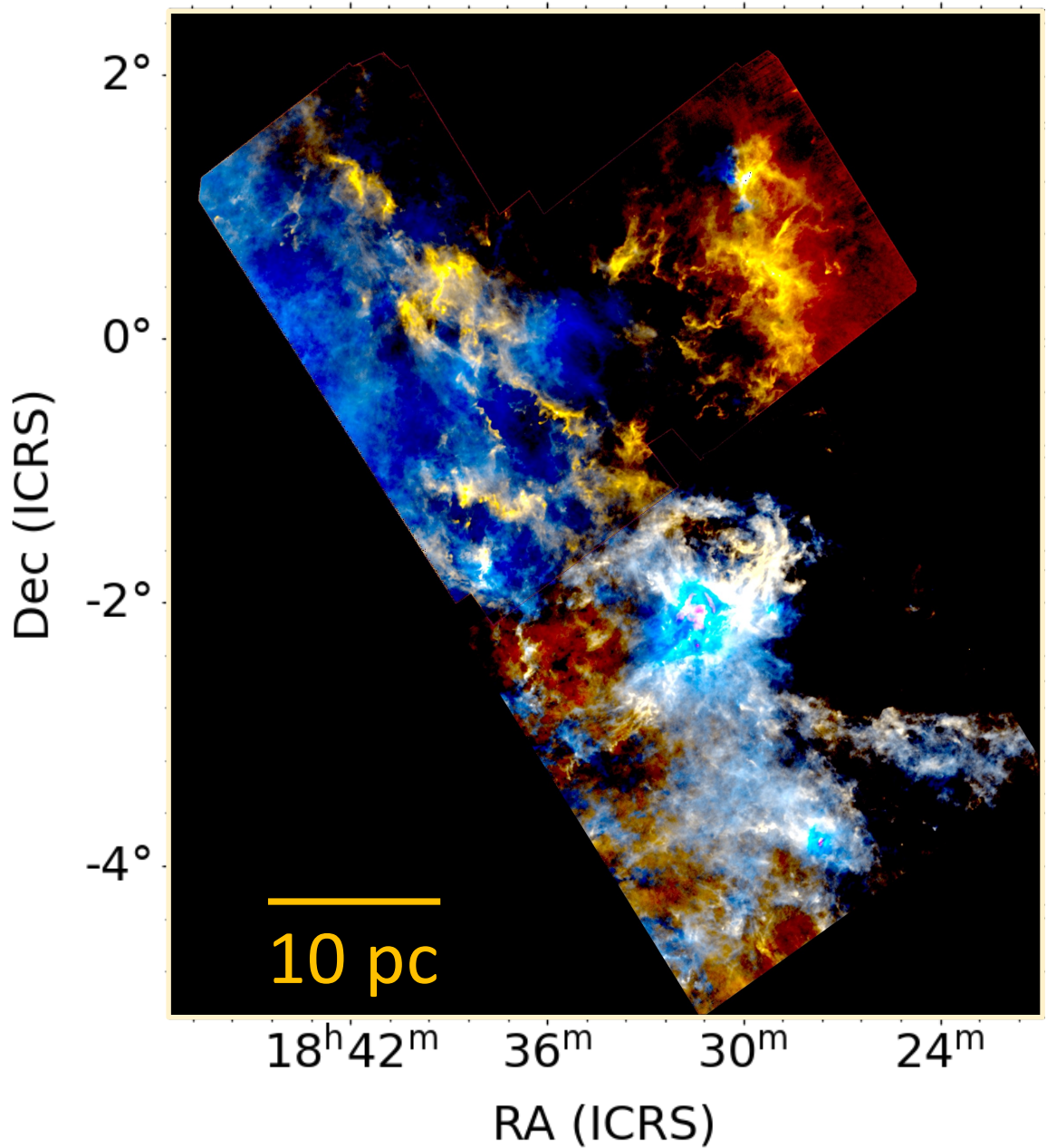}
\vspace{-1in}
\caption{A combined column density-temperature map for the Aquila cloud. The intensity of the map represents the column density of hydrogen and the color represents the dust temperature. Blue regions are comparatively hotter ($>$20 K) than red areas ($<$10 K). The figure shows the structural complexity of the cloud where currently star-forming regions such as Serpens Main and Serpens South are more dense, filamentary, and cold compared to the more evolved regions such as W40. We show the maps for column density and temperature separately in Figure \ref{fig:aquila_NH2} and Appendix \ref{app:tdust}, respectively.
\label{fig:aquila_NH2_T}}
\end{figure}

Figure \ref{fig:aquila_NH2_T} shows a column density map weighted by the dust temperature for the Aquila clouds. The column density in Figure \ref{fig:aquila_NH2_T} is shown in terms of the intensity of the map, whereas temperature is shown in terms of color. Blue regions are comparatively hotter ($>20$ K) than the red regions ($<10$ K). Figure \ref{fig:aquila_NH2_T} depicts a variety of gas structures in different star-forming regions. Recently star-forming regions such as Serpens Main and Serpens South are dense and filamentous with lower dust temperature compared to the more evolved W40 which is more diffuse and has higher dust temperature.

\section{Positional offset between Spitzer and Herschel sources}
\label{app:positions}

\begin{figure}[ht!]
\includegraphics[scale=0.65]{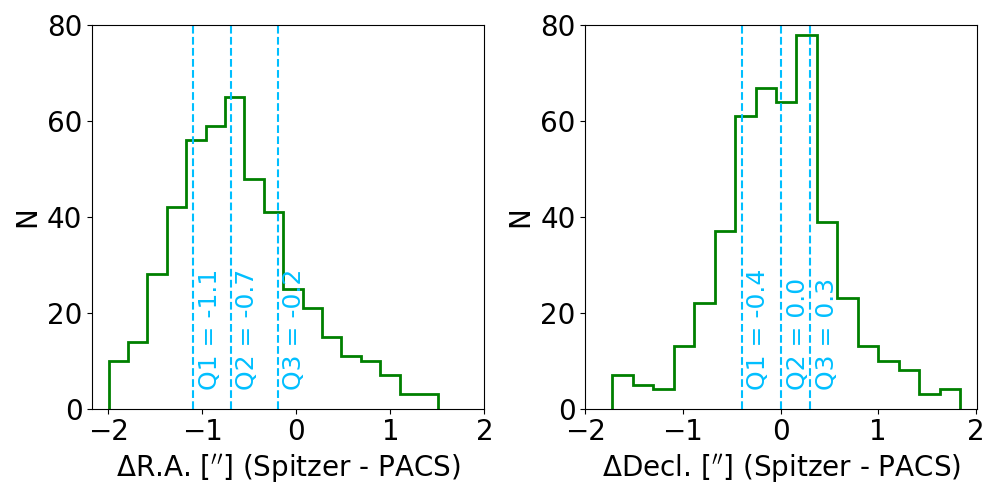}
\caption{Positional offset in arc-seconds (R.A. in the left panel and Decl. in the right panel) in the sources between Spitzer and Herschel/PACS wavelengths. 
\label{fig:coorddiff}}
\end{figure}

We use multi-wavelength photometry that spans a wide range of wavelengths to construct SEDs. Since the wavelengths range from near/mid-IR to far-IR region, we estimate the positional offset of the sources to check if the far-IR positions need adjusting. Figure \ref{fig:coorddiff} shows the offset in the average Spitzer (IRAC+MIPS) coordinates and Herschel/PACS coordinates for the same sources. The left panel shows the difference in R.A. and the right panel shows the difference in Decl. The median offset in R.A. between Spitzer and Herschel/PACS is -0.7$\arcsec$, with first and third quartiles at -1.1$\arcsec$ and -0.2$\arcsec$, respectively. The median offset in Decl. is 0.0$\arcsec$, with first and third quartiles at -0.4$\arcsec$ and 0.3$\arcsec$, respectively. The coordinate offsets for all the sources are less than the source matching tolerance limit of 2$\arcsec$. Thus we conclude that position adjustment is not required for the far-IR coordinates.

\section{SED fitting with extragalactic templates} \label{app:extgalfit}

We devise a fitting routine to fit SEDs for the sources that are not initially identified as high-confidence protostars and pre-ms stars with disks with a set of SED templates produced by \cite{Kirkpatrick15}. The extragalactic SED templates are given in terms of luminosity density ($L_{\nu})$ as a function of wavelength.  To compare the template SEDs with observed SEDs, first, we take the product of luminosity and frequency for the template SEDs to estimate $\nu L_{\nu}$. We interpolate the template SEDs to get $\nu L_{\nu}$ estimates at the same wavelengths as for the observed SEDs. 
The extragalactic SED templates are free of reddening, but the sources may suffer reddening due to the intervening gas and dust in the Milky Way. We redden the templates using the equation for foreground attenuation,
\begin{equation} \label{eq:red}
    L_{\nu}^{\rm{ext}} = L_{\nu}^{\rm{int}}~\text{exp}[-\kappa_{\nu} \mu_{H_2} m_H N(H_2)],
\end{equation}

\noindent
where $L_{\nu}^{\rm{int}}$ and $L_{\nu}^{\rm{ext}}$ are the fluxes before and after applying dust extinction, respectively. $m_H$=1.67 $\times$ 10$^{-24}$ g is the mass of a hydrogen atom, and $\mu_{H_2}$ is the mean molecular weight per hydrogen molecule (2.8 for a cloud with 71\% molecular hydrogen, 27\% helium, and 2\% metals; \citealt{Kauffmann08}). $N(H_2)$ is the beam-averaged column density at the location of the source. We use the Herschel-derived column density maps \citep{Konyves15, Fiorellino21} that are smoothed to the beam of SPIRE 500 $\micron$ (36$''$). Finally,  $\kappa_{\nu}$ is the extinction opacity from \cite{Ormel11} for their dust model ``icsgra3". The model includes graphite grains without ice coating and ice-coated silicates, with a size distribution that assumes growth of aggregates for 3 $\times$ 10$^{5}$ yr. Particle sizes in clouds range from 0.1 to 3 $\micron$, with a number density that is roughly proportional to $a^{-2.3}$ (where $a$ is the particle radius). See \citetalias{Furlan16} for a comparison and preference of the opacity law from \cite{Ormel11} over other existing models such as \cite{Ossenkopf94} and \cite{Draine03}. Then we take a ratio of both the observed and template SEDs, the median of which we use as a scaling parameter to scale the luminosities ($\nu L_{\nu}$) of the template SEDs to the fluxes ($\nu F_{\nu}$) of the observed SEDs.

We use the filter response function of the detectors at each wavelength in the reddened, scaled template SEDs to compute model photometry corresponding to the observed wavelengths. The spectral response calibration for IRAC assumes a flat-spectrum point source such that $\nu F_{\nu}$ = constant, while the MIPS calibration is not significantly dependent on spectral slope. For a photon-counting detector IRAC, \cite{Reach05} provides the details of absolute calibration and the zero-point flux estimation for a flat-spectrum source. Similarly, see \cite{Engelbracht07} for the flux density calibration of MIPS. In summary, we estimate the absolute flux densities in model templates using the following relation,

\begin{equation} \label{eq:flux_sp}
    F_{\rm{zero}} = \frac{\int F_{\nu} (\nu / \nu_{0})^{-1}~R_{\nu}~d\nu}{\int (\nu / \nu_{0})^{-2}~R_{\nu}~d\nu}
\end{equation}

\noindent
where $R_{\nu}$ is the spectral response function of the detector; $\nu_0$ = $c$/$\lambda_0$, $c$ being the velocity of light and $\lambda_0$ = $\frac{\int \lambda \nu^{-1}R_{\nu}d\nu }{\int \nu^{-1}R_{\nu}d\nu }$. We use Equation \ref{eq:flux_sp} to estimate flux densities for the filters corresponding to 2MASS, Spitzer, and WISE.

PACS, SPIRE, and SCUBA-2 detectors are bolometric detector arrays instead of photon-collectors. PACS and SPIRE are broad-band bolometric detector arrays whereas SCUBA-2 has narrower passbands that are dictated by the widths of the atmospheric windows in which they observe. For broad-band photometry, accurate knowledge of the instrument's relative spectral response function and beam properties is essential. In general, bolometers do not respond to the absorbed photon rate but to the amount of power that they absorb. The absorbed power is proportional to the spectral response function weighted flux density, which is the flux density-weighted by the overall response function and integrated across the passband \citep{Griffin13, Bendo13}.

\begin{equation} \label{eq:flux_bolo}
    F_{\rm{zero}} = \frac{\int F_{\nu}~R_{\nu}~\eta_{\nu}~d\nu}{\int R_{\nu}~\eta_{\nu}~d\nu},
\end{equation}

\noindent
where $\eta_{\nu}$ is the aperture efficiency for the particular filter. The details of the SPIRE photometry calibration are available on the observer's manual\footnote{http://herschel.esac.esa.int/Docs/SPIRE/pdf/spire\_om.pdf}. See \cite{Balog14} and the PACS observer's manual\footnote{https://www.cosmos.esa.int/documents/12133/996891/PACS+Observers\%27+Manual} for the details of the PACS photometry and \cite{Naylor14} for details on the spectral response of SCUBA-2 photometric bands. 

The variations in the  SEDs span orders of magnitude, both for the modeled as well as the observed SED. For this reason, we use the R goodness-of-fit parameter introduced by \cite{Fischer12} that measures the logarithmic deviation of observed SEDs from models. Such parameter parallels the visual comparison of the model and observed SEDs on a plot of log($\nu$F$_{\nu}$) versus log($\nu$).  The goodness-of-fit parameter is called $R$, and is defined as,

\begin{equation} \label{eq:Rfactor}
    R = \frac{1}{N} \sum_{i=1}^{N}w_{i}\left \lvert \rm{ln} \left( \frac{\it{F}_{\rm{obs, \lambda_i}}}{\it{F}_{\rm{mod, \lambda_i}}} \right) \right \lvert
\end{equation}

\noindent
where $N$ is the total number of observed photometric points in the SED; $F_{\rm{obs, \lambda_i}}$ and $F_{\rm{mod, \lambda_i}}$ are observed and modeled photometry at wavelength $\lambda_i$, respectively. Finally, $w_i$ is the weighting parameter that sets the weight of each photometric point in the SED. Following \cite{Fischer12} and \citetalias{Furlan16}, we define $w_i$ as the inverse of the approximate fractional uncertainty in each data point. For fluxes corresponding to 2MASS wavelengths, $w_i$=1/0.1, for Spitzer $w_i$=1/0.05, for WISE $w_i$=1/0.05, for Herschel/PACS 70 and 100 $\micron$ $w_i$=1/0.04, for Herschel/PACS 160 $\micron$ $w_i$=1/0.07, for Herschel/SPIRE $w_i$=1/0.4, and for SCUBA-2 $w_i$=1/0.3. We assign more uncertainty (less weight) towards the sub-mm data since the wavelengths $>200$ $\micron$ region can be strongly affected by dust emission external to the envelope. We select a model with the minimum $R$ as the best fitting model. We use $R$ to select the best fit model not only for the fitting of the SED with the grid of extragalactic templates but also for the grid of protostar models.

\section{Morphologically identified galaxies} \label{app:morph}

\begin{figure}[ht!]
\centering
\includegraphics[scale=0.35]{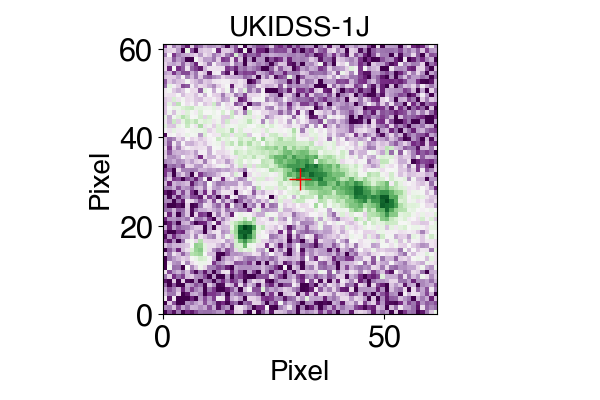}
\includegraphics[scale=0.35]{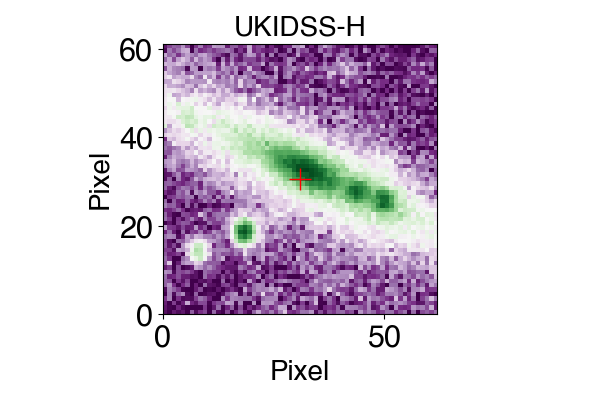}
\includegraphics[scale=0.35]{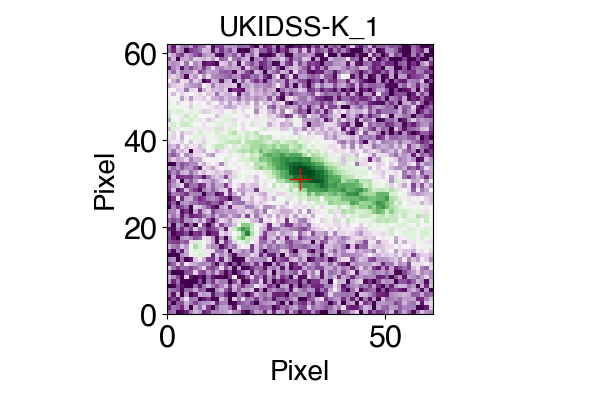}\\
\includegraphics[scale=0.35]{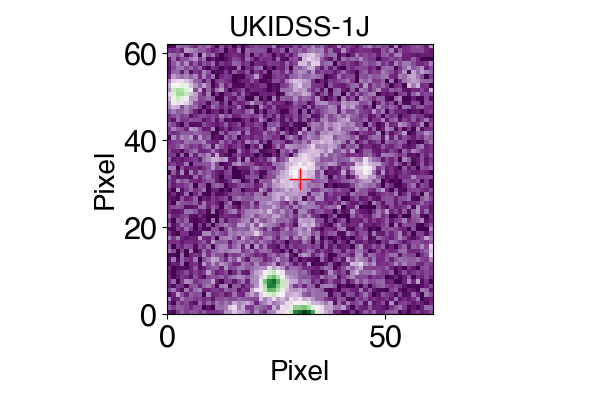}
\includegraphics[scale=0.35]{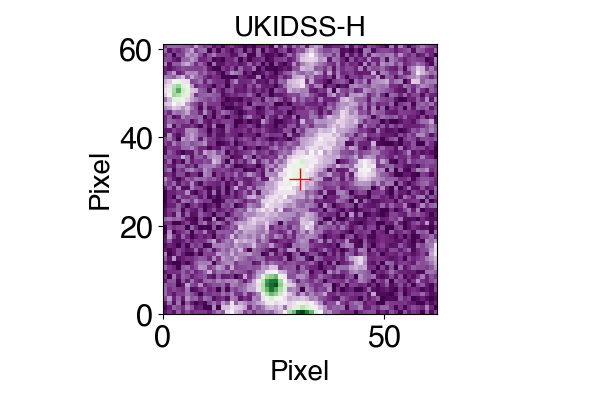}
\includegraphics[scale=0.35]{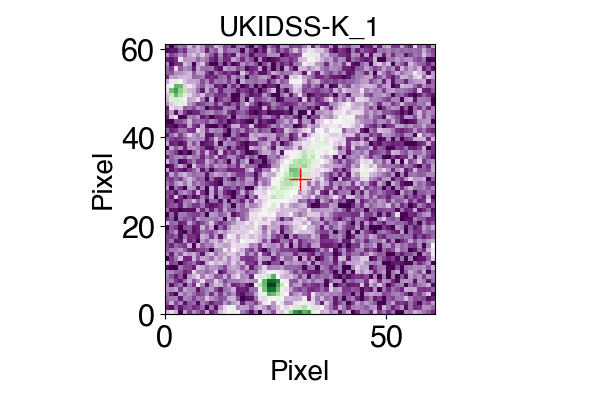}
\caption{A few examples of morphologically identified galaxies. The top three panels and the bottom three panels are the UKIDSS J, H, and K band maps of galaxies \#140000 and \#1521071, respectively (see Table \ref{tab:aquila_phot_gal}).
\label{fig:morph}}
\end{figure}

Figure \ref{fig:morph} shows a few examples of galaxies that are identified using higher-resolution UKIDSS maps. Due to the elongated morphology, these galaxies have limited detections in IRAC wavebands and thus the IRAC-color based criteria cannot be used to identify them even if we assume that they are star-forming galaxies and higher 8 $\micron$ emissions. However, these sources can be morphologically identified in the UKIDSS images regardless of their star-forming nature.

\begin{figure}[ht!]
\centering
\includegraphics[scale=0.5]{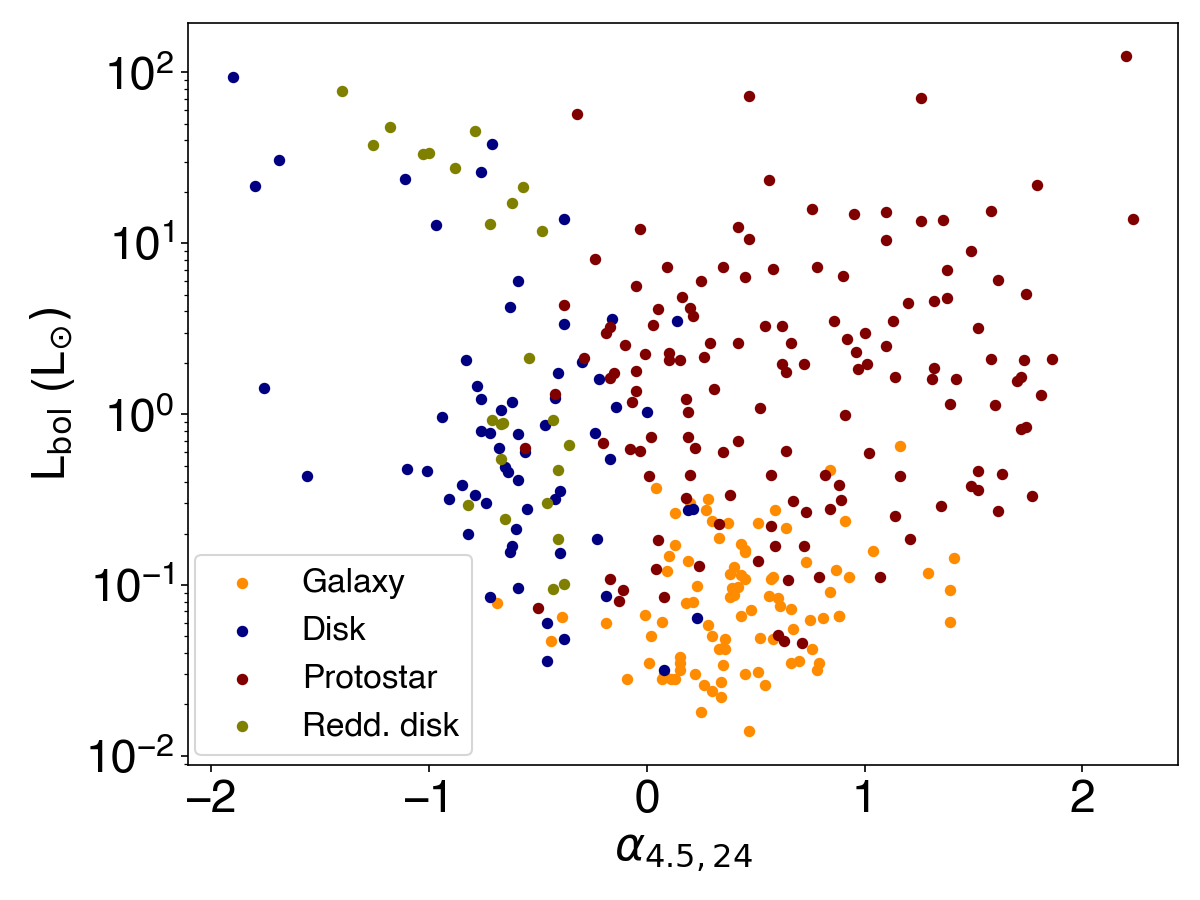}
\caption{Variation of L$_{\rm{bol}}$ with $\alpha_{4.5, 24}$ for all the sources in Aquila.
\label{fig:lbol_alpha}}
\end{figure}

In Figure \ref{fig:lbol_alpha}, show the variation of the bolometric luminosity \Lbol~with the spectral index $\alpha_{4.5, 24}$. The galaxies mostly have low luminosities and are clustered below $\Lbol<\sim0.2~L_{\odot}$ for the distance of the Aquila cloud. The protostars are mostly found at $\Lbol>\sim0.2~L_{\odot}$ and $\alpha_{4.5, 24}>0$. Similarly, pre-ms stars with disks occupy the $\Lbol>\sim0.2 L_{\odot}$ and $\alpha_{4.5, 24}<-0.3$ region, tentatively.

\section{Pre-ms stars with disks in Aquila} \label{app:disk}

Table \ref{tab:aquila_phot_disk} contains photometry for the pre-ms stars with disks (including reddened pre-ms stars) from 1.2 $\micron$ to 850 $\micron$. We only include the sources that have at least one PACS detection with signal-to-noise greater than 5 and that are included in both the Spitzer and Herschel coverage fields. We find a total of 85 pre-ms stars with disks (including 22 reddened pre-ms stars) in the surveyed region. Note that we define the ``eHOPS-aql-" designation only for the protostars in our sample. The source names listed for all other categories such as pre-ms stars with disks and galaxies are arbitrary index numbers.

\begin{longrotatetable} 
\begin{splitdeluxetable*}{c|c|c|c|cc|cc|cc|ccc|ccc|cccBccc|ccc|cc|cc|cc|cc|ccc|cccBccc|ccc|ccc|ccc|ccc|c} 
\tablecaption{Fluxes (in mJy) for the pre-ms stars with disks in Aquila. Only 5 sources are included in the table. Full list of sources is available online. \label{tab:aquila_phot_disk}} 
\setlength{\tabcolsep}{4pt} 
\tablehead{ 
\multicolumn{1}{c|}{Source \#} & \multicolumn{1}{c|}{R.A.} & \multicolumn{1}{c|}{Decl.} & \multicolumn{1}{c|}{Type} & \multicolumn{6}{|c|}{2MASS} & \multicolumn{15}{|c}{Spitzer} & \multicolumn{8}{|c}{WISE} & \multicolumn{18}{|c}{Herschel} & \multicolumn{3}{|c|}{JCMT}  & \nocolhead{} \\ 
\hline 
\multicolumn{1}{c|}{} & \multicolumn{1}{c|}{($^{\circ}$)} & \multicolumn{1}{c|}{($^{\circ}$)} & \multicolumn{1}{c|}{} & \multicolumn{6}{|c|}{} & \multicolumn{12}{|c}{IRAC} & \multicolumn{3}{|c}{MIPS} & \multicolumn{8}{|c|}{} & \multicolumn{9}{|c}{PACS} & \multicolumn{9}{|c|}{SPIRE} & \multicolumn{3}{|c|}{SCUBA2} & \nocolhead{} \\ 
\hline 
\multicolumn{1}{c|}{} & \multicolumn{1}{c|}{} & \multicolumn{1}{c|}{} & \multicolumn{1}{c|}{} & \colhead{$F_{\rm{J}}$} & \colhead{$\Delta F_{\rm{J}}$} & \multicolumn{1}{|c}{$F_{\rm{H}}$} & \colhead{$\Delta F_{\rm{H}}$} & \multicolumn{1}{|c}{$F_{\rm{K}}$} & \colhead{$\Delta F_{\rm{K}}$} & \multicolumn{1}{|c}{$F_{3.6}$} & \colhead{$\Delta F_{3.6}$} & \colhead{$F_{3.6}^C$} & \multicolumn{1}{|c}{$F_{4.5}$} & \colhead{$\Delta F_{4.5}$} & \colhead{$F_{4.5}^C$} & \multicolumn{1}{|c}{$F_{5.8}$} & \colhead{$\Delta F_{5.8}$} & \colhead{$F_{5.8}^C$} & \multicolumn{1}{c}{$F_{8.0}$} & \colhead{$\Delta F_{8.0}$} & \colhead{$F_{8.0}^C$} & \multicolumn{1}{|c}{$F_{24}$} & \colhead{$\Delta F_{24}$} & \colhead{$F_{24}^C$} & \multicolumn{1}{|c}{$F_{3.4}$} & \colhead{$\Delta F_{3.4}$} & \multicolumn{1}{|c}{$F_{4.6}$} & \colhead{$\Delta F_{4.6}$} & \multicolumn{1}{|c}{$F_{12}$} & \colhead{$\Delta F_{12}$} & \multicolumn{1}{|c}{$F_{22}$} & \colhead{$\Delta F_{22}$} & \multicolumn{1}{|c}{$F_{70}$} & \colhead{$\Delta F_{70}$} & \colhead{$F_{70}^S$} & \multicolumn{1}{|c}{$F_{100}$} & \colhead{$\Delta F_{100}$} & \colhead{$F_{100}^S$} & \multicolumn{1}{c}{$F_{160}$} & \colhead{$\Delta F_{160}$} & \colhead{$F_{160}^S$} & \multicolumn{1}{|c}{$F_{250}$} & \colhead{$\Delta F_{250}$} & \colhead{$F_{250}^S$} & \multicolumn{1}{|c}{$F_{350}$} & \colhead{$\Delta F_{350}$} & \colhead{$F_{350}^S$} & \multicolumn{1}{|c}{$F_{500}$} & \colhead{$\Delta F_{500}$} & \colhead{$F_{500}^S$} & \multicolumn{1}{|c}{$F_{850}$} & \colhead{$\Delta F_{850}$} & \colhead{$F_{850}^S$} & \multicolumn{1}{|c}{Comments} 
} 
\startdata 
30016 & 277.03533 & -0.01857 & Disk & 46.4 & 1.1 & 67.6 & 2.0 & 79.9 & 2.2 & 110.3 & 1.5 & 15.4 & 114.7 & 1.5 & 7.0 & 110.3 & 1.4 & 22.9 & 127.8 & 1.7 & 10.8 & 166.7 & 1.6 & 308.2 & 108.5 & 2.1 & 115.2 & 2.0 & 141.1 & 1.9 & 203.6 & 4.3 & 428.4 & 28.3 & 18.5 & --- & --- & --- & 374.4 & 24.0 & 18.3 & 362.2 & 39.3 & 89.4 & 303.1 & 31.3 & 148.5 & 80.1 & 56.8 & 130.6 & --- & --- & --- & Step 1--- [3.6]-[4.5] $<$ 0.65, $\alpha_{4.5,24} < -0.3$, IRS: Declining or Emission \\ 
31204 & 277.06358 & -0.04531 & Disk & --- & --- & 34.8 & 1.2 & --- & --- & 177.8 & 2.3 & 27.8 & 179.2 & 2.4 & 8.2 & 169.3 & 2.2 & 3.5 & 197.1 & 2.6 & 38.9 & 1216.6 & 13.2 & 226.7 & 152.7 & 3.2 & 159.1 & 2.8 & 317.1 & 4.4 & 1350.2 & 24.9 & 2039.0 & 110.0 & 41.4 & --- & --- & --- & 2262.4 & 115.6 & 27.3 & 1686.5 & 70.7 & 134.2 & 1616.4 & 127.9 & 198.8 & 4191.0 & 8.6 & U & --- & --- & --- & Step 3--- [3.6]-[4.5] $<$ 0.65, $\alpha_{4.5,24} < 0.3$ \\ 
33768 & 277.18998 & -0.12032 & Disk & 155.8 & 3.2 & 760.0 & 28.0 & 1834.8 & 35.5 & 2441.9 & 32.4 & 30.3 & 2690.0 & 35.5 & 117.2 & 2894.0 & 37.8 & 11.4 & 2921.9 & 38.1 & 752.1 & 2248.7 & 20.8 & 3202.7 & 2125.8 & 285.9 & 2775.9 & 125.3 & 3244.4 & 35.9 & 2372.4 & 35.0 & 324.6 & 26.6 & 20.9 & --- & --- & --- & --- & --- & 23.0 & 3712.3 & 4.8 & U & 4025.3 & 5.0 & U & 3916.9 & 5.7 & U & --- & --- & --- & Step 1--- [3.6]-[4.5] $<$ 0.65, $\alpha_{4.5,24} < -0.3$, IRS: Declining or Emission \\ 
50516 & 277.1995 & 0.14437 & Disk & 566.6 & 10.4 & 1040.5 & 54.6 & 1836.5 & 33.8 & 1975.5 & 26.1 & 92.0 & 2296.9 & 30.5 & 54.4 & 2869.5 & 37.5 & 229.5 & 3541.6 & 46.2 & 464.6 & 3411.7 & 31.7 & 5695.3 & 2624.9 & 483.5 & 3779.2 & 487.3 & 3623.5 & 36.7 & 3366.6 & 52.7 & 693.9 & 42.1 & 23.9 & 372.3 & 19.7 & 6.5 & 104.8 & 10.7 & 11.3 & 3280.5 & 4.3 & U & 3349.9 & 4.6 & U & 3260.5 & 6.3 & U & --- & --- & --- & Step 1--- [3.6]-[4.5] $<$ 0.65, $\alpha_{4.5,24} < -0.3$, IRS: Declining or Emission \\ 
50600 & 277.20929 & 0.16372 & Disk & 10.5 & 0.2 & 28.8 & 0.9 & 46.9 & 0.9 & 104.3 & 1.4 & 4.5 & 113.3 & 1.5 & 4.9 & 105.0 & 1.4 & 6.6 & 131.9 & 1.7 & 20.3 & 301.4 & 2.9 & 593.9 & 138.1 & 2.9 & 167.7 & 2.9 & 230.1 & 3.0 & 407.4 & 8.6 & 356.2 & 23.9 & 15.9 & 305.5 & 16.5 & 6.2 & 167.2 & 11.9 & 11.0 & 3885.8 & 4.8 & U & 3975.8 & 5.0 & U & 3729.1 & 6.5 & U & --- & --- & --- & Step 1--- [3.6]-[4.5] $<$ 0.65, $\alpha_{4.5,24} < -0.3$, IRS: Declining or Emission \\ 
\enddata 
\tablecomments{Column ``Type" denotes whether the pre-ms star with disk is sufficiently reddened due to foreground extinction. Other column descriptions are same as in Table \ref{tab:aquila_phot}. }
\end{splitdeluxetable*} 
\end{longrotatetable}

\section{Galaxies in the Aquila region} \label{app:gal}

We find a total of 144 galaxies in the surveyed region. This includes a diverse population from actively star-forming to AGN-dominated. Figure \ref{fig:aquila_map_gal} shows the spatial arrangement of galaxies on the column density map of Aquila. The galaxies avoid high density and are mostly found in lower density regions.
We present the photometry for galaxies that we detect in this survey in Table \ref{tab:aquila_phot_gal}.

\begin{longrotatetable} 
\begin{splitdeluxetable*}{c|c|c|cc|cc|cc|ccc|ccc|ccc|ccc|cccBcc|cc|cc|cc|ccc|ccc|cccBccc|ccc|ccc|ccc|c} 
\tablecaption{Fluxes (in mJy) for galaxies in the Aquila region. Only 5 sources are included in the table. Full list of sources is available online. \label{tab:aquila_phot_gal}} 
\setlength{\tabcolsep}{4pt} 
\tablehead{ 
\multicolumn{1}{c|}{Source \#} & \multicolumn{1}{c|}{R.A.} & \multicolumn{1}{c|}{Decl.} & \multicolumn{6}{|c|}{2MASS} & \multicolumn{15}{|c}{Spitzer} & \multicolumn{8}{c}{WISE} & \multicolumn{18}{|c}{Herschel} & \multicolumn{3}{|c|}{JCMT} & \nocolhead{} \\ 
\hline 
\multicolumn{1}{c|}{} & \multicolumn{1}{c|}{($^{\circ}$)} & \multicolumn{1}{c|}{($^{\circ}$)} & \multicolumn{6}{|c|}{} & \multicolumn{12}{|c|}{IRAC} & \multicolumn{3}{|c}{MIPS} & \multicolumn{8}{c|}{} & \multicolumn{9}{|c}{PACS} & \multicolumn{9}{c|}{SPIRE} & \multicolumn{3}{|c|}{SCUBA2} & \nocolhead{} \\ 
\hline 
\multicolumn{1}{c|}{} & \multicolumn{1}{c|}{} & \multicolumn{1}{c|}{} & \colhead{$F_{\rm{J}}$} & \colhead{$\Delta F_{\rm{J}}$} & \multicolumn{1}{|c}{$F_{\rm{H}}$} & \colhead{$\Delta F_{\rm{H}}$} & \multicolumn{1}{|c}{$F_{\rm{K}}$} & \colhead{$\Delta F_{\rm{K}}$} & \multicolumn{1}{|c}{$F_{3.6}$} & \colhead{$\Delta F_{3.6}$} & \colhead{$F_{3.6}^C$} & \multicolumn{1}{|c}{$F_{4.5}$} & \colhead{$\Delta F_{4.5}$} & \colhead{$F_{4.5}^C$} & \multicolumn{1}{|c}{$F_{5.8}$} & \colhead{$\Delta F_{5.8}$} & \colhead{$F_{5.8}^C$} & \multicolumn{1}{|c}{$F_{8.0}$} & \colhead{$\Delta F_{8.0}$} & \colhead{$F_{8.0}^C$} & \multicolumn{1}{|c}{$F_{24}$} & \colhead{$\Delta F_{24}$} & \colhead{$F_{24}^C$} & \multicolumn{1}{c}{$F_{3.4}$} & \colhead{$\Delta F_{3.4}$} & \multicolumn{1}{|c}{$F_{4.6}$} & \colhead{$\Delta F_{4.6}$} & \multicolumn{1}{|c}{$F_{12}$} & \colhead{$\Delta F_{12}$} & \multicolumn{1}{|c}{$F_{22}$} & \colhead{$\Delta F_{22}$} & \multicolumn{1}{|c}{$F_{70}$} & \colhead{$\Delta F_{70}$} & \colhead{$F_{70}^S$} & \multicolumn{1}{|c}{$F_{100}$} & \colhead{$\Delta F_{100}$} & \colhead{$F_{100}^S$} & \multicolumn{1}{|c}{$F_{160}$} & \colhead{$\Delta F_{160}$} & \colhead{$F_{160}^S$} & \multicolumn{1}{c}{$F_{250}$} & \colhead{$\Delta F_{250}$} & \colhead{$F_{250}^S$} & \multicolumn{1}{|c}{$F_{350}$} & \colhead{$\Delta F_{350}$} & \colhead{$F_{350}^S$} & \multicolumn{1}{|c}{$F_{500}$} & \colhead{$\Delta F_{500}$} & \colhead{$F_{500}^S$} & \multicolumn{1}{|c}{$F_{850}$} & \colhead{$\Delta F_{850}$} & \colhead{$F_{850}^S$} & \multicolumn{1}{|c}{Comments} 
} 
\startdata 
17703 & 277.19309 & -0.18626 & --- & --- & --- & --- & --- & --- & 0.9 & --- & 0.5 & 1.5 & --- & 1.2 & 2.2 & --- & 0.5 & 4.3 & 0.1 & 1.8 & 16.0 & 0.3 & 30.4 & 1.0 & --- & 1.7 & 0.1 & 5.9 & 0.3 & 15.4 & 1.2 & 288.8 & 25.8 & 21.2 & --- & --- & --- & 533.4 & 32.1 & 21.3 & 456.3 & 25.7 & 110.4 & 3538.2 & 5.0 & U & 3431.4 & 6.0 & U & --- & --- & --- & Step 2--- Literature  \\ 
31067 & 277.07324 & -0.11316 & 0.4 & 0.1 & 0.8 & 0.1 & 1.4 & 0.2 & 1.8 & --- & 0.5 & 1.5 & --- & 0.7 & 2.0 & 0.1 & 0.3 & 15.5 & 0.2 & 3.7 & 18.8 & 0.4 & 9.0 & 2.3 & 0.1 & 1.9 & --- & 14.3 & 0.3 & 23.0 & 1.9 & 303.0 & 23.5 & 18.0 & --- & --- & --- & 799.2 & 43.6 & 21.2 & 505.2 & 20.7 & 115.0 & 3613.5 & 5.2 & U & 3429.2 & 5.9 & U & --- & --- & --- & Step 2--- Literature; $\alpha_{3.6,4.5} < 0.5$, $\alpha_{5.8,8.0} > 3$  \\ 
63602 & 277.07327 & 0.26846 & --- & --- & --- & --- & --- & --- & 0.5 & --- & 0.2 & 0.8 & --- & 0.3 & 1.1 & --- & 0.2 & 2.0 & 0.1 & 0.3 & 9.0 & 0.2 & 3.6 & --- & --- & --- & --- & --- & --- & --- & --- & --- & --- & 16.4 & 57.9 & 5.7 & 4.9 & --- & --- & 10.5 & 2725.3 & 4.4 & U & 2864.8 & 4.3 & U & 2746.1 & 5.5 & U & --- & --- & --- & Step 2--- Literature  \\ 
65911 & 277.18147 & 0.25539 & --- & --- & --- & --- & --- & --- & 0.4 & --- & 0.2 & 0.3 & --- & 0.4 & 0.4 & --- & 0.3 & 2.8 & 0.1 & 0.8 & 3.9 & 0.1 & 5.7 & --- & --- & --- & --- & --- & --- & --- & --- & --- & --- & 18.3 & 173.5 & 10.3 & 5.5 & 244.9 & 18.5 & 11.2 & 4246.3 & 5.0 & U & 4803.0 & 5.7 & U & 4745.5 & 8.8 & U & --- & --- & --- & Step 2--- $\alpha_{3.6,4.5} < 0.5$, $\alpha_{5.8,8.0} > 3$  \\ 
67054 & 277.20625 & 0.38793 & --- & --- & --- & --- & --- & --- & 0.1 & --- & 0.7 & 0.2 & --- & 0.6 & 0.2 & --- & 0.4 & 0.6 & --- & 0.7 & 2.2 & 0.1 & 2.9 & --- & --- & --- & --- & --- & --- & --- & --- & --- & --- & 14.8 & 78.2 & 6.3 & 4.9 & --- & --- & 10.6 & 4750.3 & 5.1 & U & 5777.8 & 6.4 & U & 5926.3 & 8.3 & U & --- & --- & --- & Step 2--- $\alpha_{3.6,4.5} < 0.5$, $\alpha_{5.8,8.0} > 3$  \\ 
\enddata 
\tablecomments{Column description is same as in Table \ref{tab:aquila_phot}. }
\end{splitdeluxetable*} 
\end{longrotatetable}

\section{Candidate Protostars} \label{app:cp}

The candidate protostar (cp) includes sources that could not be robustly identified as protostars with the available data. While some sources in this list may very well be protostars, some may be low luminosity galaxies or other contaminations. The photometry of CPs are tabulated in Table \ref{tab:aquila_phot_cp}.

\begin{longrotatetable} 
\begin{splitdeluxetable*}{c|c|c|cc|cc|cc|ccc|ccc|ccc|ccc|cccBcc|cc|cc|cc|ccc|ccc|cccBccc|ccc|ccc|ccc|c} 
\tablecaption{Fluxes (in mJy) for the candidate protostars (CP) in the Aquila region. Only 5 sources are included in the table. Full list of sources is available online. \label{tab:aquila_phot_cp}} 
\setlength{\tabcolsep}{4pt} 
\tablehead{ 
\multicolumn{1}{c|}{Source \#} & \multicolumn{1}{c|}{R.A.} & \multicolumn{1}{c|}{Decl.} & \multicolumn{6}{|c|}{2MASS} & \multicolumn{15}{|c}{Spitzer} & \multicolumn{8}{c}{WISE} & \multicolumn{18}{|c}{Herschel} & \multicolumn{3}{|c|}{JCMT} & \nocolhead{} \\ 
\hline 
\multicolumn{1}{c|}{} & \multicolumn{1}{c|}{($^{\circ}$)} & \multicolumn{1}{c|}{($^{\circ}$)} & \multicolumn{6}{|c|}{} & \multicolumn{12}{|c|}{IRAC} & \multicolumn{3}{|c}{MIPS} & \multicolumn{8}{c|}{} & \multicolumn{9}{|c}{PACS} & \multicolumn{9}{c|}{SPIRE} & \multicolumn{3}{|c|}{SCUBA2} & \nocolhead{} \\ 
\hline 
\multicolumn{1}{c|}{} & \multicolumn{1}{c|}{} & \multicolumn{1}{c|}{} & \colhead{$F_{\rm{J}}$} & \colhead{$\Delta F_{\rm{J}}$} & \multicolumn{1}{|c}{$F_{\rm{H}}$} & \colhead{$\Delta F_{\rm{H}}$} & \multicolumn{1}{|c}{$F_{\rm{K}}$} & \colhead{$\Delta F_{\rm{K}}$} & \multicolumn{1}{|c}{$F_{3.6}$} & \colhead{$\Delta F_{3.6}$} & \colhead{$F_{3.6}^C$} & \multicolumn{1}{|c}{$F_{4.5}$} & \colhead{$\Delta F_{4.5}$} & \colhead{$F_{4.5}^C$} & \multicolumn{1}{|c}{$F_{5.8}$} & \colhead{$\Delta F_{5.8}$} & \colhead{$F_{5.8}^C$} & \multicolumn{1}{|c}{$F_{8.0}$} & \colhead{$\Delta F_{8.0}$} & \colhead{$F_{8.0}^C$} & \multicolumn{1}{|c}{$F_{24}$} & \colhead{$\Delta F_{24}$} & \colhead{$F_{24}^C$} & \multicolumn{1}{c}{$F_{3.4}$} & \colhead{$\Delta F_{3.4}$} & \multicolumn{1}{|c}{$F_{4.6}$} & \colhead{$\Delta F_{4.6}$} & \multicolumn{1}{|c}{$F_{12}$} & \colhead{$\Delta F_{12}$} & \multicolumn{1}{|c}{$F_{22}$} & \colhead{$\Delta F_{22}$} & \multicolumn{1}{|c}{$F_{70}$} & \colhead{$\Delta F_{70}$} & \colhead{$F_{70}^S$} & \multicolumn{1}{|c}{$F_{100}$} & \colhead{$\Delta F_{100}$} & \colhead{$F_{100}^S$} & \multicolumn{1}{|c}{$F_{160}$} & \colhead{$\Delta F_{160}$} & \colhead{$F_{160}^S$} & \multicolumn{1}{c}{$F_{250}$} & \colhead{$\Delta F_{250}$} & \colhead{$F_{250}^S$} & \multicolumn{1}{|c}{$F_{350}$} & \colhead{$\Delta F_{350}$} & \colhead{$F_{350}^S$} & \multicolumn{1}{|c}{$F_{500}$} & \colhead{$\Delta F_{500}$} & \colhead{$F_{500}^S$} & \multicolumn{1}{|c}{$F_{850}$} & \colhead{$\Delta F_{850}$} & \colhead{$F_{850}^S$} & \multicolumn{1}{|c}{Comments} 
} 
\startdata 
133408 & 277.41307 & 1.20937 & --- & --- & --- & --- & --- & --- & --- & --- & 0.2 & --- & --- & 0.2 & --- & --- & 0.3 & 0.2 & --- & 0.2 & --- & --- & 1.2 & --- & --- & --- & --- & --- & --- & --- & --- & --- & --- & 16.9 & 150.6 & 9.8 & 6.3 & 492.7 & 26.2 & 11.1 & 4857.2 & 10.0 & U & 5184.0 & 10.6 & U & 5149.2 & 12.6 & U & --- & --- & --- & Step 5--- Extended resolved emission at $\ge$100 $\micron$ \\ 
136261 & 277.48827 & 1.24477 & 5.6 & 0.5 & --- & --- & --- & --- & --- & --- & 54.6 & --- & --- & 35.0 & --- & --- & 34.0 & --- & --- & 370.3 & --- & --- & 4320.4 & --- & --- & --- & --- & --- & --- & --- & --- & 11929.3 & 605.4 & 103.8 & 9665.6 & 483.6 & 18.9 & --- & --- & 44.8 & 35817.0 & 121.3 & U & 34353.8 & 162.5 & U & 33074.0 & 170.1 & U & 734.2 & 48.3 & U & Step 5--- Using completeness limits for one or more Spitzer or Herschel/PACS photometry \\ 
136301 & 277.5818 & 1.09293 & --- & --- & --- & --- & --- & --- & 0.3 & --- & 0.4 & 0.3 & --- & 1.3 & 0.2 & --- & 0.3 & 0.3 & --- & 0.9 & 9.0 & 0.2 & 15.4 & --- & --- & --- & --- & --- & --- & --- & --- & --- & --- & 17.7 & 57.3 & 6.8 & 6.1 & --- & --- & 11.0 & 206.1 & 17.1 & 110.3 & 59.1 & 14.9 & 172.9 & 2908.8 & 5.3 & U & --- & --- & --- & Step 3--- [3.6]-[4.5] $<$ 0.65, $\alpha_{4.5,24} > 0.3$ \\ 
647743 & 277.15377 & -3.59109 & --- & --- & --- & --- & --- & --- & 0.1 & --- & 0.8 & 0.1 & --- & 0.3 & --- & --- & 0.3 & --- & --- & 0.2 & 2.5 & 0.1 & 2.9 & --- & --- & --- & --- & --- & --- & --- & --- & --- & --- & 19.9 & 106.7 & 7.1 & 4.8 & --- & --- & 10.5 & 231.0 & 34.6 & 205.3 & 6135.2 & 5.3 & U & 5870.4 & 6.4 & U & --- & --- & --- & Step 3--- [3.6]-[4.5] $<$ 0.65, $\alpha_{4.5,24} > 0.3$ \\ 
855983 & 277.42673 & -3.08261 & --- & --- & --- & --- & --- & --- & --- & --- & 0.9 & --- & --- & 0.5 & --- & --- & 0.4 & 0.4 & --- & 0.6 & --- & --- & 1.4 & --- & --- & --- & --- & --- & --- & --- & --- & --- & --- & 16.5 & 93.0 & 7.0 & 5.2 & --- & --- & 10.6 & 4411.2 & 4.0 & U & 4460.1 & 4.7 & U & 4140.2 & 5.8 & U & --- & --- & --- & Step 5--- Using completeness limits for one or more Spitzer or Herschel/PACS photometry \\ 
\enddata 
\tablecomments{Column description is same as in Table \ref{tab:aquila_phot}. }
\end{splitdeluxetable*} 
\end{longrotatetable}

\section{Sources outside the Spitzer coverage region} \label{app:out}

We list the available photometry for the sources outside the Spitzer coverage region in Table \ref{tab:aquila_phot_out}. As explained in Sec.~\ref{sec:criteria}, we divide the sources into 5 groups according to the coverage maps of Spitzer and Herschel detectors. In this study, we classify and model only the sources that are mapped by both Spitzer and Herschel (Group A). The coordinates, group (B, C, D and E; see section \ref{sec:criteria}), and photometry for the sources belonging to the rest of the groups are presented in Table \ref{tab:aquila_phot_out}. 


\begin{longrotatetable} 
\begin{splitdeluxetable*}{c|c|c|c|cc|cc|cc|ccc|ccc|ccc|cccBccc|cc|cc|cc|cc|ccc|cccBccc|ccc|ccc|ccc|ccc} 
\tablecaption{Fluxes (in mJy) for sources that are outside the Spitzer coverage region. Only 5 sources are included in the table. Full list of sources is available online. \label{tab:aquila_phot_out}} 
\setlength{\tabcolsep}{4pt} 
\tablehead{ 
\multicolumn{1}{c|}{Source \#} & \multicolumn{1}{c|}{R.A.} & \multicolumn{1}{c|}{Decl.} & \multicolumn{1}{c|}{Group} & \multicolumn{6}{|c|}{2MASS} & \multicolumn{15}{|c}{Spitzer} & \multicolumn{8}{|c}{WISE} & \multicolumn{18}{|c}{Herschel} & \multicolumn{3}{c}{JCMT} \\ 
\hline 
\multicolumn{1}{c|}{} & \multicolumn{1}{c|}{($^{\circ}$)} & \multicolumn{1}{c|}{($^{\circ}$)} & \multicolumn{1}{c|}{} & \multicolumn{6}{|c|}{} & \multicolumn{12}{c}{IRAC} & \multicolumn{3}{c}{MIPS} & \multicolumn{8}{|c|}{} & \multicolumn{9}{|c}{PACS} & \multicolumn{9}{|c}{SPIRE} & \multicolumn{3}{c}{SCUBA2} \\ 
\hline 
\multicolumn{1}{c|}{} & \multicolumn{1}{c|}{} & \multicolumn{1}{c|}{} & \multicolumn{1}{c|}{} & \colhead{$F_{\rm{J}}$} & \colhead{$\Delta F_{\rm{J}}$} & \multicolumn{1}{|c}{$F_{\rm{H}}$} & \colhead{$\Delta F_{\rm{H}}$} & \multicolumn{1}{|c}{$F_{\rm{K}}$} & \colhead{$\Delta F_{\rm{K}}$} & \multicolumn{1}{|c}{$F_{3.6}$} & \colhead{$\Delta F_{3.6}$} & \colhead{$F_{3.6}^C$} & \multicolumn{1}{|c}{$F_{4.5}$} & \colhead{$\Delta F_{4.5}$} & \colhead{$F_{4.5}^C$} & \multicolumn{1}{|c}{$F_{5.8}$} & \colhead{$\Delta F_{5.8}$} & \colhead{$F_{5.8}^C$} & \multicolumn{1}{|c}{$F_{8.0}$} & \colhead{$\Delta F_{8.0}$} & \colhead{$F_{8.0}^C$} & \multicolumn{1}{c}{$F_{24}$} & \colhead{$\Delta F_{24}$} & \colhead{$F_{24}^C$} & \multicolumn{1}{|c}{$F_{3.4}$} & \colhead{$\Delta F_{3.4}$} & \multicolumn{1}{|c}{$F_{4.6}$} & \colhead{$\Delta F_{4.6}$} & \multicolumn{1}{|c}{$F_{12}$} & \colhead{$\Delta F_{12}$} & \multicolumn{1}{|c}{$F_{22}$} & \colhead{$\Delta F_{22}$} & \multicolumn{1}{|c}{$F_{70}$} & \colhead{$\Delta F_{70}$} & \colhead{$F_{70}^S$} & \multicolumn{1}{|c}{$F_{100}$} & \colhead{$\Delta F_{100}$} & \colhead{$F_{100}^S$} & \multicolumn{1}{c}{$F_{160}$} & \colhead{$\Delta F_{160}$} & \colhead{$F_{160}^S$} & \multicolumn{1}{|c}{$F_{250}$} & \colhead{$\Delta F_{250}$} & \colhead{$F_{250}^S$} & \multicolumn{1}{|c}{$F_{350}$} & \colhead{$\Delta F_{350}$} & \colhead{$F_{350}^S$} & \multicolumn{1}{|c}{$F_{500}$} & \colhead{$\Delta F_{500}$} & \colhead{$F_{500}^S$} & \multicolumn{1}{c}{$F_{850}$} & \colhead{$\Delta F_{850}$} & \colhead{$F_{850}^S$}  
} 
\startdata 
60169 & 277.7465 & 0.08704 & D & --- & --- & 0.6 & 0.1 & 1.0 & 0.1 & --- & --- & --- & --- & --- & --- & --- & --- & --- & --- & --- & --- & --- & --- & --- & 1.0 & --- & 0.9 & --- & 8.7 & 0.3 & 24.6 & 1.4 & 335.3 & 23.8 & 16.9 & --- & --- & --- & 238.6 & 19.1 & 19.1 & 39.1 & 20.5 & 107.4 & 3088.6 & 4.2 & U & 2938.5 & 4.7 & U & --- & --- & --- \\ 
61202 & 276.93355 & 0.42019 & B & 98.4 & 2.1 & 458.0 & 22.4 & 1329.2 & 28.2 & --- & --- & --- & --- & --- & --- & --- & --- & --- & --- & --- & --- & 4946.5 & 46.3 & --- & 2567.5 & 394.9 & 5299.0 & 863.9 & 8684.2 & 232.0 & 8120.6 & 104.7 & 863.8 & 51.5 & 28.1 & 449.3 & 23.2 & 5.6 & 207.7 & 13.5 & 11.3 & 2487.8 & 4.6 & U & 2571.2 & 4.3 & U & 2427.9 & 5.7 & U & --- & --- & --- \\ 
92177 & 277.51705 & 0.57309 & B & 0.9 & 0.1 & 2.9 & 0.1 & 9.0 & 0.2 & --- & --- & --- & --- & --- & --- & --- & --- & --- & --- & --- & --- & 215.8 & 2.2 & 432.0 & 19.4 & 0.4 & 29.7 & 0.5 & 155.6 & 2.1 & 191.9 & 3.4 & 784.1 & 45.6 & 23.4 & 785.2 & 40.1 & 8.0 & 766.4 & 41.2 & 12.7 & 4308.8 & 8.4 & U & 4373.1 & 7.8 & U & 4053.8 & 7.0 & U & 37.3 & 6.3 & 0.2 \\ 
93268 & 277.6245 & 0.58343 & B & 20.0 & 0.5 & 28.2 & 0.8 & 25.5 & 0.5 & --- & --- & --- & --- & --- & --- & --- & --- & --- & --- & --- & --- & 25.0 & 0.3 & 43.8 & 15.4 & 0.3 & 9.3 & 0.2 & 2.3 & 0.2 & 26.0 & 1.5 & 410.8 & 28.9 & 20.2 & 447.8 & 23.2 & 6.2 & 464.9 & 27.8 & 11.2 & 401.2 & 40.2 & 110.0 & 380.2 & 69.5 & 177.3 & 3254.3 & 6.5 & U & --- & --- & --- \\ 
108735 & 277.59182 & 0.72683 & B & 3.2 & 0.1 & 10.5 & 0.3 & 17.9 & 0.4 & --- & --- & --- & --- & --- & --- & --- & --- & --- & --- & --- & --- & 117.2 & 1.1 & 226.7 & 26.5 & 0.6 & 31.0 & 0.6 & 55.7 & 0.9 & 138.4 & 3.6 & 206.1 & 19.9 & 17.0 & 120.4 & 7.9 & 5.1 & --- & --- & 10.8 & 4833.0 & 4.8 & U & 5042.2 & 5.1 & U & 4817.2 & 7.4 & U & --- & --- & --- \\ 
\enddata 
\tablecomments{Column description is same as in Table \ref{tab:aquila_phot}. The column ``Group" is explained in Section \ref{sec:criteria}.} 
\end{splitdeluxetable*} 
\end{longrotatetable}

\section{Comparision with previous catalogs} \label{app:prevCat}

In section \ref{sec:compare}, we compare the eHOPS-Aquila catalog with catalogs in literature. In this section, we provide a detailed source by source comparison with each of the catalog discussed in section \ref{sec:compare}. These comparisons are found in Tables \ref{tab:winston07} to \ref{tab:dunham15}.

\begin{deluxetable*}{ccccc} 
\tablecaption{Source comparison between eHOPS catalog and \cite{Winston07} for Aquila \label{tab:winston07}} 
\tablewidth{0pt} 
\tablehead{ 
\colhead{S-ID$^1$} & \colhead{R.A.} & \colhead{Decl.} & \colhead{eHOPS Name} & \colhead{Comments} \\ 
}\decimalcolnumbers 
\startdata 
1 & 18h30m02.10s & +01d13m58.86s & eHOPS-aql-100 & --- \\ 
2 & 18h29m57.83s & +01d12m51.62s & --- & No PACS, IN \\ 
3 & 18h29m59.58s & +01d11m58.78s & eHOPS-aql-88 & --- \\ 
4 & 18h29m58.79s & +01d14m26.05s & eHOPS-aql-84 & --- \\ 
5 & 18h29m47.01s & +01d16m26.82s & --- & No PACS, IN \\ 
\enddata 
\tablecomments{$^1$Spitzer ID numbers from Tables 3 and 4 in \cite{Winston07}. In ``Comments" column, ``No PACS, IN" refers the sources that are included in the PACS coverage but the sources are not detected and ``No PACS, OUT" refers to the sources that do not have PACS detection because they are outside the PACS coverage. If a protostar is reclassified  to a different Class or category then the reclassified category is mentioned in Comments column. For example, ``Redd. disk" means that the prototars is classified as a reddenned disk in eHOPS catalog. Only the protostars (Class 0, I, FS) in Table 4 of \cite{Winston07} are compared. Coordinates are from the eHOPS catalog. Full table is available in the online version.} 
\end{deluxetable*}

\begin{deluxetable*}{ccccc} 
\tablecaption{Source comparison between eHOPS catalog and \cite{Enoch09} for Aquila \label{tab:enoch09}} 
\tablewidth{0pt} 
\tablehead{ 
\colhead{ID$^1$} & \colhead{R.A.} & \colhead{Decl.} & \colhead{eHOPS Name} & \colhead{Comments} \\ 
}\decimalcolnumbers 
\startdata 
Ser-emb 1 & 18h29m09.10s & +00d31m30.74s & eHOPS-aql-40 & --- \\ 
Ser-emb 2 & 18h29m52.53s & +00d36m11.30s & eHOPS-aql-73 & --- \\ 
Ser-emb 3 & 18h28m54.89s & +00d29m52.42s & eHOPS-aql-19 & --- \\ 
Ser-emb 4 & 18h30m05.72s & +00d39m31.38s & --- & Redd. disk \\ 
Ser-emb 5 & 18h28m54.93s & +00d18m32.32s & eHOPS-aql-21 & --- \\ 
\enddata 
\tablecomments{$^1$Source ID from Table 3 in \cite{Enoch09}. Coordinates are from the eHOPS catalog. Full table is available in the online version.} 
\end{deluxetable*}

\begin{deluxetable*}{cccc} 
\tablecaption{Source comparison between eHOPS catalog and \cite{Gutermuth09} for Aquila \label{tab:gutermuth09}} 
\tablehead{ 
\colhead{R.A.} & \colhead{Decl.} & \colhead{eHOPS Name} & \colhead{Comments} \\ 
}\decimalcolnumbers 
\startdata 
18h27m36.7s & -03d50m04.7s & --- & No PACS, IN \\ 
18h27m50.22s & -03d49m14.01s & eHOPS-aql-5 & --- \\ 
18h28m05.44s & -03d46m59.75s & eHOPS-aql-8 & --- \\ 
18h30m01.54s & +01d15m14.9s & --- & No PACS, IN \\ 
18h29m31.97s & +01d18m42.59s & eHOPS-aql-50 & --- \\ 
\enddata 
\tablecomments{Only the protostars (Class I and I*) in Table 4 of \cite{Gutermuth09} are compared. Coordinates are from the eHOPS catalog. Full table is available in the online version.} 
\end{deluxetable*}

\begin{deluxetable*}{cccc} 
\tablecaption{Source comparison between eHOPS catalog and \cite{Kryukova12} for Aquila \label{tab:kryukova12}} 
\tablehead{ 
\colhead{R.A.} & \colhead{Decl.} & \colhead{eHOPS Name} & \colhead{Comments} \\ 
}\decimalcolnumbers 
\startdata 
18h28m27.11s & +00d44m45.04s & --- & No PACS, IN \\ 
18h28m41.86s & +00d03m21.33s & --- & No PACS, IN \\ 
18h28m44.03s & +00d53m38.03s & eHOPS-aql-13 & --- \\ 
18h28m44.81s & +00d51m25.53s & eHOPS-aql-14 & --- \\ 
18h28m45.05s & +00d52m02.59s & eHOPS-aql-15 & --- \\ 
\enddata 
\tablecomments{Protostars are from Table 1 of \cite{Kryukova12}. Coordinates are from the eHOPS catalog. Full table is available in the online version.} 
\end{deluxetable*}

\begin{deluxetable*}{ccc} 
\tablecaption{Source comparison between eHOPS catalog and \cite{Dunham15} for Aquila \label{tab:dunham15}} 
\tablehead{ 
\colhead{ID$^1$} & \colhead{eHOPS Name} & \colhead{Comments} 
}\decimalcolnumbers 
\startdata 
J182511.3-025853 & eHOPS-aql-1 & --- \\ 
J182513.3-025954 & eHOPS-aql-2 & --- \\ 
J182750.1-034913 & eHOPS-aql-5 & --- \\ 
J182754.7-034238 & eHOPS-aql-7 & --- \\ 
J182805.4-034659 & eHOPS-aql-8 & --- \\ 
\enddata 
\tablecomments{$^1$Spitzer ID from Table 2 in \cite{Dunham15}. Coordinates are from the eHOPS catalog. Full table is available in the online version.} 
\end{deluxetable*}

\section{Parameter uncertainties and degeneracy in Model grids} \label{app:unc}

The $R$  statistic  can pick out the model in the grid that is closest to the observed SED after applying a best-fit extinction and scale factor. Due to degeneracies in the model parameters, differences between the adopted models and actual protostars, and uncertainties in the data, there are likely multiple models that have only slightly higher $R$ values, particularly if the number of photometry points is small. The question is whether the parameters of these models have values close to those of the best-fit, $R_{\rm{min}}$ model, or whether certain parameters can accomodate a wide range of models while still providing reasonable fits. To assess the robustness of the best-fit parameters, we examine the distribution of the parameters for models within a given range of $\Delta R$ from the best fit $R$ value, $R_{\rm{min}}$. Ideally, the distributions of the  parameters for models within $\Delta R$ should have a peak around the parameter corresponding to the best fit $R$. 



We follow \citetalias{Furlan16} to  check the robustness of the best-fit parameters by comparing them with the mode of the parameter within a given range of $\Delta R$. 
Due to the possible parameter degeneracies (see Figure 56 in F16) and uncertainties involved, the models in a range of $\Delta R$ likely span a range of values. The mode estimates from models do not capture the entire extent of the parameter values but provide the most common values within $\Delta R$. Ideally, if a best fit parameter value is well-constrained, the mode of the parameter values in the range $\Delta R$ should be the same as or very close to the best fit parameter value. Similarly, if the best fit parameter is poorly constrained, there will be larger differences between the best fit and mode values. 

\begin{figure}[ht!]
\includegraphics[scale=0.35]{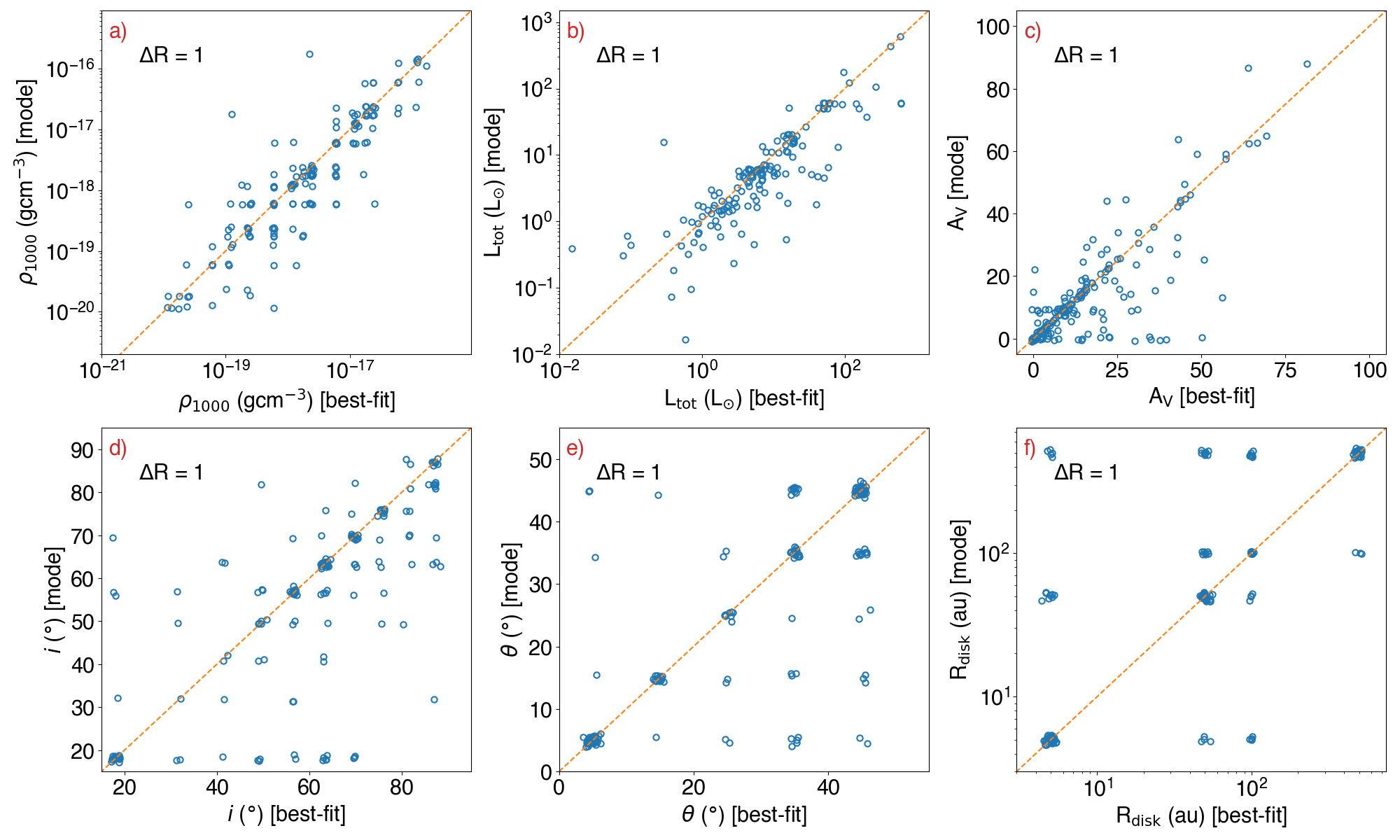}
\caption{Mode vs. best fit values for envelope reference density (panel a), total luminosity (panel b), foreground extinction (panel c), inclination angle (panel d), cavity opening angle (panel e), and disk outer radius (panel f) for the closest 50 models in R $\le R_{\rm{min}}$ + $\Delta$ R, where $\Delta$ R = 1. Random offsets are added to the values to display superimposed markers which would otherwise be hidden.  
\label{fig:gof_ltot}}
\end{figure}



Figure \ref{fig:gof_ltot} shows the correlation of the best fit values with the mode of the parameter for the envelope reference density,  total luminosity, foreground extinction, inclination angle, cavity opening angle, and disk outer radius for the closest 50 models with R $\leq$ R$_{\rm{min}}$ + $\Delta$R, where $\Delta$R = 1.
Figure \ref{fig:gof_ltot}a shows the variation of the envelope reference density. Here, we see good agreement between the mode and best fit models for  $\Delta$R = 1.  
Most of the best fit and mode values agree for the total luminosity values too (Figure \ref{fig:gof_ltot}b). 
Figure \ref{fig:gof_ltot}c shows the correlation between the best fit and the mode values for the foreground extinction. For few sources, the mode of $A_{\rm{V}}$ is small ($\approx$0), even when the best fit $A_{\rm{V}} >> 0$. Similarly there are sources with best fit $A_{\rm{V}} \sim 0$. These are the sources that lack mid-IR data so that the foreground extinction is not well-constrained by the model fits.

Figure \ref{fig:gof_ltot}d shows the variation of mode and best fit values for inclination angles. Although there is general agreement between the best fit and mode values, we find more sources with little correlation between the mode and best fit values. 
Figure \ref{fig:gof_ltot}e shows variation for the cavity opening angles. The lower cavity opening angles ($\theta \leq 25^{\circ}$) seem to be better constrained than the larger cavity angles. For large best fit opening angles, however, there is a large range in the mode values. 
For example, for $\theta = 5^{\circ}$, only three protostars have mode values different from the best fit value.
Similarly, Figure \ref{fig:gof_ltot}f shows the variation for the disk outer radius. Although there is some agreement between the best fit value and the mode, there are sources for which the variation between the best fit and mode values are more than a few step sizes. For example, there are sources for which the best fit $R_{\rm{disk}}$ is 5 au (lowest parameter value in models for $R_{\rm{disk}}$) but the mode is 500 au (highest parameter value in models for $R_{\rm{disk}}$). The mode of $R_{\rm{disk}}$ is usually larger than the best fit $R_{\rm{disk}}$ for larger $\Delta R$ bins. $R_{\rm{disk}}$ seems to be the least constrained parameter in our models.




We conclude that the envelope reference density and total luminosity are the best constrained parameters in our model fits. Other parameters such as the inclination angle, cavity opening angle and disk radius may not be well-constrained due to model degeneracies.

\section{Potentially variable protostars}\label{app:variables}

The sources are observed with the IRS and IRAC/MIPS instruments at different epochs. Discrepancies between the IRAC/MIPS photometry and corresponding IRS fluxes are expected for potentially variable sources. The IRS spectrum covers the $\sim$5-35 $\micron$ wavelength region. The corresponding Spitzer photometry in this range is at 5.8 and 8.0 $\micron$ (IRAC) and 24 $\micron$ (MIPS). We compare the Spitzer/IRAC photometry to the corresponding wavelengths in the IRS data to find potentially variable protostars.

\begin{figure}[ht!]
\centering
\includegraphics[scale=0.4]{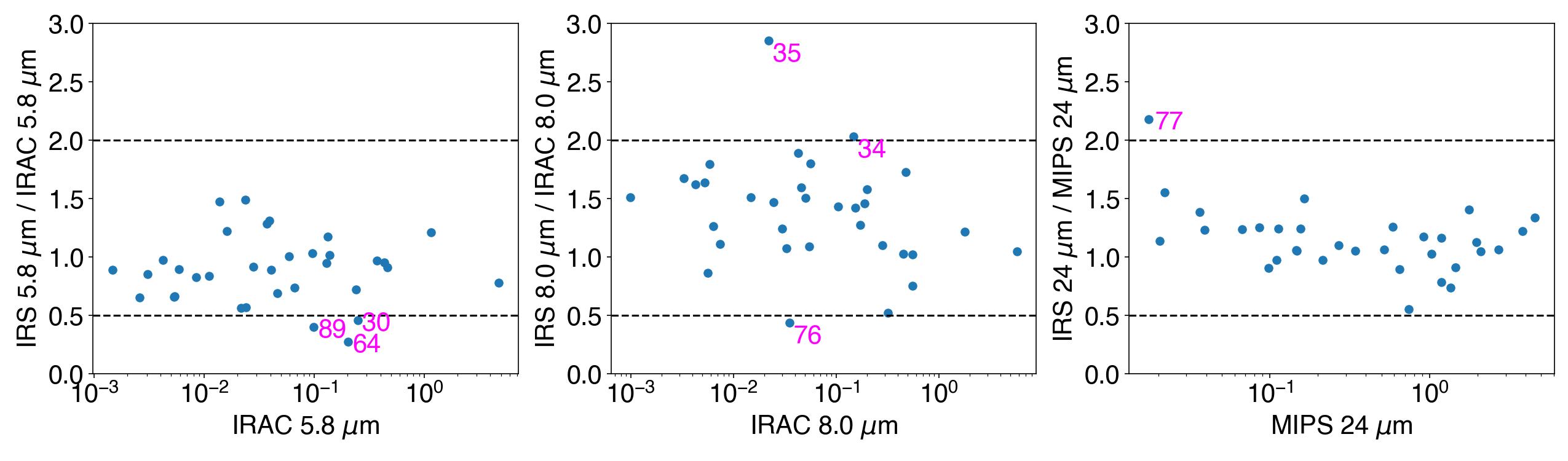}
\caption{The ratio of the IRS flux to the corresponding IRAC or MIPS photometry for 5.8, 8.0, and 24 $\micron$ wavelengths for protostars with IRS data.
\label{fig:var}}
\end{figure}

Figure \ref{fig:var} shows the ratio of the IRS equivalent flux for 5.8, 8.0, and 24 $\micron$ to the photometric fluxes, with the IRAC and MIPS fluxes at those wavelengths. The horizontal dash lines show regions where the ratio is $>2$ or $<0.5$, illustrating the sources with at least a factor of 2 change in flux between the IRS and IRAC or MIPS fluxes. We find 7 protostars for the ratio is $>2$ or $<0.5$. 

\begin{deluxetable*}{ccccc} 
\tablecaption{Observing dates of potential variable sources \label{tab:variables}} 
\tablehead{ 
\colhead{Source} & \colhead{IRAC/MIPS} & \colhead{IRS} & \colhead{Variable} & \colhead{Class} \\ 
\nocolhead{} & \colhead{Obs. date} & \colhead{Obs. date} & \colhead{Wavelength ($\mu$m)} & \nocolhead{} 
} 
\decimalcolnumbers 
\startdata 
eHOPS-aql-30 & 2004-04-05 & 2006-04-21 & 5.8 & Flat-spectrum \\ 
eHOPS-aql-34 & 2004-04-05 & 2006-04-19 & 8.0 & Class 0 \\ 
eHOPS-aql-35 & 2004-04-05 & 2005-09-09 & 8.0 & Class 0 \\ 
eHOPS-aql-64 & 2004-04-05 & 2008-04-24 & 5.8 & Class I \\ 
eHOPS-aql-76 & 2004-04-05 & 2006-10-23 & 8.0 & Class I \\ 
eHOPS-aql-77 & 2004-04-05 & 2007-04-27 & 24 & Class 0 \\ 
eHOPS-aql-89 & 2004-04-05 & 2008-04-25 & 5.8 & Class I \\ 
\enddata 
\end{deluxetable*}

Table \ref{tab:variables} shows the sources for candidates of potentially variable protostars that we find using Figure \ref{fig:var}. The table lists the dates of observation for both the photometry and the spectra for these sources, the wavelengths in which the source varies, and the SED classification of the source. 
It is worth noting that the seven sources show a flux variability of more that a factor of two at only one wavelength, therefore they may not be robust variables. Also, note that the data are taken with different instruments which can cause some biases. For example, IRS spectra can have issues with background subtractions and so have higher or lower fluxes. The aperture size of IRAC versus MIPS can matter, too, especially if there are extended emissions.
Thus, we mention the sources in Table \ref{tab:variables} only as potential variables that can be followed up by other time-domain surveys for studying photometric variability over multiple epochs.

\newpage
\bibliography{sample631}{}
\bibliographystyle{aasjournal}

\end{document}